\documentclass[final]{yorkthesis}
\usepackage{thesisoptions}

\usepackage{xr-hyper} 
\usepackage{subfiles} 
\makeatletter
\newcommand*{\addFileDependency}[1]{
  \typeout{(#1)}
  \@addtofilelist{#1}
  \IfFileExists{#1}{}{\typeout{No file #1.}}
}
\makeatother

\usepackage{verbatim} 
\usepackage[math]{blindtext} 

\usepackage{amsmath, amsthm}
\usepackage{amssymb, amsfonts}
\usepackage{dsfont}
\usepackage[final]{hyperref} 
\usepackage{xspace} 
\usepackage{xcolor} 
\usepackage[obeyDraft, textsize=small, colorinlistoftodos]{todonotes}

\usepackage{mathrsfs} 
\usepackage{bbm} 
\usepackage{url}
\usepackage{physics}

\usepackage{nicematrix}
\usepackage[numbers, sort&compress]{natbib}

\def\M{\ensuremath\mathcal}
\newtheorem{lemma}{Lemma}
\newtheorem{theorem}{Theorem}
\newtheorem{conjecture}{Conjecture}
\newtheorem{corollary}{Corollary}

\theoremstyle{definition}
\newtheorem{definition}{Definition}

\newtheorem{protocol}{Protocol}

\newcommand{\proj}[1]{\ket{#1}\!\bra{#1}}
\newcommand{\id}{\mathds{1}}
\newcommand{\idmap}{\mathcal{I}}
\newcommand{\cD}{\mathcal{D}}
\newcommand{\cC}{\mathcal{C}}
\newcommand{\cE}{\mathcal{E}}
\newcommand{\cF}{\mathcal{F}}
\newcommand{\cH}{\mathcal{H}}
\newcommand{\cI}{\mathcal{I}}
\newcommand{\cM}{\mathcal{M}}
\newcommand{\cN}{\mathcal{N}}
\newcommand{\cP}{\mathcal{P}}
\newcommand{\cQ}{\mathcal{Q}}
\newcommand{\cS}{\mathcal{S}}
\newcommand{\cT}{\mathcal{T}}
\newcommand{\cR}{\mathcal{R}}
\newcommand{\B}{\mathbf}
\newcommand{\ot}{\otimes}
\newcommand{\bin}{\mathrm{bin}}
\newcommand{\scorefunction}{S}
\newcommand{\overlapfunction}{\mathcal{O}}
\newcommand{\overlap}{\Theta}
\newcommand{\score}{\omega}
\newcommand{\Ext}{\mathrm{Ext}}
\newcommand{\Var}{\mathrm{Var}}
\newcommand{\Min}{\mathrm{Min}}
\newcommand{\Max}{\mathrm{Max}}
\newcommand{\Freq}{\mathrm{Freq}}
\newcommand{\rate}{\mathrm{rate}}
\newcommand{\entropy}{\mathcal{G}}

\newcommand{\entropytwo}{\mathcal{P}_1}

\DeclareMathOperator{\e}{e}
\newcommand{\uniteffect}{I}
\newcommand{\ii}{\mathrm{i}} 
\newcommand{\RC}[1]{{\color{black}  #1 }}
\newcommand{\RutC}[1]{{\color{black}#1}}

\newcommand{\bnkp}{B(n , \mathbf{k} , \mathbf{p})}
\newcommand{\bnkprime}{B(n' , \mathbf{k}' , \mathbf{p}')}

\newcommand{\normone}[1]{|| #1 || }

\RequirePackage{amsmath}
\RequirePackage{amsfonts}
\RequirePackage{amssymb}
\RequirePackage{amsthm}
\RequirePackage{makeidx}
\RequirePackage{graphicx}
\RequirePackage{color}
\RequirePackage{xspace}
\makeatletter
\@ifclassloaded{beamer}
	{
	}
	{
		\RequirePackage{enumitem}
	}
\makeatother
\definecolor{mpGray}{RGB}{236, 239, 241}		
\definecolor{mpBlue}{RGB}{21, 101, 192}			
\definecolor{mpLightBlue}{RGB}{144, 202, 249}	
\definecolor{mpRed}{RGB}{198, 40, 40}			
\definecolor{mpLightRed}{RGB}{239, 154, 154}	
\definecolor{mpGreen}{RGB}{46, 125, 50}			
\definecolor{mpLightGreen}{RGB}{165, 214, 167}	

\relax
\makeatletter
\@ifpackagewith{mplmacros}{no-hyperref}{
}{
	\RequirePackage{hyperref}
	\hypersetup{citecolor=magenta}
	\hypersetup{colorlinks=true}
	\hypersetup{linkcolor=blue}
	\hypersetup{urlcolor=blue}
        \hypersetup{linktocpage=true}
}
\usepackage{tikz}
\usetikzlibrary{diatikz}

\usepackage{tikz,pgfplotstable}
\usetikzlibrary{fit,calc}
\tikzstyle{porte} = [fill=gray!25, draw]

\newcommand{\cleardoublepagewithnumber}{
  \clearpage
  \if@twoside
    \ifodd\c@page
    \else
      \thispagestyle{plain}
      \hbox{}
      \newpage
      \if@twocolumn
        \hbox{}
        \newpage
      \fi
    \fi
  \fi
}

\title{Improvements on Device Independent and Semi-Device Independent Protocols of Randomness Expansion}
\author{Rutvij Bhavsar}
\department{Mathematics} 
\qualification{PhD}
\submitdate{August 2023}  

\begin{document}

\frontmatter* 

\makethesistitle

\renewcommand{\thepage}{\arabic{page}}
\pagenumbering{roman}

\chapter{Abstract}
\vspace*{-2em}\begin{quote}
\centering 
\end{quote}
{To generate genuine random numbers, random number generators based on quantum theory are essential. However, ensuring that the process used to produce randomness meets desired security standards can pose challenges for traditional quantum random number generators. This thesis delves into Device Independent (DI) and Semi-Device Independent (semi-DI) protocols of randomness expansion, based on a minimal set of experimentally verifiable security assumptions. The security in DI protocols relies on the violation of Bell inequalities, which certify the quantum behavior of devices. The semi-DI protocols discussed in this thesis require the characterization of only one device -- a power meter. These protocols exploit the fact that quantum states can be prepared such that they cannot be distinguished with certainty, thereby creating a randomness resource. In this study, we introduce enhanced DI and semi-DI protocols that surpass existing ones in terms of \RutC{output randomness rate}, security, or in some instances, both. Our analysis employs the Entropy Accumulation Theorem \RutC{(EAT)} to determine the extractable randomness for finite rounds. A notable contribution is the introduction of randomness expansion protocols that recycle input randomness, significantly enhancing finite round \RutC{randomness} rates for DI protocols based on the CHSH inequality violation. In the final section of the thesis, we delve into Generalized Probability Theories (GPTs), with a focus on Boxworld, the largest GPT capable of producing correlations consistent with relativity. A tractable criterion for identifying a Boxworld channel is presented.}

\cleardoublepagewithnumber
\chapter*{} 
\begin{center}
\large{To Hemi Foi}
\end{center}
\cleardoublepagewithnumber

\cleardoublepagewithnumber
\tableofcontents
\cleardoublepagewithnumber


\clearpage
\listoffigures
\cleardoublepagewithnumber



\chapter{Acknowledgments}

\bigskip
My deepest thanks go to my supervisor, Prof. Roger Colbeck, whose mentorship during my PhD journey has been invaluable. His profound understanding of mathematics and physics has significantly deepened my own knowledge and academic growth. I am especially grateful for his role in making my PhD journey rewarding and truly delightful, even against the backdrop of the challenges presented by the COVID-19 pandemic.

Thanks are due to Dr. Stefan Weigert and Dr. Gustav Delius for their consistent mentorship and guidance as members of my TAP. I also thank Prof. Serge Massar and Dr. Stefan Weigert for taking time to review my thesis and provide a detailed feedback on the thesis. I must acknowledge the enriching discussions with my collaborators Dr. Hamid Tebyanian and Dr. Sammy Raggy. 

I am extremely grateful for the insights and camaraderie of fellow PhD students in Quantum Information theory group including Vilasini, Vincenzo, Lewis, Shashaank, Max, Vasilis, Alistair, and Kuntal. Special thanks to Jintao, Lewis, Vincenzo, Ambroise, and Hamid for reviewing sections of this thesis and helping with LaTeX issues.

Immense gratitude is extended to my family—-my parents, Manisha and Vihang, along with my grandparents, Sarojben, Chandrikaben, Natvarlal, and Jagdishchandra—-who have been pillars of support and wisdom throughout this journey. I owe special thanks to Jintao for his unwavering support and patience. I am also indebted to my circle of friends, including Alba, Diego, Eva, Brad, Peiyun, David, Matt, Macey, Cordelia, Jenny, Beth, Ding, Ambroise, Andrew, Esther, Samantha, Sam, Jack, Jade, Guy, Ross, Emily, Simen, Firat, Vidul, Sapna, Drishti, Hardik, Akshita, Nekhel, Anjali, Ayan, Shefali, Ashvani, Shubham, Laura, Shreya, Aditya, Khushboo, and Varun, whose camaraderie and encouragement have significantly enriched my PhD experience.

Finally, I acknowledge the Quantum Communications Hub and the WW Smith Fund for financial support, which was instrumental in completing my PhD. 
\cleardoublepagewithnumber
\chapter{Author's declaration}
I declare that the work presented in this thesis, except where 
otherwise stated, is based on my own research carried out at 
the University of York under the supervision of Prof Roger Colbeck and has not been submitted previously 
for any degree at this or any other university. Sources are 
acknowledged by explicit references. 
The sections on randomness expansion protocols are largely based on the research works stated below: \\ 

Chapters ~\ref{chap: DI protocols}, \ref{chap: DI upper bounds}, \ref{chap: DI lower bounds} and \ref{chap : actual protocols for DI} are based on the following joint works: \\ 
(1) R. Bhavsar, S. Ragy, and R. Colbeck. ``Improved device-independent randomness expansion rates using two sided randomness''. New Journal of Physics 25.9 (2023): 093035.

(2) R. Bhavsar and R. Colbeck. ``Jordan's Lemma assisted lower-bounds on von-Neumann entropies for DIRNE/DIQKD protocols''. In preparation. 

The works in Chapters \ref{chap: semi-DI}, \ref{chap: semi-DI lower bounds} and \ref{chap: semi-DI protocols} are based on the joint work:

(3) R. Bhavsar, H. Tebyanian, and R. Colbeck. ``A Semi-Device independent protocol for randomness expansion''. In preparation.  \\ 

Parts of Chapter \ref{chap: DI upper bounds} including \ref{app:Proof1}, \ref{app: convex combinations of qubit strategies}, \ref{sec: reduction to pure states in DI}, \ref{app:simp} and \ref{sec: HAg00E} are extensions of the work: \\ 
S. Pironio, A Acin, N Brunner, N. Gisin, S. Massar, and V. Scarani. ``Device-independent quantum key distribution secure against collective attacks''. New Journal of Physics 11.4 (2009): 045021.\\
A discussion of the relationship of this work and the above work can be found in Chapter \ref{chap: DI upper bounds}. 

I have contributed to \textbf{all} aspects of research works (1), (2), and (3).  Portions of this work were proofread with the assistance of online tools, Grammarly and ChatGPT, as well as the valuable input from Mr Jintao Shuai, Mr Lewis Wooltorton, and Mr Vincenzo Fiorentino. All such assistance is limited to proofreading and identification of grammatical errors. I declare that no violations of the University's Guidance on Proofreading and Editing have been made.

\cleardoublepagewithnumber

\mainmatter*
\renewcommand{\thepage}{\arabic{page}}
\pagenumbering{arabic}  

\chapter{Introduction to protocols of randomness expansion}
Random numbers have a wide range of applications including cryptography, scientific experiments, games, lotteries, and gambling. Their use in these contexts is to ensure fairness and unpredictability of outcomes. 

The process of generating random numbers is often associated with a certain level of suspicion. This suspicion stems from the question of whether the numbers generated are truly random. If the randomness is compromised, the fairness of the numbers, which is one of their strengths, could be significantly weakened. For example, consider a bet based on the outcome of a coin toss. If one participant, a resourceful scientist, could account for all the forces acting on the coin, they could predict the outcome of the toss. This would make the bet unfair, even though it appears fair on the surface. This scenario, although impractical to achieve using everyday technology, demonstrates that a process that appears random to one party may be completely deterministic to another party. Therefore, any process that appears to be random may not be truly random, and could be used to gain an unfair advantage.
 
With the advent of advanced techniques such as machine learning, which can analyze large sets of data and predict outcomes, the vulnerability of randomness-based systems increases. Everyday activities such as making payments with a card or conducting transactions online depend on the security provided by random numbers. The potential gain from cracking such systems could be substantial, thus providing significant incentives for individuals or organizations to attempt guessing these random numbers. Similarly, in our data-driven world, where access to data is often considered key, governments and companies may be incentivized to break such random-number based security protocols to gain unauthorized access to private information. 

A random number generator (RNG) is a device engineered to produce a sequence of numbers following a uniform probability distribution. However, as discussed above, this definition does not entirely capture the complexities and practical necessities associated with RNGs. A desired characteristic of RNGs is their ability to generate``genuine'' or``true'' randomness. This means that the sequence of random numbers they produce is not only unpredictable to the party using the RNG but also inaccessible or unknown to any third party, including the manufacturer of the device. In other words, an RNG should be capable of producing a sequence of numbers following a uniform probability distribution for every possible agent who wants to determine it. Such a robust definition is critical in various applications that utilize random numbers, especially when used in applications such as in cryptographic protocols.

Based on our \RutC{current understanding of physics}, it is not possible to attain randomness via classical processes. The reason being, classical systems are fundamentally deterministic, meaning an extremely powerful adversary could, in \RutC{principle}, determine the \RutC{outcomes of a classical process with certainty}. Moreover, even incomplete knowledge regarding the mechanism of randomness generation could be \RutC{used} to make informed guesses about the generated numbers. Consequently, we should focus \RutC{on} quantum processes as potential candidates for constructing a reliable RNG. 

Realizing such a robust RNG presents a significant challenge even when using a quantum process. Consider, for example, a RNG designed to use a single photon passing through a 50:50 beam splitter. Theoretically, according to quantum mechanics, performing this process would result in \RutC{a} photon either going to one port of the beam splitter or the other with a 50 percent probability. Therefore, in principle, this process can be used as a quantum random number generator. However, in the real world, there are no guarantees that the outputs are genuinely uniformly distributed, due to inevitable practical imperfections in the beam splitter and the laser source. 

Moreover, if complex processes are incorporated within the RNG, the entire device must be meticulously modeled for two primary reasons. First, this is to ensure there has been no interference or tampering by an adversary, \RutC{which might even be the manufacturer of the RNG}. Second, \RutC{detailed modelling of each individual component of the RNG} is crucial to identify and account for any device imperfections. \RutC{Any} imperfections \RutC{in the RNG} could skew the distribution of the generated numbers or introduce predictability, both of which would undermine the randomness and security of the RNG.

Device Independent (DI) protocols aim to get around this. They base security on minimal and easily verifiable details about pairs of devices and without \RutC{making} any assumption on the inner working of the devices. Instead, they rely on verifying that the input-output statistics of the devices exhibit non-locality~\cite{ColbeckThesis,CK2,PAMBMMOHLMM}. In essence, Bell's theorem is used to assure the privacy of the outputs.

Bell's theorem implies that if two (or more) devices that cannot communicate are supplied with random inputs, \RutC{and} violate a Bell inequality they must be generating randomness \emph{no matter what their internal operations are}. This suggests using a Bell inequality violation to construct a Device Independent randomness expansion protocol. In such a protocol, inputs are repeatedly \RutC{provided} to two separated (non-communicating) devices and their outputs are stored. We call the inputs $X_i$ and $Y_i$ and the respective outputs $A_i$ and $B_i$, where $i$ runs from $1$ to $n$ (see chapter~\ref{chap : actual protocols for DI} for examples of such protocols). After making $n$ inputs we can estimate the average value of some Bell inequality; if there is no violation (or the violation is too small) then the protocol aborts. If the violation is large enough, then the raw outputs are run through an extractor to generate the final output randomness.

\RutC{Unfortunately, the practical application of DIRNE protocols poses challenges, primarily because achieving violations of Bell inequalities in experiments is very difficult. Although Bell's theorem was introduced in the 1960s, it took decades before the first loophole-free Bell inequality violations were experimentally verified~\cite{Giustina&, Hensen&, Shalm&}, and even these showed only modest deviations. For example, a 2021 study~\cite{LZL&gen} reported a CHSH score of 0.752484, a value merely 0.024 above the local bound of 0.75. This violation is significantly smaller than the maximum possible violation, which is roughly 0.853. As we will delve into later in this thesis,  a high CHSH score is crucial to implement a robust (CHSH-based) DIRNE protocol.}

Although achieving a Bell violation experimentally currently poses a challenge for DI protocols, significant advancements have been made in this area. For example, the violation of the local bound increased from 0.00027 in 2018 as reported by~\cite{LZL&}, to 0.024 in less than three years. As technology evolves to accommodate these needs, it is worthwhile to explore scenarios where we can trust parts of the system. Such protocols are termed semi-Device Independent (semi-DI) protocols, which are simpler to implement experimentally compared to fully device independent protocols.

A novel randomness expansion protocol was recently proposed, which operates based on energy and overlap bounds~\cite{VSoriginal}. Fundamentally, this protocol is based on a prepare-and-measure scenario consisting of a source and a measurement device. In the protocol, a source generates specific states, either $\rho_0$ or $\rho_{1}$, and the measurement device is \RutC{used} to determine if $\rho_0$ or $\rho_1$ was sent. The states $\rho_{0}$ and $\rho_{1}$ are prepared in such a way that they are near-identical (yet not entirely identical) to the unique ground state of the system (the vacuum state in the case of a laser being the source). Being \RutC{almost} identical, these states cannot be perfectly distinguished according to quantum mechanics\footnote{Note that when an arbitrarily large number of copies of two quantum states are available, perfect distinction between them becomes possible. However, here we consider the case when only a single copy of either of the two states \RutC{is sent to the measurement device at a given time}.}, which forms the basis for \RutC{randomness generation in this protocol}.

However, a major assumption in this scenario is that the prepared states are near identical to the ground state. \RutC{Unfortunately, this assumption} cannot be verified in a device independent manner. To address this, the protocol assumes \RutC{the availability of a trusted} power meter, which can measure the energy of the states prepared by the source. For the protocol that we describe \RutC{in this work}, the power meter is the only trusted component in the protocol. \RutC{Thus, this approach is appealing because} it only requires the characterization of a single component of the protocol.

\RutC{Similar to the DI protocols, this protocol should abort if the outputs do not contain sufficient extractable randomness. In particular, the protocol should abort if either of two conditions is met: the energy of the states is too high, or the detector is unable to distinguish the states $\rho_0$ and $\rho_1$ with the desired accuracy. }

\RutC{In this work, we study randomness expansion protocols based on the key ideas outlined above}. These protocols make minimal assumptions on the inner workings of the devices used to generate randomness. 

\RC{While asserting that the protocols discussed in this thesis are secure, we assume that the quantum theory is correct and complete \cite{CR_ext}. However, it is worth noting that the security of the Device Independent protocols does not rely solely on this assumption; their security has also been established under a significantly weaker condition: that the eavesdropper is bound only by the non-signalling principle \cite{BHK} (see Chapter \ref{chap: GPTs} for an explanation. Here, the non-signalling principle simply means that no superluminal signalling is possible, without necessarily asserting the correctness of the quantum theory). We also assume that the protocol is being carried out in a secure laboratory from which no information can leak out. It is crucial that humans conducting in the protocol do not cause any data leaks. In other words, having a laboratory that is well-shielded from the outside world is crucial for the security of the protocol. These assumptions are however essential for any protocol of randomness expansion. If such a protective measure is not in place, any randomness expansion protocol can be compromised, regardless of the strategy used.} 

\begin{figure}
    \centering
    \includegraphics[width=0.9\linewidth]{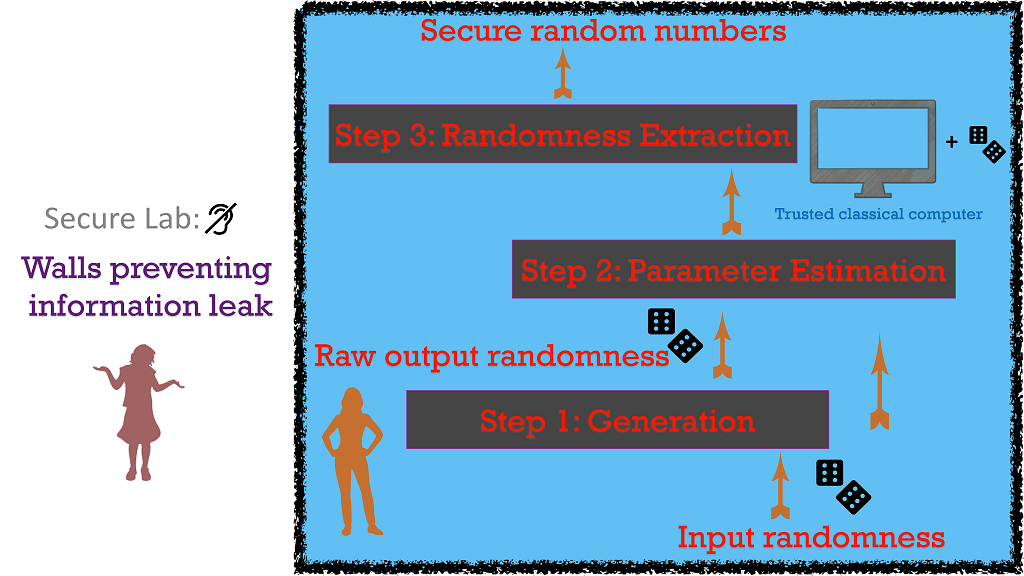}
    \caption[A typical protocol of randomness expansion.]{A typical protocol consists of three steps - randomness generation, parameter estimation and randomness extraction. Randomness extraction is typically done using a secure classical computer, using known algorithms. The lab is assumed to be secure from which no information can leak out.}
    \label{Fig:typical_protocol}
\end{figure}

Typically, \RutC{Device Independent and semi-Device Independent protocols are carried out in} three different stages as shown in Figure ~\ref{Fig:typical_protocol}. These different stages are: generation, parameter estimation, and randomness extraction.
The generation stage is the \RutC{main} part of the protocol where a certain sub-protocol is repeatedly executed. In the context of DI protocols, this step involves performing a CHSH test, while in the semi-DI protocol, it \RutC{is a ``prepare and measure round''}. 

The parameter estimation stage involves analyzing the input and output statistics collected in the generation stage. \RutC{The protocol is aborted} if these statistics are not suitable. In DI protocols, \RutC{the parameter estimation step }involves computation of the Bell score. If the \RutC{observed} Bell score is too low \RutC{(for example a CHSH score less than $3/4$)}, then the protocol is aborted. If the protocol does not abort, then the amount of uniform randomness that can be extracted from the outputs of the protocol is computed using \textit{only} the input-output statistics of the protocol (such as the Bell score). 

The final stage is the randomness extraction stage, where the \RutC{random} strings \RutC{obtained} in the \RutC{generation stage} are processed to produce a string of uniformly distributed random numbers that are secure against any adversary. This \RutC{stage also consumes} some randomness. There is a rich literature on randomness extraction (see for example~\cite{TSSR}); however, this part of the protocol is beyond the scope of this thesis.

After developing these protocols, the primary challenge lies in calculating the amount of extractable randomness as a function of the observed statistics. The difficulty of this task is amplified by the lack of structure in the problem: minimal assumptions are made on how the devices operate, requiring accounting for arbitrary pre-shared entanglement, arbitrary measurements, and potentially adaptive strategies between the rounds. Two techniques to address this exist in the literature: the quantum probability estimation framework~\cite{ZFK} and the entropy accumulation theorem (EAT)~\cite{DFR, ARV}. In this work, we employ the latter. The EAT, informally speaking, suggests that the amount of extractable randomness in the `n' bits long full string of outputs is predominantly `n' times the von Neumann entropy of a single-round strategy that would yield the observed score if used in an independent and identically distributed (i.i.d.) way. This implies that if we can solve the problem for an i.i.d. adversary, we can obtain a bound for the general case. 

Consequently, the task of determining the randomness \RutC{rate (i.e. amount of randomness generated per round) in a DI and our semi-DI protocol reduces to calculating the least value of the von Neumann entropy in a single representative round of the protocol}. \RutC{Computing lower bounds on the von Neumann entropy } generally necessitates sophisticated mathematical techniques in optimization theory, \RutC{especially in the areas of convex optimization}. In this work, we develop techniques that can be employed \RutC{to compute lower bounds on the conditional von Neumann entropy when the inputs and outputs in each round of the protocol are binary}. Essentially, this is a situation where we can apply Jordan's Lemma, leading to a significant reduction in the complexity of the problem.

This thesis focuses on the first two stages of the DI and semi-DI protocol discussed above: the generation stage and the parameter estimation stage. \RutC{ Our work focuses on two main aspects:}

\begin{itemize}
    \RutC{\item Providing new protocols for randomness expansion that use the same setup as traditional protocols, but promise more efficient and secure ways of generating randomness (for example: by reducing experimental assumptions).
    \item Finding mathematical techniques to compute bounds on randomness rates in DI and semi-DI protocols is introduced.}
\end{itemize}

We now present a brief outline of the thesis: 

The subsequent chapter (Chapter \ref{chap: maths}) discusses some optimization techniques, including the entropy accumulation theorem, which are crucial in randomness expansion protocols. We then move to Chapter \ref{chap: DI protocols}, where we provide a detailed  introduction to DI protocols for randomness expansion. We also deliberate on various entropic quantities of interest in the context of DI protocols for Randomness Expansion and Quantum Key Distribution. 

Next, in Chapter \ref{chap: DI upper bounds}, we utilize Jordan's Lemma to calculate numerical upper bounds on these entropies. In the \RutC{following} chapter (Chapter \ref{chap: DI lower bounds}), we derive the lower bounds for these entropies.

The final chapter in the section on DI protocols for randomness expansion (Chapter \ref{chap : actual protocols for DI}) presents all the protocols and exhibits the randomness rates achieved for finite rounds using the entropy accumulation theorem. Chapters \ref{chap: DI upper bounds, chap: DI lower bounds , chap : actual protocols for DI} are based on the published work \cite{BhRC}.

Following our discussion on DI protocols, we shift our focus to semi-DI protocols. We divide our work into $3$ parts. The first part introduces basic framework for the semi-DI (Chapter \ref{chap: semi-DI}) and explores different protocols. Subsequently, in the next chapter (Chapter \ref{chap: semi-DI lower bounds}), we calculate the rates for these semi-DI protocols and in the final chapter on semi-DI protocols, we discuss two different protocols along with results and discussion. We summarize our results on randomness expansion protocols.

Upon concluding our discussion on randomness expansion protocols, we transition to the part of the thesis dealing with Generalized Probability Theories (GPTs) (Chapters ~\ref{chap: GPTs} and ~\ref{chap: Boxworld}). 

\cleardoublepagewithnumber

\chapter{Preliminaries}\label{chap: maths}
\section{Optimization problems in quantum information}
The theory of optimization problems is a vast field that has been extensively studied in various disciplines such as mathematics, computer science, physics, and economics. As we shall explore further in this chapter, these optimization problems play a key role in quantum information theory as well. The main goal for the study of optimization theory is to compute the value for the following:  
\begin{equation}\label{eqn: general optimzaition problem}
    \RutC{\min_{x \in \M{D}} g(x) . }
\end{equation} 
In this case, $\M{D}$ is the domain of the optimization problem often also referred to as the feasible set, while $g: \M{D} \mapsto \mathbb{R}$ is known as the objective function. Because of the abstract nature of the problem, there are no universal algorithms available to solve a given optimization problem, as $g$ could be any function and $\M{D}$ could be any set.\par

Nevertheless, there exists a particular category of optimization problems, known as convex optimization problems, that have gained considerable interest due to their simplicity and the fact that they \RutC{often} have a unique solution. \RutC{The optimization problem \ref{eqn: general optimzaition problem} is a convex optimization problem if the objective function $g$ is a convex function and the domain $\M{D}$ is a convex set.}\par

Let's define some key terms related to convex optimization:\par

\begin{definition}[Convex Set] A set $\M{D}$ \RutC{is a} convex \RutC{set} if for any points $x$ and $y$ in $\M{D}$, the point $\mu x + (1 - \mu) y$ also belongs to $\M{D}$ \RutC{for every $\mu \in [0 , 1]$}.
\end{definition}
\begin{definition}[Convex Function]
Let $\M{D}$ be a convex set, and $g: \M{D} \mapsto \mathbb{R}$ be a function. The function $g$ is convex if, for any points $x$ and $y$ in $\M{D}$ and \RutC{any} value \RutC{$\mu \in [0 ,1]$}, $g(\mu x + (1 - \mu) y) \leq \mu g(x) + (1 - \mu) g(y)$. 
\end{definition}
\begin{definition}[Concave function]
A function $f$ is called concave if $-f$ is convex. 
\end{definition}
Convex optimization problems frequently appear in quantum information theory. This is primarily due to two fundamental convex sets in quantum theory: the set of states (state space) and the set of effects (effect space).\par

In the quantum theory, a physical system is associated with a Hilbert space $\M{H}$. A state is a \RutC{non-negative} linear map $\rho : \M{H} \mapsto \M{H}$, with a bounded trace $\tr(\rho) \leq 1$. We denote the set of all states on $\M{H}$ by the set $\M{S}(\M{H})$. This set can be easily shown to be a convex set. \par

Meanwhile, an effect in quantum theory is defined as a linear map $E: \M{S}(\M{H}) \mapsto \mathbb{R}$ such that, for any state $\rho \in \M{S}(\M{H})$, $E(\rho) = \tr(E \rho) \in [0, 1]$. The set of all effects is also easily shown to be a convex set. We will further revisit and generalize these concepts in the chapter discussing Generalized Probability Theories (GPTs) later \RutC{in Chapter \ref{chap: GPTs}}.\par

For completeness, we define a POVM (Positive Operator-Valued Measure) as a collection of effects $\{ E_i \}_{i = 1}^{N}$ that sum up to the identity $\id$. A POVM describes a measurement in quantum theory, where the probability of obtaining outcome $i$ in an experiment is given by $P[i | \rho] = \tr(E_i \rho)$. \par

Various convex and concave functions arise naturally in quantum theory. One example of a convex function on the state space is the probability of an outcome corresponding to an effect, denoted as $E(\cdot) := \tr(E \cdot)$. Another concave function that features in quantum information theory is the von Neumann entropy, denoted by $H(\rho)$. Hence, the theory of convex optimization is frequently \RutC{used} in the study of quantum information theory. In the forthcoming sections, we will encounter a range of intriguing convex optimization problems within the context of quantum information theory. The scope of the thesis for discussion of the role of optimization problems in quantum information is limited, and those unfamiliar with the topic are encouraged to go through excellent lecture notes for a good overview of this topic~\cite{Richard}.\par
\RC{As will become clear at the end of this chapter, a key quantity of interest that will appear as the objective function in most of the optimization problems in this thesis is the conditional von Neumann entropy:}
\begin{definition}[von Neumann entropy] \RC{The von Neumann entropy of a state $\rho \in \M{S}(\M{H}_{A})$ is defined as }
\begin{equation}
    \RutC{H(\rho) \equiv H(A)_{\rho} := - \tr\left( \rho \log_{2} \rho \right).}
\end{equation}
\RC{Here $\log_{2}$ is the matrix logarithm with base $2$.}
\end{definition}
\begin{definition}[conditional von Neumann entropy] \RC{For a state $\rho_{AB} \in \M{S}(\M{H}_{A} \otimes \M{H}_{B})$ the conditional von Neumann entropy $H(A|B)_{\rho_{AB}}$ is given by}
\begin{eqnarray}
    \RutC{H(A|B)_{\rho_{AB}} := H(AB)_{\rho_{AB}} - H(B)_{\rho_{B}},}
\end{eqnarray}
\RC{where $ \rho_{B} := \tr_{A} (\rho_{AB}) \in \M{S}(\M{H}_{B})$ is the state on the subsystem $B$. }
\end{definition}
\section{Semi-Definite Programs}\label{sec: sdps}
Even though a general convex optimization problem often holds the appeal of having a unique solution, this solution can be quite challenging to find. The complexity stems from the lack of efficient numerical algorithms capable of reliably solving such general convex optimization problems. Nevertheless, certain optimization problems that we frequently encounter belong to specific classes that can be solved using efficient numerical algorithms. Prominent examples of such problems include Linear Programs and Semi-Definite Programs (SDPs).
\begin{definition}[Semi-Definite Program] An optimization problem is an SDP if it is of the following form 
\begin{eqnarray}
\begin{aligned}
   g   &= \quad \inf_\RutC{{X \in \mathbb{H}^{m}}} \tr (C X)  \\ 
    \mathrm{s.t.}   & \quad  X \succeq 0 \\ 
                    & \RutC{\quad \forall i: \quad \mathcal{A}_i(X) = B_i} .
\end{aligned}
\end{eqnarray}
where $C , X \in \mathbb{H}^{m}$ (the set of $m \times m$ Hermitian matrices) , \RutC{$B_i \in \mathbb{H}^{n_i}$} and \RutC{$\mathcal{A}_i: \mathbb{H}^{m} \mapsto \mathbb{H}^{n_i}$} is a linear map.
\end{definition}
As stated above, efficient numerical algorithms exist that can solve SDPs numerically. Popular software packages like Mosek~\cite{mosek} and CVXPY~\cite{CVXPY} have integrated these algorithms \RutC{in a very user-friendly fashion}. These algorithms and details on how to program them efficiently are beyond the scope of the thesis.\par    
SDPs are particularly relevant to the study of quantum information theory, as we often need to maximize or minimize a linear function \RutC{involving arbitrary quantum states or effects}. 
Consider, for example, the problem of operationally distinguishing two quantum states. Suppose we are given a uniform mixture of two states, $\rho_1$ and $\rho_2$, and we aim to perform a single measurement to determine whether $\rho_1$ or $\rho_2$ was prepared. We are interested in determining the best possible strategy that allows us to distinguish these states. This problem can be directly formulated as an optimization problem.\par
To do so, we establish a decision rule by employing a two-outcome measurement $ \{ E_1$, $E_2 = \id - E_1 \}$. If we observe the outcome corresponding to $E_1$, we guess that $\rho_1$ was prepared, and vice versa. The success probability \RutC{$p_{\text{succ}}$} of our strategy can be computed as:
\begin{eqnarray}
    p_{\text{succ}} &=& \frac{1}{2} \mathbb{P}[E_1 | \rho_1] + \frac{1}{2} \mathbb{P}[E_2 | \rho_2] \\ 
                    &=& \frac{1}{2} \left( \tr\left(E_1 \rho_1 \right) + \tr\left(E_2 \rho_2 \right) \right) \nonumber \\
                    &=& \frac{1}{2} + \frac{1}{2}\tr \left( E_1 (\rho_1 - \rho_2) \right) \nonumber .
\end{eqnarray}
\RutC{Here $\mathbb{P}[E_i|\rho_i]$ is the probability of observing the outcome $i \in \{ 0 ,1\}$ given that the state $\rho_{i}$ has been prepared}.
In order to find the optimal measurement, we would like to maximize the \RutC{success probability}. As \RutC{$\{ E_1 , E_2 \}$} is a POVM, \RutC{we have the constraint} $ 0 \leq E_1 \leq \id$. This means that our best chance of the guessing the correct state in this scenario is given by the optimization problem 
\begin{eqnarray}
\begin{aligned}
   p_{\text{succ}}^{\max}   &= \quad \sup  \left(\frac{1}{2} + \frac{1}{2} \tr \left(E_1 (\rho_1 - \rho_2)  \right) \right)  \\ 
    \mathrm{s.t.}   & \quad  E_1 \succeq 0 \\ 
                    & \quad  \id - E_1 \succeq 0 .
\end{aligned}
\end{eqnarray}
The above problem can be easily identified as a SDP. Notably, this optimization problem also has an analytical solution known as the Holevo-Helstrom theorem~\cite{helstrom1969quantum, holevo1973optimal}, which states that the maximum success probability is given by:
\begin{eqnarray}
    p_{\text{succ}}^{\max} =  \frac{1}{2}  + \frac{1}{4}|| \rho_1 - \rho_2 ||_1  ,
\end{eqnarray}
where $|| . ||_1$ is the trace norm defined as 
\begin{eqnarray*}
    || \tau ||_{1} := \tr \left( \sqrt{\tau^2} \right).
\end{eqnarray*}
\RutC{The POVM $\{E_1 , E_2 \}$ which leads to the maximum success probability is the POVM $\{ P_+ , P_{-} \}$ where $P_+$ and $P_{-}$ are the positive part and the negative part of the operator $\rho - \sigma$\footnote{Every Hermitian operator $X$ has a unique deposition $X = P_+ - P_{-}$, where the operators $P_{+} \succeq 0$ and $P_{-} \succeq 0$ are the called the positive and negative part of $X$ respectively. If the operator $X$ is traceless then $P_{+} + P_{-} = \id$.}. }
In the next sections, we will explore the different SDPs that can arise in studying protocols of randomness expansion.

\section{Polynomial optimization problems}\label{sec: polysdp}
Another interesting set of optimization problems are the polynomial optimization problems. The polynomial optimization problems take the form 
\begin{eqnarray}\label{eqn: constrained optimization}
\begin{aligned}
   \min &\quad   p(x_1 , \cdots x_n)  \\ 
    \mathrm{s.t.}   & \quad  g_i(x_1 , \cdots x_n) \geq 0  \\ 
                    & \quad  h_j(x_1, \cdots x_n) =  0 ,
\end{aligned}
\end{eqnarray}
with $p,q_i$ and $h_j$ all being polynomials \RutC{of degree less than or equal to $d \in \mathbb{N}$} and $x_1 ,\cdots x_n$ being real valued variables - i.e. 
\begin{eqnarray}
p \in \mathbb{R}[x_1 , \cdots , x_n]_{\leq d} \nonumber ,\\  
g_i \in \mathbb{R}[x_1 , \cdots ,  x_n]_{\leq d} \nonumber ,\\
h_j \in \mathbb{R}[x_1 , \cdots , x_n]_{\leq d} \nonumber . 
\end{eqnarray}
Note that such optimization problems are not necessarily convex. \RutC{Due to the lack of structure in such problems}, there are no direct techniques for efficiently solving them. However, we shall demonstrate that any polynomial optimization problem can be cast into a converging sequence of SDPs, each of which can be solved using established algorithms. This approach allows us to \RutC{compute reliable upper and lower bounds} for these optimization problems. It is important to mention, though, as we go higher up in the hierarchy, the SDPs become increasingly more time and resource consuming to solve. 

Our discussion in this section is primarily based on the lectures on semi-definite programming by Prof. Hamza Fawzi (refer to~\cite{Hamza} for the slides). We have provided here a concise overview of the technique, with a focus on its core principles. For a comprehensive exploration, references like~\cite{Hamza_book} are highly recommended.
\subsection{Proving non-negativity of a polynomial}
To better understand how to construct such a hierarchy of SDPs that approximately solve a polynomial optimization problem, we can consider a simpler problem: demonstrating the non-negativity of a given polynomial:
\begin{eqnarray}
\forall x_1, x_2 , \cdots x_{n}: \quad p(x_1 , \cdots x_n) \geq 0 .
\end{eqnarray}
We use the notation $\text{POS}$ to represent the set of all (globally) non-negative polynomials.

\subsubsection{Proving if a function is sum of squares}\label{subsec: sos}
A simple \RutC{method} to determine the non-negativity of a polynomial is to show that it can be expressed as a sum of squares. We define $\text{SOS}_{d}$ as the set of all polynomials of degree less than or equal to $d$ that have a sum of squares decomposition. \\

The problem of determining whether a given polynomial can be expressed as a sum of squares - i.e. $p \in \text{SOS}_{d}$, can be cast as an SDP. To understand why, consider a polynomial $p \in \text{SOS}_{d}$  of degree $d$. Assuming it has a sum of squares decomposition, $p = \sum_{i} p_i^2$, where each $p_i$ is a polynomial of degree less than or equal to $d/2$. Each $p_i$ can be expressed in matrix form:
$$ p_i = [1 , x_1 , x_2 , \cdots , x_1 x_2 , \cdots] \begin{bmatrix}
    a_{0}\\
    a_{1} \\
    \vdots\\
    a_{12} \\ 
    \vdots
\end{bmatrix} ,$$
i.e, $p_i = \mathbf{x}^{T}\mathbf{a}$, where $\mathbf{x}$ is \RutC{a} vector consisting of monomials and $\mathbf{a}$ is the vector corresponding coefficients of the monomial. Then, it is easy to see that $$ p_i^2 = \mathbf{x}^T \mathbf{a}^T \mathbf{a} \mathbf{x}.$$ 
Note that the matrix $\mathbf{a}^T \mathbf{a}$ is positive semi-definite. We can immediately deduce that if a polynomial $p \in \text{SOS}_{d}$, then there exists a matrix $\mathbf{A}$ such that
 $\mathbf{A}$ is a sum of matrices \RutC{$\mathbf{a}^T \mathbf{a}$}, and is itself positive semi-definite. \RutC{It is} also evident from straightforward reasoning that if $p = \mathbf{x}^T \mathbf{A} \mathbf{x}$ for some $\mathbf{A} \succeq 0$, then $p$ can be expressed as a sum of squares of polynomials. The simplest approach do prove this is to express $\mathbf{A}$ through its spectral decomposition to explicitly construct the sum-of-squares polynomials. Hence, a polynomial $p$ belongs to the set $\text{SOS}_{d}$ if and only if there exists a positive semi-definite matrix $\mathbf{A}$ such that $p = \mathbf{x}^T \mathbf{A} \mathbf{x}$. 

Let's now consider a scenario where we have an arbitrary polynomial \(g\), and our objective is to determine whether \(g \in \text{SOS}_{d}\). To achieve this, we aim find a suitable $D \succeq 0$ such that  \(g = \mathbf{x}^{T} D \mathbf{x}\). 

Since \(g\) is a polynomial with a degree at most \(d\), we can uniquely determine it through a finite set of constraints. These constraints, for instance can take the form \(g(c_0^{(0)}, c_1^{(0)}, \cdots, c_{d}^{(0)}) = b_{0}\), \(g(c_0^{(1)}, c_1^{(1)}, \cdots, c_{d}^{(1)}) = b_{1}\), and so forth. These constraints can \RutC{be expanded as} linear conditions on the matrix \(D\):
\[ \mathbf{c}_{i}^{T} D \mathbf{c}_{i} = b_{i} .\]
Thus, if there is a solution to the following problem:
\begin{equation}\label{eqn: trivial optimization problem}
\begin{aligned}
\max \quad & 0 \\
\text{s.t.} \quad & \forall i\in \{ 1 , 2 , \cdots , k \}: \mathbf{c}_{i}^{T} D \mathbf{c}_{i} = b_{i} \\
& D \succeq 0 .
\end{aligned}
\end{equation}
Then we have successfully decomposed \(g\) into a sum of squares of polynomials. Conversely, if no solution exists for the above problem, we have demonstrated that \(g\) cannot be expressed as a sum of squares. \\
\RutC{Optimization problems with trivial objective functions such as \eqref{eqn: trivial optimization problem} are often used to prove the existence (or lack of existence) of solutions of simultaneous equations and inequalities. The solution to such an optimization problem is $0$ iff there the set of constraints can be simultaneously satisfied. Having no solution to such an optimization problem is proof that the constraints can not be simultaneously satisfied. }

\subsubsection{Using SOS condition to determine positivity} 
\RutC{In the previous subsection \ref{subsec: sos}, we have seen an efficient method to determine if a given polynomial $p$ can be expressed as a sum of squares. It is immediately clear that any polynomial that can be expressed as a sum of squares is non-negative - i.e. $\text{SOS} \subseteq \text{POS}$. }

However, the intriguing question arises: Is it possible that \(\text{SOS} = \text{POS}\), or is it the case that \text{SOS} $\subset$ \text{POS}? If the former holds true, then indeed, we would have already found an efficient way to determine the positivity of any polynomial. 

\begin{figure}[h!]
  \centering
  \includegraphics[width=0.8\textwidth]{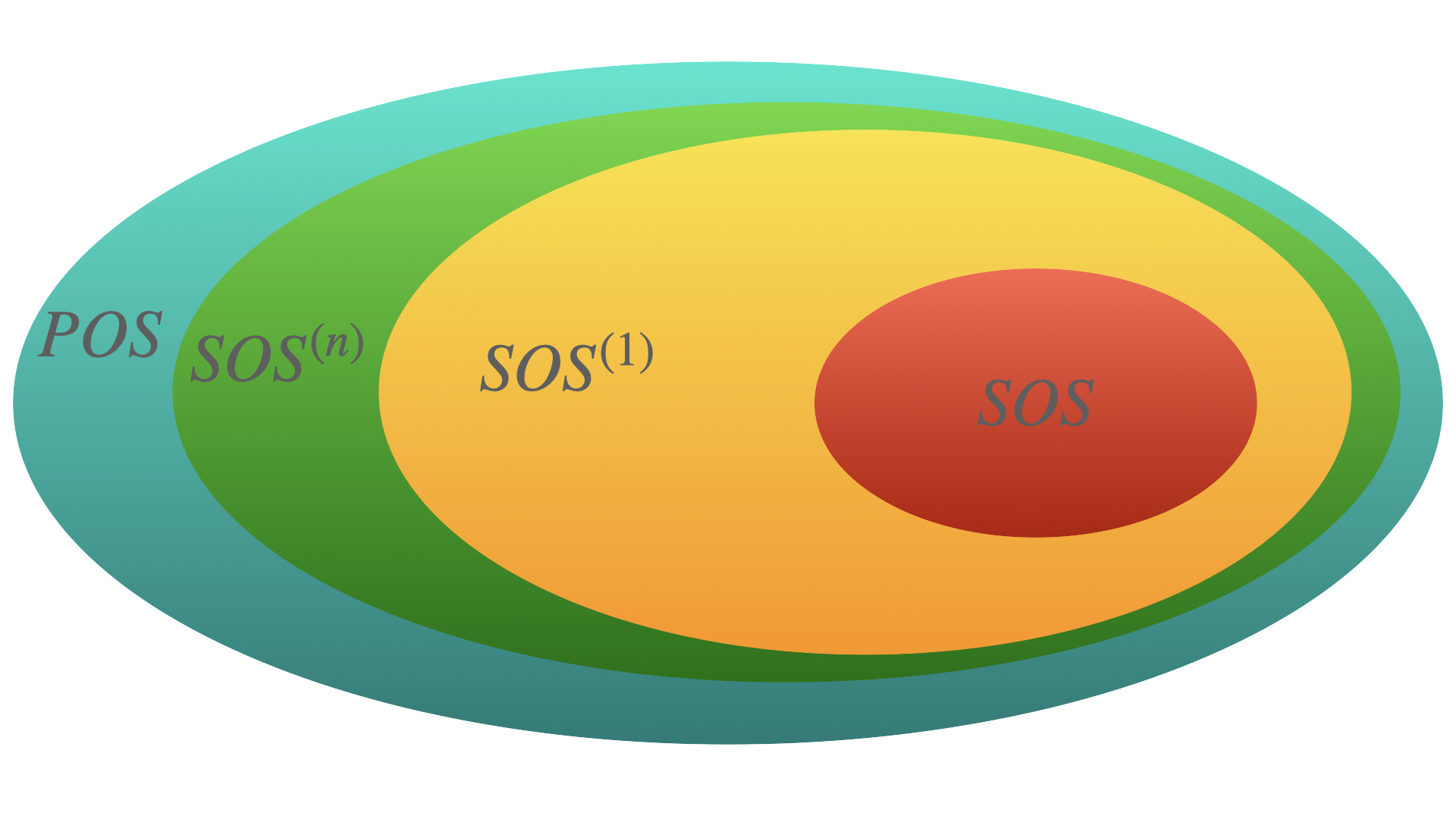}
  \caption[A schematic depiction of the sets SOS and POS.]{A \RutC{schematic depiction of the} converging sets $\text{SOS}^{(n)}$ \RutC{and the} set $\text{POS}$. }
  \label{SOS}
\end{figure}

The answer to this problem is shown to be unfortunately in the negative. However, \RutC{it turns out that there is a very interesting relationship between the sets $\text{SOS}$ and $\text{POS}$}. It can be shown that 
\begin{eqnarray}
p \text{ is non-negative } \quad \iff \exists \text{ polynomial }  q \text{ such that } qp \in \text{SOS} .\nonumber    
\end{eqnarray}
In fact, we can reduce the search of polynomials $q$ to be of the form $q_n = ( x_1^2 + \cdots + x_d^2)^{n}$. \\
Note that now there is a possibility of coming up with a way to show if a polynomial is positive. Consider the set $\text{SOS}^{(n)}$ defined as 
$$ \{p : \quad ( x_1^2 + x_2^2 + \cdots x_d^2)^{n} p \in \text{SOS} \} .$$
Then $p \in \text{POS}$ implies that there exists $n \in \mathbb{N}$ such that $p \in \text{SOS}^{(n)}$. Thus, there is a hierarchy of sets, which will eventually cover all the positive polynomials. Note that we have now a hierarchy of sets $\text{SOS}^{(n)}$ that converge \RutC{to} the set of positive polynomials from inside 
$$ \text{SOS} \subseteq \text{SOS}^{(1)} \subseteq \text{SOS}^{(2)} \subseteq \cdots = \text{POS} .$$
See Figure~\ref{SOS} for a \RutC{schematic} illustration of the hierarchy. 
It is also possible to come up with a converging hierarchy of semi-definite programs that converge \RutC{to} the set \text{POS} from outside. This discussion is beyond the scope of the thesis. 

\subsection{Solving an unconstrained polynomial optimization problem}
Now that we possess a method for determining whether $p \in \text{POS}$, we can also extend this same argument to address the unconstrained polynomial optimization problem 
$$\min_{x \in \mathbb{R}^{d}} p(x) ,$$
by observing that it can be equivalently reformulated as the subsequent optimization problem:
\begin{equation}
\begin{aligned}
\RutC{\max_{\gamma \in \mathbb{R}}} \quad & \gamma \\
\text{s.t.} \quad & p - \gamma \in \text{POS} .
\end{aligned}
\end{equation}
This reformulation can be further transformed into a progressively convergent hierarchy of SDPs problems, expressed as follows:
\begin{equation}
\begin{aligned}
\RutC{\max_{\gamma \in \mathbb{R}}} \quad & \gamma \\
\text{s.t.} \quad & p - \gamma \in \text{SOS}^{(n)} .
\end{aligned}
\end{equation}
\subsection{Solving a constrained polynomial optimization problem} 
We saw above that determining global non-negativity of a polynomial can be cast as a hierarchy of converging semi-definite programs. In order to solve the constrained polynomial optimization problem such as  \eqref{eqn: constrained optimization}, we construct the following polynomial 
\begin{eqnarray}\label{eqn: f function}
     f(x_1\cdots x_{k}) =  p(x_1  .. x_n) + \sum_i \sigma_i(x_1 .. x_{r}) g_i(x_1 .. x_n) +  \sum_{j} \rho_j(x_1 
     \cdots x_{l}) h_j(x_1 .. x_{n}),
\end{eqnarray}
where $\sigma_i(x_1 , \cdots x_{r}) \in \text{POS}$ are known non-negative polynomials, and $\rho_{j}(x_1 , \cdots x_{r})$ are any polynomials. Note that $f(x_1 , \cdots x_{k})$ is non-negative whenever the constraints are satisfied - i.e. when $ g_i(x_1, \cdots x_n) \geq 0$ and $h_j(x_1 ,\cdots x_{n}) = 0$. 
We can now find a set of solutions to our global polynomial optimization problems by solving the following problem 
\begin{eqnarray}
\begin{aligned}
   \RutC{\max_{\gamma \in \mathbb{R}}} &\quad   \gamma  \\ 
    \mathrm{s.t.}  & \exists f \text{ of the form \eqref{eqn: f function}} \\ 
     &\quad f  - \gamma \in  \text{POS}. 
\end{aligned} 
\end{eqnarray}
We can again find suitable relaxations of this problem in terms of the semi-definite programs as discussed in the previous subsection. \\ 
These techniques are well known in the convex optimization literature. There are standard packages for many programming languages such as Python and Matlab that solve such optimization problems \RutC{using techniques similar to the one discussed in this section}. The package used to solve the optimization problems in this work is the NCPol2SDPA package~\cite{NCPOL2SDPA}. Other semi-definite programs are solved using solvers PICOS~\cite{PICOS}. 

\section{Operational meaning of the min-entropy}\label{sec: min-entropy}
In a randomness expansion protocol, the main objective is to quantify the amount of randomness \RutC{produced by the }protocol. The key question then arises: How can we quantify the amount of randomness? To address this query, we must accurately characterize the broader situation in which we are operating.\\
Consider a secure laboratory where we have a random variable, denoted $A$. Intuitively, we would consider the random variable $A$ as a source of randomness if its outcome cannot be accurately predicted. However, since we are dealing with a cryptographic scenario, we must be thorough and also account for the potential existence of an adversary for whom this random variable should also remain unpredictable. Thus, we will envision an imaginary adversary, whom we will call Eve or \RutC{the} Eavesdropper (or adversary), possessing certain information related to the random variable $A$. This information could include a correlated random variable or even a quantum state. However, Eve does not have direct access to \RutC{the} variable $A$. 

In the most general case we allow that adversary to hold a quantum state $\rho^{a}_{E}$ for every outcome $a$ of the random variable $A$. This scenario can be effectively described using the \RutC{Classical-Quantum (CQ)} state:
$$\rho_{CQ} = \sum_{a} p_{A}(a)\ketbra{a}{a} \otimes \rho^{a}_{E} .$$ 

Now, imagine Eve attempts to determine the value of the random variable \(A\). To achieve this, she performs a POVM, denoted as \(\{ \mathcal{F}_{a} \}_{a}\), on her state, where each outcome corresponds to a possible value of \(A\). The probability that Eve correctly guesses the outcome is expressed as:

\begin{equation*}
    p_{\text{guess}} = \sum_{a} p_{A}(a) \M{F}_{a}(\rho^{a}_{E}).
\end{equation*}
Given that the choice of measurement used by Eve can be arbitrary, it is reasonable to assume that she would select the optimal measurement strategy to maximize her guessing probability. With this in mind, quantifying the amount of randomness in a cryptographic scenario is related an optimization problem, which can be formulated as follows:
\begin{eqnarray}
     \begin{aligned}
   p_{\text{guess}}^* =  & \quad \sup \sum_{a} p_{A}(a)\M{F}_{a}(\rho^{a}_{E})  \\ 
    \mathrm{s.t.}   & \quad  \{ \M{F}_{a} \}_{a} \text{ is a POVM.}    
\end{aligned}
\end{eqnarray}

It is easy to see that the optimization problem described above can be written in terms of an SDP. This is due to the linearity of the objective function with respect to the variables $\{\M{F}_{a}\}_{a}$, and $\{\M{F}_{a}\}_{a}$ being effects can be represented in terms positive semi-definite operators \RutC{$\{F_{a}\}_{a}$} by the relation $\M{F}_{a}(.) = \tr(F_{a}(.)) $. Furthermore, the constraints \RutC{that $\sum_{a} \M{F}_{a} = \M{I}$ (where $\M{I}$ is the identity map)} is a linear constraint $\sum_{a} F_{a} = \id$. 

\begin{figure}[h!]
  \centering
  \includegraphics[width=0.8\textwidth]{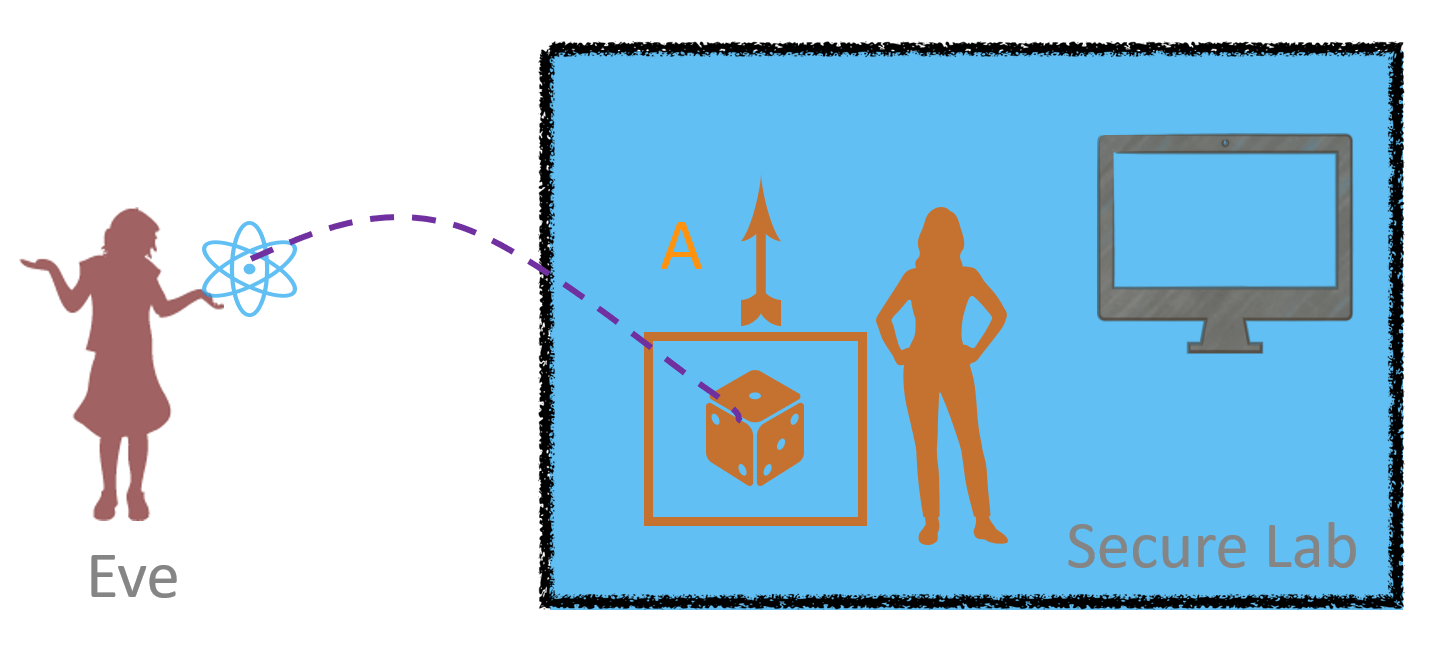}
  \caption[A diagram depicting the scenario in which an adversary is trying to guess the output random variable of a protocol.]{A diagram depicting the scenario in which an adversary is trying to guess the random variable $A$ in the secure lab.}
  \label{fig:min-entropy}
\end{figure}

Intuitively, the amount of randomness should be a function of this optimal guessing probability $p_{\text{guess}}^*$. 
Formally, in quantum information theory, the amount of randomness in a protocol is quantified using the min-entropy~\cite{KRS}: 
\begin{definition}[Min-entropy] The min-entropy of a state  $\rho_{AB} \in \mathcal{H}_{A} \otimes \mathcal{H}_{B}$ is given \RutC{by} 
\begin{eqnarray}
 H_{\min}(A|B)_{\rho_{AB}} &=& -\inf_{\sigma_{B}} \quad \{ \lambda | \quad \rho_{AB} \preceq 2^{\lambda} (\id_{A} \otimes \sigma_{B}) \},
\end{eqnarray}
where $\sigma_{B} \in \cS(\M{H}_{B})$.
\end{definition}

\RC{To understand the significance of why min-entropy is an appropriate measure for randomness in a cryptographic scenario, we must understand that the primary objective of such a protocol is to generate random bits that are uniformly (or almost uniformly) distributed. Moreover, these bits should be independent and uncorrelated with any side information held by an adversary. Hence, at the end of the protocol, we require a string of uniformly distributed random numbers $\B{R}$ uncorrelated with any system held by a potential adversary. Randomness extractors are procedures that "extract" such a random string $\B{R}$ from a random-variable $A$ (obtained as an output of a protocol) that could, in general, be correlated with the side-information of the adversary. Roughly speaking, a quantum-proof strong extractor $\mathrm{Ext}$ is a function that takes a random variable $A$ and an input random seed $U$, outputting the desired string: $\B{R} = \mathrm{Ext}(A, U)$. Technical results, like the generalized left over-hashing lemma~\cite{DPVR}, show that it is possible to extract approximately $H_{\min}(A|E)_{\rho_{AE}}$ such uniformly random bits from the random variable $A$ given a particular classical-quantum state $\rho_{AE}$ (that describes the protocol). Further details on randomness extractors are beyond the scope of the thesis.}

However, as it stands, the definition of min-entropy is not very insightful or intuitive. We sketch the proof from~\cite{KRS} to show that the min-entropy and the guessing probability are in-fact related to each other. 

To begin, let us introduce a new variable $\tilde{\sigma}_{B} = 2^{\lambda} \sigma_{B}$. Exploiting the monotonicity property of the logarithms, it becomes apparent that the min-entropy is given by $-\log_2(g)$, where $g$ can be computed through the following optimization problem: 
\begin{eqnarray}
\begin{aligned}
   g   &= \quad \inf \tr (\tilde{\sigma}_{B})  \\ 
    \mathrm{s.t.}   & \quad  (\id_{A} \otimes \tilde{\sigma}_{B}) - \rho_{AB} \succeq 0 \\ 
                    &\quad \tilde{ \sigma}_{B} \succeq 0 .
\end{aligned}
\end{eqnarray}
We will now show that this function $g$ is \RutC{, in fact,} the best guessing probability $p_{\text{guess}}^*$. 

Using the duality theory of optimization~\cite{boyd2004convex}, it is known that for every SDP, a corresponding dual SDP can be constructed, and the solutions to both problems are identical. In the case of the problem above, the dual for the SDP above is:
\begin{eqnarray}
     \begin{aligned}
      & \quad \sup \tr (Y \rho_{AB})  \\ 
    \mathrm{s.t.}   & \quad  Y \succeq 0 \\ 
                    &\quad \tr_{A}(Y) = 1 .
\end{aligned}
\end{eqnarray}
We can now use the Choi--Jamiołkowski isomorphism~\cite{jamiolkowski1972linear} to relate the non-negative operator $Y$ in terms of a channel $\M{E}$ as
\begin{eqnarray}
    Y = d_{A}(\id_{A} \otimes \M{E}) (\ketbra{\Phi_{AB}}{\Phi_{AB}}),
\end{eqnarray}
where $\ket{\Phi_{AB}} = \sum_{x}\frac{1}{d_{A}}\ket{x,x}$ is the maximally entangled state and $d_{A} = \dim(\M{H}_{A})$. \\  As $\tr(\id_{A} \otimes \M{E}(\ketbra{\Phi_{AB}}{\Phi_{AB}}) \rho_{AB}) = \tr(\ketbra{\Phi_{AB}} {\Phi_{AB}} \id_{A} \otimes \M{E}^{\dagger}(\rho_{AB}) ) $, we can re-write the \RutC{trace} $\tr(Y\rho_{AB})$ as: 
\begin{eqnarray}
\tr(\ketbra{\Phi_{AB}}{\Phi_{AB}}  \M{I} \otimes \M{E}^{\dagger}(\rho_{AB}) ) = \sum_{x} \bra{x, x} (\id \otimes \M{E}^{\dagger}(\rho_{AB})) \ket{x, x}.
\end{eqnarray}
This gives the optimization problem, 
\begin{eqnarray}\label{eqn: operational meaning}
\begin{aligned}
   g   &= \quad \sup  \sum_{x} \bra{x, x} (\id \otimes \M{F}(\rho_{AB})) \ket{x, x}      \\ 
    \mathrm{s.t.}   & \quad  \M{F} \text{ is a quantum channel}  ,
\end{aligned}
\end{eqnarray} 
where $\M{F} = \M{E}^{\dagger}$ is any channel. We return to the case of computing the min-entropy for the CQ states of the form\RutC{
\begin{eqnarray}
    \rho_{AB} = \sum_{a} p_{A}(a) \ketbra{a}{a} \otimes \rho^{a}_{B}. 
\end{eqnarray}}
Substituting the explicit form of $\rho_{AB}$ in the objective function of eqn. ~\eqref{eqn: operational meaning} gives the  following objective function: 
\begin{eqnarray}
    \sum_{a} p_{A}(a) \bra{a} \M{F}(\rho_{B}^{a}) \ket{a} \equiv p_{\text{guess}}.
\end{eqnarray}
The collection of maps $\{ \M{F}_{a} \equiv  \bra{a} \M{F}( . ) \ket{a} \}_{a}$ without any loss of generality is an arbitrary POVM, due to the fact that \RutC{ 
\begin{eqnarray}
  \forall \rho \in \M{S}(\M{H}_{A} \ot \M{H}_{B}):\qquad  \sum_{a} \bra{a} \M{F}(\rho) \ket{a}= \tr(\M{F} \rho_{AB}) = \tr(\rho).
\end{eqnarray}}
Note that above we have used the fact that $\M{F} = \M{E}^{\dagger}$ can be taken to be trace-preserving. 

\RutC{For consistency, unless stated otherwise, throughout the thesis, we have reserved the system $E$ for the system of the Eavesdropper or the adversary. }

\section{Quantifying randomness in a protocol}
\begin{figure}
    \centering
    \includegraphics[width=0.9\linewidth]{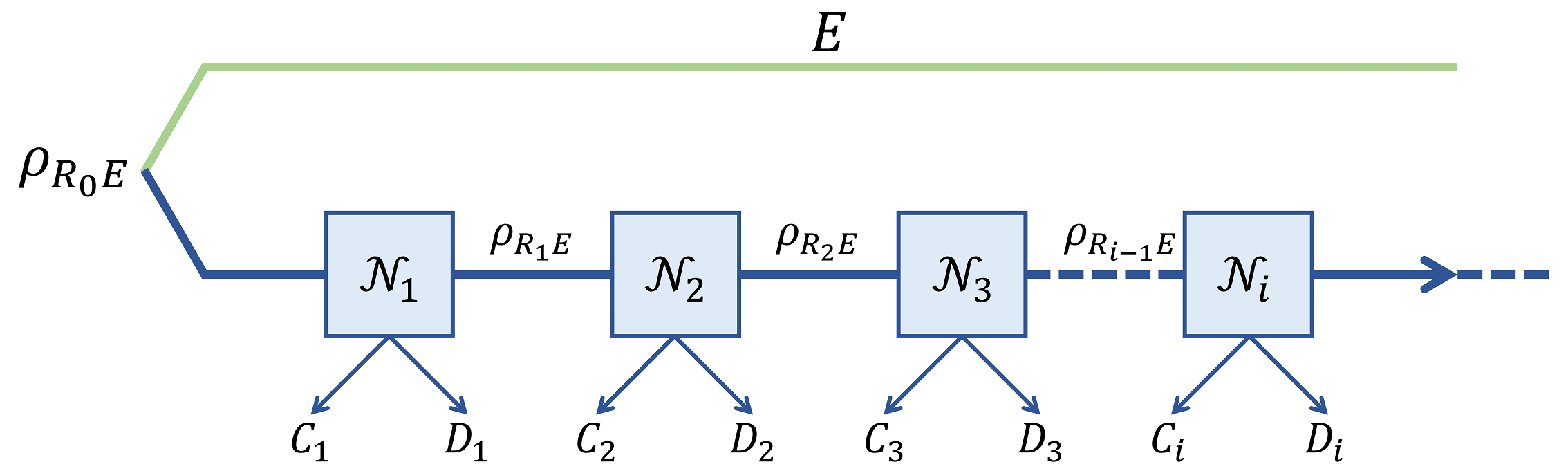} 

    \caption[A protocol of randomness expansion in terms of sub-protocols.]{A protocol of randomness expansion in terms of sub-protocols. $\rho_{R_0 E}$ is the initial state of the protocol and EAT channels.}
    \label{fig: EAT channel}
\end{figure}

The framework described in the previous section is very general and can be useful for any cryptographic scenario. In this thesis, our primary interest is to quantify randomness in randomness expansion protocols, which has more structure. Randomness expansion protocols are inherently sequential by design; that is, these protocols are typically carried out over multiple rounds, each identified by \(i \in \{ 1 , \cdots , n \}\). In each round, a sub-protocol is performed, and this process is repeated \(n\) times. For instance, in the Device Independent scenario, the sub-protocol might correspond to a single CHSH test. The randomness expansion protocols, in general, may be described using an initial state shared by the lab and Eve along with a sequence of channels, each representing a sub-protocol as shown in Figure~\ref{fig: EAT channel}. 

In the protocols we consider in this thesis, each round \(i \in \{ 1 , \cdots , n \}\) requires generating a random number \(D_i\) from an an existing source of randomness. By the end of round \(i\), the protocol produces another random number, \(C_i\). When considering CHSH Device Independent protocols, \(D_i \equiv X_i , Y_i\) are the inputs for Alice and Bob in each round, and \(C_i \equiv A_i B_i\) represent the outputs of a round of the CHSH test (see Section~\ref{sec: Bells theorem} for more details). Each round is characterized by a channel \(\mathcal{N}_i\), and the overall protocol is defined by the collection of channels \( \{ \mathcal{N}_i\}_{i} \) where $i$ runs from $1$ to $n$.

The protocol starts with the initial state \(\rho^{0}_{R_{A} E}\), where \(R_{A}\) represents the laboratory system, and \(E\) is the system in possession of Eve. In randomness expansion protocols, this is often an unknown, pre-programmed quantum resource. In each round, the channel \(\mathcal{N}_{i}\) acts on the state \(\rho^{i-1}_{R_{i-1} E}\) to output classical random variables \(C_i , D_i \) and a state \( \rho^{i}_{R_{i} , E}\) for the next round\footnote{In reality, devices may receive new states each round. However, it is equally valid to assume that the devices have pre-shared entangled quantum resources needed for the entire protocol.}. For CHSH tests, \(R_0\) denotes the initial pre-shared quantum state between Alice, Bob, and Eve. We further make the assumption that inputs \(D_{i}\) are independent random variables, and are uncorrelated with any other input random variables for any round other than round $i$. Moreover, \( D_{i} \) cannot be correlated with any outputs \( A_{1} , A_{2} , \cdots , A_{i-1} \) generated before round $i$.\par 

After the protocol is finished, the objective is to determine the randomness in the collection \(\mathbf{C} = ( C_1 , C_{2} , \cdots , C_{n}) \) (or \(\mathbf{AB} = (A_{0}, B_{0} , \cdots , A_{n} , B_{n})\) in the DI protocol). This calculation is done conditioned upon a specific event, \(\Omega\), taking place. In our context, \(\Omega\) is the event that protocol does not abort. Recall that, a DIRNE protocol aborts if the CHSH score \(\score \) does not exceed a given threshold score. In general, the event \(\Omega\) (abort condition) is determined by the input-output statistics \( \{C_{i} , D_{i}\}_{i} \). Note that computing the randomness conditioned only on the input and outputs is in the spirit of Device Independence.  The input-output statistics can be directly observed in the laboratory, and therefore are known quantities to the party conducting the protocol. The internal mechanism of the devices used to generate the experimentally observed statistics is dictated by the pre-shared initial state \(\rho_{R_{0} , E}\) and the channels \(\{\mathcal{N}_{i}\}_{i} \) of the protocol. As the state and the channels are treated as unknown in the protocol, we make no assumption on the inner workings of the devices other than the fact that $\rho_{R_0 E}$ is a valid state in Quantum Theory and $\{\M{N}_i\}_{i}$ are valid quantum channels. 

We say that the protocol generates randomness if output string \(\mathbf{C}\) contains secure randomness; that is, given access to the system \(E\) and the input string \(\mathbf{D} = (D_{1} , D_{2} , \cdots , D_{n})\), the output string \( \mathbf{C} = (C_{1} , C_{2} , \cdots , C_{n}) \) cannot be determined with certainty. Consequently, using the discussion in the previous section, our main aim is to determine \(H_{\min}(\mathbf{C}| \mathbf{D} E)_{\rho}\) in a protocol that does not abort.

This calculation needs to be done considering the worst-case scenario, where \(\rho_{R_{0} E}\) is unknown to the party executing the protocol but fully accessible to Eve. Moreover, the channels \(\mathcal{N}_{i}\) are also known to Eve. The main assumption here is that Eve cannot access the laboratory system after the protocol begins. Given this immense power granted to Eve, tackling this optimization problem is extremely challenging. Therefore instead of finding the exact min-entropy in a given protocol, it is acceptable to determine a lower bound on the min-entropy. This lower bound tells us the minimum randomness we can ``safely'' extract from the protocol. However, it is crucial that this lower bound is not too far off from the actual value or else we will waste randomness.

In the literature, incremental progress was made towards getting lower bounds on the min-entropy for a cryptographic protocol. It began with the asymptotic equipartition theorem~\cite{TCR}, then advanced with the Entropy Accumulation Theorem (EAT), and subsequent proofs catering to more generalized settings~\cite{MFR}. In this thesis, we use the EAT to compute randomness in a protocol. Informally, the EAT states that 
\begin{eqnarray}
    H_{\min}(\mathbf{C}| \mathbf{D} E)_{\rho} \geq n\inf H(C|DE) - \sqrt{n} v,
\end{eqnarray}
where \(H(C |DE)\) is the single round von Neumann entropy. Here \(\inf\) is taken over all single-round strategies, which would reproduce the observed statistics if executed in an i.i.d. manner.

EAT significantly simplifies the challenge of determining the extractable randomness for the protocol. Without EAT, one would need to account for every individual and potentially undefined channel \(\mathcal{N}_i\) and any arbitrary initial state \(\rho_{R_0 E}\), whereas EAT permits us to focus solely on the von Neumann entropy of a representative single round of the protocol. Moreover, as \(n \rightarrow \infty\), this lower bound becomes tight, implying that we can use the quantity \(\inf H(C|DE)\) as the asymptotic randomness rate (amount of extractable randomness per round) in a protocol. The error term \(\sqrt{n} v\) is intricate and defined via the min-tradeoff function. These technical details will be elaborated on in the next section of this thesis. 

However, the EAT does not fully resolve the problem of determining rates, leaving the crucial task of optimizing single-round von Neumann entropies. This optimization becomes extremely crucial when developing these protocols and is the central theme of many chapters in this thesis. 

\section{Entropy Accumulation Theorem}\label{app:EAT}
In this section we state the Entropy Accumulation Theorem (EAT) more formally with all relevant details. The theorem is phrased in terms of a set of channels $\{\cM_i\}_i$ called EAT channels, where $\cM_i: $ $\cS(R_{i-1}) $ $\to \cS(C_iD_iU_iR_i)$ \RutC{(here $i \in \{ 1 , 2 , \cdots, n \}$)}.
\begin{definition}[EAT Channels]
Let $\{R_i\}_{i=0}^n$  be arbitrary quantum systems and $\{C_i\}_{i=1}^n$, $\{D_i\}_{i=1}^n$, and $\{U_i\}_{i=1}^n$ be finite dimensional classical systems. Suppose that $U_i$ is a deterministic function of $C_i$, $D_i$ and that $\{\cM_i\}_{i=1}^n$, $\cM_i:\cS(R_{i-1})\to\cS(C_iD_iU_iR_i)$ are a set of quantum channels. These channels form a set of EAT channels if for all $\rho_{R_0E}\in\cS(R_0E)$ the state $\rho_{\B{ CDU}R_nE}=(\cM_n\circ\ldots\cM_1)(\rho_{R_0E})$ after applying the channels satisfies $I(C_1^{i-1}:D_i|D_1^{i-1}E)=0$, where $I$ is the mutual information, and $C_1^{i-1}$ is shorthand for $C_1C_2\ldots C_{i-1}$.
\end{definition} 
In the context of protocols, the register $U_i$ records the score \RutC{for the round  $i$}. Each EAT channel for the randomness expansion protocols is a set of maps of the form
\begin{eqnarray}\label{EAT channel}
\cM_i(\rho) = \sum_{c,d} \proj{c}\ot\proj{d}\ot\proj{u(c,d)}\ot\cM_i^{c,d}(\rho)\,,
\end{eqnarray}
where $u(c,d)$ records the score in the protocol, and each $\M{M}_i^{c,d}$ is a subnormalized quantum channel from $\cS(R_{i-1})$ to $\cS(R_i)$. The joint distribution of the classical variables $C_i$ and $D_i$ is
\begin{equation}
    p_{C_iD_i}(c,d):=\tr(\M{M}_i^{c,d}(\rho))\,.
\end{equation}
\begin{definition}[Frequency distribution function]
Let $\B{ U}=U_1U_2\ldots U_n$ be a string of variables. The associated frequency distribution is 
\begin{equation}\label{eqn: eat freq}
    \Freq_{\B{ U}}(u) := \frac{|\{i\in\{1,\ldots,n\}:U_i=u\}|}{n}\,.
\end{equation}
\end{definition}
\RutC{In the above equation (eqn. \ref{eqn: eat freq}), the notation $|.|$ is used to represent the cardinality of a set. We use this notation to denote the cardinality of sets throughout this thesis. }
\begin{definition}
Given a set of channels $\mathfrak{G}$ whose outputs have a register $U$, the set of achievable score distributions is
\begin{equation}
    \cQ_{\mathfrak{G}}:=\{p_U:\cM(\rho)_U=\sum_up_U(u)\proj{u}\text{ for some }\cM\in\mathfrak{G}\}.
\end{equation}
\RutC{For the spot checking protocol (introduced in Chapter \ref{chap : actual protocols for DI}), there is an additional quantity of interest  }
\begin{equation}
    \cQ_{\mathfrak{G}}^\gamma:=\{p_U:p_U(\bot)=(1-\gamma)\text{ and }p_U(u)=\gamma\tilde{p}_U(u)\text{ with }\tilde{p}_U\in\cQ_{\mathfrak{G}}\}.
\end{equation}
\end{definition} 
\RutC{The above definition is given for completeness here. The variable $\perp$ is defined in Protocol \ref{prot:spotcheck} (see chapter \ref{chap : actual protocols for DI} for further details).  }
\begin{definition}[Rate function]
Let $\mathfrak{G}$ be a set of EAT channels. A \emph{rate function} $\rate:\cQ_\mathfrak{G}\to\mathbb{R}$ is any function that satisfies
\begin{eqnarray}
\rate(q) \leq \inf_{(\M{M},\rho_{RE})\in\Gamma_{\mathfrak{G}}(q)} H(C|DE)_{(\M{M}\ot\cI_E)(\rho_{RE})}\,,
\end{eqnarray}
where
\begin{equation}
\Gamma_{\mathfrak{G}}(q):=\{(\cM,\rho_{RE}):(\cM\ot\cI_E)(\rho_{RE})_U=\sum_uq(u)\proj{u}\text{ for some }\cM\in\mathfrak{G}\}
\end{equation}
is the set of states and channels that can achieve distribution $q$.
\end{definition}
\begin{definition}[Min-tradeoff function]
A function $f:\cQ_\mathfrak{G}\to\mathbb{R}$ is a \emph{min-tradeoff function} if $f$ is an affine rate function. Since min-tradeoff functions are affine, we can naturally extend their domain to all probability distributions on $U$, denoted $\cP$.
\end{definition}
The entropy accumulation theorem then can be stated as follows (this is Theorem~2 of~\cite{LLR&}, which is a generalization of the results of~\cite{DF}).

\begin{theorem}\label{thm:EAT}
 Let $f$ be a min-tradeoff function for a set of EAT channels $\mathfrak{G}=\{\cM_i\}_i$ and $\rho_{\B{ CDU}E}$ be the output after applying these channels to initial state $\rho_{RE}$. In addition let $\epsilon_h\in(0,1)$, $\alpha\in(1,2)$ and $r\in\mathbb{R}$ and $\Omega$ be an event on $\B{ U}$ that implies $f(\Freq_\B{ U})\geq r$. We have
 \begin{align}
     H_{\min}^{\epsilon_h}(\B{ C}|\B{ D}E)_{\rho_{\B{ CD}E|\Omega}}>&nr-\frac{\alpha}{\alpha-1}\log\left(\frac{1}{p_\Omega(1-\sqrt{1-\epsilon_h^2})}\right)+\nonumber\\
     &n\inf_{p\in\cQ_{\mathfrak{G}}}\left(\Delta(f,p)-(\alpha-1)V(f,p)-(\alpha-1)^2K_{\alpha}(f)\right)\,,
 \end{align}
where $\Delta(f,p)=\rate(p)-f(p)$, and
\begin{align*}
    V(f,p)&=\frac{\ln2}{2}\left(\log(1+2d_C)+\sqrt{2+\Var_p(f)}\right)^2\\
    K_\alpha(f)&=\frac{1}{6(2-\alpha)^3\ln 2}2^{(\alpha-1)(\log(d_C)+\Max(f)-\Min_{\cQ_{\mathfrak{G}}}(f))}\ln^3\left(2^{\log(d_C)+\Max(f)-\Min_{\cQ_{\mathfrak{G}}}(f)}+\e^2\right)\,,
\end{align*}
and we have also used
\begin{align*}
\Max(f)&=\max_{p\in\cP} f(p)\\
\Min_{\cQ_{\mathfrak{G}}}(f)&=\inf_{p\in\cQ_{\mathfrak{G}}}f(p)\\
\Var_p(f)&=\sum_up(u)\left(f(\delta_u)-\mathbb{E}(f(\delta_u))\right)^2\,,
\end{align*}
and $\delta_u$ is the deterministic distribution with outcome $u$.
\end{theorem}
\RutC{To use this theorem we have to assign the variables $C_i$ and $D_i$ to the parameters in the protocol. For example, as stated in the previous section, in the context of CHSH-based Device Independent Protocols, the variables $C_i$ can be output variables $A_i B_i$ and the variable $D_i$ can be taken to be the input variables $X_i Y_i$.}

\cleardoublepagewithnumber
\part{Device Independent Protocols}

\chapter{Introduction to Device Independent Protocols}\label{chap: DI protocols}
\section{Introduction}
Bell's theorem states that quantum mechanics is not compatible with local hidden variable theories. A Bell test is a non-local experiment that consists of two (or more) space-like separated (or non-communicating) observers who share an entangled quantum state. At the beginning of the experiment, each observer generates a discrete random input (with a finite number of possible inputs). Based on this input, these observers perform a measurement on their shared state to generate an output. The observers repeat this process for a large number of rounds to get some input-output statistics. A Bell inequality is a relation on the joint input-output statistics of both these parties that is satisfied if these statistics can be obtained by a local-hidden variable theory. Therefore, if the obtained input-output statistics indicate the violation of a Bell inequality, then those statistics cannot have arisen from a local deterministic behaviour. 

In the context of Device Independent randomness expansion (DIRNE), the main idea is that the ability to violate a Bell inequality implies that the devices doing so must be generating randomness~\cite{ColbeckThesis,CK2} (see section~\ref{sec: Bells theorem} for further details). Thus, in a sense, the protocol self-tests\footnote{\RutC{Roughly speaking, a protocol is called a self testing protocol if we are able to infer the underlying physical process solely from the observable outcomes of the protocol.}}~\cite{MayersYao} the devices during its operation, leading to enhanced security. Although challenging to accomplish, recently the first experimental demonstrations of DIRNE were performed~\cite{LZL&,Shalm_rand,LLR&}, following earlier experiments considering randomness generation~\cite{PAMBMMOHLMM,BKGZM&,LZL&gen}.

On the theoretical side, the main difficulty is then to calculate how much extractable randomness there is as a function of the Bell violation. This task is made more challenging by the lack of structure of the problem: no assumption is made on how the devices operate and so one has to account for arbitrary pre-shared entanglement, arbitrary measurements, and strategies that may be adaptive between the rounds. An increasingly sophisticated series of proofs~\cite{VV,MS1,MS2} leads to two techniques for dealing with this exist in the literature: the quantum probability estimation framework~\cite{ZFK} and the entropy accumulation theorem (EAT)~\cite{DFR,ARV} (\RutC{see} Section~\ref{app:EAT} for details). We use the latter in this work. Informally speaking, the EAT states that the amount of extractable randomness in the full string of outputs is to leading order $n$ times the von Neumann entropy of a single-round strategy that would give the observed score if used in an i.i.d.\ way. In other words, the EAT implies that if we can solve the problem for an i.i.d.\ adversary, then we can get a bound for the general case.

In a DIRNE protocol, randomly chosen inputs are made to two separated devices so that each device cannot determine the input of the other device. We use \(X\) and \(Y\) to label the inputs, and \(A\) and \(B\) to label the outputs, taking into account an adversary with side information \(E\). This side information could be quantum; the general strategy allows for the adversary holding the \(E\) part of a state \(\rho_{A'B'E}\), with the \(A'\) and \(B'\) systems retained by the devices. Each input \(X\) to the first device corresponds to a measurement on \(A'\) yielding outcome \(A\); similarly, each input \(Y\) to the other device corresponds to a measurement on \(B'\) resulting in outcome \(B\). Two quantities of interest arise, both dependent on the post-measurement state: the first is the score in a non-local game -- a function of the conditional distribution \(p_{AB|XY}\)—and the second, the von Neumann entropy of either one or both of the outputs. Specifically, we aim to express the minimum von Neumann entropy in terms of the score. In this work, we study six von Neumann entropies: \(H(AB|X=0,Y=0,E)\), \(H(AB|XYE)\), \(H(AB|E)\), \(H(A|X=0,Y=0,E)\), \(H(A|XYE)\), and \(H(A|E)\) \footnote{In the one-sided cases, conditioning on the variable \(Y\) is unnecessary, but it is retained for notational symmetry.}.

In essence, \RutC{using the EAT}, the problem \RutC{of computing extractible randomness} is reduced to finding the smallest von Neumann entropy of the outputs conditioned on the adversary's side information and the inputs, with the property that the state and measurements used would give a particular Bell value. More precisely, a strategy (for the adversary) corresponds to picking a quantum state $\rho_{A'B'E}$ and for each possible input $X=x$ a POVM $\{M_{a|x}\}_a$ on $A'$, and for each $Y=y$ a POVM $\{N_{b|y}\}_b$ on $B'$. From this, the implementation of the protocol corresponds to repeated actions of the channel:
\begin{eqnarray*}
\cN&:&\cS(\cH_{A'}\ot\cH_{B'})\to\cS(\cH_A\ot\cH_B\ot\cH_X\ot\cH_Y): \\ 
 \sigma&\mapsto& \sum_{abxy}p_{XY}(x,y)\proj{a}\ot\proj{b}\ot\proj{x}\ot\proj{y}\tr\left((M_{a|x}\ot N_{b|y})\sigma\right) ,   
\end{eqnarray*}
where $p_{XY}$ is the input distribution used in the protocol. Applying this channel to $A'B'$ and acting as identity on $E$ generates the final state $\tau_{ABXYE}=(\cN\ot\cI_E)(\rho_{A'B'E})$. The Bell value is a linear function on this state, and we seek to minimize an entropy (e.g., $H(AB|XYE)$) for this state, over all strategies that have a given Bell value (see later for a discussion of different entropies).

While the minimization of the single round entropic quantities above proves simpler than the direct optimization of the min-entropy, this minimization remains challenging. \RutC{The main challenge is that} the von Neumann entropies are non-linear. Additionally, there is no preliminary upper bound on the dimensions of the systems \(A'\), \(B'\), and \(E\). For instance, some Bell inequalities suggest that the maximum quantum violation cannot be realized if \(A'\) and \(B'\) are finite dimensional~\cite{Slofstra}. Moreover, evidence suggests this remains true even when \(X\) and \(Y\) are binary, and \(A\) and \(B\) have only three possible outcomes~\cite{PV}. However, when \(A\), \(B\), \(X\), and \(Y\) are all binary, Jordan's lemma~\cite{Jordan} asserts that there is no loss in generality when considering a convex combination of strategies wherein \(A'\) and \(B'\) are two-dimensional. This observation paved the way for establishing a tight lower bound on the one-sided entropy based on the CHSH score~\cite{PABGMS} and is pivotal for this study.

In Chapter~\ref{chap: DI upper bounds}, we discuss computing the von Neumann entropy bounds, later presenting numerically generated upper bound curves for each of the six quantities for the protocols based on violation of the CHSH inequality~\ref{sec:num}. By employing Jordan's lemma and other technical maneuvers~\cite{PABGMS}, we can reduce the problem to seven real parameters (three for state specification and four for measurement selection), making it suitable for numerical optimization. Using heuristic numerical techniques we determined \RutC{bounds} which serve as useful references for the optimal values von Neumann entropic quantities for the six different cases as a function of the CHSH score, \(\score\) (notably, an analytic bound exists for the first case~\cite{PABGMS}). For \(H(A|XYE)\) and \(H(AB|XYE)\), we also propose conjectured analytic forms for the curves. 

In Chapter~\ref{chap: DI lower bounds}, rigorous lower bounds on the entropic quantities \(H(A|XYE)\) and \(H(AB|X=0,Y=0,E)\) as a function of the CHSH score, \(\score\), are also derived. We show that lower bounds for these quantities can be calculated by solving an optimization problem over three real parameters. This realization lets us partition the domain of these new optimization problems into cuboids and compute the objective function on their edges. The objective function's value within each cuboid is underestimated using Taylor's theorem, allowing us to obtain a lower bound on the entropies by taking the minimum over all the function's under-estimations on each cube.

Using \RutC{a} similar method of partitioning the domain into (hyper) cuboids, combined with techniques for lower bounding polynomial optimization problems, we have also found reliable lower bounds for the entropy \(H(AB|XYE)\).

The derived entropy lower bounds can be made arbitrarily tight by refining the partition, albeit with increased computation time (see Chapter~\ref{chap: DI lower bounds}). If one needs to compute the min-tradeoff function for application to the EAT, a more refined partition can be used since entropy only needs computation for a limited CHSH score range. Thus, our methodology \RutC{has} relevance in practical applications \RutC{of DIRNE protocols}. For \RutC{the} entropy \(H(AB|X=0,Y=0,E)\), our findings are compared with the recent breakthrough technique~\cite{BFF2022} for optimizing von Neumann entropies. \RutC{We find that our }the lower bounds \RutC{are very close to the numerical upper} bounds for both \(H(A|XYE)\) and \(H(AB|X=0,Y=0,E)\). \RutC{For the case \(H(AB|X=0,Y=0,E)\), our lower bounds surpass the one ones obtained using the technique in \cite{BFF2022}.} Consequently, we conjecture that the numerical bounds established for all six entropic quantities are tight. \RutC{Hence, in our work,} the conjectured bounds are used to determine randomness rates across various protocols for randomness expansion.

Given the challenges associated with optimizing von Neumann entropies in randomness expansion protocols, the one-sided quantity \(H(A|X=0,Y=0,E)\) has often been used due to its existing analytic bound~\cite{PABGMS}. However, as this omits Bob's output, it can be deemed wasteful in terms of generating randomness. With our new bounds available, we can now employ the corresponding two-sided quantities in randomness expansion protocols, such as the spot-checking protocol (refer to Chapter~\ref{chap : actual protocols for DI}). This bound also facilitates the calculation of randomness expansion in protocols using (heavily) biased input randomness, allowing for the closure of the locality loophole present in the spot-checking protocol. Furthermore, the \RutC{lower bounds for the }entropy \(H(AB|XYE)\) \RutC{allow} for the recycling of input randomness, ensuring a more efficient utilization of all resources.

Our newly derived entropies are employed in a protocol for Device Independent randomness generation, allowing two key improvements over prior works:

\begin{itemize}
    \item The protocol exploits two-sided randomness.
    \item It recycles input randomness.
\end{itemize}

These modifications not only \RutC{improve the randomness rate} but also \RutC{close the locality loophole associated with the spot-checking protocol}. In Chapter~\ref{chap : actual protocols for DI}, we outline rate curves with finite round numbers (employing the EAT with the updated numerical bounds on the single-round von Neumann entropy). As anticipated, the gains from the i.i.d. case transition seamlessly to the finite regime. For example, when taking the experimental conditions from~\cite{LLR&}, using two-sided randomness coupled with randomness recycling culminates in a rate increase of multiple orders of magnitude.

\section{Bell's theorem and randomness}\label{sec: Bells theorem} 
Before delving into the intricacies of DI protocols for randomness expansion, we will first discuss the Bell's theorem and its role in the randomness expansion protocols. 

\begin{figure}[h!]
  \centering
  \includegraphics[width=0.8\textwidth]{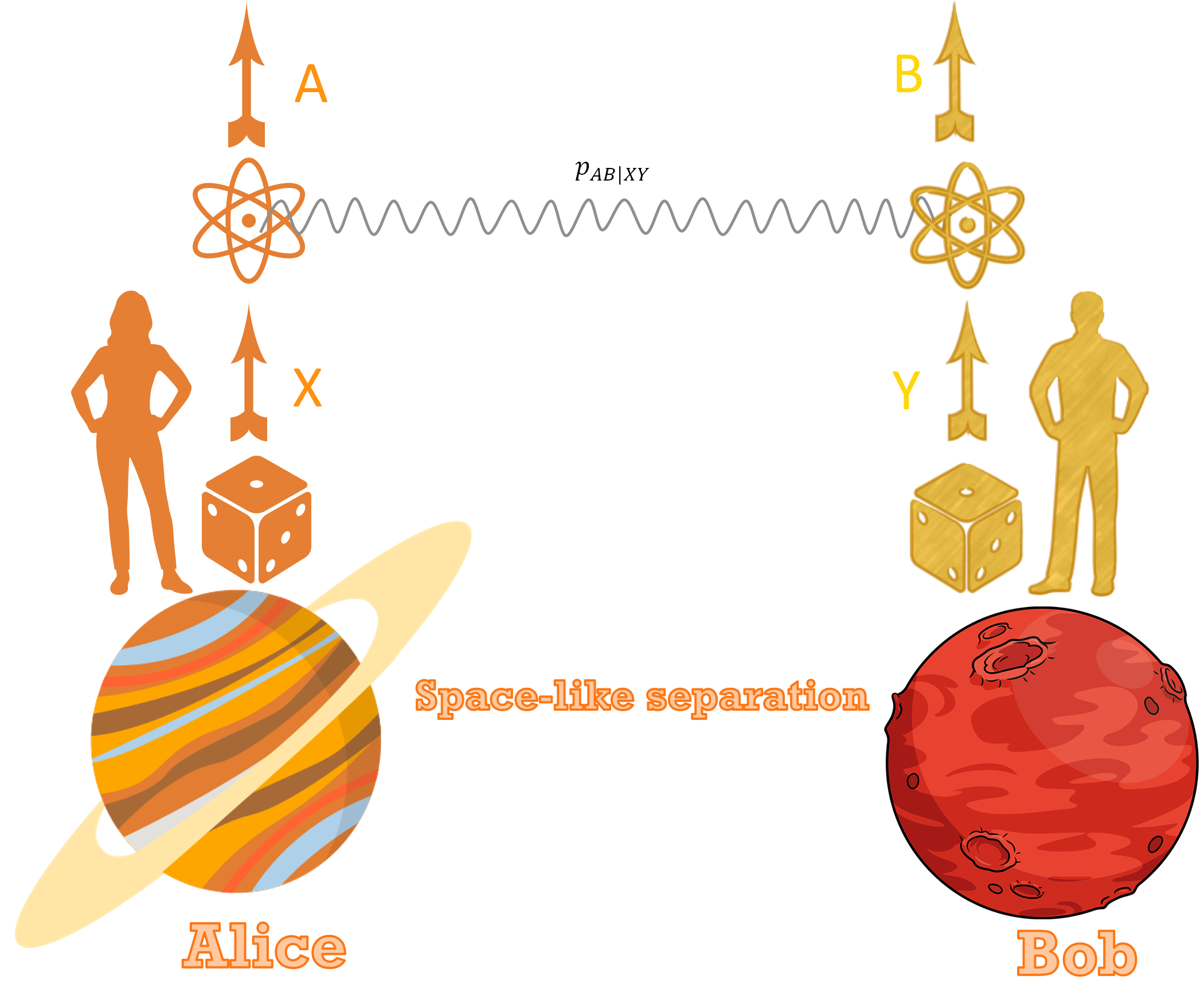}
  \caption{A diagram depicting a typical Bell scenario involving two parties.}
  \label{fig: Bells theorem}
\end{figure}

As depicted in Figure~\ref{fig: Bells theorem}, the simplest scenario of the Bell setup involves two spatially separated \textit{or non communicating} \RutC{parties}, say Alice and Bob, sharing a common resource. This resource could be, in general, a quantum state.

A Bell test is conducted over several rounds. In each round, the procedure remains the same: an input \(X\) is chosen randomly and dispatched to Alice's device, and similarly, an input \(Y\) is selected and sent to Bob. When Alice and Bob conduct measurements based on their randomly chosen inputs \(X\) and \(Y\), their outcomes are denoted as \(A\) and \(B\), respectively. After numerous repetitions, we can determine the conditional probability distribution \(p_{AB|XY}\).

John Bell posed the question: can this probability distribution be derived ``classically''\footnote{Here the word ``classical'' is used to refer to any theory which obeys local determinism.} -- in other words, can the outputs \( A , B\)  be solely predetermined by the input random variables \(X, Y\), and a classical random variable \(\Lambda\) that could, in theory, be known to any entity within the past light cone of both Alice and Bob?

For this scenario, we consider the simplest situation where the inputs \(X, Y\) and the outputs \(A, B\) are binary. The CHSH score, represented as \(\score\), can then be defined as:
\begin{equation}
    \RutC{\score := \frac{1}{4}\left(\sum_ap_{AB|00}(a,a)+\sum_ap_{AB|01}(a,a)+\sum_ap_{AB|10}(a,a)+\sum_ap_{AB|11}(a,a\oplus1)\right),}
\end{equation} 
which can be determined solely using collected input and output statistics $p_{AB|XY}$. Bell's theorem states that if $\score > \frac{3}{4}$, the outputs must have been generated by a non-classical resource. This means that the outcomes of the given measurement cannot be achieved in a local deterministic fashion. Thus they cannot be determined with certainty by anyone including any adversary, who has the full knowledge of the inner workings of the device. This principle lays the groundwork for the generation of random numbers within the Bell type setup.

\begin{figure}[h!]
  \centering
  \includegraphics[width=0.8\textwidth]{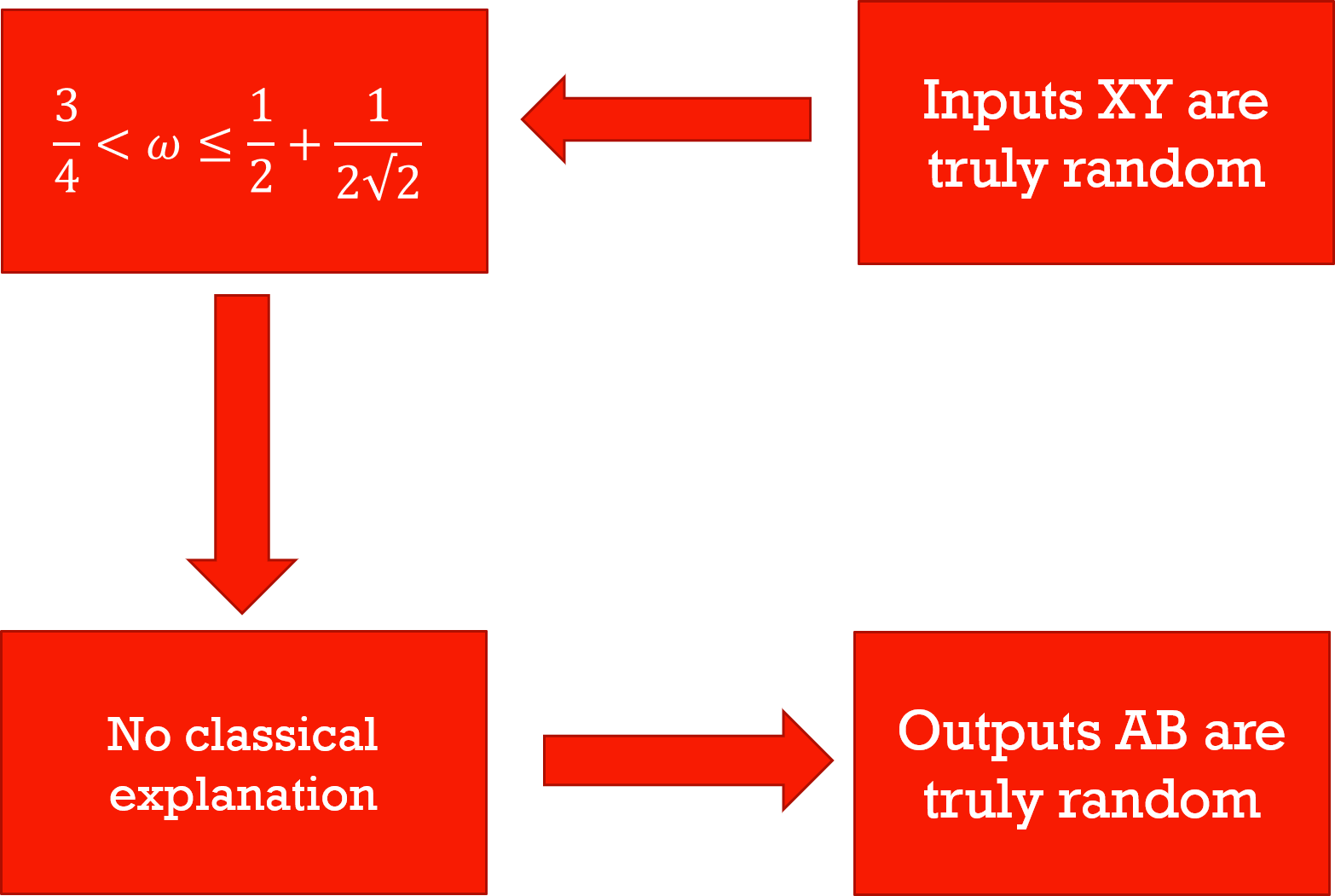}
  \caption{A diagram depicting an informal summary of Bell's theorem. }
  \label{fig: Bells theorem explanatio }
\end{figure}

Bell's theorem is often also described using the scenario where two non-communicating parties -- Alice and Bob \RutC{are} playing a \RutC{non-local} game. In each round, they are posed a question represented by the value of the input of random variables \(X \in \{ 0, 1 \}\) and \(Y \in \{ 0, 1 \}\). They are then expected to output \(A, B \in \{ 0, 1 \}\). Alice and Bob are considered to have won a round of the game if \(XY = A \oplus B\), and they lose if this condition is not met. If Alice and Bob play this game (referred to as the CHSH game) using local deterministic strategies, their maximum winning probability is \(\leq \frac{3}{4}\). If they wish to win with a probability higher than this, they must employ a non-classical strategy.

With the understanding that the outputs $A, B$ are random given the random inputs $X,Y$ and a CHSH score $\omega \geq \frac{3}{4}$, the next question \RutC{is} to consider how much randomness can be extracted from the outputs. As discussed in Chapter~\ref{chap: maths}, the randomness per CHSH test is roughly characterized by the von Neumann entropy $ \inf H(AB|XYE)$. It is important to note that this value represents the randomness per round in the asymptotic limit (i.e., the randomness when the CHSH test is repeated indefinitely). The \RutC{infimum} is taken over all strategies that achieve a CHSH score $\score$, assuming the strategy is performed in an i.i.d. manner (see Chapter~\ref{chap: DI upper bounds} and~\ref{chap: DI lower bounds} for details). Note that other entropic quantities such \RutC{as $\inf H(AB|X=0,Y=0,E)$} may also be of interest depending upon the protocol being performed. In the next section, we discuss the role of different entropic quantities that are useful in Device Independent protocols based on the violation of the (generalized) CHSH inequality.

In the literature, the inequality $\score > 3/4$ is called a CHSH inequality. In a more general setting, if $p_{AB\cdots | XY \cdots}$ is a valid $n$ partite probability distribution, \RutC{a} Bell inequality can be defined \RutC{as} $\M{B}(p_{AB\cdots | XY \cdots}) > l$, where $\M{B}$ is a linear functional on the probability distribution $p_{AB\cdots | XY \cdots}$ and $l$ is the maximum value of $\M{B}(p_{AB\cdots | XY \cdots})$ that can be achieved by a probability distribution corresponding to any local deterministic strategy. 

It is worth highlighting that the CHSH inequality is not the sole Bell inequality, even when considering the two-input, two-output, two-party scenario. Distinct classes of Bell inequalities exist for this simplest scenario, such as the tilted Bell inequalities~\cite{AMP-rand, WAP}. Another category pertains to optimal inequalities tailored for randomness expansion~\cite{WBC}. Yet, the CHSH inequality violation remains the most extensively researched in context of DIRNE protocols. The collection of CHSH inequalities, defined by permutations of inputs $X,Y$ and outputs $A,B$ in the CHSH score above, stands out as, in the vector space of all probability distributions, the CHSH inequalities  serve as facets of the polytope formed by all the local deterministic distributions~\cite{brunner2014bell}. In this dissertation, while determining randomness rates for our protocols, we employ generalized Bell inequalities expressed as:
\RutC{\begin{eqnarray}\label{eqn: genCHSHscore}
    \score := \sum_a \left(\gamma_{00}p_{AB|00}(a,a)+ \gamma_{01} p_{AB|01}(a,a)+\gamma_{10}p_{AB|10}(a,a)+ \gamma_{11}p_{AB|11}(a,a\oplus1) \right),
\end{eqnarray}}

where $\gamma_{ij} \in \mathbb{R}$ are some coefficients. 
However when it comes to analysis of the protocols, we restrict ourselves to the protocols based only on the CHSH tests. 

\section{The progress of DIRNE protocols: a brief review} 
\RC{
Before discussing the DIRNE protocols, let us briefly highlight some foundational literature on their development.

The first DIRNE protocol was introduced by Colbeck \cite{ColbeckThesis, CK2} based on the GHZ test. This work introduced the idea of private randomness generation certified by non-local games. Building on this work, Pironio et al. \cite{PAMBMMOHLMM} presented a construction of DIRNE protocol along with its experimental demonstration (this was not a loophole-free experimental demonstration). Further, it was then proven that unbounded randomness can be produced from a finite initial seed of random numbers using many copies of entangled states \cite{MS1, CY}. Other works have focused on amplification of randomness from an initial source of weak randomness \cite{CR_free}.

Meanwhile, improving the security proof of the DIRNE protocols started to attract attention. Pironio and Massar demonstrated that the DIRNE protocol is secure against classical adversaries~\cite{PM}. This was followed by Vazirani and Vidick, who proved the security of the protocol against an entangled adversary~\cite{VV}; however, this was without any noise tolerance. Miller and Shi then proved the security in the presence of errors~\cite{MS2}.

However, challenges remained for computing reasonable tight bounds on the min-entropy (which quantifies the randomness in a DIRNE protocol -- see Chapter \ref{chap: maths}) for the outputs of the protocol. The difficulty stems from the fact that one needs to account for different collective attacks that a potential adversary may attempt to tamper with the protocol. Assuming the protocol is carried out in an i.i.d. fashion, the min-entropy can be bound using the single round min-entropy \cite{BRC} at a finite level of the NPA hierarchy \cite{NPA, NPA2}. However, these bounds are generally not very tight. This issue was addressed by the Asymptotic Equipartition Theorem (AEP) \cite{TCR}, which bounds the min-entropy of the protocol in terms of the single round von Neumann entropy. This bound holds for i.i.d. protocols and also in the limit that the number of rounds is large. As von Neumann entropy is larger than the min-entropy, the AEP improves the randomness rates of DIRNE protocols. Therefore, the main challenge was then reduced to getting bounds on the min-entropy of the protocol, keeping two aspects in mind: the i.i.d. assumption needs to be relaxed and the second aspect is to get bounds for finitely many rounds.

Two approaches were developed to address this -- the quantum probability estimation framework \cite{ZFK} and the entropy accumulation theorem (EAT) \cite{DFR, ARV}. The EAT bounds the min-entropy in terms of the von Neumann entropy, just like AEP does -- however it also provides bounds for finite rounds and for the non-i.i.d. scenario by introducing a penalty term that vanishes as the number of rounds increases. The penalty term in the original EAT was not tight and this EAT bound was further refined in \cite{DF} and subsequently in \cite{LLR&}. A generalized version of the EAT was published very recently \cite{MFR}, which extends EAT to cases where the side information of the adversary can be updated after every EAT round.

Although the single-round von Neumann entropy is significantly easier to compute compared to the min-entropy of the entire protocol, the bounds of the single-round von Neumann entropy are hard to compute even for a single round. The challenge arises due to the non-linearity of von Neumann entropy and the lack of assumptions about the states and measurements used. For a vast majority of research up until recently, the randomness rate was determined using known tight bounds on the randomness of just one of the devices. In their work, Pironio et al. \cite{PABGMS} calculated bounds on the randomness of a single device in the context of computing key rates for Device Independent Quantum Key Distribution (DIQKD). Since this represents a bound on the randomness of one device, we refer to such a bound as a one-sided rate. As the  output of one device is a binary, at most one bit of randomness could be certified using this method. One-sided rates, as opposed to two-sided rates, are useful for DIQKD because the secret key is formed from the outputs of only one party. The other party then performs error correction to agree with the secret key held by the first party. In the literature, this DIQKD bound was re-used to determine a lower bound on the output randomness of the protocol. The advantage was that bounds existed for all values of the CHSH score, allowing for a non-zero randomness rate as long as there was a violation of the Bell inequality. This result was employed in one of the first loophole-free experimental demonstrations of DIRNE protocols \cite{LLR&}, which was conducted recently.

Later, it was shown that in principle, up to 2 bits of randomness could be achieved in a (2-input, 2-output, 2-party) DIRNE protocol using tilted Bell inequalities \cite{AMP-rand}. This result showed the existence of such strategies, but did not compute the randomness rates for all values of Bell violation. One-sided rates for such tilted Bell inequalities (introduced in the context of DIQKD protocols) for all values of Bell violation were presented by Woodhead et al. \cite{WAP}. A significant limitation of using such tilted Bell inequalities is that 2 bits can only be certified while being arbitrarily close to the local set (i.e., the CHSH score will be near 2), making high randomness rates less robust against noise.

Therefore, determining bounds on randomness rates for DIRNE protocols (i.e., two-sided bounds) instead of re-using existing one-sided bounds for DIQKD is one the main motivation for this thesis. While working on this thesis, several other important works were also done in this direction, each attempting to address the challenge of finding lower bounds on one-sided and two-sided randomness rates. This led to important advances in computing these bounds \cite{SBVTRS, TSBSRSL, TSGPL}. More recently, a breakthrough paper \cite{BFF, BFF2022} was published, presenting a method to compute reliable lower bounds on the von Neumann entropy for the most general non-local games. This technique relies on semi-definite programming, leveraging the NPA hierarchy and methods from non-commutative polynomial optimization theory to derive reliable bounds on von Neumann entropy. Building on this, the recent work by Wooltorton et al. \cite{WBC} demonstrated that up to two bits of randomness can be extracted even when the observed correlations are far from the local bound (up to a CHSH score of about 2.6). Hence, up to 2 bits of randomness can be extracted from a DIRNE protocol in a more robust manner.

The intriguing question of the quantity of randomness that can be extracted from a single source through multiple measurements was tackled by Curchod et al. Their study~\cite{CJAHWA} revealed that by employing the same initial state and conducting repeated measurements, an unbounded amount of randomness can be extracted. Their method utilized tilted Bell inequalities to certify this randomness. Even more recently, Brown and Colbeck~\cite{BC} demonstrated that the CHSH inequality can be violated an unbounded number of times, suggesting the potential for extracting unbounded randomness through numerous violations of the CHSH inequality.

Ongoing advancements in theoretical analysis have yielded considerable enhancements in DIRNE protocols. Coupled with the rapid progress observed in experiments~\cite{LLR&, LZL&, LZL&gen}, DIRNE protocols are getting increasingly more practical.}
\section{The significance of various entropic quantities}\label{sec:signif}
In this section, we discuss the significance of the six entropic quantities given above in the context of DIRNE, noting that the one-sided quantities are also useful for DIQKD (Device Independent Quantum Key Distribution). To do so, we first describe the general structure of the raw randomness generation part of a spot-checking and non-spot-checking DIRNE protocol. A more complete description of the protocols is in Chapter~\ref{chap : actual protocols for DI}. 

In a protocol without spot-checking \RutC{(such as the protocol which recycles input randomness)}, two untrusted devices are used. In every round, their inputs $X_i$ and $Y_i$ are generated according to some distribution $p_{XY}$. Often two independent random number generators are used for this, so that $p_{XY}=p_Xp_Y$. The generated numbers are used as inputs to the devices, which return two outputs $A_i$ and $B_i$ respectively. This is repeated for $n$ rounds generating the raw randomness $\B{A}$, $\B{B}$, where the bold font denotes the concatenation of all the outputs.

In a spot-checking protocol, there is an added step. In this step, each round is declared either a test round ($T_i=1$) or a generation round ($T_i=0$). Test rounds occur with a typically small probability $\gamma$. \RutC{In} test rounds, $X_i$ and $Y_i$ are generated according to some distributions $p_{XY}$. \RutC{In} generation rounds, $X_i$ and $Y_i$ are set according to some other distributions -- in this work we use the deterministic distribution $X_i=Y_i=0$. These are used as inputs to the devices, which return two outputs $A_i$ and $B_i$ respectively. The rationale behind using a spot-checking protocol is that randomness is required to perform a Bell test and it is desirable to be able to run the protocol with a smaller requirement on the amount of input randomness required. Choosing whether to test or not requires roughly $H_\bin(\gamma)$ bits of randomness per round\footnote{Here \RutC{$H_\bin: [0 ,1] \mapsto [0, 1]$ }denotes the binary entropy \RutC{defined as $H_{\bin}(x) = - x \log_2(x) - (1 -x) \log_2(1 -x)$. Here $\log_2$ is the logarithm with base $2$. Further, $0 \log_2 0$ is taken to be $0$.}}, so choosing $\gamma$ small enough leads to an overall saving. Furthermore, protocols often discard the input randomness, in which case for many Bell tests spot-checking is necessary in order to achieve expansion. In the CHSH game, for instance, if $p_{XY}$ is chosen uniformly, each test round requires $2$ bits of randomness, but the amount of two-sided randomness output by the quantum strategy with the highest possible winning probability is only $1+H_\bin(\frac{1}{2}(1+\frac{1}{\sqrt{2}}))\approx1.60$ bits. However, as we discuss later, the input randomness need not be discarded.

In the case of small $\gamma$, almost every round is a generation round hence an eavesdropper wishes to guess the outputs for the inputs $X=0$ and $Y=0$. The entropy $H(AB|X=0,Y=0,E)$ is thus the relevant quantity for spot-checking DIRNE protocols. The one-sided quantity $H(A|X=0,Y=0,E)$ has often been used instead because of the existing analytic bound for this~\cite{PABGMS,WAP}, but, because this ignores one of the outputs, it is wasteful as an estimate of the generated randomness. For DIQKD protocols, on the other hand, the one-sided entropy is the relevant quantity. \RutC{This is because in order to make a key, the random strings held by Alice and Bob, $\B{A}$ and $\B{B}$ respectively should match. Thus, only one of the strings $\B{A}$ can be used as a key, and the other string $\B{B}$ should be corrected to match the with the string $\B{A}$}. We also remark that these quantities can be useful bounds for protocols without spot-checking if the distribution $p_{XY}$ is heavily biased towards $X=Y=0$.

The quantities $H(AB|XYE)$ and $H(A|XYE)$ are useful for protocols without spot checking. One might imagine, for example, using a source of public randomness, such as a randomness beacon to choose the inputs to the protocol, in which case $X$ and $Y$ become known to the adversary (but are not known before the devices are prepared). In this case, rather than being interested in randomness expansion, the task is to turn public randomness into private randomness in a Device Independent way. One can also use $H(AB|XYE)$ and $H(A|XYE)$ in protocols when the input randomness is recycled. In this case we are really interested in $H(ABXY|E)$, but, because $X$ and $Y$ are chosen independently of $E$, this can be expanded as $H(XY)+H(AB|XYE)$. Hence $H(AB|XYE)$ is the relevant quantity in this case as well. The one-sided quantity $H(A|XYE)$ could also be used for DIQKD without spot-checking.

In addition we consider the quantities $H(AB|E)$ and $H(A|E)$. The second of these could be useful for QKD protocols in which the key generation rounds do not have a fixed input and where Alice and Bob do not publicly reveal their measurement choices during the protocol. For instance, the sharing of these choices could be encrypted using an initial key, analogously to a suggested defence against memory attacks~\cite{bckone}\footnote{Note that such protocols would only be useful if more key is generated than is required, so the protocol we are thinking of here is really quantum key expansion. Furthermore, the results presented in Figure~\ref{fig:rates} show that the use of $H(A|E)$ only gives a minor advantage over $H(A|XYE)$.}. $H(AB|E)$ would be a useful quantity for randomness generation in a protocol without spot-checking and in which $X$ and $Y$ are kept private after running the protocol and not used in the overall output. When such protocols are based on the CHSH inequality, they cannot allow expansion. These quantities can also be thought of as quantifying the fundamental amount of randomness obtainable from a given Bell violation. Although we have computed the graphs for $H(AB|E)$ and $H(A|E)$, existing versions of the EAT cannot be directly applied to them --- see Appendix~\ref{appendix: EAT}.

\cleardoublepagewithnumber

\chapter{Upper bounds on the entropic quantities}\label{chap: DI upper bounds} 

In the previous chapter, we noted the necessity for computing lower bounds on entropic quantities $H(.|. E)$ for any strategy that achieves a CHSH score $\score$ when carried out in an i.i.d. manner. This computation is crucial for determining the rates of CHSH-based protocols. In this chapter, we proceed with a rigorous definition of such a strategy, formulating it as an optimization problem for the entropic quantities. Techniques such as Jordan's lemma will be employed to simplify these optimization problems, enabling us to calculate the asymptotic rates of the protocols (i.e., the randomness per round). Interestingly, we observe that our techniques have a broader applicability and can be extended to accommodate a wider class of Bell inequalities, of which the CHSH inequality is a special case. We refer to these inequalities as CHSH-type inequalities and will define them in the subsequent section. Nonetheless, for our analysis, we primarily focus on the CHSH inequality due to its widespread use in literature concerning randomness expansion protocols. 

\RC{Parts of Chapter \ref{chap: DI upper bounds}, including Sections \ref{app:Proof1}, \ref{app: convex combinations of qubit strategies}, \ref{sec: reduction to pure states in DI}, \ref{app:simp}, and \ref{sec: HAg00E}, build upon techniques presented in the work by Pironio et al. \cite{PABGMS}. Their work was dedicated to finding tight bounds on the one-sided rate \(H(A|X=0,Y=0,E)\) as a function of the CHSH score \(\score\) and was derived for a DIQKD protocol. In this thesis, we extend their approach and present it within a more mathematically rigorous framework to make the approach taken compatible with the framework of the entropy accumulation theorem. Moreover, we extend these results to compute the bounds for all one and two-sided entropies \(H(.|.E)\) as a function of the CHSH score \(\score\) (refer to Section \ref{sec:signif} for the appropriate definitions). We also show that these techniques can further be extended to find tight bounds for all the \(6\) one- and two-sided rates as a function of the generalized CHSH score as defined in equation \ref{eqn: genCHSHscore}.}

\section{Rates for (generalized) CHSH-based protocols}\label{sec:rates}
We calculate various one-sided and two-sided rates for protocols based on the (generalized) CHSH game, which involves trying to violate the CHSH Bell inequality~\cite{CHSH}. Recall that in this game, each party uses a binary input and receives a binary output and the game is won if $A\oplus B=XY$. We define the generalized CHSH score by
\begin{eqnarray}\label{eqn: Generalized CHSH score}
  \score &:=& \gamma_{00} \sum_ap_{AB|00}(a,a)+  \gamma_{01} \sum_ap_{AB|01}(a,a)+ \gamma_{10} \sum_ap_{AB|10}(a,a)  \nonumber \\ 
      & & \hspace{7.5cm}  + \gamma_{11}\sum_ap_{AB|11}(a,a\oplus1) ,
\end{eqnarray}
which is the probability of winning the generalized CHSH game when the inputs are chosen at random\footnote{Note that even if nonuniform distributions of inputs are used when running protocols, in this work the CHSH score is always defined \RutC{as \eqref{eqn: Generalized CHSH score}}.}. Note here that \RutC{$\gamma_{ij}$} are some real coefficients. A special feature of this class of scores is that they remain invariant under the re-labelling \RutC{of the output variables }$(a, b) \rightarrow (a \oplus 1 , b \oplus 1)$. \RutC{As we shall see later in this chapter, having such a relabelling symmetry implies that there are always two different strategies (related by a deterministic local operation) that yield the same score. This property will be leveraged to simplify the optimization problem for computing the randomness rates as a function of the score $\score$.}

The CHSH score is the special case when all the coefficients \RutC{$\gamma_{ij} = \frac{1}{4}$}. Classical strategies \RutC{the} in case of the protocols based on the CHSH inequality can win this game with probability at most $3/4$, while quantum strategies can get as high as $\frac{1}{2}\left(1+\frac{1}{\sqrt{2}}\right)\approx0.85$. 

For a fixed generalized CHSH score, $\score$, we wish to compute the minimum von Neumann entropy over all strategies achieving that score. In this context, a strategy comprises three Hilbert spaces $\cH_{A'}$, $\cH_{B'}$ and $\cH_E$, POVMs $\{M_{a|x}\}_a$ on $\cH_{A'}$ for both $x=0$ and $x=1$, POVMs $\{N_{b|y}\}_b$ on $\cH_{B'}$ for both $y=0$ and $y=1$, and a state $\rho_{A'B'E}$ on $\cH_{A'}\ot\cH_{B'}\ot\cH_E$. Given a strategy and a distribution $p_{XY}$ there is an associated channel $\cN$ that acts on $A'B'$ and takes the state to the post-measurement state, i.e.,
\begin{eqnarray}
\tau_{ABXYE}&=&(\cN\ot\cI_E)(\rho_{A'B'E})  \nonumber\\ 
        &=& \sum_{abxy}p_{XY}(x,y)\proj{abxy}_{A,B,X,Y} \ot\tr_{A'B'}\left((M_{a|x}\ot N_{b|y}\ot \id_E)\rho_{A'B'E}\right)\,, \nonumber 
\end{eqnarray}
where $\cI_E$ is the identity channel on $E$. The entropic quantities we consider all pertain to this state\footnote{In the cases where we condition on $X=0$ and $Y=0$, we can project this state onto $\proj{0}_X\ot\proj{0}_Y$ and renormalize --- see section ~\ref{app:simp} for more detail.}. Note also that the generalized CHSH score that is achieved using the strategy (if it is performed in an i.i.d. fashion) is a function of $\tau$, which we denote \RutC{by} $\scorefunction(\tau)$ -- in particular $\scorefunction(\tau)$ is defined using Equation~\eqref{eqn: Generalized CHSH score} and substituting $p_{AB|xy}(a,b)$ by the expression $\tr\left((M_{a|x}\ot N_{b|y}\ot \id_E)\rho_{A'B'E}\right)$. 

For each of the six entropic quantities previously discussed, we consider the infimum over all strategies that achieve a given score. We use this to define a set of curves \RutC{corresponding to each of the $6$ entropic quantities}. We write $F_{AB|XYE}(\score,p_{XY})=\inf H(AB|XYE)_{\tau}$, where the infimum is over all strategies for which $\scorefunction(\tau)=\score$. In the same way we define $F_{AB|E}(\score,p_{XY})$, $F_{A|XYE}(\score,p_{XY})$ and $F_{A|E}(\score,p_{XY})$ \RutC{by} replacing the objective function \RutC{with} the corresponding entropy. We also define $F_{AB|00E}(\score)$ and $F_{A|00E}(\score)$ analogously, noting that these are independent of $p_{XY}$. Furthermore, if we write $F_{AB|XYE}(\score)$ etc.\ (i.e., leaving out the $p_{XY}$), we refer to the case where $p_{XY}$ is uniform over $X$ and $Y$. For a more precise writing of these optimizations, see Equation~\eqref{opt_1}.

We also consider a related set of functions $G_{AB|XYE}(\score,p_{XY})$, $G_{AB|E}(\score,p_{XY})$ etc.\ that are defined analogously, but while optimizing over a smaller set of allowed strategies. More precisely, the $G$ functions are defined by restricting $\cH_{A'}$ and $\cH_{B'}$ to be two dimensional and $\cH_E$ to have dimension 4, taking $\rho_{A'B'E}$ to be pure with $\rho_{A'B'}$ diagonal in the Bell basis, and taking the POVMs to be projective measurements onto states of the form $\cos(\alpha)\ket{0}+\sin(\alpha)\ket{1}$ (see Equation~\eqref{opt_simp}).

\RutC{Note that ideally, the family of functions $F_{.|.E}(\score,p_{XY})$ and $G_{.|,E}(\score, p_{XY})$ should explicitly indicate the coefficients $\gamma_{ij}$ used to define the score. However, for the sake of brevity and to avoid unnecessary complexity in notation, we will omit this explicit dependence. In the later parts of this chapter and the entirety of the next chapter, we will use $\score$ to refer exclusively to the CHSH score.}

As we shall see later in this chapter, it turns out that $F_{AB|00E}(\score)=G_{AB|00E}(\score)$, $F_{A|00E}(\score)=G_{A|00E}(\score)$, and that in each of the other four cases $F$ \RutC{and $G$ are related by}
\begin{eqnarray}\label{eqn: conenv}
    \RutC{F = \text{convenv}(G)}.
\end{eqnarray} 
\RutC{Here $\text{convenv}(G)$ represents the convex envelope (or the convex lower bound) of $G$.  Roughly speaking, the convex envelope of a function $g$ is the smallest convex function $f$ that is not greater than $g$. The rigorous definition of this function in its most general form can be found in Section \ref{app: LF tranform}. The underlying reason why the above equation (eqn. \ref{eqn: conenv}) holds is due to the Jordan's lemma. In our context,} Jordan's lemma~\cite{Jordan} implies that in the case of Bell inequalities with two inputs and two outputs, any strategy is equivalent to a convex combination of strategies in which $A'$ and $B'$ are qubit systems. This means that if we solve the qubit case, the general case follows by taking the convex lower bound. Such an argument was made in~\cite{PABGMS} and we now proceed to show this in the next few sections.

A note on notation: in this work we measure entropies in bits, taking $\log$ to represent the logarithm base 2, and $\ln$ for the natural logarithm where needed.

\section{Simplifying the strategy}\label{app:Proof1}
Given a Hilbert space $\cH$, we \RutC{define} $\cP(\cH)$ to be the set of positive semi-definite operators on $\cH$, and $\cS(\cH)$ to be the set of density operators, i.e., elements of $\cP$ with trace 1. The pure states on $\cH$ (elements of $\cS(\cH)$ with rank 1) will be denoted $\cS_P(\cH)$. A POVM on $\cH$ is a set of positive operators $\{E_i\}_i$ with $E_i\in\cP(\cH)$  for all $i$ and $\sum_i E_i=\id_\cH$, where $\id_\cH$ is the identity operator on $\cH$. A projective measurement on $\cH$ is a POVM on $\cH$ where $E_i^2=E_i$ for all $i$. We define the Bell states
\begin{eqnarray}
\ket{\Phi_0}&=&\frac{1}{\sqrt{2}}\left(\ket{00}+\ket{11}\right)\label{bell1}\\
\ket{\Phi_1}&=&\frac{1}{\sqrt{2}}\left(\ket{00}-\ket{11}\right)\\
\ket{\Phi_2}&=&\frac{1}{\sqrt{2}}\left(\ket{01}+\ket{10}\right)\\
\ket{\Phi_3}&=&\frac{1}{\sqrt{2}}\left(\ket{01}-\ket{10}\right)\,,\label{bell4}
\end{eqnarray}
and use $\sigma_1=\ketbra{1}{0}+\ketbra{0}{1}$, $\sigma_2=\ii\ketbra{1}{0}-\ii\ketbra{0}{1}$ and $\sigma_3=\proj{0}-\proj{1}$ as the three Pauli operators.

In this section we make a series of simplifications of the form of the optimization. The argument given broadly follows the logic of~\cite{PABGMS} (see also~\cite{WAP} for an alternative).

\begin{definition}
  A \emph{single-round $2-2-2$ measurement strategy} is a tuple  \\ $(\cH_{A'},\cH_{B'},\{M_{a|x}\}_{x,a}, \{N_{b|y}\}_{y,b})$, where $\cH_{A'}$ and $\cH_{B'}$ are Hilbert spaces, and $\{M_{a|x}\}_a$ is a POVM on $\cH_{A'}$ for each $x\in\{0,1\}$ and likewise $\{N_{b|y}\}_b$ is a POVM on $\cH_{B'}$ for each $y\in\{0,1\}$. In the case that all the POVMs are projective we will call this a \emph{single-round $2-2-2$ projective measurement strategy}.
\end{definition}

Note that here $2-2-2$ stands for $2$ possible values of inputs, $2$ possible values for the outputs and $2$ parties. 

\begin{definition}
  A \emph{single-round $2-2-2$ strategy} is a single-round $2-2-2$ measurement strategy together with a state $\rho_{A'B'E}\in\cH_{A'}\ot\cH_{B'}\ot\cH_E$, where $\cH_E$ is an arbitrary Hilbert space.
\end{definition}
Note that in a device-independent scenario, such a strategy can be chosen by the adversary.

\begin{definition}
  Given a single-round $2-2-2$ measurement strategy and a distribution $p_{XY}$ over the settings $X$ and $Y$, the \emph{associated $2-2-2$ channel} is defined by
\begin{eqnarray*}
\cN:\cS(\cH_{A'}\ot\cH_{B'})\to\cS(\cH_A\ot\cH_B\ot\cH_X\ot\cH_Y):  \\ \sigma\mapsto\sum_{abxy}p_{XY}
(x,y)\proj{a}\ot\proj{b}\ot\proj{x}\ot\proj{y}\tr\left((M_{a|x}\ot N_{b|y})\sigma\right)\,,
\end{eqnarray*}
where $\cH_A$, $\cH_B$, $\cH_X$ and $\cH_Y$ are two dimensional Hilbert spaces.
The union of the sets of associated $2-2-2$ channels for all single-round $2-2-2$ measurement strategies for some fixed input distribution $p_{XY}$ is denoted $\cC(p_{XY})$. The union of the sets of associated $2-2-2$ channels for all single-round $2-2-2$ projective measurement strategies is denoted $\cC_{\Pi}(p_{XY})$.
\end{definition}
The output of the associated $2-2-2$ channel is classical, and $\cH_A\ot\cH_B\ot\cH_X\ot\cH_Y$ stores the outcomes and the chosen measurements. We will usually apply this channel to the $AB$ part of a tripartite system, giving
\begin{equation}\label{eq:post-meas}
(\cN\ot\cI_E)(\rho_{A'B'E})=\sum_{abxy}p_{XY}(x,y)p_{AB|xy}(a,b)\proj{a}\ot\proj{b}\ot\proj{x}\ot\proj{y}\ot\tau^{a,b,x,y}_E\,,
\end{equation}
where $\tau^{a,b,x,y}_E\in\cS(\cH_E)$ for each $a,b,x,y$ (it is the normalization of \\  $\tr_{A'B'}\left((M_{a|x}\ot N_{b|y}\ot \id_E)\rho_{A'B'E}\right)$).

Note that the generalized CHSH score, which we denote $\scorefunction((\cN\ot\cI_E)(\rho_{A'B'E}))$ does not depend on the distribution $p_{XY}$ of input settings.

We will be interested in optimization problems of the form
\begin{align}
  F(\score,p_{XY})=&\inf_{\cR}  \bar{H}((\cN\ot\cI_E)(\rho_{A'B'E}))\text{, where  }\label{opt_1}\\\nonumber
  &\cR=\{(\cN,\rho_{A'B'E}):\cN\in\cC(p_{XY}),
               \,\scorefunction((\cN\ot\cI_E)(\rho_{A'B'E}))=\score\} ,
\end{align}
where $\cH_E$ is an arbitrary Hilbert space and the spaces $\cH_{A'}$ and $\cH_{B'}$ are those from the chosen element of $\cC(p_{XY})$, i.e., the set $\cR(\score)$ runs over all possible dimensions of these spaces, and $\omega$ is some fixed real number. Here $\bar{H}$ can be any one of the  following entropic quantities defined on the state $(\cN\ot\cI_E)(\rho_{A'B'E})$: $H(AB|X=0,Y=0,E)$, $H(AB|XYE)$, $H(AB|E)$, $H(A|X=0,Y=0,E)$, $H(A|XYE)$ or $H(A|E)$. We consider the family of optimizations in this work and many of the arguments that follow \RutC{to be} independent of this choice.

\subsection{Reduction to projective measurements}
In this section, we conclude that there is no loss in generality in assuming that the devices perform projective measurements. More precisely, we prove the following lemma.
\begin{lemma}\label{lem:proj}
  The sets
  \begin{eqnarray*}
    \cT_1&:=&\{(\cN\ot\cI_E)(\rho_{A'B'E}):\cN\in\cC(p_{XY}), \rho_{A'B'E}\in\cS(\cH_{A'}\ot\cH_{B'}\ot\cH_E)\}\text{  and}\\
    \cT_2&:=&\{(\cN\ot\cI_E)(\rho_{A'B'E}):\cN\in\cC_\Pi(p_{XY}), \rho_{A'B'E}\in\cS(\cH_{A'}\ot\cH_{B'}\ot\cH_E)\}
    \end{eqnarray*}
are identical.
\end{lemma}

This is a corollary of Naimark's theorem, which we state in the following way.
\begin{theorem}[Naimark's theorem]\label{thm:naimark}
  Let $\{E_i\}_i$ be a POVM on $\cH$. There exists a Hilbert space $\cH'$ and a projective measurement $\{\Pi_i\}_i$ on $\cH\ot\cH'$ such that for any $\rho\in\cS(\cH)$
  $$\sum_i\proj{i}\tr(\rho E_i)=\sum_i\proj{i}\tr(\Pi_i(\rho\ot\proj{0}))\,.$$
\end{theorem}
\begin{proof}
  We can directly construct this measurement as follows. Consider the isometry $V:\cH\to\cH\ot\cH'$ given by $V=\sum_i\sqrt{E_i}\ot\ket{i}$, and let $U$ be the extension of $V$ to a unitary with the property that $U(\ket{\psi}\ot\ket{0})=\sum_i\sqrt{E_i}\ket{\psi}\ot\ket{i}$ for any $\ket{\psi}\in\cH$. This construction ensures that the channels
  \begin{eqnarray*}
    \cE:\rho&\mapsto&\sum_i\proj{i}\tr(E_i\rho)\text{  and}\\
\cE':\rho&\mapsto&\sum_i\proj{i}\tr\left((\id\ot\proj{i})U(\rho\ot\proj{0})U^\dagger\right)
  \end{eqnarray*}
  are identical. The second of these can be rewritten
  $$\rho\mapsto\sum_i\proj{i}\tr\left(\Pi_i(\rho\ot\proj{0})\right)\,,$$
where we take $\Pi_i=U^\dagger(\id\ot\proj{i})U$, as required.
\end{proof}

\begin{proof}[Proof of Lemma~\ref{lem:proj}]
By definition $\cT_2\subseteq\cT_1$. For the other direction, consider a state $\rho_{A'B'E}$ and POVMs $\{M_{a|x}\}_{a,x}$ and $\{N_{b|y}\}_{b,y}$ forming a single-round CHSH strategy in $\cT_1$. We use the construction in the proof of Theorem~\ref{thm:naimark} to generate the projectors $\Pi^A_{a|x}$ and $\Pi^B_{b|y}$ as Naimark extensions of the POVMs. Instead of creating the state $\rho_{A'B'E}$, the state $\rho_{A'B'E}\ot\proj{0}_{A''}\ot\proj{0}_{B''}$ is created, where the projectors $\Pi^A_{a|x}$ act on $A'A''$ and $\Pi^B_{b|y}$ act on $B'B''$. Since the latter is a strategy in $\cT_2$ leading to the same post-measurement state~\eqref{eq:post-meas}, we have $\cT_1\subseteq\cT_2$, which completes the proof.
\end{proof}

\subsection{Reduction to convex combinations of qubit strategies}
This is a consequence of Jordan's lemma~\cite{Jordan} and is a special feature that applies only because the Bell inequality has two inputs and two outputs for each party.

\begin{lemma}[Jordan's lemma]\label{lem:Jordan}
Let $A_1$  and $A_2$ be two Hermitian operators on $\cH$ with eigenvalues $\pm1$, then we can decompose $\cH=\bigoplus_{\alpha}\cH_{\alpha}$ such that $A_1$ and $A_2$ preserve the subspaces $\cH_\alpha$, and where each $\cH_\alpha$ has dimension at most 2.
\end{lemma}
\begin{corollary}\label{cor:proj}
Let $\Pi_1$ and $\Pi_2$ be two projections on $\cH$. We can decompose $\cH=\bigoplus_{\alpha}\cH_{\alpha}$ such that $\Pi_1$, $\id-\Pi_1$, $\Pi_2$ and $\id-\Pi_2$ preserve the subspaces $\cH_\alpha$, and where each $\cH_\alpha$ has dimension at most 2.
\end{corollary}
\begin{proof}
Apply Jordan's lemma to the Hermitian operators $A_1=2\Pi_1-\id$ and $A_2=2\Pi_2-\id$ with eigenvalues $\pm1$, and consider $\ket{\psi}\in\cH_\alpha$ for some $\alpha$. By construction $A_1\ket{\psi}\in\cH_\alpha$ from which it follows that $\Pi_1\ket{\psi}\in\cH_\alpha$, and hence also $(\id-\Pi_1)\ket{\psi}\in\cH_\alpha$. Thus, $\Pi_1$ and $\id-\Pi_1$ preserve the subspace; likewise $\Pi_2$ and $\id-\Pi_2$.
\end{proof}

This implies the following
\begin{lemma}
  Let $\cC_{2\times2}(p_{XY})$ be the set of CHSH channels associated with the single-round CHSH projective measurement strategies where each of the four projectors $M_{a|x}$ is block diagonal with $2\times2$ blocks, and each of the four projectors $N_{b|y}$ is block diagonal with $2\times2$ blocks.
  The sets
  \begin{eqnarray*}
    \cT_2&:=&\{(\cN\ot\cI_E)(\rho_{A'B'E}):\cN\in\cC_\Pi(p_{XY}), \rho_{A'B'E}\in\cS(\cH_{A'}\ot\cH_{B'}\ot\cH_E)\}\text{  and}\\
    \cT_3&:=&\{(\cN\ot\cI_E)(\rho_{A'B'E}):\cN\in\cC_{2\times2}(p_{XY}), \rho_{A'B'E}\in\cS(\cH_{A'}\ot\cH_{B'}\ot\cH_E)\}
    \end{eqnarray*}
are identical.
\end{lemma}
\begin{proof}
This follows by applying Corollary~\ref{cor:proj} to the projectors $M_{0|0}$ and $M_{0|1}$ to get the blocks on $\cH_{A'}$ and to the projectors $N_{0|0}$ and $N_{0|1}$ to get the blocks on $\cH_{B'}$. Although some of the blocks may be $1\times1$, we can collect these together and treat them as a $2\times2$ block, or add an extra dimension to the space (on which the state has no support) to achieve all $2\times2$ blocks.
\end{proof}
We can also make the state only have support on the $2\times2$ blocks.
\begin{lemma}
  The sets
  \begin{eqnarray*}
    \cT_3&:=&\{(\cN\ot\cI_E)(\rho_{A'B'E}):\cN\in\cC_{2\times2}(p_{XY}), \rho_{A'B'E}\in\cS(\cH_{A'}\ot\cH_{B'}\ot\cH_E)\}\text{  and}\\
    \cT_4&:=&\{(\cN\ot\cI_E)(\rho_{A'B'E}):\cN\in\cC_{2\times2}(p_{XY}), \rho_{A'B'E}\in\cS_{2\times2}(\cH_{A'}\ot\cH_{B'}\ot\cH_E)\}
    \end{eqnarray*}
are identical. Here $\cS_{2\times2}(\cH_{A'}\ot\cH_{B'}\ot\cH_E)$ is the subset of $\cS(\cH_{A'}\ot\cH_{B'}\ot\cH_E)$ such that $\rho_{A'B'E}\in\cS_{2\times2}(\cH_{A'}\ot\cH_{B'}\ot\cH_E)$ implies
  $$\rho_{A'B'E}=\sum_{\alpha,\beta}(\Pi^\alpha_{A'}\ot \Pi^\beta_{B'}\ot\id_E)\rho_{A'B'E}(\Pi^\alpha_{A'}\ot \Pi^\beta_{B'}\ot\id_E)\,,$$
where $\{\Pi^{\alpha}\}_\alpha$ are projectors onto the $2\times2$ diagonal blocks.
\end{lemma}
\begin{proof}
Consider a state $\rho_{A'B'E}$ and sets of projectors $\{M_{a|x}\}_{a,x}$ and $\{N_{b|y}\}_{b,y}$ from the set $\cT_3$.
  For brevity, write $\Pi^{\alpha,\beta}=\Pi^\alpha_{A'}\ot \Pi^\beta_{B'}$. Then, since
  $$M_{a|x}\ot N_{b|y}=\sum_{\alpha,\beta}(\Pi^\alpha_{A'}\ot \Pi^\beta_{B'})(M_{a|x}\ot N_{b|y})(\Pi^\alpha_{A'}\ot \Pi^\beta_{B'})\,,$$
  we have
  \begin{eqnarray*}
    \tr((M_{a|x}\ot N_{b|y}\ot\id)\rho_{A'B'E})&=&\tr\left(\sum_{\alpha,\beta}(\Pi^{\alpha,\beta}\ot\id_E)(M_{a|x}\ot N_{b|y}\ot\id)(\Pi^{\alpha,\beta}\ot\id_E)\rho_{A'B'E}\right)\\
                                                     &=&\tr\left(\sum_{\alpha,\beta}(M_{a|x}\ot N_{b|y}\ot\id)(\Pi^{\alpha,\beta}\ot\id_E)\rho_{A'B'E}(\Pi^{\alpha,\beta}\ot\id_E)\right)\\
    &=&\tr_{A'B'}((M_{a|x}\ot N_{b|y}\ot\id)\rho'_{A'B'E})\,,
  \end{eqnarray*}
where $\rho'_{A'B'E}=\sum_{\alpha,\beta}(\Pi^{\alpha,\beta}\ot\id_E)\rho_{A'B'E}(\Pi^{\alpha,\beta}\ot\id_E)$ and $\tr$ is over the registers $A'$ and $B'$. Thus, if we replace $\rho_{A'B'E}$ by $\rho'_{A'B'E}$ we obtain the same post-measurement state~\eqref{eq:post-meas}. Hence $\cT_3\subseteq\cT_4$, and, since the other inclusion is trivial, $\cT_3=\cT_4$.
\end{proof}

\begin{lemma}
Let $\cN\in\cC_{2\times2}(p_{XY})$ and $\rho_{A'B'E}\in\cS_{2\times2}(\cH_{A'}\ot\cH_{B'}\ot\cH_E)$. The state $(\cN\ot\cI_E)(\rho_{A'B'E})$ can be formed as a convex combination of states $(\cN_\lambda\ot\cI_E)(\rho^\lambda_{A''B''E})$, where for each $\lambda$, the channel $\cN_\lambda$ is that associated with a single-round measurement strategy with two 2-dimensional Hilbert spaces and distribution $p_{XY}$.
\end{lemma}
\begin{proof}
Since $\rho_{A'B'E}\in\cS_{2\times2}(\cH_{A'}\ot\cH_{B'}\ot\cH_E)$ the $2\times2$ block structure means we can write
$\rho_{A'B'E}=\sum_{\alpha,\beta}p_{\alpha,\beta}\rho^{\alpha,\beta}_{A'B'E}$, where $p_{\alpha,\beta}\rho^{\alpha,\beta}_{A'B'E}=(\Pi^{\alpha,\beta}\ot\id_E)\rho_{A'B'E}(\Pi^{\alpha,\beta}\ot\id_E)$ and $\tr(\rho^{\alpha,\beta}_{A'B'E})=1$ for all $\alpha$ and $\beta$. Likewise, taking $M^\alpha_{a|x}=\Pi_{A'}^\alpha M_{a|x}\Pi_{A'}^\alpha$ and $N^\beta_{b|y}=\Pi^\beta_{B'} N_{b|y}\Pi^\beta_{B'}$ we can write $M_{a|x}=\sum_\alpha M^\alpha_{a|x}$ and $N_{b|y}=\sum_\beta N^\beta_{b|y}$. In terms of these we have
\begin{eqnarray*}
  \tr_{A'B'}((M_{a|x}\ot N_{b|y}\ot\id)\rho_{A'B'E})&=&\sum_{\alpha,\beta}p_{\alpha,\beta}\tr_{A'B'}((M^\alpha_{a|x}\ot N^\beta_{b|y}\ot\id)\rho^{\alpha,\beta}_{A'B'E}).
\end{eqnarray*}
We can then associate a value of $\lambda$ with each pair $(\alpha,\beta)$, replace each $\rho^{\alpha,\beta}_{A'B'E}$ by a state on $A''B''E$ in which $A''$ and $B''$ are two-dimensional (the support of $\rho^{\alpha,\beta}_{A'B'}$ has dimension at most 4), and likewise replace the projectors by qubit projectors. In terms of these we have
\begin{equation*}
  (\cN\ot\cI_E)(\rho_{A'B'E})=\sum_\lambda p_\lambda(\cN_\lambda\ot\cI_E)(\rho^\lambda_{A''B''E})\,.\qedhere
  \end{equation*}
\end{proof}
In other words, any post-measurement state~\eqref{eq:post-meas} that can be generated in the general case, can also be generated if Eve sends a convex combination of two qubit states, and where the measurements used by the separated devices depend on the state sent. Eve could realise such a strategy in practice by using pre-shared randomness. We can proceed to consider strategies in which qubits are shared between the two devices, and then consider the mixture of such strategies after doing so.

\section{Qubit strategies}~\label{app: convex combinations of qubit strategies}
\noindent In this section we consider the single-round $2-2-2$ measurement strategies in which $\cH_{A'}$ and $\cH_{B'}$ are two-dimensional and the measurements are rank-1 projectors. Given a distribution $p_{XY}$ we use $\cC_{\Pi_1,2}(p_{XY})$ to denote the set of associated $2-2-2$ channels. We restrict to rank-1 projectors because if one of the projectors is \RutC{equal to the} identity it is not possible to achieve a non-classical generalized CHSH score, and the non-classical scores are the ones of interest.
\begin{lemma}
Consider a single-round $2-2-2$ measurement strategy for which one of the POVM elements is identity and let $\cN$ be the associated $2-2-2$ channel. For any state $\rho_{A'B'}$ on which $\cN$ can act we have $\scorefunction(\cN(\rho_{A'B'}))\leq l$, where $l$ is the local bound for the generalized CHSH inequality.
\end{lemma}
\begin{proof}
  Suppose the identity element corresponds to $M_{0|0}$ (the other cases follow symmetrically). \RutC{Any} conditional distribution $p_{AB|XY}$ \RutC{that obeys the non-signalling conditions takes the form}\\
  \begin{tabular}{ |cc| c  c| c c| }
\hline
        && $X=0$ &  & $X=1$ &\\
        && $A=0$ & $A=1$  & $A=0$ & $A=1$\\
\hline
 $Y=0$&$B=0$   & $\mu$ &  $0$ & $\nu$  & $\mu-\nu$\\  
  &$B=1$ & $1-\mu$ & $0$ & $\zeta$ &  $1-\mu-\zeta$     \\ 
  \hline
  $Y=1$&$B=0$  & $\gamma$ &  $0$ & $\xi$ & $\gamma-\xi$ \\
  &$B=1$ &  $1-\gamma$  &  $0$ & $\nu+\zeta-\xi$ & $1+\xi-\gamma-\nu-\zeta$\\ 
  \hline
  \end{tabular}\\
where \RutC{$\mu, \nu , \zeta$ and $\gamma$ are any arbitrary non-negative numbers that do not exceed $1$}. In order to prove the lemma, it suffices to show that this distribution can be achieved using a local strategy. For this, we show that the CHSH score for this strategy $\leq \frac{3}{4}$. 
The associated CHSH  score is
\begin{eqnarray*} \frac{1}{4}\left(\mu+\nu+(1-\mu-\zeta)+\gamma+(\gamma-\xi)+(\nu+\zeta-\xi)\right)&=&\frac{1}{4}\left(1+2\nu+2\gamma-2\xi\right).
 \end{eqnarray*}
 Since every element of the distribution must be between $0$ and $1$ we have  $1+\xi-\gamma-\nu-\zeta\geq0$, and hence $1+2\nu+2\gamma-2\xi\leq 3-2\zeta\leq3$, from which the claim follows.
\end{proof}

We will then consider an optimization of the form~\eqref{opt_1}, but restricting to $\cC_{\Pi_1,2}$, i.e. 
\begin{align}
  h(\score)&=\inf_{\cR(\score)}  \bar{H}((\cN\ot\cI_E)(\rho_{A'B'E}))\text{, where  }\label{opt_2}\\\nonumber
  \cR(\score)&=\{(\cN,\rho_{A'B'E}):\cN\in\cC_{\Pi_1,2}(p_{XY}),
               \,\scorefunction((\cN\ot\cI_E)(\rho_{A'B'E}))=\score\}.
\end{align}

The next step is to show that without loss of generality we can reduce to states that are invariant under application of $\sigma_2\ot\sigma_2$ on $A'B'$.
\begin{lemma}\label{lem:marginals}
Let $p_{XY}$ be a distribution, $\cN\in\cC_{\Pi_1,2}(p_{XY})$ and $\rho_{A'B'E}\in\cS(\cH_{A'}\ot\cH_{B'}\ot\cH_E)$ be such that $\scorefunction((\cN\ot\cI_E)(\rho_{A'B'E}))=\score$. There exists a state $\tilde{\rho}_{A'B'EE'}\in\cS(\cH_{A'}\ot\cH_{B'}\ot\cH_E\ot\cH_{E'})$ such that $\scorefunction((\cN\ot\cI_{EE'})(\tilde{\rho}_{A'B'EE'}))=\score$, $\tilde{\rho}_{A'B'EE'}=(\sigma_2\ot\sigma_2\ot\id_{EE'})\tilde{\rho}_{A'B'EE'}(\sigma_2\ot\sigma_2\ot\id_{EE'})$ and $\bar{H}((\cN\ot\cI_{EE'})(\tilde{\rho}_{A'B'EE'}))=\bar{H}((\cN\ot\cI_E)(\rho_{A'B'E}))$ for all six of the entropic functions given earlier.
\end{lemma}
Note that this implies that $p_{A|X}$ and $p_{B|Y}$ can be taken to be uniform.

This is a consequence of the following lemmas.
\begin{lemma}\label{lem:real}
  Let $\{\Pi_{0|0},\Pi_{1|0}\}$ and $ \{\Pi_{0|1} ,\Pi_{1|1}\}$ be two rank-one projective measurements on a two dimensional Hilbert space $\cH$. There exists a basis $\{\ket{e_i}\}_{i=1}^{2}$ such that $\bra{e_l}\Pi_{i|j}\ket{e_k} \in \mathbb{R}$ for all $i,j,k,l$.
  \end{lemma}
\begin{proof}
Without loss of generality, we can take $\Pi_{0|0}=\proj{0}$ and $\Pi_{1|0}=\proj{1}$, and then write $\Pi_{0|1}=\proj{\alpha_{0|1}}$ and $\Pi_{1|1}=\proj{\alpha_{1|1}}$, where $\ket{\alpha_{0|1}}=\cos(\lambda) \ket{0} + \e^{\ii\chi} \sin(\lambda) \ket{1}$ and $\ket{\alpha_{1|1}}=\sin(\lambda) \ket{0} - \e^{\ii\chi} \cos(\lambda) \ket{1}$. Then, we can re-define $\ket{1} \rightarrow \e^{\ii\chi} \ket{1}$ so that $\ket{\alpha_{0|1}} = \cos(\lambda) \ket{0} + \sin(\lambda) \ket{1}$ and $\ket{\alpha_{1|1}} = \sin(\lambda) \ket{0} - \cos(\lambda) \ket{1}$, with $\lambda \in \mathbb{R}$.
\end{proof}

\begin{lemma}\label{lem:Uproj}
Let $\{\Pi_{0|0},\Pi_{1|0}\}$ and $ \{\Pi_{0|1} ,\Pi_{1|1}\}$ be two rank-one projective measurements on a two dimensional Hilbert space $\cH$, then, there exists a unitary transformation $U$ such that $U\Pi_{j|i}U^\dagger=\Pi_{j\oplus1|i}$ for all $i,j$. 
\end{lemma}
\begin{proof}
Let $\Pi_{j|i}=\proj{\alpha_{0|1}}$ for all $i,j$. From Lemma~\ref{lem:real}, we can change basis such that $\ket{\alpha_{0|0}}=\ket{0}$, $\ket{\alpha_{1|0}}=\ket{1}$, $\ket{\alpha_{0|1}}=\cos(\lambda)\ket{0}+\sin(\lambda)\ket{1}$ and $\ket{\alpha_{1|1}}=\sin(\lambda)\ket{0}-\cos(\lambda)\ket{1}$ for some $\lambda\in\mathbb{R}$. Any unitary of the form $U=\e^{\ii\phi}(\ketbra{0}{1}-\ketbra{1}{0})$, with $\phi\in\mathbb{R}$ then satisfies the desired relations. 
\end{proof}

The following lemma is well-known (it follows straightforwardly from e.g.,~\cite[Section~11.3.5]{Nielsen&Chuang})
\begin{lemma}\label{lem:N&C}
For $\rho_{CZEE'}=\sum_ip_i\rho^i_{CZE}\ot\proj{i}_{E'}$ we have $H(C|ZEE')_\rho=\sum_ip_iH(C|ZE)_{\rho^i}$.
\end{lemma}

We now prove Lemma~\ref{lem:marginals}.
\begin{proof}[Proof of Lemma~\ref{lem:marginals}]
  Let $U_A$ and $U_B$ be the unitaries formed by applying Lemma~\ref{lem:Uproj} to respective measurements of each device and using the choice of basis specified in the proof of Lemma~\ref{lem:Uproj} we can take $U_A=\sigma_2$ and $U_B=\sigma_2$. Then define $\rho'_{A'B'E}=(\sigma_2\ot\sigma_2\ot\id_E)\rho_{A'B'E}(\sigma_2\ot\sigma_2\ot\id_E)$. The states $(\cN\ot\cI_E)(\rho'_{A'B'E})$ and $(\cN\ot\cI_E)(\rho_{A'B'E})$ are related by
  $$(\cN\ot\cI_E)(\rho'_{A'B'E})=(\id_{XYE}\ot\sigma_1\ot\sigma_1)(\cN\ot\cI_E)(\rho_{A'B'E})(\id_{XYE}\ot\sigma_1\ot\sigma_1)\,.$$
In other words $(\cN\ot\cI_E)(\rho'_{A'B'E})$ is identical to $(\cN\ot\cI_E)(\rho_{A'B'E})$, except that the outcomes of each device have been relabelled $(a,b) \rightarrow (a \oplus 1 , b \oplus 1)$. Note that our choice of the generalized CHSH score is invariant under the re-labelling of the outputs. It follows that $\scorefunction((\cN\ot\cI_E)(\rho'_{A'B'E}))=\score$ and $\bar{H}((\cN\ot\cI_E)(\rho'_{A'B'E}))=\bar{H}((\cN\ot\cI_E)(\rho_{A'B'E}))$.

Now consider the state $\tilde{\rho}_{A'B'EE'}=(\rho_{A'B'E}\ot\proj{0}_{E'}+\rho'_{A'B'E'}\ot\proj{1}_{E'})/2$. We have $(\cN\ot\cI_{EE'})(\tilde{\rho}_{A'B'EE'})=((\cN\ot\cI_E)(\rho_{A'B'E})\ot\proj{0}_{E'}+(\cN\ot\cI_E)(\rho'_{A'B'E'})\ot\proj{1}_{E'})/2$. Since the score is linear, we have $\scorefunction((\cN\ot\cI_{EE'})(\tilde{\rho}_{A'B'EE'}))=\score$. By construction, $\tilde{\rho}_{A'B'E}=(\sigma_2\ot\sigma_2\ot\id_E)\tilde{\rho}_{A'B'E}(\sigma_2\ot\sigma_2\ot\id_E)$. Finally, as a consequence of Lemma~\ref{lem:N&C}, for any of the entropy functions $H$ we have $\bar{H}(\tilde{\rho}_{A'B'EE'})=\bar{H}(\rho_{A'B'E})$. 
\end{proof}

\begin{corollary}\label{cor:sigma2}
Any optimization of the form~\eqref{opt_2} is equivalent to an optimization of the same form but where each of the projectors are onto states of the form $\alpha\ket{0}+\beta\ket{1}$ with $\alpha,\beta\in\mathbb{R}$ and $\rho_{A'B'E}=(\sigma_2\ot\sigma_2\ot\id)\rho_{A'B'E}(\sigma_2\ot\sigma_2\ot\id)$.
\end{corollary}

Next we consider the form of the reduced state $\rho_{A'B'}$ in the Bell basis.
\begin{lemma}
  Let $\cN$ be the channel associated with a single-round CHSH strategy in which each POVM element is a projector of the form $\cos(\alpha)\ket{0}+\sin(\alpha)\ket{1}$ with $\alpha\in\mathbb{R}$. 
  The state $\rho^P_{A'B'E}$ satisfies $(\cN\ot\cI_E)(\rho^P_{A'B'E})=(\cN\ot\cI_E)(\rho_{A'B'E})$, where $\rho^P_{A'B'E}$ is formed from $\rho_{A'B'E}$ by taking the partial transpose on $A'B'$ in the Bell basis.
\end{lemma}
\begin{proof}
By definition, the partial transpose generates the state 
\begin{eqnarray}
    \RutC{\rho^P_{A'B'E}=\sum_{ij}(\ketbra{\Psi_i}{\Psi_j}\ot\id_E)\rho_{A'B'E}(\ketbra{\Psi_i}{\Psi_j}\ot\id_E)\,.}
\end{eqnarray}
Writing the partial trace out in the Bell basis, for any two projectors $\Pi_1$ and $\Pi_2$ on $\cH_{A'}$ and $\cH_{B'}$ we have
\begin{eqnarray}
  \tr((\Pi_1\ot\Pi_2\ot\id_E)\rho^P_{A'B'E})&=&\sum_i(\bra{\Psi_i}(\Pi_1\ot\Pi_2)\ot\id_E)\rho^P_{A'B'E}(\ket{\Psi_i}\ot\id_E)\nonumber\\
                                                   &=&\sum_{ijk}((\bra{\Psi_i}(\Pi_1\ot\Pi_2)\ketbra{\Psi_j}{\Psi_k})\ot\id_E)\rho_{A'B'E} \nonumber\\& & \hspace{5.5cm}(\ketbra{\Psi_j}{\Psi_k}\ket{\Psi_i}\ot\id_E)\nonumber\\
                                                   &=&\sum_{ij}\bra{\Psi_i}(\Pi_1\ot\Pi_2)\ket{\Psi_j}(\bra{\Psi_i}\ot\id_E)\rho_{A'B'E} \nonumber\\
                                                   & & \hspace{6cm}(\ket{\Psi_j}\ot\id_E)\,.\label{eq:sw}
\end{eqnarray}
When $\Pi_1$ and $\Pi_2$ are each projectors onto states of the form $\cos(\alpha)\ket{0}+\sin(\alpha)\ket{1}$ a short calculation reveals $\bra{\Psi_i}(\Pi_1\ot\Pi_2)\ket{\Psi_j}=\bra{\Psi_j}(\Pi_1\ot\Pi_2)\ket{\Psi_i}$. Using this in~\eqref{eq:sw} we can conclude that
$$\tr_{A'B'}((\Pi_1\ot\Pi_2\ot\id_E)\rho^P_{A'B'E})=\tr_{A'B'}((\Pi_1\ot\Pi_2\ot\id_E)\rho_{A'B'E})\,,$$
from which it follows that $(\cN\ot\cI_E)(\rho^P_{A'B'E})=(\cN\ot\cI_E)(\rho_{A'B'E})$.
\end{proof}

\begin{corollary}\label{cor:trans}
Any optimization of the form~\eqref{opt_2} is equivalent to an optimization of the same form but where each of the projectors are onto states of the form $\alpha\ket{0}+\beta\ket{1}$ with $\alpha,\beta\in\mathbb{R}$, $\rho_{A'B'E}=(\sigma_2\ot\sigma_2\ot\id)\rho_{A'B'E}(\sigma_2\ot\sigma_2\ot\id)$ and $\rho_{A'B'E}=\rho^P_{A'B'E}$.
\end{corollary}
\begin{proof}
We established the invariance under $(\sigma_2\ot\sigma_2\ot\id)$ in Corollary~\ref{cor:sigma2}. Since $(\cN\ot\cI_E)(\rho^P_{A'B'E})=(\cN\ot\cI_E)(\rho_{A'B'E})$, if Eve uses the state $(\rho_{A'B'E}\ot\proj{0}_{E'}+\rho^P_{A'B'E}\ot\proj{1}_{E'})/2$, then, by the same argument used at the end of the proof of Lemma~\ref{lem:marginals}, the entropy and scores are unchanged while the state satisfies the required conditions.
\end{proof}

The next step is to show that the state on $A'B'$ can be taken to come from the set of density operators that are diagonal in the Bell basis. We define
\begin{align}
\cS_B&:=&\{\lambda_0\proj{\Phi_0}+\lambda_1\proj{\Phi_1}+\lambda_2\proj{\Phi_2}+\lambda_3\proj{\Phi_3}  :1\geq\lambda_0\geq\lambda_3\geq0, \nonumber \\ 
& &\,1\geq\lambda_1\geq\lambda_2\geq0,\vphantom{\sum_i} \lambda_0-\lambda_3\geq\lambda_1-\lambda_2,\sum_i\lambda_i=1 \}\,,\label{eq:bellset}
\end{align}
where the states $\{\ket{\phi_i}\}_i$ are defined by~\eqref{bell1}--\eqref{bell4}.

\begin{lemma}
Any optimization of the form~\eqref{opt_2} is equivalent to an optimization of the same form but where each of the projectors are onto states of the form $\cos(\alpha)\ket{0}+\sin(\alpha)\ket{1}$ with $\alpha\in\mathbb{R}$ and $\rho_{A'B'}\in\cS_B$.
\end{lemma}
\begin{proof}
  From Corollary~\ref{cor:trans}, we have that $\rho_{A'B'}$ can be taken to be invariant under $\sigma_2\ot\sigma_2$. Hence we can write
  \begin{equation}\label{eq:bell_form}
\rho_{A'B'} = \left(\begin{array}{cccc}\lambda'_0&0&0&r_1\\0&\lambda'_1&r_2&0\\0&r_2^*&\lambda'_2&0\\r_1^*&0&0&\lambda'_3\end{array}\right)\,
    \end{equation}
where the matrix is expressing the coefficients in the Bell basis. \RutC{From corollary \ref{cor:trans} we can impose that} $\rho_{A'B'}=\rho_{A'B'}^T$, \RutC{which} then implies that $r_1$ and $r_2$ are real. Note that in order that $\rho_{A'B'}$ is a positive operator we require $r_1^2\leq\lambda'_0\lambda'_3$ and $r_2^2\leq\lambda'_1\lambda'_2$.

    Let $U_\theta=\cos(\theta/2)\proj{0}+\sin(\theta/2)\ketbra{0}{1}-\sin(\theta/2)\ketbra{1}{0}+\cos(\theta/2)\proj{1}$, so that $U_\theta$ preserves the set $\{\cos(\alpha)\ket{0}+\sin(\alpha)\ket{1}:\alpha\in\mathbb{R}\}$. We proceed to show that for any state of the form~\eqref{eq:bell_form} with $r_1$ and $r_2$ real, there exist values of $\theta_A$ and $\theta_B$ such that
  $$\rho'_{A'B'} = (U_{\theta_A}\ot U_{\theta_B})\rho_{A'B'}(U^\dagger_{\theta_A}\ot U^\dagger_{\theta_B})$$
  is diagonal in the Bell basis. We can compute the form of $\rho'_{A'B'}$ in the Bell basis. This has the same form as~\eqref{eq:bell_form}, but with $r_1$ replaced by $r_1\cos(\theta_A-\theta_B)+\frac{\lambda'_0-\lambda'_3}{2}\sin(\theta_A-\theta_B)$ and $r_2$ replaced by
  $r_2\cos(\theta_A+\theta_B)+\frac{\lambda'_2-\lambda'_1}{2}\sin(\theta_A+\theta_B)$. To make these zero we need to choose $\theta_A$ and $\theta_B$ such that $\cos^2(\theta_A-\theta_B)=\frac{(\lambda'_0-\lambda'_3)^2}{(\lambda'_0-\lambda'_3)^2+4r_1^2}$ and $\cos^2(\theta_A+\theta_B)=\frac{(\lambda'_1-\lambda'_2)^2}{(\lambda'_1-\lambda'_2)^2+4r_2^2}$. If we write
  \begin{align*}
    \phi_1=\cos^{-1}\left(\frac{\lambda'_0-\lambda'_3}{\sqrt{(\lambda'_0-\lambda'_3)^2+4r_1^2}}\right),\quad&\phi_2=\cos^{-1}\left(\frac{\lambda'_1-\lambda'_2}{\sqrt{(\lambda'_1-\lambda'_2)^2+4r_2^2}}\right),\\
    \zeta_A=\frac{\phi_1+\phi_2}{2}\quad\text{and}\quad&\zeta_B=\frac{\phi_1-\phi_2}{2}
\end{align*}    
then we can express the four solutions
$$(\theta_A,\theta_B)=(\zeta_A,\zeta_B),\,(\zeta_A+\pi/2,\zeta_B-\pi/2),\,(\zeta_A+\pi/2,\zeta_B+\pi/2),\,(\zeta_A+\pi,\zeta_B)\,.$$
Each of these brings the state into the form $\rho_{A'B'}=\lambda_0\proj{\Phi_0}+\lambda_1\proj{\Phi_1}+\lambda_2\proj{\Phi_2}+\lambda_3\proj{\Phi_3}$.
The difference between the first two of these is an exchange of $\lambda_0$ with $\lambda_3$, the difference between the first and the third is an exchange of $\lambda_1$ with $\lambda_2$ and the difference between the first and the fourth is an exchange of $\lambda_0$ with $\lambda_3$ and of $\lambda_1$ with $\lambda_2$. It follows that we can ensure $\lambda_0\geq\lambda_3$ and $\lambda_1\geq\lambda_2$. Finally, if $\lambda_0-\lambda_3<\lambda_1-\lambda_2$ we can apply $\sigma_3\ot\id$ to the resulting state, which simultaneously switches $\lambda_0$ with $\lambda_1$ and $\lambda_2$ with $\lambda_3$, while again preserving the set $\{\cos(\alpha)\ket{0}+\sin(\alpha)\ket{1}:\alpha\in\mathbb{R}\}$.
\end{proof}

The culmination of this section is the following.
\begin{lemma}\label{lem:bellred}
  For given $p_{XY}$, let
  \begin{eqnarray*}
    \cR_1(\score)&:=&\{(\cN,\rho_{A'B'E}):\cN\in\cC_{\Pi_1,2}(p_{XY}),\, \\
           & & \hspace{0.3cm} \rho_{A'B'E}\in\cS(\cH_{A'}\ot\cH_{B'}\ot\cH_E),\,\scorefunction((\cN\ot\cI_E)(\rho_{A'B'E}))=\score\}\quad\text{and}\\
    \cR_2(\score)&:=&\{(\cN,\rho_{A'B'E}):\cN\in\cC_{\Pi_1,2}(p_{XY}),\, \\
           & & \hspace{0.3cm}\rho_{A'B'E}\in\cS(\cH_{A'}\ot\cH_{B'}\ot\cH_E),
\,\rho_{A'B'}\in\cS_B,\,\scorefunction((\cN\ot\cI_E)(\rho_{A'B'E}))=\score\}\,.
  \end{eqnarray*}
We have $\inf_{\cR_2(\score)}\bar{H}((\cN\ot\cI_E)(\rho_{A'B'E}))=\inf_{\cR_1(\score)}  \bar{H}((\cN\ot\cI_E)(\rho_{A'B'E}))$ for any of the six entropy functions $\bar{H}$.
\end{lemma}

\section{Reduction to pure states}\label{sec: reduction to pure states in DI}
Here we show that it is sufficient to restrict any of the optimizations we are interested in to pure states.
\begin{lemma}\label{lem:pure}
  For given $p_{XY}$ let $\cR_2(\score)$ be as in Lemma~\ref{lem:bellred} and consider
  \begin{eqnarray*}
    \cR_3(\score)&:=&\{(\cN,\rho_{A'B'E}):\cN\in\cC_{\Pi_1,2}(p_{XY}),\, \rho_{A'B'E}\in\cS_P(\cH_{A'}\ot\cH_{B'}\ot\cH_E),\, \\ 
                & & \hspace{0.3 cm} \rho_{A'B'}\in\cS_B,\,\scorefunction((\cN\ot\cI_E)(\rho_{A'B'E}))=\score\}\,.
    \end{eqnarray*}
We have $\inf_{\cR_3(\score)}  \bar{H}((\cN\ot\cI_E)(\rho_{A'B'E}))=\inf_{\cR_2(\score)}  \bar{H}((\cN\ot\cI_E)(\rho_{A'B'E}))$ for any of the six entropy functions $\bar{H}$.
\end{lemma}
\begin{proof}
Since $\cR_3\subset\cR_2$, we have $\inf_{\cR_3(\score)}  \bar{H}((\cN\ot\cI_E)(\rho_{A'B'E}))\geq\inf_{\cR_2(\score)}  \bar{H}((\cN\ot\cI_E)(\rho_{A'B'E}))$. For the other direction, consider a state $\rho_{A'B'E}$ from the set $\cR_2$, and let $\rho_{A'B'EE'}$ be its purification. Using the strong subadditivity of the von-Neumann entropy, $H(C|ZEE')\leq H(C|ZE)$, a new strategy in which the only change is that Eve holds a purification of $\rho_{A'B'E}$ cannot increase any of the entropic quantities of interest and makes no change to the score. Thus, $\inf_{\cR_3(\score)}  \bar{H}((\cN\ot\cI_E)(\rho_{A'B'E}))\leq\inf_{\cR_2(\score)}  \bar{H}((\cN\ot\cI_E)(\rho_{A'B'E}))$.
\end{proof}

Note that this also means that we can restrict $\cH_E$ to be 4 dimensional.

\section{Simplifications of qubit strategies for specific entropic quantities}\label{app:simp}
In this section, we compute expressions for each of the entropies of interest, based on the simplifications from the previous section. In other words, we are considering the optimizations
\begin{align}
G(\score,p_{XY}):=\min\ &\bar{H}((\cN\ot\cI_E)(\rho_{A'B'E}))\label{opt_simp}\\
        &\cH_{A'}=\cH_{B'}=\mathbb{C}^2,\ \cH_E=\mathbb{C}^4,\\ \nonumber    
        &\rho_{A'B'E}\in\cS_P(\cH_{A'}\ot\cH_{B'}\ot\cH_E),\ \rho_{A'B'}\in\cS_B\nonumber\\
        &\cN: \text{ is of the form \ref{eqn: Channel N}}\nonumber\\
                         &\ket{\phi^A_{a|x}}=\cos(\alpha_{a|x})\ket{0}+\sin(\alpha_{a|x})\ket{1}\quad \text{and} \\
                         &\ket{\phi^B_{b|y}}=\cos(\beta_{b|y})\ket{0}+\sin(\beta_{b|y})\ket{1}\\
&\alpha_{1|x}=\pi/2+\alpha_{0|x}\quad\text{and}\quad\beta_{1|x}=\pi/2+\beta_{0|x}\\
                         &\scorefunction((\cN\ot\cI_E)(\rho_{A'B'E}))=\score.
\end{align}
For convenience we sometimes use $\alpha_x=\alpha_{0|x}$ and $\beta_x=\beta_{0|x}$ and 
\begin{eqnarray}\label{eqn: Channel N}
    \sigma_{A'B'}\mapsto\sum_{abxy}p_{XY}(x,y)\proj{abxy}\tr\left( \left(\proj{\phi^A_{a|x}} \ot\proj{\phi^B_{b|y}})\sigma_{A'B'} \right)\right) 
\end{eqnarray}
Let
\begin{eqnarray}
  \tau_{ABXYE}&=&(\cN\ot\cI_E)(\rho_{A'B'E})\nonumber\\
&=&\sum_{abxy}p_{XY}(x,y)\proj{abxy}\ot\tr\left(\left(\proj{\phi^A_{a|x}}\ot\proj{\phi^B_{b|y}}\ot\id_E\right)\rho_{A'B'E}\right)\nonumber\\
&=&\sum_{abxy}p_{XY}(x,y)p_{AB|xy}(a,b)\proj{abxy}\ot\tau_E^{abxy}\label{eq:tau}\,,
\end{eqnarray}
where $\{\tau_E^{abxy}\}$ are normalized and $\tr$ is with respect to the systems $A'B'$.

We make a few initial observations.

Consider $p_{AB|xy}(a,b)\tau_E^{abxy}=\tr_{A'B'}\left(\left(\proj{\phi^A_{a|x}}\ot\proj{\phi^B_{b|y}}\ot\id_E\right)\rho_{A'B'E}\right)$. Since $\rho_{A'B'E}$ is pure, we can use the Schmidt decomposition to write $\rho_{A'B'E}=\proj{\Phi}_{A'B'E}$, where
$$\ket{\Phi}_{A'B'E}=\sum_i\sqrt{\lambda_i}\ket{\Psi_i}\ot\ket{i}\,$$
where $\{\ket{i}\}$ is an orthonormal basis for $\cH_E$. We have
\begin{eqnarray*}
p_{AB|xy}(a,b)\tau_E^{abxy}&=&\sum_{ij}\sqrt{\lambda_i\lambda_j}\left(\bra{\phi^A_{a|x}}\ot\bra{\phi^B_{b|y}}\right)\ket{\Psi_i}\bra{\Psi_j}\left(\ket{\phi^A_{a|x}}\ot\ket{\phi^B_{b|y}}\right)\ketbra{i}{j} \\ 
&=& \proj{\zeta^{abxy}}\,,
\end{eqnarray*}
where
\begin{eqnarray}\label{eq:zeta}
\ket{\zeta^{abxy}}=\!\sum_i\sqrt{\lambda_i}(\bra{\phi^A_{a|x}}\ot\bra{\phi^B_{b|y}})\ket{\Psi_i}\ket{i}\!\quad\!  \text{and} \\ \!\quad\! p_{AB|xy}(a,b)=\!\sum_i\lambda_i\left|(\bra{\phi^A_{a|x}}\ot\bra{\phi^B_{b|y}})\ket{\Psi_i}\right|^2\,.
\end{eqnarray}
Hence $\tau_E^{abxy}$ is pure for each $a,b,x,y$. Note also that
\begin{eqnarray}
  \left(\bra{\phi^A_{a|x}}\ot\bra{\phi^B_{b|y}}\right)\ket{\Psi_0}&=&\frac{\cos(\beta_{b|y}-\alpha_{a|x})}{\sqrt{2}}\label{eq:innerp1}\\
  \left(\bra{\phi^A_{a|x}}\ot\bra{\phi^B_{b|y}}\right)\ket{\Psi_1}&=&\frac{\cos(\beta_{b|y}+\alpha_{a|x})}{\sqrt{2}}\\
  \left(\bra{\phi^A_{a|x}}\ot\bra{\phi^B_{b|y}}\right)\ket{\Psi_2}&=&\frac{\sin(\beta_{b|y}+\alpha_{a|x})}{\sqrt{2}}\\
  \left(\bra{\phi^A_{a|x}}\ot\bra{\phi^B_{b|y}}\right)\ket{\Psi_3}&=&\frac{\sin(\beta_{b|y}-\alpha_{a|x})}{\sqrt{2}}\label{eq:innerp4}
\end{eqnarray}

Because $\tau_{ABXYE}$ is formed from $\rho_{A'B'E}$ without acting on $E$, we have $H(E)_\tau=H(E)_\rho$, and because $\rho_{A'B'E}$ is pure, $H(E)_\rho=H(A'B')_\rho=H(\{\lambda_0,\lambda_1,\lambda_2,\lambda_3\})$.\footnote{We use $H$ for both the von Neumann and Shannon entropies; if a list of probabilities is given as the argument to $H$ it signifies the Shannon entropy.} For the same reason, $\sum_{ab}p_{AB}(a,b)\tau_E^{abxy}=\tau_E$ for all $x,y$.

\begin{lemma}\label{lem:ent_simp}
For $\sigma_{AXE}=\sum_{ax}p_{AX}(a,x)\proj{a}\ot\proj{x}\ot\sigma^{a,x}_E$, we have
$$H(A|XE)=
H(A|X)+\sum_{ax}p_{AX}(a,x)H(\sigma^{a,x}_E)-\sum_xp_X(x)H\left(\sum_ap_{A|x}(a)\sigma^{a,x}_E\right)\,.$$
\end{lemma}
\begin{proof}
We have 
\begin{align*}
    H(A|XE)&=H(AXE)-H(XE) \\
    &=H(AX)+\sum_{ax}p_{AX}(a,x)H(E|A=a,X=x)-H(X)- \\
    & \hspace{8cm}\sum_xp_X(x)H(E|X=x)\\
    &=H(A|X)+\sum_{ax}p_{AX}(a,x)\left(H(\sigma^{a,x}_E)- H\left(\sum_{a'}p_{A|x}(a')\sigma^{a',x}_E\right)\right)\,.\qedhere
\end{align*}
\end{proof}

We can parameterize the Bell diagonal state in the following way:
\begin{eqnarray}
  \lambda_0&=&\frac{1}{4}+\frac{R\cos(\theta)}{2}+\delta\label{param1}\\
  \lambda_1&=&\frac{1}{4}+\frac{R\sin(\theta)}{2}-\delta\label{param2}\\
  \lambda_2&=&\frac{1}{4}-\frac{R\sin(\theta)}{2}-\delta\label{param3}\\
  \lambda_3&=&\frac{1}{4}-\frac{R\cos(\theta)}{2}+\delta\label{param4}
\end{eqnarray}
where $0\leq R\leq1$, $0\leq\theta\leq\pi/4$ if $R\leq1/\sqrt{2}$, or $0\leq\theta\leq\pi/4-\cos^{-1}(1/(R\sqrt{2}))$ if $R>1/\sqrt{2}$ and $-1/4+R\cos(\theta)/2\leq\delta\leq1/4-R\sin(\theta)/2$.

\begin{lemma}\label{lem:delta_opt}
For $R>1/\sqrt{2}$, $\max_\delta H(\{\lambda_0,\lambda_1,\lambda_2,\lambda_3\})$ is achieved when $\delta=\delta^*=\frac{R^2\cos(2\theta)}{4}$.
\end{lemma}
\begin{proof}
One can compute the derivative of $H(\{\lambda_0,\lambda_1,\lambda_2,\lambda_3\})$ with respect to $\delta$ to see that it is $0$ only for $\delta=\delta^*:=\frac{R^2\cos(2\theta)}{4}$. 

We next check that $\delta^*$ is in the valid range of $\delta$. The condition $\delta^*\leq1/4-R\sin(\theta)/2$ rearranges to
  $2R^2\sin^2(\theta)-2R\sin(\theta)+1-R^2\geq0$. The roots of the quadratic equation $2R^2x^2-2Rx+1-R^2$ are at $x=\frac{1}{2R}(1\pm\sqrt{2R^2-1})$. For $R\geq\frac{1}{\sqrt{2}}$ the roots are real\footnote{If $R<\frac{1}{\sqrt{2}}$ there are no real roots and the condition always holds.}. Our condition on $\theta$ implies that $0\leq\sin(\theta)\leq\frac{1}{2R}(1-\sqrt{2R^2-1})$, hence taking $x=\sin(\theta)$ we are always to the left of the first root and so $\delta^*\leq1/4-R\sin(\theta)/2$. A similar argument shows $\delta^*\geq-1/4+R\cos(\theta)/2$.

We can then compute the double derivative of $H(\{\lambda_0,\lambda_1,\lambda_2,\lambda_3\})$ with respect to $\delta$ and evaluate it at $\delta^*$. This gives $-\frac{32}{\ln(2)(2-4R^2+R^4+R^4\cos(4\theta))}$, which can be shown to be negative for $R>1/\sqrt{2}$ and any valid $\theta$ using a similar argument to that above.
\end{proof}

Using this state and the specified measurements, the probability table for the observed distribution has the form given in the table below (whose entries correspond to $p_{AB|XY}$):\smallskip

\begin{tabular}{|c |c| c  c| c c| }
  \hline
  &&$Y=0$&&$Y=1$&\\
       & & $B=0$ & $B$ = 1  & $B=0$ & $A=1$\\
\hline
$X=0$& $A=0$   & $\epsilon_{00}$ &  \vphantom{$\frac{\sum^a}{f}$} $\frac{1}{2} -\epsilon_{00} $ & $\epsilon_{01}$ &  $\frac{1}{2} -\epsilon_{01} $\\  
&  $A=1$ & \vphantom{$\frac{\sum^a}{f}$} $\frac{1}{2} - \epsilon_{00}$ &   $\epsilon_{00}$ & $\frac{1}{2} - \epsilon_{01}$ &   $\epsilon_{01}$     \\ 
  \hline
$X=1$&  $A=0$  & $\epsilon_{10}$ &  \vphantom{$\frac{\sum^a}{f}$} $\frac{1}{2} -\epsilon_{10} $ & $\frac{1}{2} - \epsilon_{11}$ &   $\epsilon_{11}$ \\
&  $A=1$ &  \vphantom{$\frac{\sum^a}{f}$}$\frac{1}{2} - \epsilon_{10}$ &   $\epsilon_{10}$& $\epsilon_{11}$ &  $\frac{1}{2} -\epsilon_{11} $\\ 
  \hline
\end{tabular}\smallskip\\
where
\begin{align*}
  \epsilon_{00}&=\frac{1}{4}\left(1+R\cos(\theta)\cos(2(\alpha_0-\beta_0))+R\sin(\theta)\cos(2(\alpha_0+\beta_0))\right)\\
  \epsilon_{01}&=\frac{1}{4}\left(1+R\cos(\theta)\cos(2(\alpha_0-\beta_1))+R\sin(\theta)\cos(2(\alpha_0+\beta_1))\right)\\
  \epsilon_{10}&=\frac{1}{4}\left(1+R\cos(\theta)\cos(2(\alpha_1-\beta_0))+R\sin(\theta)\cos(2(\alpha_1+\beta_0))\right)\\
  \epsilon_{11}&=\frac{1}{4}\left(1-R\cos(\theta)\cos(2(\alpha_1-\beta_1))-R\sin(\theta)\cos(2(\alpha_1+\beta_1))\right)\,.
\end{align*}

Note that
\begin{align}\label{eq:score}
\scorefunction(\tau_{ABXY})=&   2\sum_{ij} \RutC{\gamma_{ij}} \epsilon_{ij}
\end{align}
is independent of $\delta$.
In the upcoming sections, we use these simplifications to obtain analytic forms for entropic quantities for the qubit case.
\section{H(AB|X=0,Y=0,E)}\label{app:AB00E}
For $H(AB|X=0,Y=0,E)$ we are interested in the state
$$\tau'_{ABE}=\sum_{ab}p_{AB|00}(a,b)\proj{a}\ot\proj{b}\ot\tau_E^{ab00}$$
since $H(AB|X=0,Y=0,E)_\tau=H(AB|E)_{\tau'}$. Note that, as above, $H(E)_{\tau'}=H(E)_\rho$. Using Lemma~\ref{lem:ent_simp} we have
\begin{align*}
H(AB|E)_{\tau'}&=H(AB)_{\tau'}+\sum_{ab}p_{AB}(a,b)H(\tau_E^{ab00})-H\left(\sum_{ab}p_{AB}(a,b)\tau_E^{ab00}\right)\\
  &=H(AB)_{\tau'}+\sum_{ab}p_{AB}(a,b)H(\tau_E^{ab00})-H(E)_{\tau'}\,.
\end{align*}
However, since $\tau_E^{ab00}$ is pure for each $a,b$, $H(\tau_E^{ab00})=0$ and we find
\begin{align*}
  H(AB|E)_{\tau'}&=H(AB)_{\tau'}-H(E)_{\rho}\\
&=H(\{\epsilon_{00},\epsilon_{00},1/2-\epsilon_{00},1/2-\epsilon_{00}\})-H(\{\lambda_0,\lambda_1,\lambda_2,\lambda_3\})\\
  &=1+H_\bin(2\epsilon_{00})-H(\{\lambda_0,\lambda_1,\lambda_2,\lambda_3\})\,.
\end{align*}
Lemma~\ref{lem:delta_opt} shows that $\max_\delta H(\{\lambda_0,\lambda_1,\lambda_2,\lambda_3\})$ is achieved for $\delta=\delta^*=\frac{R^2\cos(2\theta)}{4}$. Since the score is independent of $\delta$ we can take the state to satisfy $\delta=\delta^*$ and remove $\delta$ from the optimization.

\section{H(AB|XYE)}\label{app:ABgXYE}
In this case we again use Lemma~\ref{lem:ent_simp} to obtain
\begin{align*}
H(AB|XYE)&=H(AB|XY)+\sum_{abxy}p_{ABXY}(a,b,x,y)H(\tau_E^{abxy})- \\ 
  & \hspace{5.5cm}\sum_{xy}p_{XY}(x,y)H\left(\sum_{ab}p_{AB|xy}(a,b)\tau_E^{abxy}\right)\\
  &=H(AB|XY)-H(E)\,,
\end{align*}
where we again use that $H(\tau_E^{abxy})=0$, and note that $\sum_{ab}p_{AB|xy}(a,b)\tau_E^{abxy}=\rho_E$ for all $x,y$. Note that
\begin{align*}
  H(AB|XY)&=\sum_{xy}p_{XY}(x,y)H(AB|X=x,Y=y)\\
          &=1+\sum_{xy}p_{XY}(x,y)H_\bin(2\epsilon_{xy})\,,
\end{align*}
and so we have
\begin{equation}\label{eq:HABgXYE}
    H(AB|XYE)=1+\sum_{xy}p_{XY}(x,y)H_\bin(2\epsilon_{xy})-H(\{\lambda_0,\lambda_1,\lambda_2,\lambda_3\})\,.
\end{equation}
This is again independent of $\delta$, so, like in the case of $H(AB|X=0,Y=0,E)$ we can take $\delta=\delta^*$ and remove $\delta$ from the optimization.

\section{H(AB|E)}
We first trace out $XY$ to give
$\tau_{ABE}=\sum_{ab}p_{AB}\proj{a}\ot\proj{b}\ot\sum_{xy}p_{XY|ab}(x,y)\tau_E^{abxy}$. For this state, Lemma~\ref{lem:ent_simp} gives
\begin{align*}
H(AB|E)&=H(AB)+\sum_{ab}p_{AB}(a,b)H\left(\sum_{xy}p_{XY|ab}(x,y)\tau_E^{abxy}\right)- \\ 
       & \hspace{7cm}H\left(\sum_{abxy}p_{AB}(a,b)p_{XY|ab}(x,y)\tau_E^{abxy}\right)\\
       &=H(AB)+\sum_{ab}p_{AB}(a,b)H\left(\sum_{xy}p_{XY|ab}(x,y)\tau_E^{abxy}\right)-H(E)\\
       &=H(AB)+\sum_{ab}p_{AB}(a,b)H\left(\sum_{xy}\frac{p_{XY}(x,y)p_{AB|xy}(a,b)\tau_E^{abxy}}{p_{AB}(a,b)}\right)-H(E)\,.
\end{align*}
In this case we cannot remove the middle term, and the middle term is not independent of $\delta$. The optimization in this case is hence significantly more complicated. Note that
\begin{align*} H(AB)=1+H_\bin\left(2\left(\sum_{(x , y) \ne (1,1)}p_{XY}(x,y)\epsilon_{xy}+p_{XY}(1,1)\left(\frac{1}{2}-\epsilon_{11}\right)\right)\right)\,.
\end{align*}

\section{H(A|X=0,Y=0,E)}\label{sec: HAg00E}
For this section, we only focus on the case when $\score$ is the CHSH score. This case was already covered in~\cite{PABGMS} where it was solved analytically (see also~\cite{WAP} for a slight generalization).
\begin{lemma}\label{lem:Ag00E}
For $3/4\leq\omega\leq\frac{1}{2}(1+\frac{1}{\sqrt{2}})$ the solution to the optimization problem~\eqref{opt_simp} when $\bar{H}=H(A|X=0,Y=0,E)$ is $1-H_\bin\left(\frac{1}{2}(1+\sqrt{16\score(\score-1)+3})\right)$.
\end{lemma}
For completeness we give a proof here as well. We first show that the maximum CHSH score for a Bell diagonal state depends only on $R$.
\begin{lemma}\label{lem:Ag00Escore}
Given a state $\rho_{A'B'E}$ with $\rho_{A'B'}$ parameterized as in~\eqref{param1}--\eqref{param4}, if $\cN$ satisfies the requirements of the optimization problem~\eqref{opt_simp}, then $\tau_{ABXYE}=(\cN\ot\cI_E)(\rho_{A'B'E})$ satisfies $\scorefunction(\tau_{ABXYE})\leq\frac{1}{2}+\frac{R}{2\sqrt{2}}$, and there exists a channel $\cN$ achieving equality.
\end{lemma}
\begin{proof}
Consider the score function~\eqref{eq:score}. Collecting all the terms involving $\alpha_0$ and $\alpha_1$ and some manipulation gives
  \begin{align*}
\scorefunction(\tau_{ABXYE})=&\frac{1}{2}+\frac{R}{2\sqrt{2}}\cos(\beta_0-\beta_1) A_{1} +\frac{R}{2\sqrt{2}}\sin(\beta_0-\beta_1) B_1,
  \end{align*}
where we have used $\cos(\theta)+\sin(\theta)=\sqrt{2}\sin(\frac{\pi}{4}+\theta)$ and $\cos(\theta)-\sin(\theta)=\sqrt{2}\cos(\frac{\pi}{4}+\theta)$ and 
\begin{align*}
  A_1 &=    \left[\sin(2\alpha_0)\sin(\beta_0+\beta_1)\cos(\frac{\pi}{4}+\theta)+\cos(2\alpha_0)\cos(\beta_0+\beta_1)\sin(\frac{\pi}{4}+\theta)\right]\\
  B_1  &= \left[\sin(2\alpha_1)\cos(\beta_0+\beta_1)\cos(\frac{\pi}{4}+\theta)-\cos(2\alpha_1)\sin(\beta_0+\beta_1)\sin(\frac{\pi}{4}+\theta)\right]
\end{align*}.
  For brevity we write $\bar{\theta}=\frac{\pi}{4}+\theta$. We then use that for $r,t,\phi\in\mathbb{R}$ we have $r\cos(\phi)+t\sin(\phi)\leq\sqrt{r^2+t^2}$ with equality if $r\cos(\phi)+t\sin(\phi)\geq0$ and $r\sin(\phi)=t\cos(\phi)$. This allows us to form the bound
  \begin{align*}
\scorefunction(\tau_{ABXYE})\leq&\frac{1}{2}+\frac{R}{2\sqrt{2}}\left(|\cos(\beta_0-\beta_1)|\sqrt{\sin^2(\beta_0+\beta_1)\cos^2(\bar{\theta})+\cos^2(\beta_0+\beta_1)\sin^2(\bar{\theta})}\right.\\
                           &\left.+|\sin(\beta_0-\beta_1)|\sqrt{\cos^2(\beta_0+\beta_1)\cos^2(\bar{\theta})+\sin^2(\beta_0+\beta_1)\sin^2(\bar{\theta})}\right)\\
    \phantom{\scorefunction}\leq&\frac{1}{2}+\frac{R}{2\sqrt{2}}\,.
  \end{align*}
Choosing $\alpha_0=0$, $\alpha_1=\pi/4$, $\beta_0=\frac{\pi}{8}-\frac{\theta}{2}$, $\beta_1=-\frac{\pi}{8}+\frac{\theta}{2}$ achieves equality (for instance).
\end{proof}
It follows that $\score>3/4$ is only possible if $R>1/\sqrt{2}$.

We now turn to the entropy. In this case we trace out $B$ from the state $\tau'$ in Section~\ref{app:AB00E} to give
$\tau'_{AE}=\sum_ap_{A|00}(a)\proj{a}\ot\sum_bp_{B|a00}(b)\tau_E^{ab00}$, so that $H(A|X=0,Y=0,E)_\tau=H(A|E)_{\tau'}$. Using Lemma~\ref{lem:ent_simp} we have
\begin{align*}
H(A|E)_{\tau'}&=H(A)_{\tau'}+\sum_ap_{A|00}(a)H\left(\sum_bp_{B|a00}(b)\tau_E^{ab00}\right)-H\left(\sum_{ab}p_{AB|00}(a,b)\tau_E^{ab00}\right)\\
&=1+\sum_a\frac{1}{2}H\left(\sum_bp_{B|a00}(b)\tau_E^{ab00}\right)-H(E)_{\rho}\\
              &=1+\sum_a\frac{1}{2}H\left(\sum_b2p_{AB|00}(a,b)\tau_E^{ab00}\right)-H(E)_{\rho}\,,
\end{align*}
where we have used the fact that $p_{A|00}(a)=1/2$ for $a=0,1$. 
The eigenvalues of $\sum_b2p_{AB|00}(a,b)\tau_E^{ab00}$ turn out to be
$$\frac{1}{2}\left(1\pm\sqrt{2(\lambda_0-\lambda_3)(\lambda_1-\lambda_2)\cos(4\alpha_0)+(\lambda_0-\lambda_3)^2+(\lambda_1-\lambda_2)^2}\right)\,,$$
independently of $a$. Hence, we can write $H(A|E)_{\tau'}$ in terms of the Bell diagonal state using
\begin{align*}
\sum_a\frac{1}{2}H\left(\sum_b2p_{AB|00}(a,b)\tau_E^{ab00}\right)&=\Phi_{s}\Big(2(\lambda_0-\lambda_3)(\lambda_1-\lambda_2)\cos(4\alpha_0) \\
                                                                 & \quad \quad \quad \quad \quad \quad \quad \quad +(\lambda_0-\lambda_3)^2+(\lambda_1-\lambda_2)^2\Big)\\
  H(E)_\rho&=H(\{\lambda_0,\lambda_1,\lambda_2,\lambda_3\})
\end{align*}
where $\Phi_{s}(x) = H_{\bin}(\frac{1}{2} + \frac{\sqrt{x}}{2})$ is a short-hand notation.
Having established this, we show the following.
\begin{lemma}\label{lem:Ag00Eent}
Let $\rho_{A'B'E}$ be pure with $\cH_{A'}=\cH_{B'}=\mathbb{C}^2$, and let $\rho_{A'B'}$ be a Bell diagonal state parameterized by~\eqref{param1}--\eqref{param4} with $R>1/\sqrt{2}$. Let $\tau$ be the state defined by~\eqref{eq:tau} $H(A|X=0,Y=0,E)_\tau\geq1+H_\bin\left(\frac{1}{2}\left(1+\sqrt{2R^2-1}\right)\right)$ where equality is achievable for $\alpha_0=0$.
\end{lemma}
\begin{proof}
We note that
  \begin{align*}
\sum_a\frac{1}{2}H\left(\sum_b2p_{AB|00}(a,b)\tau_E^{ab00}\right)&\geq \Phi_{s}\left(2(\lambda_0-\lambda_3)(\lambda_1-\lambda_2)+(\lambda_0-\lambda_3)^2+(\lambda_1-\lambda_2)^2\right)\\
  &=H_\bin(\lambda_0+\lambda_1)\,,
\end{align*}
with equality for $\alpha_0=0$.
Hence
\begin{equation}\label{eq:ent45}
  H(A|E)_{\tau'}\geq 1+H_\bin(\lambda_0+\lambda_1)-H(\{\lambda_0,\lambda_1,\lambda_2,\lambda_3\})\,.
  \end{equation}

Using the parameterization of~\eqref{param1}--\eqref{param4} we have $\lambda_0+\lambda_1=1/2(1+R(\cos(\theta)+\sin(\theta)))$. Thus, the minimum of $H(A|E)_{\tau'}$ over $\delta$ is achieved for $\delta=\delta^*$ (as in the case $H(AB|X=0,Y=0,E)$). Taking $\delta=\delta^*$ and differentiating the resulting expression with respect to $\theta$ yields
$$\frac{R}{2}(\cos(\theta)+\sin(\theta))\log\left(\frac{1-R\cos(\theta)+R\sin(\theta)}{1+R\cos(\theta)-R \sin(\theta)}\right)\,.$$
Since $\cos(\theta)+\sin(\theta)=\sqrt{2}\sin(\pi/4+\theta)$, the leading factor is always positive over our range of $\theta$. The logarithm term is always negative, except for $\theta=\pi/4$ where it reaches zero. Thus, the minimum over $\theta$ is always obtained at the largest possible $\theta$, i.e., $\theta=\pi/4-\cos^{-1}\left(1/(R\sqrt{2})\right)$.

With this substitution the right hand side of~\eqref{eq:ent45} reduces to
$$1+H_\bin\left(\frac{1}{2}\left(1+\sqrt{2R^2-1}\right)\right)\,,$$
establishing the claim.
\end{proof}

Lemma~\ref{lem:Ag00E} is then a corollary of Lemmas~\ref{lem:Ag00Escore} and~\ref{lem:Ag00Eent}.
\begin{proof}[Proof of Lemma~\ref{lem:Ag00E}]
  From Lemma~\ref{lem:Ag00Eent} we have
  $$H(A|X=0,Y=0,E)_\tau\geq1+H_\bin\left(\frac{1}{2}\left(1+\sqrt{2R^2-1}\right)\right)\,.$$
  However, Lemma~\ref{lem:Ag00Escore} then implies
  \begin{eqnarray*}
    H(A|X=0,Y=0,E)_\tau&\geq&1+H_\bin\left(\frac{1}{2}\left(1+\sqrt{4(2\score-1)^2-1}\right)\right)\\
    &=&1+H_\bin\left(\frac{1}{2}\left(1+\sqrt{16\score(\score-1)+3)}\right)\right)\,,
  \end{eqnarray*}
where we use the fact that $H_\bin(p)$ is decreasing and concave for $p\geq1/2$.
  Equality is achievable by taking $\alpha_0=0$, $\alpha_1=\pi/4$, $\beta_0=\frac{\pi}{8}-\frac{\theta}{2}$, $\beta_1=-\frac{\pi}{8}+\frac{\theta}{2}$. 
\end{proof}

We use this case to gain confidence in our numerics, since we can make a direct comparison to the analytic curve.

\section{H(A|XYE)}\label{app:AgXYE}
In this case Lemma~\ref{lem:ent_simp} gives
\begin{eqnarray*}
  H(A|XYE)&=&H(A|XY)+\sum_{axy}p_{AXY}(a,x,y)H\left(\sum_bp_{B|axy}(b)\tau_E^{abxy}\right)\\
  &&-\sum_{xy}p_{XY}(x,y)H\left(\sum_{ab}p_{AB|xy}(a,b)\tau_E^{abxy}\right)\\
        &=&1+\sum_{axy}p_{XY}(x,y)p_{A|xy}(a)H\left(\sum_b2p_{AB|xy}(a,b)\tau_E^{abxy}\right)-H(E)\,.
\end{eqnarray*}
The eigenvalues of $\sum_b2p_{AB|xy}(a,b)\tau_E^{abxy}$ can be computed to be
$$\frac{1}{2}\left(1\pm\sqrt{2(\lambda_0-\lambda_3)(\lambda_1-\lambda_2)\cos(4\alpha_x)+(\lambda_0-\lambda_3)^2+(\lambda_1-\lambda_2)^2}\right)\,,$$
independently of $a,y$. If we define $$g(\alpha):=\frac{1}{2}\left(1+\sqrt{2(\lambda_0-\lambda_3)(\lambda_1-\lambda_2)\cos(4\alpha)+(\lambda_0-\lambda_3)^2+(\lambda_1-\lambda_2)^2}\right)$$
then
\begin{eqnarray}\label{eqn: optimization for AgXYE objective function}
    H(A|XYE)&=1+\sum_xp_X(x)H_\bin(g(\alpha_x))-H(E)\,.
\end{eqnarray}
Note that using the parameterization~\eqref{param1}--\eqref{param4} we have
$$g(\alpha)=\frac{1}{2}\left(1+R\sqrt{1+\sin(2\theta)\cos(4\alpha)}\right)\,.$$
Since this is independent of $\delta$, we can again use $\delta=\delta^*$ to remove one parameter when minimizing $H(A|XYE)$.
  
\section{H(A|E)}
In this case Lemma~\ref{lem:ent_simp} gives
\begin{align*}
H(A|E)&=H(A)+\sum_ap_A(a)H\left(\sum_{bxy}p_{BXY|a}(b,x,y)\tau_E^{abxy}\right)-\\ 
      & \hspace{6cm} H\left(\sum_{abxy}p_{AB}(a,b)p_{XY|ab}(x,y)\tau_E^{abxy}\right)\\
      &=1+\frac{1}{2}\sum_aH\left(\sum_{bxy}2p_{ABXY}(a,b,x,y)\tau_E^{abxy}\right)-H(E)\\
      &=1+\frac{1}{2}\sum_aH\left(\sum_{bxy}2p_{XY}(x,y)P_{AB|xy}(a,b)\tau_E^{abxy}\right)-H(E)\\
      &=1+\frac{1}{2}\sum_aH\left(\sum_x2p_X(x)\tr_{A'}\left(\left(\proj{\phi^A_{a|x}}\ot\id_E\right)\rho_{A'E}\right)\right)-H(E)\,.
\end{align*}
Like in the case $H(AB|E)$ the middle term cannot be removed and this term is not independent of $\delta$.

\section{Numerically computing upper bounds on rates}\label{sec:num}
As we have shown, the optimizations that define the $G$ functions can be expressed in terms of at-most $7$ real parameters ($3$ to specify the state and $4$ to choose the measurements). They are hence amenable to numerical optimizations. We note also that except in the cases $G_{AB|E}(\score,p_{XY})$ and $G_{A|E}(\score,p_{XY})$ we can remove an additional parameter. In this section, we present a heuristic method to estimate the rates. We restrict ourselves to the case when $\score$ is the CHSH score (i.e. \RutC{$\gamma_{ij} = \frac{1}{4}$}). This is also the case for the next chapter. 

We obtain upper bounds by using numerical solvers that attempt to compute $G$ (these give upper bounds because the computations are not guaranteed to converge). Our program for computing $G$ runs in $N$ iterations. In each iteration the program starts by making a random guess for the parameters from within the valid range. It then uses sequential quadratic programming to minimize $\bar{H}((\cN\ot\cI_E)(\rho_{A'B'E}))$ subject to the CHSH score being fixed [here $\bar{H}$ is a placeholder for one of the entropic quantities of interest]. On each iteration, the program arrives at a candidate for the minimum value, and we run $N\approx10^4$ iterations to arrive at the conjectured minimum value for $\bar{H}((\cN\ot\cI_E)(\rho_{A'B'E}))$. The numerical optimization is performed in Python using the sequential least squares programming (SLSQP) solver in SciPy. The curves obtained were found to match those generated by solving numerically in Mathematica and Matlab.

Since these optimizations are not guaranteed to converge, the generated curves are upper bounds on the infima. Some confidence of their tightness comes from the smoothness of the curves, the consistency across different numerical solvers, and that the generated points match the known analytic tight bound in the case $H(A|X=0,Y=0,E)$. They also closely match the numerical lower bounds we computed for $G_{A|XYE}$ and $G_{AB|X=0,Y=0,E}$ discussed in the next chapter~\ref{chap: DI lower bounds}.

$G_{AB|00E}$ and $G_{A|00E}$ are convex functions, and hence $F=G$ for these. For the other cases we generate the graphs in the case where $p_{XY}$ is uniform, observing that each of the $G$ curves starts with a concave part and switches to convex for larger CHSH scores. Since the minimum entropy is always zero for classical scores, each of the $G$ curves approach $0$ as $\score$ approaches $3/4$. Each of the $F$ curves can be found from $G$ by finding the tangent to $G$ that passes through $(3/4,0)$. We call the score at which this tangent is taken $\score^*$, defined by $(\score^*-3/4)G'(\score^*)=G(\score^*)$. We then have
\begin{equation}
    F(\score) =     \begin{cases}
    G'(\score^*) (\score -\frac{3}{4}) & \text{if}\ \score \leq \score^* \\
    G(\score) & \text{otherwise} \\
    \end{cases}\,.
\end{equation}
We give estimates for $\score^*$ for each of the cases below. In essence, what this means is that for $\score<\score^*$ the optimal strategy for Eve is to either use a deterministic classical strategy with score $3/4$ or a strategy that achieves score $\score^*$, mixing these such that the average score is $\score$. Eve can remember which strategy she used, and hence the entropy from her perspective is also the convex mixture of the entropies of the endpoints.

\begin{figure}
\includegraphics[width=\textwidth]{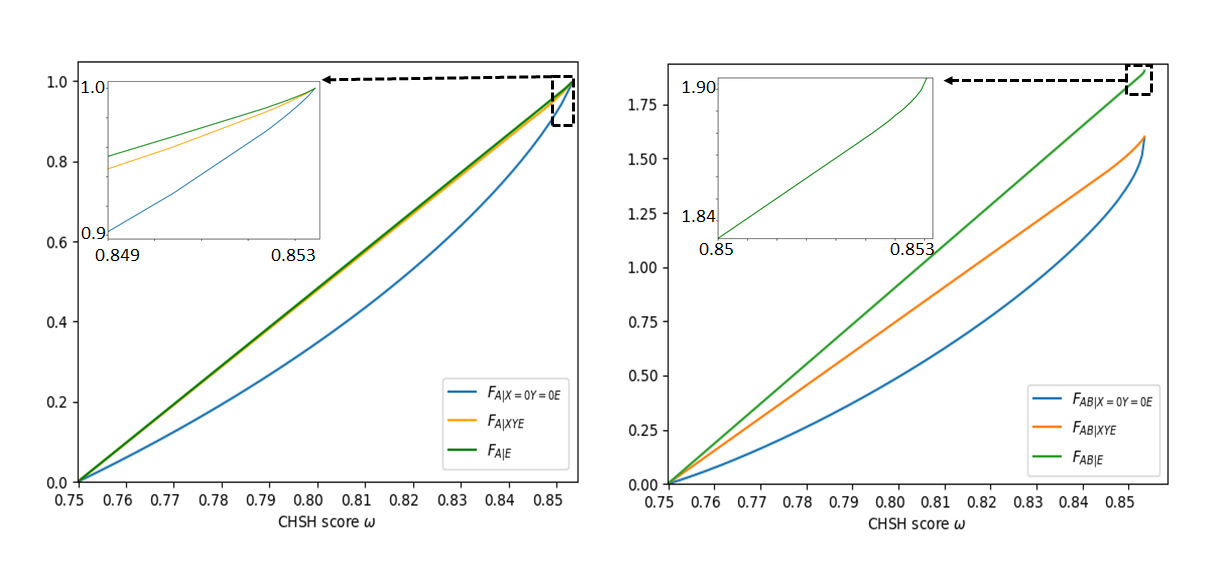}\\
\phantom{m}\hspace{8em}(a)\hspace{17em}(b)
\caption{Graphs of the rates for (a) the one-sided and (b) the two-sided randomness with uniformly chosen inputs. Each of these curves has a non-linear part and the blue curves do not have a linear part.}
\label{fig:rates}
\end{figure}

Figure~\ref{fig:rates} shows the curves we obtained for the functions $F$ in each of the six cases. Note that, except in the cases where we condition on $X=0$ and $Y=0$, the graphs all have linear sections as a result of taking the convex lower bound. In the figure~\ref{fig:rates} we show the graphs for $G$ together with those for $F$. The approximate coordinate of the top of the linear segment for $F_{AB|E}$ is $(0.8523,1.8735)$ and for $F_{A|E}$ it is $(0.8505,0.967)$. Note also that $1+H_\bin\left(\frac{1}{2}+\frac{1}{\sqrt{32}}\right)\approx1.908$ is the maximum value on the graph $F_{AB|E}(\score)$.

By examining the parameters that come out of the numerical optimizations we have the following.

\begin{lemma}\label{lem:ABgXYE}
Consider the curve $g_1(\score)=1+H_\bin(\score)-2H_\bin(\frac{1}{2}+\frac{2\score-1}{\sqrt{2}})$. $F_{AB|XYE}(\score)$ can be upper bounded in terms of $g_1$ as follows
\begin{equation}\label{eq:main_result}
F_{AB|XYE}(\score)\leq\begin{cases}g_1(\score)&\score_{AB|XYE}^*\leq\score\leq\frac{1}{2}\left(1+\frac{1}{\sqrt{2}}\right) \\
g_1'(\score_{AB|XYE}^*)(\score-3/4)&3/4\leq\score\leq\score_{AB|XYE}^*\end{cases}\,.
\end{equation}
where $\score_{AB|XYE}^*\approx0.84403$ is the solution to $g'_1(\score)(\score-3/4)=g_1(\score)$. Note that $g_1(\score_{AB|XYE}^*)\approx1.4186$ and the maximum value reached is $1+H_\bin(1/2+1/(2\sqrt{2}))\approx1.601$.
\end{lemma}
\begin{proof}
We first consider an upper bound on $G_{AB|XYE}(\score)$. In section ~\ref{app:simp} we give a parameterization of a two-qubit state (with parameters $R$, $\theta$ and $\delta$) and measurements (with parameters $\alpha_0$, $\alpha_1$, $\beta_0$ and $\beta_1$) before computing an expression for $H(AB|XYE)$ in terms of these (see~\eqref{eq:HABgXYE}). We also obtain an expression for the CHSH score (see~\eqref{eq:score}). Choosing $R=\sqrt{2}(2\score-1)$, $\theta=0$, $\delta=R^2/4$, $\alpha_0=0$, $\alpha_1=\pi/4$, $\beta_0=\pi/8$, $\beta_1=-\pi/8$ we find a score $\score$, and calculating $H(AB|XYE)$ we obtain $H(AB|XYE)=g_1(\score)$ and hence $G_{AB|XYE}(\score)\leq g_1(\score)$. Since, $G_{AB|XYE}(3/4)=0$, and $F_{AB|XYE}$ is formed from $G_{AB|XYE}$ by taking the convex lower bound, we establish the claim.
\end{proof}

\begin{lemma}\label{lem:AgXYE}
Consider the curve $g_2(\score)=1-H_\bin\left(\frac{1}{2}+\frac{2\score-1}{\sqrt{2}}\right)$.
$F_{A|XYE}(\score)$ is upper bounded by the convex lower bound of $g_2(\score)$. In other words,
\begin{equation}
F_{A|XYE}(\score)\leq\begin{cases}g_2(\score)&\score_{A|XYE}^*\leq\score\leq\frac{1}{2}\left(1+\frac{1}{\sqrt{2}}\right) \\
g_2'(\score_{A|XYE}^*)(\score-3/4)&3/4\leq\score\leq\score_{A|XYE}^*\end{cases}\,.
\end{equation}
where $\score_{A|XYE}^*\approx0.84698$ is the solution to $g'_2(\score)(\score-3/4)=g_2(\score)$. Note that $g_2(\score_{A|XYE}^*)\approx0.92394$.
\end{lemma}
\begin{proof}
The proof is the same as for Lemma~\ref{lem:ABgXYE}, except that $g_1$ is replaced by $g_2$ --- the choice of state and measurements remains the same.
\end{proof}
\section{Upper bounds for other entropic quantities}\label{app:graphs}
\begin{figure}[h!]
\includegraphics[scale=0.50]{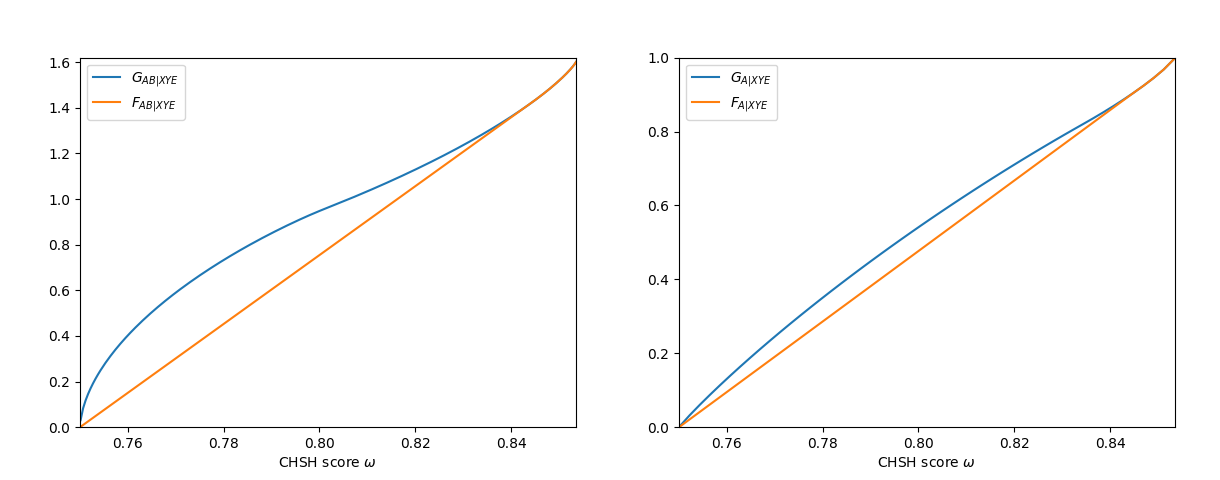} \\
\phantom{m}\hspace{8em}(a)\hspace{17em}(b)
\caption[(a) Two-sided and (b) one-sided entropy curves conditioned on X, Y and E with uniform input distribution.]{ (a) Two-sided and (b) one-sided entropy curves conditioned on $X$, $Y$ and $E$ with uniform input distribution.}
\label{fig:gXYE}
\end{figure}

\begin{figure}
\includegraphics[scale=0.50]{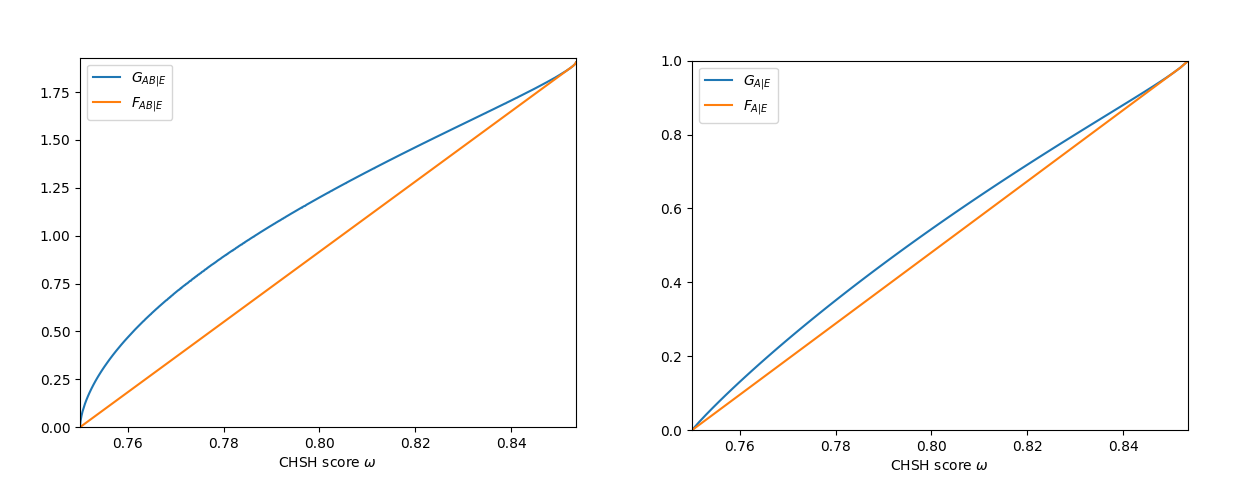} \\
\phantom{m}\hspace{8em}(a)\hspace{17em}(b)
\caption[(a) Two-sided and (b) one-sided entropy curves conditioned on the system E with uniform input distribution.]{(a) Two-sided and (b) one-sided entropy curves conditioned on $E$ with uniform input distribution.}
\label{fig:gE}
\end{figure}
\RC{Figure~\ref{fig:gXYE} gives one-sided and two-sided randomness rates as a function of CHSH score when inputs are chosen uniformly at random. Recall that the curves for $F_{A|E}(\score)$ and $F_{AB|E}(\score)$ (Fig~\ref{fig:gE}) are fundamental upper bounds on one-sided and two-sided output randomness rates as a function of the CHSH score. The curves $F_{.|.E}$ are obtained by taking the convex lower bound (of convex envelope) of $G_{.|.E}$, which is the randomness rates if only qubit-strategies were used -- i.e. Alice and Bob are only allowed to share qubit strategies.}

Figure~\ref{fig:gXYE}(a) gives the graphs of $F_{AB|XYE}(\score)$ and $G_{AB|XYE}(\score)$, while Figure~\ref{fig:gXYE}(b) shows those for $F_{A|XYE}(\score)$ and $G_{A|XYE}(\score)$. In each case the $G$ graphs have a concave and convex part and the $F$ graphs are formed by taking the convex lower bound. For these cases the points at which the tangents are taken are $\score^*_{AB|XYE}\approx0.8440$ and $\score^*_{A|XYE}\approx0.8470$.

Figure~\ref{fig:gE}(a) gives the graphs of $F_{AB|E}(\score)$ and $G_{AB|E}(\score)$, while Figure~\ref{fig:gE}(b) shows those for $F_{A|XYE}(\score)$ and $G_{A|XYE}(\score)$. Again, in each case the $G$ graphs have a concave and convex part and the $F$ graphs are formed by taking the convex lower bound. For these cases the points at which the tangents are taken are $\score^*_{A|E}\approx0.8505$ and $\score^*_{AB|E}\approx0.8523$.

\RC{These curves are generated using heuristic numerical optimizations, and therefore can be only be treated as good estimates for the randomness rate. In the next chapter, we shall derive some of these curves using reliable and rigorous techniques.}

\chapter{Lower bounds on the entropies}\label{chap: DI lower bounds}
\section{Lower bounds}\label{sec:lower}
Lemma~\ref{lem:AgXYE} gives an upper bound on the one-sided randomness using an explicit strategy. However, for security proofs a lower bound is needed. In this chapter, we compute such lower bounds for $G_{A|XYE}$, $G_{AB|X=0,Y=0,E}$ and $G_{AB|XYE}$. Note that for this chapter, we restrict to the case when $\score$ is the CHSH score instead of the generalized CHSH score. 

The idea behind our lower bounds is as follows. We first show that for every fixed value of $\score$ the functions $G_{A|XYE}(\score)$ and $G_{AB|X=0,Y=0,E}(\score)$ can each be expressed as a minimization over $3$ real parameters. For the function $G_{AB|XYE}$, we use polynomial optimization techniques to form a lower bound. For fixed $\score$ we compute the values of the objective function on a grid of points comprising these parameters. By bounding the derivative of the objective function within the cuboids generated by the grid we establish a lower bound on the function over the possible parameters. The lower bound we generate can in principle be made arbitrarily good by decreasing the grid spacing (at the expense of taking more time to evaluate).

Given lower bounds on $G_{A|XYE}(\score), G_{AB|X=0,Y=0,E}(\score)$ and $G_{AB|XYE}(\score)$ for a finite set of values of $\score$, we can get lower bounds for all values of $\score$ by using that the $G$ functions are monotonically increasing in $\score$, so we have $G(\score)\leq G(\score + \nu)$, where $\nu$ is the spacing between the finite set of values of $\score$. Hence, \RutC{we can only compute the lower bounds for finitely many values} $\M{W}=\{\score_1,\score_2,\ldots\}$ in the range $(3/4,(1/2)(1+(1/2)^{1/2})]$. We can then consider the points $\{(\score_1,0),(\score_2,G(\score_1)),(\score_3,G(\score_2)),\ldots\}$, i.e., where each is shifted one place. Taking the convex lower bound of these shifted points gives a convex lower bound for $G_{A|XYE}$ and $G_{AB|X=0,Y=0,E}$. By taking more points in the set $\M{W}$ tighter lower bounds can be obtained. 

In the next few sections of this chapter, we shall find ways to compute reliable lower bounds on $G_{A|XYE}(\score) , G_{AB|X=0,Y=0,E}(\score)$ and $G_{AB|XYE}(\score)$ for any arbitrary score $\score \in [\frac{3}{4} , \frac{1}{2} + \frac{1}{2 \sqrt{2}}]$. 
\section{H(A|XYE)}\label{app:AgXYElb}
Recall from \ref{eqn: optimization for AgXYE objective function} that 
\begin{eqnarray}
    H(A|XYE)&=1+\sum_xp_X(x)H_\bin(g(\alpha_x))-H(E)\,.
\end{eqnarray}
where we define $$g(\alpha):=\frac{1}{2}\left(1+\sqrt{2(\lambda_0-\lambda_3)(\lambda_1-\lambda_2)\cos(4\alpha)+(\lambda_0-\lambda_3)^2+(\lambda_1-\lambda_2)^2}\right)$$
Using the parameterization~\eqref{param1}--\eqref{param4} we have
$$g(\alpha)=\frac{1}{2}\left(1+R\sqrt{1+\sin(2\theta)\cos(4\alpha)}\right).$$ 

We now restrict to the case where $p_X(x)=1/2$ for $x=0,1$. Since $H(E)$ is independent of $\{\alpha_x\}$, we can consider the optimization
\begin{align}
  \min_{\alpha_0,\alpha_1,\beta_0,\beta_1} &\ H_\bin(g(\alpha_0))+H_\bin(g(\alpha_1)) \nonumber\\
\text{subject to } &\ \scorefunction(\tau_{ABXYE})=\score\label{eq:redopt}
\end{align}
for some fixed values of $\score$, $R$ and $\theta$.

We proceed to make a series of simplifications of this optimization.
\begin{lemma}
  The optimization~\eqref{eq:redopt} is equivalent to
\begin{align}
  \min_{u,v} &\ H_\bin(g((v+u)/4))+H_\bin(g((v-u)/4))\nonumber\\
  \mathrm{s.t.}\ &\ S(u,v):=\frac{1}{2}+\frac{R}{4}\Big(\cos(\frac{u}{2})\sqrt{1+\cos(v)\sin(2\theta)} \nonumber\\ 
                 & \hspace{6cm}+\sin(\frac{u}{2})\sqrt{1-\cos(v)\sin(2\theta)}\Big)=\score\label{eq:redopt2}\\
  &0\leq u\leq\pi\nonumber
\end{align}
\end{lemma}
\begin{proof}
Noting that the objective function in~\eqref{eq:redopt} is independent of Bob's angles ($\beta_0$ and $\beta_1$), analogously to the derivation of Lemma ~\eqref{lem:Ag00Escore} we can bound the score function using
\begin{align*}
S(\tau_{ABXYE})\leq&\ \frac{1}{2}+\frac{R}{2\sqrt{2}}\left(\left|\cos(\alpha_0-\alpha_1)\right|\sqrt{\sin^2(\alpha_0+\alpha_1)\cos^2(\bar{\theta})+\cos^2(\alpha_0+\alpha_1)\sin^2(\bar{\theta})}\right.\\                           &\left.+\left|\sin(\alpha_0-\alpha_1)\right|\sqrt{\cos^2(\alpha_0+\alpha_1)\cos^2(\bar{\theta})+\sin^2(\alpha_0+\alpha_1)\sin^2(\bar{\theta})}\right).
\end{align*}
We now substitute $v/2=\alpha_0+\alpha_1$ and $u/2=\alpha_0-\alpha_1$ and rearrange (recalling that $\bar{\theta}=\pi/4+\theta$) to give
\begin{eqnarray*}
S(\tau_{ABXYE})&\leq& \frac{1}{2}+\frac{R}{4}\Big(\left|\cos(\frac{u}{2})\right|\sqrt{1+\cos(v)\sin(2\theta)} \\ 
                &   & \hspace{5cm} +\left|\sin(\frac{u}{2})\right|\sqrt{1-\cos(v)\sin(2\theta)}\Big). 
\end{eqnarray*}

Since $G_{A|XYE}(\score)$ is monotonically increasing in $\score$ (see \ section~\ref{app:mono}), it follows that we wish to choose the angles to achieve the largest possible score function.

We first note that if $\cos(u/2)<0$ we can make the substitution $\alpha_0\mapsto \pi/2+\alpha_1$ and $\alpha_1\mapsto\alpha_0-\pi/2$ which maintains the objective function, constraint, $v$ and $\sin(u/2)$, while changing the sign of $\cos(u/2)$. In addition, if $\sin(u/2)<0$, the substitution $\alpha_0\mapsto\alpha_1$ and $\alpha_1\mapsto\alpha_0$ maintains the objective function, constraint, $v$ and $\cos(u/2)$ while changing the sign of $\sin(u/2)$. It follows that the maximum of $H_\bin(g((v+u)/4))+H_\bin(g((v-u)/4))$ for fixed score is obtained when 
\begin{eqnarray*}
 S(\tau_{ABXYE})=S(u,v)&:=& \frac{1}{2}+\frac{R}{4}\Big(\cos(u/2)\sqrt{1+\cos(v)\sin(2\theta)} \\ 
                    & &\hspace{4.5cm} +\sin(u/2)\sqrt{1-\cos(v)\sin(2\theta)}\Big)   
\end{eqnarray*}
and when both $\cos(u/2)\geq0$ and $\sin(u/2)\geq0$, or, alternatively $0\leq u\leq\pi$.
\end{proof}
\begin{lemma}
  In the optimization~\eqref{eq:redopt2} we can restrict to $0\leq u\leq\pi/2$ and $0\leq v\leq\pi/2$ without affecting the result.
\end{lemma}
\begin{proof}
Consider a $u$ that satisfies $0\leq u\leq\pi$. If $\cos(u/2)\geq\sin(u/2)$ then $0\leq u\leq\pi/2$. Otherwise, consider $u\mapsto\pi-u$, $v\mapsto\pi-v$. This maintains the constraint and the value of the objective function and hence the optimal value can still be obtained, but now with $\cos(u/2)\geq\sin(u/2)$, so we can assume $0\leq u\leq\pi/2$.

For the restriction on $v$, first note that transforming $v\mapsto-v$ has no affect on either the objective function or the constraint so we can take $\sin(v)\geq0$, or $0\leq v\leq\pi$. If $v>\pi/2$, then $\cos(v+u)<0$. Furthermore, $\cos(v-u)+\cos(v+u)=2\cos(v)\cos(u)\leq0$ and hence $\cos(v-u)\leq|\cos(v+u)|$. Let $\bar{v}=\pi-v$. We have 
\begin{eqnarray*}
 S(u,v)-S(u,\bar{v}) &=& \frac{R}{4}\Big(\left(\sin(u/2)-\cos(u/2)\right)\big(\sqrt{1-\cos(v)\sin(2\theta)} \\ 
                        & & \hspace{5.5cm}   - \sqrt{1+\cos(v)\sin(2\theta)}\big)\Big)\leq0\,,   
\end{eqnarray*}

where the inequality follows from $\cos(v)<0$, $\sin(2\theta)\geq0$ and $\cos(u/2)\geq\sin(u/2)$. Hence, the mapping $v\mapsto\pi-v$ increases $S(u,v)$.

Consider now the effect on the objective function $$J(u,v):=\Phi\left(R\sqrt{1+\sin(2\theta)\cos(v+u)}\right)+\Phi\left(R\sqrt{1+\sin(2\theta)\cos(v-u)}\right),$$
where we have used the notation $$\Phi(x) := H_{\bin}\left(\frac{1}{2} + \frac{x}{2} \right),$$ which will be useful shorthand throughout this thesis.
Note that each binary entropy term decreases as its cosine term increases. We have
$$J(u,\bar{v}):=\Phi\left(R\sqrt{1-\sin(2\theta)\cos(v+u)}\right)+\Phi\left(R\sqrt{1-\sin(2\theta)\cos(v-u)}\right).$$

If $\cos(v-u)\leq0$ then $J(u,\bar{v})\leq J(u,v)$, so the transformation decreases the objective function.

On the other hand, if $\cos(v-u)\geq0$ we must have $\cos(v-u)\leq|\cos(v+u)|$ and hence 
\begin{eqnarray*}
\sqrt{1-\sin(2\theta)\cos(v+u)}&\geq& \sqrt{1+\sin(2\theta)\cos(v-u)} \\ 
                               &\geq& \sqrt{1-\sin(2\theta)\cos(v-u)} \\ 
                               &\geq& \sqrt{1+\sin(2\theta)\cos(v+u)}.
\end{eqnarray*}
Thus, $J(u,\bar{v})\leq J(u,v)$ and the transformation decreases the objective function.

Thus, in both cases the transformation decreases the objective function while increasing the score. Using the monotonicity of $G_{A|XYE}(\score)$ with the score $\score$ (see \ section~\ref{app:mono}), it follows that we can further reduce the objective function while bringing the score back to its original level.
\end{proof}

Let us turn to the constraint. We have
\begin{eqnarray*}
    \score &=& \frac{1}{2}+\frac{R}{4}\left(\cos(u/2)\sqrt{1+\cos(v)\sin(2\theta)}+\sin(u/2)\sqrt{1-\cos(v)\sin(2\theta)}\right)\\ 
           &=& \frac{1}{2}+\frac{R}{2\sqrt{2}}\cos(u/2-\phi)\,,
\end{eqnarray*}
where $\cos(\phi)=\sqrt{(1+\cos(v)\sin(2\theta))/2}$ and $\sin(\phi)=\sqrt{(1-\cos(v)\sin(2\theta))/2}$. We can rearrange this to $\cos(u/2-\phi)=\sqrt{2}(2\score-1)/R$ and hence there are two possibilities for $u$: \begin{equation}\label{eq:uforms}
u_{\pm}=2\cos^{-1}\sqrt{(1+\cos(v)\sin(2\theta))/2}\pm2\cos^{-1}(\sqrt{2}(2\score-1)/R).
\end{equation}
We can hence remove the constraint and consider the optimizations
\begin{align*}
  \min_v\ J(u_{\pm}(v),v)
\end{align*}

Summarizing the above analysis we have the following.
\begin{corollary}\label{cor:onesidelower}
Let $\score \in(3/4,(1+1/\sqrt{2})/2]$ and $$K(R,\theta)=1-H(\{\lambda_0(R,\theta),\lambda_1(R,\theta),\lambda_2(R,\theta), \lambda_3(R,\theta)\}),$$ where $\{\lambda_i(R,\theta)\}_{i=0}^3$ are given by~\eqref{param1}--\eqref{param4} with $\delta=\delta^*$. Defining $$\M{D}_\score=\{(R,\theta,v):R\in[\sqrt{2}(2\score-1),1],\ \theta\in[0,\frac{\pi}{4}-\cos^{-1}\left(1/(\sqrt{2}R)\right)],\ v\in[0,\frac{\pi}{2}]\},$$ we have
\begin{eqnarray}\label{One-sided-final opt}
  G_{A|XYE}(\score) &=&\min_{\M{D}_\score , u\in\{u_+,u_-\}}J(u(v),v)/2+K(R,\theta)\,.
\end{eqnarray}
\end{corollary}

\subsection{Monotonicity properties of the function K}\label{app: Monotonicty of K}
The following monotonicity properties of the function $K(R,\theta)$ will be useful later. 
\begin{lemma}\label{lem: monotonicity of K(R)}
For any $\score\in(3/4,(1+1/\sqrt{2})/2]$, and $(R,\theta)\in \M{D}_{\score}$ we have $\partial_{R} K(R,\theta)\geq0$.
\end{lemma}
\begin{proof}
Note that 
\begin{equation*}
    \frac{\lambda_0 \lambda_3}{\lambda_1 \lambda_2} = 1 
\end{equation*}
and $\lambda_0 > \lambda_3$ and $\lambda_1 > \lambda_2$. We differentiate $K(R, \theta)$ with respect to $R$ 
\begin{eqnarray}
\partial_{R} K(R , \theta) &=&  \sum_{i} \Big( \log(\lambda_i) + \frac{1}{\ln2} \Big) \frac{\partial \lambda_i}{\partial R} \nonumber \\ 
&=& \frac{1}{2}\left(\left(\cos(\theta)+R\cos(2\theta)\right)(\log\lambda_0+\frac{1}{\ln2})+\left(\sin(\theta)-R\cos(2\theta)\right)(\log\lambda_1+\frac{1}{\ln2})-\right.\nonumber\\
&&\left.\left(\sin(\theta)+R\cos(2\theta)\right)(\log\lambda_2+\frac{1}{\ln2})-\left(\cos(\theta)-R\cos(2\theta)\right)(\log\lambda_3+\frac{1}{\ln2})\right)   \nonumber \\ 
&=& \frac{1}{2}\left(\log\Big(\frac{\lambda_0}{\lambda_3} \Big) \cos(\theta) + \log\Big(\frac{\lambda_1}{\lambda_2} \Big)\sin(\theta)+ R\cos(2 \theta)\log\Big(\frac{\lambda_0 \lambda_3}{\lambda_1 \lambda_2} \Big)\right) \nonumber \\
&=& \log\Big(\frac{\lambda_0}{\lambda_3} \Big) \cos(\theta) + \log\Big(\frac{\lambda_1}{\lambda_2} \Big)\sin(\theta) \geq0 \nonumber 
\end{eqnarray}
as claimed.
\end{proof}
We derive a similar result for monotonicity of $K(R,\theta)$ with respect to $\theta$
\begin{lemma}\label{lem: monotonicity of K(th)}
For any $\score\in(3/4,(1+1/\sqrt{2})/2]$, and $(R,\theta)\in \M{D}_{\score}$ we have $\partial_{\theta} K(R,\theta)\geq0$.
\end{lemma}
\begin{proof}
We differentiate $K(R,\theta)$ with respect to $\theta$ 
\begin{eqnarray}
\partial_{\theta} K(R , \theta) &=&  \sum_{i} \Big( \log(\lambda_i) + \frac{1}{\ln2} \Big) \frac{\partial \lambda_i}{\partial \theta} \nonumber \\ 
&=&  \sum_{i} \Big( \log(\lambda_i) \frac{\partial \lambda_i}{\partial \theta}\Big) + \frac{1}{\ln2}\sum_{i} \frac{\partial \lambda_i}{\partial \theta}  \nonumber \\ 
&=& \sum_{i} \Big( \log(\lambda_i) \frac{\partial \lambda_i}{\partial \theta}\Big)  \nonumber \\ 
&=&\frac{1}{2}\left(-\log\Big(\frac{\lambda_0}{\lambda_3} \Big) R\sin(\theta) + \log\Big(\frac{\lambda_1}{\lambda_2} \Big)R\cos(\theta) - R^2 \sin(2 \theta)\log\Big(\frac{\lambda_0 \lambda_3}{\lambda_1 \lambda_2} \Big)\right) \nonumber \\
&=& -\frac{R\sin(\theta)}{2} \log\Big(\frac{\lambda_0}{\lambda_3} \Big) + \frac{R\cos(\theta)}{2} \log\Big(\frac{\lambda_1}{\lambda_2} \Big)   \nonumber
\end{eqnarray}
We now consider the function 
\begin{eqnarray}
f(R, \theta) = \frac{2\ln2}{R}\partial_\theta K(R,\theta)=- \sin(\theta)\ln\left(\frac{\lambda_0(R, \theta)}{\lambda_3(R, \theta)}\right)+\cos(\theta)\ln\left(\frac{\lambda_0(R, \theta)}{\lambda_3(R, \theta)}\right)
\end{eqnarray}
Note that $\score\in(3/4,(1+1/\sqrt{2})/2]$ implies $1/\sqrt{2}<R\leq1$ and $0\leq\theta\leq\pi/4-\cos^{-1}(1/(R\sqrt{2}))$, or $0\leq\theta\leq\pi/4$, $1/\sqrt{2}<R\leq1/(\cos(\theta)+\sin(\theta))$. We can extend the domain of $f(R,\theta)$ to $0\leq\theta\leq\pi/4$, $0 \leq R \leq 1/(\cos(\theta) + \sin(\theta))$. \\ 
Taking derivative of $f$ with respect to $R$ gives 
\begin{eqnarray}
\partial_{R}f(R, \theta) = \frac{2 R^2 \sin (4 \theta)}{(1 - R^2(\cos(\theta) + \sin(\theta))^2) (1 - R^2(\cos(\theta) - \sin(\theta))^2 )}.
\end{eqnarray}
Thus, $\partial_{R}f(R , \theta) > 0$ whenever $\theta \in [0 ,\frac{\pi}{4}]$. We can then infer that $\partial_{\theta}K(R, \theta) = \frac{R}{2\ln2}f(R, \theta) \geq\frac{R}{2\ln2} f(0 , \theta) = 0$. 
\end{proof}

\subsection{Lower bounding the objective function}\label{app: lower bound using grids}
In this section, we propose a method to compute the lower bound of the function $G_{.|.E}$ by partitioning the domain. We start by considering an abstract version of the problem, which has the form
\begin{eqnarray}
 \min_{\B{x} \in \M{D}}  Q(\B{x})
\end{eqnarray}
where $\M{D} \subset \mathbb{R}^{n}$ is a compact set and $Q: \M{D} \mapsto \mathbb{R}$ is bounded. Furthermore, we assume we know an upper bound $M$ such that $Q(\B{x})\leq M$ for all $\B{x}\in\M{D}$.

We use the notation $\M{C}_{\B{a},\B{b}}=[a_1,b_1]\times[a_2,b_2]\times\cdots\times[a_n,b_n]$, i.e., $\M{C}_{\B{a},\B{b}}$ is a hyper-cuboid with $\B{a}$ and $\B{b}$ as two opposite vertices. Then let $\M{C}\supseteq\M{D}$ be any hyper-cuboid that completely contains $\M{D}$.
We say $\M{P}=\{\M{C}_{\B{a}^i,\B{b}^i}\}_i$ is a partition of $\M{C}$ if 
\begin{eqnarray}
\M{C} = \bigcup_{i} \M{C}_{\B{a}^{i}, \B{b}^{i}} 
\end{eqnarray}
where $\{\M{C}_{\B{a}^{i}, \B{b}^{i}}\}_i$ are cuboids whose intersection has zero volume.

The main idea behind our lower bounds is to find lower bounds on $Q(\B{x})$ that hold on each cuboid and then to take the minimum of all the lower bounds. In some cases these bounds are formed by starting from a corner and using bounds on the derivatives of $Q(\B{x})$ on the cuboid to form a bound that holds across the cuboid. In other cases, we use monotonicity arguments to imply that evaluation at one of the corners lower bounds the whole cuboid. Some of our cuboids lie entirely outside the original domain $\M{D}$. To save calculation we assign the known upper bound on the function as the upper bound on cuboids in our partition that lie outside of $\M{D}$.

\subsection{Obtaining a lower bound on $G_{A|XYE}$}
We now return to our optimization problem~\eqref{One-sided-final opt}. It is convenient to switch parameterization to use $\eta := \cos^{-1}(\sqrt{2}(2 \score - 1)/ R)$ instead of $R$. Taking $\B{x}=(\eta,\theta,v)$ we rewrite~\eqref{One-sided-final opt} as \begin{eqnarray}\label{eqn: Optimization problem on grids}
G_{A|XYE}(\score) = \min_{\mathbf{x} \in \M{D}_\score,u\in\{u_+,u_-\}} F_1(\eta , \theta, v) + F_2(\eta , \theta , v) + K(R(\eta), \theta) 
\end{eqnarray}
where
\begin{eqnarray}
F_{1}(\eta, \theta , v) &=& \frac{1}{2}H_{\bin} \Big(\frac{1}{2} + \frac{R(\eta)}{2} \sqrt{1 + \cos(u(v) + v) \sin(2 \theta)}\Big) \nonumber\\ 
F_{2}(\eta, \theta , v) &=& \frac{1}{2}H_{\bin} \Big(\frac{1}{2} + \frac{R(\eta)}{2} \sqrt{1 + \cos(u(v) - v) \sin(2 \theta)}\Big) \nonumber\\ 
R(\eta)                 &=& \frac{\sqrt{2}(2 \score - 1)}{ \cos(\eta)}\,. \nonumber
\end{eqnarray}
Here the domain $\M{D}_\score$ is the set 
\begin{eqnarray}
\label{eqn: Domain for one sided rate}
\M{D}_\score &=&  \Big\{ (\eta , \theta , v) : \eta\in[0,\cos^{-1}\big(\sqrt{2}(2 \score - 1)\big)] \nonumber \\ 
& & \hspace{2.5cm}, \theta\in[0,\frac{\pi}{4}-\cos^{-1}(\cos(\eta)/(4\score-2))] , v\in[0,\frac{\pi}{2}]\Big\}.
\end{eqnarray} 
Define a cuboid $\M{C} \supseteq \M{D}_{\score}$ as $[0 , \cos^{-1}\big(\sqrt{2}(2 \score - 1)\big)]\times[0 , \frac{\pi}{4} - \cos^{-1}\big(1/(4\score-2)\big)]\times[0,\frac{\pi}{2}]$. We then partition $\M{C}$ as follows. We take $\{\eta_{i}\}_{i=0}^{N+1}$ to be such that $0 = \eta_0<\eta_1<\eta_2 \cdots <\eta_{N+1} = \cos^{-1} \big( 
\sqrt{2}(2\score-1)\big)$. Similarly define $\{\theta^{(i)}_j\}_{j=0}^{M(i)+1}$ be such that $0=\theta^{(i)}_{0}<\theta^{(i)}_{1}<\cdots<\theta^{(i)}_{M(i)+1}=\frac{\pi}{4} - \cos^{-1}\big(1/(4\score-2)\big)$ and $\{v_{k}^{(i,j)}\}_{k = 0}^{P(i,j)+1} $ be such that $0 = v^{(i,j)}_{0}<v^{(i,j)}_{1}< v^{(i,j)}_{2} \cdots < v^{(i,j)}_{P(i,j)+ 1} = \frac{\pi}{2}$. Thus $\M{C} = \bigcup_{i,j,k}\M{C}_{i,j,k}$, where, to streamline the notation, we have used $\M{C}_{i,j,k}:=\M{C}_{\B{x}_{i, j , k}, \B{x}_{i +1, j +1, k+1}}$ with $\B{x}_{i,j,k} := (\eta_i, \theta^{(i)}_j, v^{(i,j)}_k)$.

From~\eqref{eq:uforms} there are two possible functional forms of $u_{\pm}$. Taking derivatives we find
\begin{eqnarray}
\partial_{\eta} u_{\pm} &=&  \pm 2\nonumber \\ 
\partial_{\theta} u_{\pm} &=& -2 \frac{\cos (2 \theta ) \cos (v)}{\sqrt{1-\sin ^2(2 \theta ) \cos ^2(v)}}  \in [-2 , 0] \label{eq:uderivs} \\ 
\partial_{v}u_{\pm} &=& \frac{\sin (2 \theta ) \sin (v)}{ \sqrt{1-\sin ^2(2 \theta ) \cos ^2(v)}} \in [0 ,1]. \nonumber
\end{eqnarray}

We return to the problem of deriving an upper bound on the functions $F_{1}$ and $F_{2}$. To do so, we first need bounds on the functions $\cos(u_{\pm} \pm v) \sin(2 \theta)$. Our bounds use Taylor's theorem, which we first state for convenience.
\begin{theorem}[Taylor]\label{thm:Taylor}
Let $\mathcal{D}\subseteq\mathbb{R}^n$ be compact and $f:\mathcal{D}\to\mathbb{R}$ be differentiable on $\mathcal{D}$, then for all $\B{ a,x}\in\mathcal{D}$ there exists $\B{ x'}\in\mathcal{D}$ such that
\[f(\B{ x})=f(\B{ a})+\nabla f\big|_{\B{ x'}}\cdot(\B{ x}-\B{ a}).
\]
\end{theorem}
Thus, we can find lower bounds on $f$ in the domain $\M{D}$ by computing $f$ at any point $\B{a} \in \M{D}$ and the upper-bound $\max_{\B{x}' \in \M{D}} \nabla f(\B{x}')$. 
We apply this to the functions  $\cos(u_{\pm}(\B{x} \pm v)) \sin(2\theta)$ in appendix~\ref{appendix: H(A|XYE)}, where we shall the show all the detailed calculations. For brevity, we write $g_{\pm,y}(\B{ x})=u_{\pm}(\B{x})+(-1)^yv$ with $y\in\{0,1\}$. In particular, we show that we have the following bounds for any $\B{x}\in \M{C}_{i,j,k}$:
\begin{eqnarray}
\cos(g_{+,y}(\mathbf{x})) \sin(2 \theta) \leq \zeta^{i,j,k}_{+,y} \text{ and }   \cos(g_{-,y}(\mathbf{x})) \sin(2 \theta) \leq \zeta^{i,j,k}_{-,y} 
\end{eqnarray} 
where, 
\begin{eqnarray}
\zeta^{i,j,k}_{+,y} :=  \begin{cases} 
\big(\cos(g_{+,y}(\B{x}_{i,j,k})) + \Delta_{+,y}  \big) \sin(2 \theta^{(i)}_{j}) 
& \text{ if } \big(\cos(g_{+,y}(\B{x}_{i,j,k})) + \Delta_{+,y}  \big) < 0 \\
\big(\cos(g_{+,y}(\B{x}_{i,j,k})) + \Delta_{+,y}\big) \sin(2 \theta^{(i)}_{j+1}) & 
 \text{ otherwise }
\end{cases} \nonumber 
\end{eqnarray}

\begin{eqnarray}
\zeta^{i,j,k}_{-,y} := \begin{cases}  
\big(\cos(g_{-,y}(\B{x}_{i,j,k})) + \Delta_{-,y}   \big) \sin(2 \theta^{(i)}_{j}) 
& \text{ if } \cos(g_{-,y}(\B{x}_{i,j,k})) + \Delta_{-,y} < 0 \\
\big(\cos(g_{-,y}(\B{x}_{i,j,k})) + \Delta_{-,y}   \big) \sin(2 \theta^{(i)}_{j+1}) & 
 \text{ otherwise }
\end{cases}.\nonumber  
\end{eqnarray}
and 
\begin{align*}
  \Delta_{+,0}&=\max(2(\theta^{(i)}_{j+1}-\theta^{(i)}_j),\ 2(\eta_{i+1}-\eta_i)+2(v_{k+1}^{(i,j)}-v_k^{(i,j)}))\\
\Delta_{+,1}&=\max(2(\eta_{i+1}-\eta_i),\ 2(\theta^{(i)}_{j+1}-\theta^{(i)}_j)+(v_{k+1}^{(i,j)}-v_k^{(i,j)})\\
\Delta_{-,0}&=\max(2(v_{k+1}^{(i,j)}-v_k^{(i,j)}),\ 2(\eta_{i+1}-\eta_i)+2(\theta^{(i)}_{j+1}-\theta^{(i)}_j))\\
  \Delta_{-,1}&=2(\eta_{i+1}-\eta_i)+2(\theta^{(i)}_{j+1}-\theta^{(i)}_j)+(v_{k+1}^{(i,j)}-v_k^{(i,j)}).
\end{align*}

With this established we return to the optimization problem~\eqref{eqn: Optimization problem on grids}. We define the objective function $Q(\eta,\theta,v):=F_1(\eta,\theta,v)+F_2(\eta,\theta,v)+K(R(\eta),\theta)$, which we want to optimize over $\M{D}_{\score}$ and $u\in\{u_+,u_-\}$.

\begin{lemma}
Let $\M{P} = \bigcup_{i,j,k} \M{C}_{i,j,k}$ be a partition of $\M{C}$ as specified above. Define $g_{i,j,k}$ and $h_{i,j,k}$ as follows
\begin{eqnarray}
g_{i,j,k} &:=&  \frac{1}{2}\Phi \left({R(\eta_{i+1})} \sqrt{1 + \zeta_{+,0}^{i,j,k}}\right) + \frac{1}{2} \Phi \left(R(\eta_{i+1}) \sqrt{1 + \zeta^{i,j,k}_{+,1} }\right) + K\left( R(\eta_{i}) , \theta^{(i)}_{j} \right) \nonumber\\ 
    h_{i,j,k} &:=&  \frac{1}{2}\Phi\left(R(\eta_{i+1}) \sqrt{1 + \zeta_{-,0}^{i,j,k} }\right) + \frac{1}{2} \Phi \left( R(\eta_{i+1}) \sqrt{1 + \zeta^{i,j,k}_{-,1} }\right)  + K\left(R(\eta_i) , \theta^{(i)}_{j} \right). \nonumber
\end{eqnarray}
Let $M \in \mathbb{R}$ be any upper bound on $Q$, i.e., $M \geq \max_{\B{x} \in \M{D}} Q(\B{x})$. Then
\begin{equation}\label{eq:def_f}
    Q(\B{x})\geq f_{i,j,k}:=\begin{cases}\min\{g_{i,j,k}, h_{i,j,k}\}&\text{ if }\B{x}\in\M{C}_{i,j,k}\text{ such that }\M{C}_{i,j,k} \cap \M{D} \ne \emptyset\\
 M & \text{ otherwise.}
\end{cases}
\end{equation}
\end{lemma}
\begin{proof}
From Lemmas~\ref{lem: monotonicity of K(R)} and~\ref{lem: monotonicity of K(th)} we know that $\partial_{R} K>0$ and $\partial_{\theta} K>0$. In addition, $\partial_{\eta} K(R(\eta), \theta)  = \frac{\sqrt{2}(2 \score - 1) \sin(\eta )}{ \cos^2(\eta)} \partial_{R} K(R , \theta)$. Positivity of $\partial_{\eta} K$ and $\partial_{\theta} K$, implies  $K( R(\eta) , \theta) \geq K( R(\eta_{i}) , \theta^{(i)}_{j})$ within $\M{C}_{i,j,k}$. Furthermore, $H_{\bin}(\frac{1}{2} + \frac{x}{2})$ is decreasing for $x \geq 0$. Since $R(\eta) \sqrt{1 + \cos(v \pm u) \sin(2 \theta)} >0$,
\begin{align*}
\Phi\left(R(\eta) \sqrt{1 + \cos(u_{+} \pm v) \sin(2 \theta)}\right) &\geq\Phi \left(R(\eta_{i+1}) \sqrt{1 + \cos(u_{+} \pm v) \sin(2 \theta)}\right)\\
&=\Phi\left(R(\eta_{i+1}) \sqrt{1 + \cos(g_{+,(1\mp1)/2}) \sin(2 \theta)}\right)\\
&\geq \Phi\left(R(\eta_{i+1}) \sqrt{1 + \zeta^{i,j,k}_{+,(1\mp1)/2}}\right).
\end{align*}

Similarly,
$$\Phi\left(R(\eta_{i+1}) \sqrt{1 + \cos(u_{-} \pm v) \sin(2 \theta)}\right) \geq \Phi\left(R(\eta_{i+1}) \sqrt{1 + \zeta^{i,j,k}_{-,(1\mp1)/2}}\right),$$
which establishes the claim.
\end{proof}
Combining the results in this section we obtain the following corollary. 
\begin{corollary}
Let $\score \in (\frac{3}{4}, \frac{1}{2} + \frac{1}{2 \sqrt{2}}]$ be fixed. Let $\M{D}_{\score}$ be defined as in \eqref{eqn: Domain for one sided rate} and $\M{P} = \bigcup_{i,j,k} \M{C}_{i,j,k}$ be any partition of the cuboid $\M{C} = [0 , \cos^{-1}\big(\sqrt{2}(2 \score - 1)\big)]\times[0 , \frac{\pi}{4} - \cos^{-1}\big((4 \score -2)^{-1}\big)]\times[0 , \frac{\pi}{2}]$. Then 
\begin{eqnarray}
G_{A|XYE}(\score) \geq \min_{i,j,k} f_{i,j,k}.
\end{eqnarray}
where $f_{i,j,k}$ are defined in~\eqref{eq:def_f}.
\end{corollary}
This means that for fixed $\score$ we can lower bound the randomness by evaluating $f_{i,j,k}$ at all grid points in the relevant cuboid and taking the minimum. This is how our numerical algorithm works (note that the lower bound gets tighter as the number of grid points is increased).

\subsection{Lower bounding the randomness rate}\label{sec: lower bounding the randomness rate}
In the previous section, we derived a technique to lower bound the function $G_{A|XYE}(\score)$ for a fixed value of the score, $\score$. In Section~\ref{app: convex combinations of qubit strategies}, we showed that the asymptotic rate $F_{A|XYE}$ can be computed by taking the convex lower bound on $G_{A|XYE}$. In this section, we construct a lower bound on the function $F_{A|XYE}$ using a lower bound on $G_{A|XYE}$. 

We start with a general lemma.
\begin{lemma}\label{lem:FLB}
Let $a$ and $b$ be real numbers, $a<b$ and $\Tilde{G}:[a,b]\to\mathbb{R}$ be a lower bound on $G:[a,b]\to\mathbb{R}$. Let $\Tilde{F}[a,b]\to\mathbb{R}$ and $F[a,b]\to\mathbb{R}$ be convex lower bounds on $\Tilde{G}$ and $G$ respectively. Then $\Tilde{F}$ is a lower bound on $F$.
\end{lemma}
\begin{proof}
Let $M_{\score_{0}}$ be the set of probability measures on the interval $[a,b]$ satisfying $\int d \mu(\score) \score = \score_0$.
\begin{eqnarray}
F(\score_0) = \inf_{\mu \in M_{\score_0}} \int d \mu(\score) G(\score)  
\end{eqnarray}
Since $G(\score) \geq \Tilde{G}(\score)$ for every value of $\score\in[a,b]$, for every measure $\mu \in M_{\score}$ we must have that $\int d \mu(\score) G(\score) \geq \int d \mu(\score) \Tilde{G}(\score)$. Thus
\begin{equation*}
F(\score_0) := \inf_{\mu \in \score_{0}}\int d \mu(\score) G(\score)  \geq \inf_{\mu \in \score_0} \int d \mu(\score) \Tilde{G}(\score) \geq \Tilde{F}(\score_0).\qedhere
\end{equation*}
\end{proof}

Since we can only compute our lower bound $G^{\M{P}}_{A|XYE}$ on $G_{A|XYE}$ for a finite set of values of $\score$, to form a lower bound that holds for all values of $\score$, we construct a function $\tilde{G}_{A|XYE}$ as follows. Let $\{\score_i\}_{i=1}^N$ be an ordered set of values in $[\frac{3}{4},\frac{1}{2}+\frac{1}{2 \sqrt{2}}]$ with $\score_1=3/4$ at which we have computed $G^{\M{P}}_{A|XYE}$. We define $\tilde{G}_{A|XYE}(\score)$ to be equal to $G^{\M{P}}_{A|XYE}(\score_i)$ for $\score\in[\score_i,\score_{i+1})$, and equal to $G^{\M{P}}_{A|XYE}(\score_N)$ for $\score\geq\score_N$. Because $G_{A|XYE}$ is monotonically increasing in $\score$ (see Lemma~\ref{lemm: Monotonicity of one sided rates}), it follows that for $\score\in[\score_i,\score_{i+1})$, $G_{A|XYE}(\score)\geq G_{A|XYE}(\score_i) \geq G^{\M{P}}_{A|XYE}(\score_i)=\Tilde{G}_{A|XYE}(\score)$.

A lower bound $\Tilde{F}_{A|XYE}$ of $F_{A|XYE}$ can then be formed by taking the convex lower bound of $\Tilde{G}^{\M{P}}_{A|XYE}$ (see\ Lemma~\ref{lem:FLB}).

\section{H(AB|X=0, Y=0, E)}
Recall from the section~\ref{app:AB00E} that 
\begin{eqnarray*}
H(AB|X=0,Y=0,E) &=& 1+H_\bin(2\epsilon_{00})-H(\{\lambda_0,\lambda_1,\lambda_2,\lambda_3\})\, \\ 
                &=& H_{\bin}(2\epsilon_{00}) + K(R,\theta)   . 
\end{eqnarray*}
Thus we need to minimize the above function with respect to the constraint: 
\begin{eqnarray*}
\sum_{ij} \epsilon_{i,j} &=& 2(2\score-1)\,.
\end{eqnarray*}
\subsection{Reparameterizing the optimisation problem}
We introduce some notation for convenience. Let $\mathbf{x} = (R , \theta , \alpha_0 ,\alpha_1 , \beta_0 , \beta_1)$ and define  
\begin{eqnarray}
\hat{\epsilon}_{00}(\mathbf{x})  &:=&  \cos(\theta) \cos(2 \alpha_0 - 2 \beta_0) + \sin(\theta) \cos(2 \alpha_0 + 2 \beta_0) \\ 
\hat{\epsilon}_{10}(\mathbf{x})  &:=&  \cos(\theta) \cos(2 \alpha_1 - 2 \beta_0) + \sin(\theta) \cos(2 \alpha_1 + 2 \beta_0) \\ 
\hat{\epsilon}_{01}(\mathbf{x})  &:=&  \cos(\theta) \cos(2 \alpha_0 - 2 \beta_1) + \sin(\theta) \cos(2 \alpha_0 + 2 \beta_1) \\ 
\hat{\epsilon}_{11}(\mathbf{x})  &:=&  -\cos(\theta) \cos(2 \alpha_1 - 2 \beta_1) - \sin(\theta) \cos(2 \alpha_1 + 2 \beta_1) \\ 
K(\mathbf{x}) &:=& K(R, \theta)\,,
\end{eqnarray}
where $K(R,\theta)$ is given in Corollary~\ref{cor:onesidelower}.
In this notation, the equation for the constraint is 
\begin{eqnarray}\label{eq:epsconstr}
\sum_{ij} \hat{\epsilon}_{i,j} &=& \frac{4(2\score-1)}{R}\,,
\end{eqnarray}
and hence the optimization problem is
\begin{equation}\label{Two-sided optimization}
\begin{aligned}
G_{AB|X=0,Y=0,E}(\score) = \min_{\mathbf{x}\in\cD_\score} \quad & \Big(H_{\bin}\left(\frac{1}{2} + \frac{R}{2} \hat{\epsilon}_{00}(\mathbf{x})\right)  + K(\mathbf{x}) \Big)\\
\textrm{s.t.} \quad & \sum_{ij} \hat{\epsilon}_{ij}(\mathbf{x}) = \frac{4(2\score-1)}{R}\,,
\end{aligned}
\end{equation}
where $\cD_\score=\{R\in[\sqrt{2}(2\score-1),1],\theta\in[0,\pi/4-\cos^{-1}(1/(R\sqrt{2}))],(\alpha_0,\alpha_1,\beta_0,\beta_1)\in\mathbb{R}^4\}$ (see Lemma~\ref{lem:Ag00Escore} for the justification of the range of $R$).

For brevity we use $P(\mathbf{x})$ for the objective function. We call $\mathbf{x}\in\cD_\score$ a solution to the optimization problem~\eqref{Two-sided optimization} if $G_{AB|X=0,Y=0,E}(\score)=P(\mathbf{x})$ and $\mathbf{x}$ satisfies the constraint. For reasons that shall be clear later, we now define the following functions on the extended domain
\begin{eqnarray}
\hat{H}_{\bin}(x) = \begin{cases}
H_{\bin}(x)               & \text{ if } x \in [\frac{1}{2} , 1 ]   \\              1    & \text{ otherwise }
\end{cases}
\end{eqnarray}
and 
\begin{eqnarray}
\hat{K}(R , \theta) = \begin{cases}
K(R , \theta)            & \text{ if } \sqrt{2} (2 \score -1 ) \leq R \leq 1 \text{ and } 0 \leq \theta \leq \frac{\pi}{4} - \cos^{-1} \left(\frac{1}{\sqrt{2}R}\right)    \\               
1    & \text{ otherwise }  
\end{cases}
\end{eqnarray}
Here $\hat{K}(R , \theta)$ and $\hat{H}_{\bin}$ both take the value $1$ when the functions $K(R, \theta)$ and $H_{\bin}(x)$ are outside the stated range. These values are chosen such that upon extension of the domain, the resulting optimization problem still has the same minimum\footnote{That $H_{\bin}(x) \leq 1$ and $K(R, \theta) \leq 1$ whenever defined justifies the choice made for defining $\hat{H}_{\bin}(x)$ and $\hat{K}(R,\theta)$.}.
\begin{lemma}\label{Claim: positive espilon}
Let $\mathbf{X}_\score$ be the set of solutions of~\eqref{Two-sided optimization} for some $\score\in(\frac{3}{4}, \frac{1}{2} + \frac{1}{2 \sqrt{2}}]$. There exists $\mathbf{x} \in \mathbf{X}_\score$ such that 
 $\hat{\epsilon}_{00}(\mathbf{x}) > 0$ and $\displaystyle\hat
 {\epsilon}_{00}(\mathbf{x}) = \max_{i,j} |\epsilon_{ij}(\mathbf{x})|$.
\end{lemma}
\begin{proof}
We first prove that we can choose
\begin{eqnarray}
|\hat{\epsilon}_{00}(\B{x})| = \max_{i,j} |\hat{\epsilon}_{ij}(\B{x})|.
\end{eqnarray}
From the symmetry of the binary entropy, $H_{\bin}(\frac{1}{2} + \frac{y_1}{2}) < H_{\bin}(\frac{1}{2} + \frac{y_2}{2})$ for $|y_1| > |y_2|$. Now consider the following cases
\begin{itemize}
    \item Suppose $|\hat{\epsilon}_{00}(\B{x})| < |\hat{\epsilon}_{10}(\B{x})|$: Perform the transformation $\alpha_0  \leftrightarrow \alpha_1$ and $\beta_1 \rightarrow \beta_1 + \frac{\pi}{2}$. Under this transformation $\hat{\epsilon}_{00}(\B{x}) \leftrightarrow \hat{\epsilon}_{01}(\B{x})$ and $\hat{\epsilon}_{10}(\B{x}) \leftrightarrow \hat{\epsilon}_{11}(\B{x})$. The CHSH score is hence preserved. This transformation also decreases the objective function, so $\B{ x}$ cannot have been an solution to~\eqref{Two-sided optimization} prior to the transformation.
    \item Suppose $|\hat{\epsilon}_{00}(\mathbf{x})| < |\hat{\epsilon}_{01}(\mathbf{x})|$: Perform the transformation $\beta_0  \leftrightarrow \beta_1$ and $\alpha_1 \rightarrow \alpha_1 + \frac{\pi}{2}$. Under this transformation $\hat{\epsilon}_{00}(\B{x}) \leftrightarrow \hat{\epsilon}_{10}(\B{x})$ and $\hat{\epsilon}_{01}(\B{x}) \leftrightarrow \hat{\epsilon}_{11}(\B{x})$. Again, this preserves the CHSH score while reducing the objective function.
    \item Suppose $|\hat{\epsilon}_{00}(\mathbf{x})| < |\hat{\epsilon}_{11}(\mathbf{x})|$: Perform the transformation $\alpha_0 \rightarrow \alpha_1 + \frac{\pi}{2}$ , $\alpha_1 \rightarrow \alpha_0$, $\beta_0 \rightarrow \beta_1$ and $\beta_1 \rightarrow \beta_0 + \frac{\pi}{2}$. Under this transformation $\hat{\epsilon}_{00}(\B{x}) \leftrightarrow \hat{\epsilon}_{11}(\B{x})$ and $\hat{\epsilon}_{01}(\B{x}) \leftrightarrow \hat{\epsilon}_{10}(\B{x})$. Again, this preserves the CHSH score while reducing the objective function.
\end{itemize}
Finally, we can show that $\hat{\epsilon}_{00}(\B{x}) > 0$ by observing that for $\score\in(\frac{3}{4}, \frac{1}{2} + \frac{1}{2 \sqrt{2}}]$ we have
\begin{eqnarray}
   R \sum_{i,j} \hat{\epsilon}_{ij}(\B{x}) &=& 4 (2 \score - 1) > 2.\label{eq:bound1}
   \end{eqnarray}
   In addition, for all $i,j$,
   \begin{eqnarray}
   R \hat{\epsilon}_{ij}(\B{x}) &\leq&R(\cos(\theta)+\sin(\theta))\leq1,\label{eq:bound2}
\end{eqnarray}
where the last inequality follows from~\eqref{param1} and~\eqref{param2} whose sum can be at most $1$.

Now suppose that $|\hat{\epsilon}_{00}(\B{x})|= \displaystyle\max_{i,j} |\hat{\epsilon}_{ij}|$ and $\hat{\epsilon}_{00}(\B{x})<0$. It follows that
\begin{eqnarray}
    R\sum_{i,j} \hat{\epsilon}_{ij} &=& R\big( \hat{\epsilon}_{00} + \hat{\epsilon}_{01} \big) + R\big( \hat{\epsilon}_{10} + \hat{\epsilon}_{11}\big) \nonumber \\ 
    &\leq& R\big( \hat{\epsilon}_{00} + \hat{\epsilon}_{01} \big) + 2 \nonumber \\
    &\leq& 2, \nonumber
\end{eqnarray}
where the first inequality uses~\eqref{eq:bound2}. This is in contradiction with~\eqref{eq:bound1}.
\end{proof}
\begin{lemma}\label{lemm: H(A|X=0,Y=0,E) on grids}
Let $\hat{P}$ be the objective function with extended domain, i.e., $\hat{P}(\mathbf{x}) := \hat{H}_\bin\left( \frac{1}{2} + \frac{R \epsilon_{00}(\B{x})}{2}\right) + \hat{K}(\mathbf{x})$, $\score\in(\frac{3}{4}, \frac{1}{2} + \frac{1}{2 \sqrt{2}}]$, and let $\M{X}$ be a set such that $\M{D}_\score\subseteq\M{X}\subseteq\mathbb{R}^6$. Then,
\begin{eqnarray}\label{Revised two sided}
\begin{aligned}
G_{AB|X=0,Y=0 ,E}(\score ) = \min_{\mathbf{x}\in\M{X}} \quad & \hat{P}(\mathbf{x}) \\
\mathrm{s.t.} \quad & \sum_{ij} \hat{\epsilon}_{ij}(\mathbf{x}) = \frac{4(2 \score - 1 )}{R},
\end{aligned}   
\end{eqnarray}
i.e., optimizing over $\hat{P}$ on an extended domain $\M{X}$ gives the same solution as the original optimization~\eqref{Two-sided optimization}.
Furthermore $\exists \mathbf{x} \in \M{D}_\score$ that is a solution to both optimization problems.
    \end{lemma}
\begin{proof}
Let $\B{x'}\in\M{D}_\score$ achieve the optimal value of $P$ and have $\hat{\epsilon}_{00}(\B{x'})>0$. [From Lemma~\ref{Claim: positive espilon} such an $\B{x'}$ exists.]  Since $\hat{P}(\B{x})=P(\B{x})\leq2$ for all $\B{x}\in\M{D}_\score$, and $\hat{P}(\B{x})=2$ for $\B{x}\in \mathbb{R}^{6}\setminus \M{D}_\score$, $\B{x'}$ must also achieve the optimal value of $\hat{P}$, where it takes the same value.
\end{proof}

\subsection{Some simplifications}
\begin{lemma}\label{lem:signs}
Let $\mathbf{X}_\score$ the set of solutions to the optimization problem~\eqref{Two-sided optimization} for some $\score\in(\frac{3}{4}, \frac{1}{2} + \frac{1}{2 \sqrt{2}}]$. There exists $\mathbf{x} = (R ,\theta , \alpha_0 , \alpha_1 , \beta_0 , \beta_1) \in \mathbf{X}_\score$ such that the following hold
\begin{itemize}
    \item $\sin(\beta_0+\beta_1) \geq 0$
    \item $\sin(\beta_0-\beta_1) \leq 0$
\end{itemize}
\end{lemma}
\begin{proof}
The expression for the CHSH score satisfies:
\begin{align*}
\sqrt{2} (2 \score - 1)=&R\cos(\beta_0 - \beta_1) A_1 +R\sin(\beta_0 - \beta_1) B_1.
\end{align*}
where,
\begin{align*}
A_1 &= \left[\sin(2\alpha_0)\sin(\beta_0 + \beta_1)\cos(\frac{\pi}{4}+\theta)+\cos(2\alpha_0)\cos(\beta_0 + \beta_1)\sin(\frac{\pi}{4}+\theta)\right]    \\ 
B_1 &= \left[\sin(2\alpha_1)\cos(\beta_0+ \beta_1)\cos(\frac{\pi}{4}+\theta)-\cos(2\alpha_1)\sin(\beta_0 + \beta_1)\sin(\frac{\pi}{4}+\theta)\right]
\end{align*}
Let $\alpha_0,\alpha_1, \beta_0, \beta_1 $ be optimal parameters. Consider the following algorithm, in which each step is performed in the order shown and depends on the previous ones.
\begin{enumerate}
    \item If $\sin(\beta_0 + \beta_1 ) <  0$, then perform the transformations $\beta_i \rightarrow -\beta_i$ and $\alpha_i \rightarrow -\alpha_i$. We get $\sin(\beta_0 + \beta_1) \geq 0$ from this step onwards. 
    \item If $\sin(\beta_0 - \beta_1) > 0$ then perform the transformations $\beta_0 \rightarrow \beta_0 + \frac{\pi}{2}$, $\beta_1 \rightarrow \beta_1 - \frac{\pi}{2}$,   $\alpha_i \rightarrow \alpha_i + \frac{\pi}{2}$. This step does not affect $\sin(\beta_0 + \beta_1)$. Thus we ensure that  $\sin(\beta_0 - \beta_1) \leq 0$ and $\sin(\beta_0 + \beta_1) \geq 0$. 
    \end{enumerate}
In each step of the algorithm, the values of $\epsilon_{ij}$ for all $i,j$ remain the same, hence the CHSH score and the objective function remains invariant throughout. Thus, the transformations maintain optimal parameters.
\end{proof}

\subsection{Reduction in parameters}\label{app: two sided proof}
To rewrite the optimization in a way that removes the constraint we introduce the following functions 
\begin{eqnarray}
\hat{\alpha}_{0}(\lambda , v ,\theta) &:=&    -2\tan^{-1} \left( \frac{1}{\tan(\lambda) \tan(\frac{\pi}{4} + \theta)}\right) +  \tan^{-1} \left( \frac{1}{\tan(v) \tan(\frac{\pi}{4} + \theta)}\right) \nonumber\\ 
\tilde{\epsilon}(\lambda, v, \theta) &:=& \cos(\theta) \cos(\hat{\alpha}_0 - 2 v + \lambda  ) + \sin(\theta) \cos(\hat{\alpha}_0 + 2 v - \lambda) \label{two sided function 1} \\ 
z(\lambda , v ,\theta) &=& \cos(\lambda\!-\!v)\!\left[\sin(\hat{\alpha}_0)\sin(v)\cos(\frac{\pi}{4}\!+\!\theta)\!+\!\cos(\hat{\alpha}_0)\cos(v)\sin(\frac{\pi}{4}\!+\!\theta)\right]\! \nonumber\\ 
 & & \hspace{6cm}+\!\frac{\sin(\lambda-v)}{\sqrt{2}} \sqrt{1\!-\!\cos(2 v)\sin(2 \theta)}  \nonumber  \\ 
\hat{R}(\lambda , v , \theta) &:=& \frac {\sqrt{2} (2 \score - 1) }{z(\lambda , v, \theta )}. \label{two sided function 2}
\end{eqnarray}

We also state the following small lemma for convenience.
\begin{lemma}\label{lem:arcsin}
Let $a,b\in\mathbb{R}$ with $a\neq0$. The values of $\gamma\in\mathbb{R}$ that form extrema of $a\cos(\gamma)+b\sin(\gamma)$ are
\begin{equation}
\gamma=\tan^{-1}(b/a)+n\pi
\end{equation}
for any $n\in\mathbb{Z}$.
If $a>0$ the maxima occur when $n$ is even and the minima when $n$ is odd, and vice-versa if $a<0$.
\end{lemma}
\begin{proof}
The problem is equivalent to maximizing
\begin{equation*}
\frac{a}{\sqrt{a^2+b^2}}\cos(\gamma)+\frac{b}{\sqrt{a^2+b^2}}\sin(\gamma).
\end{equation*}
Let $\phi$ satisfy $\cos(\phi)=\left(\frac{a}{\sqrt{a^2+b^2}}\right)$ and $\sin(\phi)=\left(\frac{b}{\sqrt{a^2+b^2}}\right)$. Thus, the expression is equivalent to $\cos(\gamma-\phi)$ which has maxima for $\gamma=\phi+2n\pi$ and minima for $\gamma=\phi+\pi+2n\pi$ for $n\in\mathbb{Z}$.

If $a>0$ then this gives maxima for $\gamma=\tan^{-1}(b/a)+2n\pi$ and minima for $\gamma=\tan^{-1}(b/a)+(2n+1)\pi$.

Alternatively, if $a<0$ then this gives maxima for $\gamma=\tan^{-1}(b/a)+(2n+1)\pi$ and minima for $\gamma=\tan^{-1}(b/a)+2n\pi$.
\end{proof}

\begin{lemma} \label{lem: twosided lower bound}
Let $\score\in(\frac{3}{4}, \frac{1}{2} + \frac{1}{2 \sqrt{2}}]$ and 
$$\M{D}_\score' = \left\{(\lambda , v , \theta) \in \mathbb{R}^{3} : \lambda \in [0 , \pi] , v \in [0 , \pi] , \theta \in [0 , \frac{\pi}{4} - \cos^{-1} \left(1/(4 \score -2) \right) ] \right\},$$ then
\begin{eqnarray}\label{eq:newopt}
G_{AB|X=0, Y=0, E}(\score) =  \inf_{\M{D}_\score'} \left( \hat{H}_{\bin}\left(\frac{1}{2} +\frac{\hat{R}(\lambda,v,\theta)\tilde{\epsilon}(\lambda,v,\theta)}{2}\right) + \hat{K}(\hat{R} , \theta) \right) 
\end{eqnarray}
\end{lemma}
\begin{proof}
Start from the form of $G$ in Lemma~\ref{lemm: H(A|X=0,Y=0,E) on grids}. The objective function $\hat{P}$ is independent of the parameters $\alpha_1$ and $\beta_1$, and, as shown in Lemma~\ref{lemm: H(A|X=0,Y=0,E) on grids}, the optimum is achieved for some $\B{ x}\in\cD_\score$. Because the function $G_{AB|X=0,Y=0,E}(\score)$ is increasing in $\score$ (see Lemma~\ref{lem:monot:w}), the optimal values of the parameters $\alpha_1$ and $\beta_1$ must maximize the CHSH score. Recall that the score can be related to $\alpha_i$ and $\beta_i$ by 
\begin{align}
\sqrt{2} (2 \score - 1) =&R\cos(\beta_0 - \beta_1) A_1 +R\sin(\beta_0 - \beta_1),\label{eq:secn}
  \end{align}
where 
\begin{align*}
    A_1 &= \left[\sin(2\alpha_0)\sin(\beta_0 + \beta_1)\cos(\frac{\pi}{4}+\theta)+\cos(2\alpha_0)\cos(\beta_0 + \beta_0)\sin(\frac{\pi}{4}+\theta)\right] \\ 
    B_1 &= \left[\sin(2\alpha_1)\cos(\beta_0+ \beta_1)\cos(\frac{\pi}{4}+\theta)-\cos(2\alpha_1)\sin(\beta_0 + \beta_1)\sin(\frac{\pi}{4}+\theta)\right].
\end{align*}
Consider maximizing this over $\alpha_1$. From Lemma~\ref{lem:signs} we can assume $\sin(\beta_0-\beta_1)\leq0$, so we want to minimize the second term in square brackets in~\eqref{eq:secn}. This has the form of the expression in Lemma~\ref{lem:arcsin}. Since the sine and cosine of $\pi/4+\theta$ are both positive, and from Lemma~\ref{lem:signs} we can assume $\sin(\beta_0+\beta_1)\geq0$, the minima of the square bracket (and hence maxima overall) occur for 
\begin{equation}\label{eq:alpha1}
2\alpha_1=-\tan^{-1}\left(\cot(\beta_0+\beta_1)\cot(\frac{\pi}{4}+\theta)\right)+2n\pi.
\end{equation}

The CHSH score is symmetric in the parameters for Alice and Bob, so we can re-write it as 
\begin{align*}
\sqrt{2} (2 \score - 1) =&  R\cos(\alpha_0 - \alpha_1) \hat{A}_1 +R\sin(\alpha_0 - \alpha_1) \hat{B}_1
\end{align*}
where, 
\begin{align*}
\hat{A}_1 &= \left[\sin(2\beta_0)\sin(\alpha_0 + \alpha_1)\cos(\frac{\pi}{4}+\theta)+\cos(2\beta_0)\cos(\alpha_0 + \alpha_0)\sin(\frac{\pi}{4}+\theta)\right] \\ 
\hat{B}_1 &= \left[\sin(2\beta_1)\cos(\alpha_0+ \alpha_1)\cos(\frac{\pi}{4}+\theta)-\cos(2\beta_1)\sin(\alpha_0 + \alpha_1)\sin(\frac{\pi}{4}+\theta)\right].
\end{align*}
If we now maximize over $\beta_1$, from Lemma~\ref{lem:signs} the solutions either satisfy 
\begin{align*}
2\beta_1&=-\tan^{-1}\left(\cot(\alpha_0+\alpha_1)\cot(\frac{\pi}{4}+\theta)\right)+2n\pi\quad\text{or}\\
2\beta_1&=-\tan^{-1}\left(\cot(\alpha_0+\alpha_1)\cot(\frac{\pi}{4}+\theta)\right)+(2n+1)\pi
\end{align*}
for $n\in\mathbb{Z}$. (Which one holds depends on the signs of $\sin(\alpha_0-\alpha_1)$ and $\sin(\alpha_0+\alpha_1)$.)  In both cases, $\tan(2\beta_1)=\cot(\alpha_0+\alpha_1)\cot(\frac{\pi}{4}+\theta)$.

By symmetry (and because we can take $\sin(\alpha_0-\alpha_1)\leq0$ and $\sin(\alpha_0+\alpha_1)\geq0$ from Lemma~\ref{lem:signs}) the maxima of this over $\beta_1$ occur for
\begin{equation}
2\beta_1=-\tan^{-1}\left(\cot(\alpha_0+\alpha_1)\cot(\frac{\pi}{4}+\theta)\right)+2n\pi.
\end{equation}
Rearranging gives
\begin{equation}
\tan(\alpha_0+\alpha_1)=-\cot(2\beta_1)\cot(\frac{\pi}{4}+\theta)\,,
\end{equation}
and hence
\begin{eqnarray*}
\alpha_0 &=& -\alpha_1 - \tan^{-1}\left(\cot(2\beta_1) \cot(\frac{\pi}{4}+\theta) \right) + n\pi
\end{eqnarray*}
for $n \in \mathbb{Z}$. Using~\eqref{eq:alpha1} we find
\begin{equation}\label{eq:alpha0}
2\alpha_0=\tan^{-1}\left(\cot(\beta_0+\beta_1)\cot(\frac{\pi}{4}+\theta)\right) - 2\tan^{-1}\left(\cot(2\beta_1) \cot(\frac{\pi}{4}+\theta) \right) + 2n\pi.
\end{equation}

The proof proceeds as follows. We use~\eqref{eq:alpha1} to eliminate $\alpha_1$ from the constraint, noting that the value of $n$ in~\eqref{eq:alpha1} does not change the value so we can take $n=0$. We then use~\eqref{eq:alpha0} to reparameterize the objective function in terms of $\beta_1$ instead of $\alpha_0$ (again the value of $n$ in~\eqref{eq:alpha0} makes no difference and we take $n=0$). The parameters that remain are hence $\beta_0$, $\beta_1$, $R$ and $\theta$. We then reparameterize using $v=\beta_0+\beta_1$ and $\lambda=2\beta_1$, so that both the constraint and objective function are written in terms of $v$, $\lambda$, $R$ and $\theta$. We then use the constraint to write $R$ in terms of the other parameters, reducing the objective function to an unconstrained optimization over $v$, $\lambda$ and $\theta$. 

At this stage $v$ and $\lambda$ range over all reals, which can readily be restricted to $[0,2\pi]$. In fact, we can restrict both to $[0,\pi]$ by noting that Lemma~\ref{lem:signs} shows that it suffices to take $\sin(v)=\sin(\beta_0+\beta_1)\geq0$ hence $v\in[0,\pi]$. We then consider the transformation $(\lambda\mapsto2\pi-\lambda,v\mapsto\pi-v)$. We find
\begin{eqnarray*}
    \hat{\alpha}_0(2\pi-\lambda,\pi-v,\theta)&=&-\hat{\alpha}_0(\lambda,v,\theta)\\
    \tilde{\epsilon}(2\pi-\lambda,\pi-v,\theta)&=&\tilde{\epsilon}(\lambda,v,\theta)\\
    \hat{R}(2\pi-\lambda,\pi-v,\theta)&=&\hat{R}(\lambda,v,\theta)\,,
\end{eqnarray*}
from which it follows that we can restrict both $\lambda$ and $v$ to the range $[0,\pi]$. Finally, the original range of $\theta$ is $[0,\pi/4-\cos^{-1}(1/(R\sqrt{2})]$, with $R\in[\sqrt{2}(2\score-1),1]$, hence the largest $\theta$ that needs to be considered for a given $\score$ is $\pi/4-\cos^{-1}(1/(4\score-2))$. Since we are using the functions with extended domain, it does not matter that we allow the range of $\theta$ to potentially be incompatible with the value of $\hat{R}$. This gives the optimization claimed in~\eqref{eq:newopt}.
\end{proof}

\subsection{Lower bounding the function} 
Consider a partition $\M{P}$ of $\M{D}_\score'$. Let $\M{C}_{i,j,k}$ be a cuboid (with $i$ label corresponding to $\lambda$, $j$ label for $v$ and $k$ label for $\theta$). Again using Taylor's theorem, we bound the objective function 

$$ \left( \hat{H}_{\bin}\left(\frac{1}{2} +\frac{\hat{R}(\lambda,v,\theta)\tilde{\epsilon}(\lambda,v,\theta)}{2}\right) + \hat{K}(\hat{R} , \theta) \right)  $$ 
in the cuboid $\M{C}_{i,j,k}$ , by upper-bounding the absolute values of the derivatives in the cuboid. Let us define $z(\lambda, v, \theta)$ to be the denominator in~\eqref{two sided function 2}, i.e., 
 \begin{eqnarray*}
z(\lambda, v, \theta)\!&:=&\!\cos(v\!-\!\lambda)\!\left[\sin(\hat{\alpha}_0)\sin(v)\cos(\frac{\pi}{4}\!+\!\theta)\!+\!\cos(\hat{\alpha}_0)\cos(v)\sin(\frac{\pi}{4}\!+\!\theta)\right]\! \nonumber\\ 
& & \hspace{4.5cm} - \quad\frac{\sin(v\!-\!\lambda)}{\sqrt{2}}\sqrt{1\!-\!\cos(2 v)\sin(2 \theta)}.
\end{eqnarray*}
In appendix \ref{appendix: H(AB|00E)} we use taylor's theorem and some monotonicity results to find the parameters $z_{\lambda}, z_{v} , z_{\theta}$ and $\epsilon_{\lambda}, \epsilon_{v} , \epsilon_{\theta}$ defined as the following upper-bounds : 
\begin{eqnarray*}
 z_{\lambda} \geq \max_{\B{x} \in \M{C}_{i,j,k}} |\partial_{\lambda} z| ,\quad z_{v} \geq \max_{\B{x} \in \M{C}_{i,j,k}}|\partial_{v} z| \quad\text{and}\quad z_{\theta} \geq  \max_{\B{x} \in \M{C}_{i,j,k}}|\partial_{\theta} z| \\ 
 \epsilon_{\lambda} \geq \max_{\B{x} \in \M{C}_{i,j,k}} |\partial_{\lambda} \tilde{\epsilon}|  , \quad  \epsilon_{v} \geq \max_{\B{x} \in \M{C}_{i,j,k}} |\partial_{v} \tilde{\epsilon}| \quad \text{and} \quad \epsilon_{\theta} \geq \max_{\B{x} \in \M{C}_{i,j,k}} |\partial_{\theta} \tilde{\epsilon}|.
\end{eqnarray*}    
Let $\Delta z = z_{\lambda}(\lambda_{i+1} - \lambda_i) + z_{v}(v^{(i)}_{j+1} - v^{(i)}_j)+z_{\theta}(\theta^{(i,j)}_{k+1} - \theta^{(i,j)}_k)$, then in $\M{C}_{i,j,k}$. 
\begin{eqnarray}
R_{\min}^{i,j,k} &:=& \frac{\sqrt{2}(2 \score -1 )}{z(\lambda_i, v_j^{(i)}, \theta_k^{(i,j)}) + \Delta z } \leq \hat{R}(\lambda , v , \theta) = \frac{\sqrt{2}(2 \score -1 )}{z(\lambda_i, v_j^{(i)}, \theta_k^{(i,j)})  }  \\ 
 R_{\max}^{i,j,k} &:=&  \frac{\sqrt{2} (2 \score - 1)}{z(\lambda_i, v_j^{(i)}, \theta_k^{(i,j)}) - \Delta z} \geq  \hat{R}(\lambda , v , \theta)
\end{eqnarray}
Also let $\Delta \epsilon := \epsilon_{\lambda}(\lambda_{i+1} - \lambda_i) + \epsilon_{v}(v_{j+1}^{(i)} - v_j^{(i)})+\epsilon_{\theta}(\theta_{k+1}^{(i,j)} - \theta_k^{(i,j)})$, then in $\M{C}_{i,j,k}$ we have
\begin{eqnarray}
\tilde{\epsilon}(\lambda,v,\theta)\leq\epsilon_{\max}^{i,j ,k} := \tilde{\epsilon}(\lambda_i, v_j, \theta_k) + \Delta \epsilon  
\end{eqnarray}
For each cuboid we define a continuous function $g_{i,j,k}: \M{C}_{i,j,k} \rightarrow \mathbb{R}$ such that $g_{i,j,k}(\B{x}) \leq \hat{P}(\B{x})$ for all $\B{x} \in \M{C}_{i,j,k}$. Then we lower bound $G_{AB|X=0,Y=0,E}$ by using the following.
\begin{lemma}\label{lemm: Lower bound om H(A|X=0,Y=0,E) on grids}
Let
\begin{eqnarray}\label{eqn: two sided lower bound}
g_{i,j,k} := \hat{H}_{\bin}\left(\frac{1}{2} + \frac{R^{i,j,k}_{\max} \epsilon_{\max}^{i,j,k}}{2}\right) + \hat{K}(R^{i,j,k}_{\min}, \theta_{k}^{(i,j)}).
\end{eqnarray}
Then $\hat{P}(\B{x})\geq g_{i,j,k}$ for all $\B{x}\in\M{C}_{i,j,k}$.
\end{lemma}
\begin{proof}
By definition, we have $R_{\max}^{i,j,k} \epsilon_{\max}^{i,j,k} \geq \hat{R}(\lambda, v, \theta) \tilde{\epsilon}(\lambda, v, \theta)$ for all $\B{x}\in\M{C}_{i,j,k}$. Using the monotonicity of the function $\hat{H}_{\bin}(\frac{1}{2} + \frac{x}{2})$, we obtain $\hat{H}_{\bin}(\frac{1}{2} + \frac{R_{\max}^{i,j,k} \epsilon_{\max}^{i,j,k}}{2}) \leq \hat{H}_{\bin}(\frac{1}{2} + \frac{\hat{R}(\lambda, v, \theta) \tilde{\epsilon}(\lambda, v, \theta)}{2})$. Similarly, the monotonicity of $\hat{K}(R,\theta)$ with respect to $R$ and $\theta$ (see Lemmas~\ref{lem: monotonicity of K(R)} and~\ref{lem: monotonicity of K(th)}) implies $\hat{K}(R(\lambda,v,\theta),\theta)\geq\hat{K}(R_{\min},\theta_{k}^{(i,j)})$ for all $\B{x}\in\M{C}_{i,j,k}$. These imply the claim.
\end{proof} 
\noindent Combining all the results in this section, we have the following
\begin{corollary}
Let $\score \in (\frac{3}{4}, \frac{1}{2} + \frac{1}{2 \sqrt{2}}]$ be fixed. Let $\M{D}_\score' = \{(\lambda, v, \theta) \in \mathbb{R}^{3} : \lambda \in [0, \pi], v \in [0, \pi], \theta \in [0, \frac{\pi}{4} - \cos^{-1} \left(\frac{1}{2(2 \score -1 )} \right) ] \}$ and $\M{P} = \bigcup_{i,j,k} \M{C}_{i,j,k}$ be a partition of any cuboid $\M{C} \supseteq \M{D}'(\score)$ as specified above. Then 
\begin{eqnarray}
G_{AB|X=0,Y=0,E}(\score) \geq \min_{i,j,k} g_{i,j,k}
\end{eqnarray}
where $g_{i,j,k}$ are defined in~\eqref{eqn: two sided lower bound}.
\end{corollary}
\begin{proof}
This is a direct consequence of Lemmas~\ref{lemm: H(A|X=0,Y=0,E) on grids} and~\ref{lemm: Lower bound om H(A|X=0,Y=0,E) on grids}.
\end{proof}

\section{H(AB|XYE)}
From section~\ref{app:ABgXYE}, recall that
\begin{equation}
    H(AB|XYE)=1+\sum_{xy}p_{XY}(x,y)H_\bin(2\epsilon_{xy})-H(\{\lambda_0,\lambda_1,\lambda_2,\lambda_3\})\,.
\end{equation}
So, we set $\delta = \delta^*$, as argued in the section~\ref{app:ABgXYE}, and arrive at the optimization problem: 
\RutC{\begin{equation*}
\begin{aligned}
G_{AB|XYE}(\score) = \min \quad &  \sum_{ij} \frac{1}{4} H_{\bin}(\epsilon_{ij}) + K(R ,\theta)\\
\textrm{s.t.} \quad & \sum_{i,j} \epsilon_{ij} = 2(2 \score - 1)  \\ 
                    &  R \cos(\theta) + R \sin(\theta) \leq  1 
\end{aligned}
\end{equation*}
}
\subsection{Optimization of H(AB|XYE)} 
We introduce some notation for convenience. Let $\mathbf{x} = (R , \theta , \alpha_0 ,\alpha_1 , \beta_0 , \beta_1)$ and define  
\begin{eqnarray*}
\hat{\epsilon}_{00}(\mathbf{x})  &:=& R \cos(\theta) \cos(2 \alpha_0 - 2 \beta_0) + R \sin(\theta) \cos(2 \alpha_0 + 2 \beta_0) \\ 
\hat{\epsilon}_{10}(\mathbf{x})  &:=& R  \cos(\theta) \cos(2 \alpha_1 - 2 \beta_0) + R \sin(\theta) \cos(2 \alpha_1 + 2 \beta_0) \\ 
\hat{\epsilon}_{01}(\mathbf{x})  &:=&  R \cos(\theta) \cos(2 \alpha_0 - 2 \beta_1) + R \sin(\theta) \cos(2 \alpha_0 + 2 \beta_1) \\ 
\hat{\epsilon}_{11}(\mathbf{x})  &:=&  -R \cos(\theta) \cos(2 \alpha_1 - 2 \beta_1) -R \sin(\theta) \cos(2 \alpha_1 + 2 \beta_1) \\ 
K(\mathbf{x}) &:=& K(R, \theta)\,,
\end{eqnarray*}
where $K(R,\theta)$ is given in Corollary~\ref{cor:onesidelower}.
In this notation, the equation for the constraint is 
\begin{eqnarray}
\sum_{ij} \hat{\epsilon}_{i,j} &=& 4(2\score-1)\,.
\end{eqnarray}
Hence we obtain the optimization problem:
\begin{equation}\label{Two-sided recyling optimization}
\begin{aligned}
G_{AB|XYE}(\score) = \min_{\mathbf{x}\in\cD_\score} \quad & \sum_{ij}\frac{1}{4}\Phi\left( \hat{\epsilon}_{ij}(\mathbf{x})\right)  + K(\mathbf{x}) \\
\textrm{s.t.} \quad & \sum_{ij} \hat{\epsilon}_{ij}(\mathbf{x}) = 4(2\score-1)\,,
\end{aligned}
\end{equation}
where, $$\cD_\score=\{R\in[\sqrt{2}(2\score-1),1],\theta\in[0,\pi/4-\cos^{-1}(1/(R\sqrt{2}))],(\alpha_0,\alpha_1,\beta_0,\beta_1)\in\mathbb{R}^4\}$$ (see Lemma~\ref{lem:Ag00Escore} for the justification of the range of $R$). Due to the monotonocity of the function $G_{AB|XYE}(\score)$ (see lemma \ref{lem: monotonicity of recyled inputs protocol}).\\ 
In the following section, we will demonstrate how to obtain reliable lower bounds for $G_{AB|XYE}$ (converging in the asymptotic limit) by considering a sequence of polynomial optimization problems. To initiate this process of finding appropriate lower bounds, we first introduce the following modified optimization problem:
\begin{equation}\label{Two-sided recyling (n) optimization}
\begin{aligned}
G_{AB|XYE}^{(n)}(\score) = \min_{\mathbf{x}\in\cD_\score} \quad & \sum_{ij}\frac{1}{4}\Phi_{n}\left(R \hat{\epsilon}_{ij}(\mathbf{x})\right)  + K(\mathbf{x}) \\
\textrm{s.t.} \quad & \sum_{ij} \hat{\epsilon}_{ij}(\mathbf{x}) \geq 4(2\score-1).\
\end{aligned}
\end{equation}
Here, \(\Phi_{n}(x)\) is a polynomial that serves as a lower bound to the function \(\Phi(x)\) in the range \([-1, 1]\). Conveniently, we have a set of polynomial lower bounds for the function \(\Phi(x)\), discussed in detail during the discussion of rates for semi Device Independent protocols in Section \ref{sec: poly-approximations of Phi(x)}. These functions possess an appealing property, in that they converge to \(\Phi(x)\) from below. Specifically, for all \(x \in [0, 1]\), the following holds:
\[
\lim_{n \rightarrow \infty} \left( \Phi(x) - \Phi_{n}(x) \right) = 0.
\]
If we chose \(\Phi_{n}(x)\) functions to be these special sequence of functions, then we obtain a sequence \(\{ G^{(n)}_{AB|XYE} \}_{n}\) that must converge to the function \(G_{AB|XYE}\) from below.

\subsection{Partitioning the domain}
Let $\M{P}$ be a partition of the set $$ \cD'_{\score} := \{ (R , \theta) \in \mathbb{R}^{2} : \quad R\in[\sqrt{2}(2\score-1),1],\theta\in[0,\pi/4-\cos^{-1}(1/(R\sqrt{2}))] \}. $$ 
Let $\M{C}_{a,b}$ be a cuboid $[R_{a} , R_{a+1}] \times [\theta_{b}^{(a)} , \theta_{b+1}^{(a)} ]$ and $\B{x}_{a,b} := (R_{a} , \theta^{(a)}_{b} )  $. Let $\Delta R := R_{a+1} - R_{a} $ and $\Delta \Theta := \theta^{(a)}_{b+1} - \theta^{(a)}_{b} $. \\ 
Furthermore define
\begin{eqnarray*}
\hat{\epsilon}_{00}^{(a,b)}(\mathbf{x})  &:=&  R_{a} \cos(\theta^{(a)}_{b}) \cos(2 \alpha_0 - 2 \beta_0) + R_{a}\sin(\theta^{(a)}_{b}) \cos(2 \alpha_0 + 2 \beta_0))\\ 
\hat{\epsilon}_{10}^{(a,b)}(\mathbf{x})  &:=& R_{a} \cos(\theta^{(a)}_{b}) \cos(2 \alpha_1 - 2 \beta_0) + R_{a}\sin(\theta^{(a)}_{b}) \cos(2 \alpha_1 + 2 \beta_0) \\ 
\hat{\epsilon}_{01}^{(a,b)}(\mathbf{x})  &:=&  R_{a}\cos(\theta^{(a)}_{b}) \cos(2 \alpha_0 - 2 \beta_1) +  R_{a} \sin(\theta^{(a)}_{b}) \cos(2 \alpha_0 + 2 \beta_1) \\ 
\hat{\epsilon}_{11}^{(a,b)}(\mathbf{x})  &:=&  -R_{a} \cos(\theta^{(a)}_{b}) \cos(2 \alpha_1 - 2 \beta_1) - R_{a} \sin(\theta^{(a)}_{b}) \cos(2 \alpha_1 + 2 \beta_1) 
\end{eqnarray*}
On a cuboid $\M{C}_{a,b}$ define the restriction of the optimization problem \eqref{Two-sided recyling (n) optimization} on the set $\M{C}_{a,b}$ as: 
\begin{eqnarray}\label{eqn: Gridding the two sided optimization problem}
\begin{aligned}
G^{(n)}_{\M{C}_{a,b}}(\score) := \min_{(R, \theta) \in \M{C}_{a,b}} & \sum_{i,j} \Phi_{n}\left( \hat{\epsilon}_{i,j} \right) + K(R, \theta)\nonumber\\
\textrm{s.t.} \quad & \sum_{ij} \hat{\epsilon}_{ij}(\mathbf{x}) \geq 4(2\score-1)\nonumber\\
              \quad & \forall i,j: \alpha_{i} , \beta_{j} \in \mathbb{R}\nonumber. 
\end{aligned}
\end{eqnarray}
We now seek a lower bound $g_{a,b}^{(n)}$ for the function $G^{(n)}_{\M{C}_{a,b}}$, which will allow us to compute a lower bound on the function $G^{(n)}_{AB|XYE}$ using 
Lemma~\ref{lemm: H(A|X=0,Y=0,E) on grids}. The following result finds appropriate lower bounds $g_{a,b}^{(n)}$. 
\begin{lemma}\label{lemm: recyled randomness on grids}
Let $\score \in (\frac{3}{4}, \frac{1}{2} + \frac{1}{2 \sqrt{2}}]$ be fixed, and $\M{P} = \bigcup_{a,b} \M{C}_{a,b}$ be a partition of any cuboid $\M{C} \supseteq \M{D}'(\score)$ as specified above. For any $\M{C}_{a,b}$ in the partition, we define:
\begin{eqnarray}
\begin{aligned}
g_{a,b} := \min_{(R, \theta) \in \M{C}_{a,b}} & \frac{1}{4}\sum_{i,j} \Phi_{n}\left( \hat{\epsilon}_{i,j}^{(a,b)} \right) + K(R_{i} , \theta_{j}^{(i)}) - 2 |\Phi'_{n}(1)|  \sqrt{\Delta R^2 + \Delta \theta^2 }\nonumber\\
\textrm{s.t.} \quad & \sum_{ij} \hat{\epsilon}_{ij}(\mathbf{x}) \geq 4(2\score-1) - 2 \sqrt{(\Delta R)^2 + (\Delta \theta)^2}\nonumber\\
              \quad & \forall i,j : \alpha_{i} , \beta_{j} \in \mathbb{R}\nonumber. 
\end{aligned}
\end{eqnarray}
Then, for every $\M{C}_{a,b}$ in the partition, we have $g_{a,b} \leq G^{(n)}_{\M{C}_{a,b}}$.
\end{lemma}
\begin{proof}
We invoke lemma~\ref{lemm: gazab book-keeping} (see appendix~\ref{app: Two sided rates on grids}) to compute the lower bound on the optimization problem for \eqref{eqn: Gridding the two sided optimization problem}. \\  
Let $\B{x} = (R , \theta)$ and $\B{y} = (\alpha_0 , \alpha_1 , \beta_0 , \beta_1)$ be vectors corresponding to the parameters in our domain. This allows us to compare optimization problem \eqref{eqn: Gridding the two sided optimization problem} to the optimization problem in lemma~\ref{lemm: gazab book-keeping} by identifying functions $g(\B{x} , \B{y}) = \sum_{a,b}\Phi_n(\hat{\epsilon}_{i,j})$ and $h(\B{x}) = K(R, \theta)$. The constraint function $f_1(\B{x} , \B{y}) = \sum_{i,j} \hat{\epsilon}_{i,j}(\B{x} , \B{y} )$. \\
The monotonicity of $\hat{K}(R,\theta)$ with respect to $R$ and $\theta$ (see Lemmas~\ref{lem: monotonicity of K(R)} and~\ref{lem: monotonicity of K(th)}) implies $\hat{K}(R,\theta) \geq\hat{K}(R_{a},\theta_{b}^{(a)})$ for all $\B{x}\in\M{C}_{a,b}$. This inspires us to choose $\B{x}_0 = (R_{a} , \theta^{(a)}_{b})$. \\ 
Now, we find the $\Delta_{\max} , g_{\max}$ and $f_{\max}$. To find $\Delta_{\max}$ note that every $\B{x} \in \M{C}_{a,b}$, the following holds: 
$$ || \Delta_{\B{x}_0}(\B{x}) || = \sqrt{(R - R_{a})^2 + (\theta - \theta^{(a)}_b )^2} \leq \sqrt{(\Delta R)^2 + (\Delta \theta)^2}. $$
Now, we can bound the gradients: 
\begin{eqnarray*}
    |\partial_{R} \hat{\epsilon}_{i,j}| &\leq&  \left( \cos(\theta) + \sin(\theta) \right) \leq \sqrt{2}    \\ 
    |\partial_{\theta} \hat{\epsilon}_{i,j}| &\leq&  R \left( \cos(\theta) + \sin(\theta) \right) \leq \sqrt{2}  \\   
    |\partial_{R} g| &\leq& | \Phi_{n}'(1) | |\partial_{R} \hat{\epsilon}_{i,j}| \leq \sqrt{2}|\Phi_{n}'(1)|   \\ 
    |\partial_{\theta} g| &\leq& | \Phi_{n}'(1) | |\partial_{\theta} \hat{\epsilon}_{i,j}| \leq \sqrt{2}|\Phi_{n}'(1)| ,
\end{eqnarray*}
where we have used the fact that the functions $\max_{x \in [- 1 , 1]} :|\Phi'_{n}(x)| = |\Phi'_{n}(1)| = |\Phi'_{n}(-1)| $. Combining the values of the derivatives gives $g_{\max} = 2 \Phi'(1)$ and $f_{\max} = 2$.
\end{proof}
\noindent Combining all the results in this section, we have the following
\begin{corollary}
Let $\score \in (\frac{3}{4}, \frac{1}{2} + \frac{1}{2 \sqrt{2}}]$ be fixed. Let $\cD'_{\score} := \{ (R , \theta) \in \mathbb{R}^{2} : \quad R\in[\sqrt{2}(2\score-1),1],\theta\in[0,\pi/4-\cos^{-1}(1/(R\sqrt{2}))] \}$ and $\M{P} = \bigcup_{a,b} \M{C}_{a,b}$ be a partition of any cuboid $\M{C} \supseteq \M{D}'(\score)$ as specified above. Then 
\begin{eqnarray}
G_{AB|XYE}(\score) \geq \min_{a,b} g_{a,b},
\end{eqnarray}
where $g_{a , b}$ are defined in Lemma~\ref{lemm: recyled randomness on grids}.
\end{corollary}
\begin{proof}
This is a direct consequence of Lemmas~\ref{lemm: H(A|X=0,Y=0,E) on grids} and~\ref{lemm: recyled randomness on grids}.
\end{proof} 
The primary motivation for going through the various steps and simplifications is to recognize that $g_{a,b}^{(n)}$ can be easily cast to a polynomial optimization problem. This realization stems from the fact that both, the objective function and the constraints involve the sine and cosine of variables $\alpha_0, \alpha_1, \beta_0, \beta_1 \in \mathbb{R}$. Therefore, we introduce new variables $\cos(\alpha_i) = x_{\alpha_i}$ and $\sin(\alpha_i) = y_{\alpha_i}$, along with analogous variables for $\beta_j$. By doing so, both the objective function and the constraints in the \RutC{optimization problem \ref{eqn: Gridding the two sided optimization problem}} can be expressed in terms of polynomials involving $x_{\alpha_i}, y_{\alpha_i}, x_{\beta_j}, y_{\beta_j}$. \RutC{Further, we should} introduce the additional constraints $x_{\alpha_i}^2 + y_{\alpha_i}^2 = 1$ and $x_{\beta_j}^2 + y_{\beta_j}^2 = 1$ to ensure that $x$ and $y$ correspond to cosine and sine of a valid angle. Consequently, we can determine reliable lower bounds for $g_{a,b}$ using the \RutC{SDP based techniques to solve polynomial optimization problems as} discussed in the section~\ref{sec: polysdp}. These lower bounds converge asymptotically to the actual value of $g_{a,b}$. Furthermore, refining the partition leads to a reduction in the size of cuboid dimensions $\Delta R$ and $\Delta \theta$, giving arbitrary tight lower bounds on $G_{AB|XYE}$. 

\section{Monotonicity of rates}\label{app:mono}
In this section we prove the monotonicity of the functions $G_{A|XYE}(\score)$, $G_{AB|00E}(\score)$ and $G_{AB|XYE}(\score)$. There is a common part to the proofs, which we first establish.
\begin{lemma}\label{Monotonicity claim 2}
Let $\lambda_0(R,\theta), \lambda_1(R, \theta) , \lambda_2(R , \theta)$ and $\lambda_3(R, \theta)$ be the eigenvalues of a Bell-diagonal state $\rho_{A'B'}$ as in~\eqref{param1}--\eqref{param4} in the case where $\delta=\frac{R^2}{4}\cos(2\theta)$. Then
\begin{eqnarray}
\frac{\partial}{\partial R} \left(H_{\bin}(\lambda_0 + \lambda_1) - H(\{\lambda_0,\lambda_1,\lambda_2,\lambda_3\}) \right) > 0. 
\end{eqnarray}
\end{lemma}
\begin{proof}
\begin{eqnarray*}
\frac{\partial}{\partial R} \Big( H_{\bin}(\lambda_0 + \lambda_1) \Big) &=& -\log(\lambda_0 + \lambda_1) \frac{\partial}{\partial R} (\lambda_0 + \lambda_1)  -\log(\lambda_2 + \lambda_3) \frac{\partial}{\partial R} (\lambda_2 + \lambda_3) 
\end{eqnarray*}
The equality above follows from the fact that $1 - \lambda_1 - \lambda_0 = \lambda_2 + \lambda_3$ and thus $H_{\bin}(\lambda_0+ \lambda_1) = -(\lambda_0 + \lambda_1) \log(\lambda_1 + \lambda_0) - (\lambda_2 + \lambda_3) \log(\lambda_2 + \lambda_3)$. We also have that
\begin{eqnarray}
\frac{\partial }{ \partial R} H(\{\lambda_0,\lambda_1,\lambda_2,\lambda_3\}) =- \sum_i\log{\lambda_i}\frac{\partial \lambda_i}{\partial R} .
\end{eqnarray}
For convenience, we write $$G = H_{\bin}(\lambda_0 + \lambda_1) - H(\{\lambda_0,\lambda_1,\lambda_2,\lambda_3\}) .$$
Adding the derivatives, we define 
\begin{eqnarray}
\frac{\partial}{\partial R} \Big( G\Big)  &=& \log\left(\frac{\lambda_0}{\lambda_0 + \lambda_1}\right) \frac{\partial \lambda_0}{\partial R} + \log\left(\frac{\lambda_1}{\lambda_0 + \lambda_1}\right) \frac{\partial \lambda_1}{\partial R}  \nonumber \\& & + \log\left(\frac{\lambda_2}{\lambda_3 + \lambda_2}\right) \frac{\partial \lambda_2}{\partial R} + \log\left(\frac{\lambda_3}{\lambda_2 + \lambda_3}\right) \frac{\partial \lambda_3}{\partial R}  \nonumber\\ 
&=& \log_{2}\Big( \frac{\lambda_0}{\lambda_0 + \lambda_1}\Big) \frac{\partial}{\partial R}(\lambda_0 + \lambda_2) + \log_{2}\Big( \frac{\lambda_1}{\lambda_0 + \lambda_1}\Big) \frac{\partial}{\partial R}(\lambda_1 + \lambda_3) \nonumber  \\  
&=& \log\left(\frac{\lambda_0}{\lambda_1}\right)\frac{\partial}{\partial R}(\lambda_0 + \lambda_2)=\log\left(\frac{\lambda_0}{\lambda_1}\right)\frac{\cos(\theta)-\sin(\theta)}{2}\nonumber\\
&\geq& 0 ,
\end{eqnarray}
where the second equality follows from the fact that for Bell-diagonal states parameterized by $\delta = \frac{R^2}{4}\cos(2\theta)$, the eigenvalues obey 
\begin{eqnarray*}
\frac{\lambda_0}{\lambda_0 + \lambda_1} = \frac{\lambda_2}{\lambda_2 + \lambda_3} \quad\text{and}\quad
\frac{\lambda_1}{\lambda_0 + \lambda_1} = \frac{\lambda_3}{\lambda_2 + \lambda_3}
\end{eqnarray*}
and the inequality comes from the parameterization.
\end{proof}

\begin{lemma}\label{lemm: Monotonicity of one sided rates}
For $\score\in(\frac{3}{4},\frac{1}{2}(1+\frac{1}{\sqrt{2}}))$ and any distribution $p_{XY}$, the function $G_{A|XYE}(\omega,p_{XY})$ is increasing in $\score$.
\end{lemma}

\begin{proof}
Let us fix the score $\omega$. From the analysis in section~\ref{app:AgXYElb} we know that the optimum value of $\delta$ is $\frac{R^2}{4}\cos(2\theta)$. Throughout this proof we take $\delta=\frac{R^2}{4}\cos(2\theta)$ and consider $\rho_{A'B'}$ to depend on two parameters $R$ and $\theta$. Let $(\M{N}^*, \rho^*) \equiv \rho(R^*, \theta^*)$ be the channel and state that that solves the optimization problem for $G_{A|XYE}(\score,p_{XY})$, i.e., such that $G_{A|XYE}(\score,p_{XY})=H(A|XYE)_{(\cN^*\ot\cI_E)(\rho_{A'B'E}(R^*,\theta^*))}$.
It suffices to show that there exists a curve $\sigma: [-1 , 0] \mapsto \M{S}(\M{H}_{A'}\ot\M{H}_{B'}\ot\M{H}_E)$, such that
\begin{enumerate}
    \item $\sigma(0) = \rho^*$ 
    \item $g(t) := H(A|XYE)_{(\M{N}^*\ot\cI_E)(\sigma(t))}$ is differentiable for all $t \in [-1,0]$. 
    \item $\left.\frac{\dd g(t)}{\dd t}\right|_{t=0} > 0$ 
    \item $\forall t: \frac{\dd}{\dd t}\scorefunction\left((\M{N}^*\ot\cI_E)(\sigma(t))\right)>0$.
\end{enumerate}
Then, if 1--4 hold, using the fact that $g(t)$ is continuous and has a positive derivative at $t = 0$, there exists $t_0 < 0$ such that for $t\in(t_0,0)$, $g(t) < g(0)$. Since the $\scorefunction\left((\M{N}^*\ot\cI_E)(\sigma(t))\right)$ is continuous function, we must have that for any $t\in(t_0,0)$
\begin{eqnarray}
H(A|X Y  E)_{(\M{N}^*\ot\cI_E)(\rho_{A'B'E}^*)} &>& H(A|X Y E)_{(\M{N}^*\ot\cI_E)(\sigma(t))} \\ 
&\geq& G_{A|XYE}\left(\scorefunction((\M{N}^*\ot\cI_E)(\sigma(t)),p_{XY}\right).
\end{eqnarray}
Since $\scorefunction((\M{N}^*\ot\cI_E)(\sigma(t))<\score$ this establishes the claim.

It remains to show that there exists a function $\sigma(t)$ such that 1--4 hold. Recall from section~\ref{app:AgXYElb} that we can write
\begin{eqnarray}\label{Just above}
H(A|XYE)=1+\sum_{x \in \{0 , 1\}}p_X(x)H_{\bin}(g(\theta,\alpha_x)) -H(\{\lambda_0,\lambda_1,\lambda_2,\lambda_3\}) ,
\end{eqnarray}
where $g(\theta, \alpha) := \frac{1}{2}\left(1+R\sqrt{1+\sin(2\theta)\cos(4\alpha)}\right)$. We then set
\begin{eqnarray}
\sigma(t)=\rho(R^*+\kappa t,\theta^*)
\end{eqnarray}
for some positive number $\kappa$ such that $R^*-\kappa>3/4$. Thus, $\sigma(0)=\rho^*$ and differentiability of $g(t)$ can be shown using the form~\eqref{Just above}. We compute the $t$ derivative:
\begin{eqnarray}
\left.\frac{\dd g(t)}{\dd t}\right|_{t=0}=\left.\kappa\frac{\partial}{\partial R} \Big(H(A|X Y E)_{(\M{N^*}\ot\cI_E)(\rho_{A'B'E}(R,\theta))} \Big) \right|_{R=R^*,\theta=\theta^*} .
\end{eqnarray}
Note that 
\begin{eqnarray*}
\frac{\partial}{\partial R} H_{\bin} (g(\theta, \alpha)) &=&  H_{\bin}'(g(\theta, \alpha)) \frac{\sqrt{1 + \sin(2 \theta) \cos(4\alpha)}}{2}    \\
&\geq& H_{\bin}'\Big(\frac{1}{2} + \frac{R}{2} (\cos(\theta) + \sin(\theta)) \Big) \frac{\cos(\theta) + \sin(\theta)}{2} \\
&=& H_{\bin}'(\lambda_0 + \lambda_1) \frac{\partial}{\partial R}(\lambda_0 + \lambda_1)\\
&=& \frac{\partial}{\partial R}H_{\bin}(\lambda_0+\lambda_1)\,,
\end{eqnarray*} 
where we have used that $H'_\bin(p)$ is decreasing in $p$ for $p>1/2$, so we take $\alpha=0$ to obtain a bound. It follows that 
\begin{eqnarray*}
\left.\frac{\dd g(t)}{\dd t}\right|_{t=0}=\kappa\frac{\partial}{\partial R} \Big(H_{\bin}(\lambda_0 + \lambda_1) - H(\{\lambda_0,\lambda_1,\lambda_2,\lambda_3\}) \Big) > 0\,,
\end{eqnarray*}
where the inequality is Lemma~\ref{Monotonicity claim 2}.

Finally, the function 
$$\scorefunction((\M{N}^*\ot\cI_E)(\sigma(t))=\frac{1}{2}\sum_{i,j} \epsilon_{ij}$$
increases linearly with $t$ (the score is linear in $R$).
\end{proof}

\begin{lemma}\label{lem:monot:w}
For $\score\in(\frac{3}{4},\frac{1}{2}(1+\frac{1}{\sqrt{2}}))$, the function $G_{AB|X=0,Y=0,E}(\score)$ is increasing in $\score$.
\end{lemma}
\begin{proof}
The proof follows the same lines as the previous lemma but with the entropy changed. From section~\ref{app:AB00E} we have
$$H(AB|X=0,Y=0,E)=1+H_\bin(2\epsilon_{00})-H(\{\lambda_0,\lambda_1,\lambda_2,\lambda_3\}).$$
We have
\begin{eqnarray}
\frac{\partial}{\partial R}  H_{\bin} (2\epsilon_{00})&=&H'_{\bin}(2\epsilon_{00})\frac{\cos(\theta)\cos(2(\alpha_0-\beta_0))+\sin(\theta)\cos(2(\alpha_0+\beta_0))}{2}\nonumber\\
&\geq&H'_{\bin}\left(\frac{1}{2}+\frac{R}{2}(\cos(\theta)+\sin(\theta))\right)\frac{\cos(\theta)+\sin(\theta)}{2}\label{eq:eps00}
\end{eqnarray}
and the remainder of the argument matches the previous proof.
\end{proof}

\begin{lemma}\label{lem: monotonicity of recyled inputs protocol}
For $\score\in(\frac{3}{4},\frac{1}{2}(1+\frac{1}{\sqrt{2}}))$ and any distribution $p_{XY}$, the function $G_{AB|XYE}(\score,p_{XY})$ is increasing in $\score$.
\end{lemma}
\begin{proof}
The proof for this again follows those above, except in this case (see section~\ref{app:ABgXYE})
\begin{eqnarray}
H(AB|XYE) &=& 1 + \sum_{xy} p_{XY}(x,y) H_{\bin}(2\epsilon_{xy}) - H(\{\lambda_0,\lambda_1,\lambda_2,\lambda_3\})\,.
\end{eqnarray}
The bound that holds for $\epsilon_{00}$ in~\eqref{eq:eps00} holds for all $\epsilon_{xy}$, and hence the rest of the argument goes through as before.
\end{proof}
\section{Results for the lower bounds}
\begin{figure}[h!]
\includegraphics[width=\textwidth]{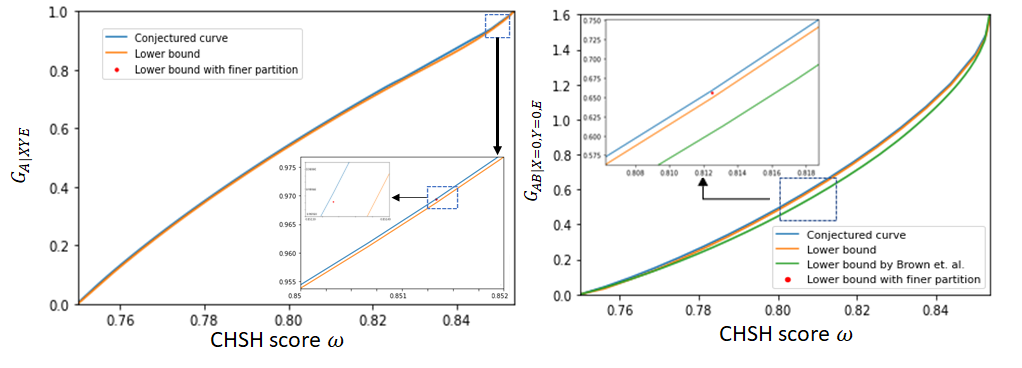}\\
\phantom{m}\hspace{7em}(a)\hspace{18em}(b)
\caption[Graphs of the conjectured rates and lower bounds for various DIRNE protocols assuming that Alice and Bob share qubits.]{Graphs of the conjectured rates and lower bounds for (a) $G_{A|XYE}$ (b) $G_{AB|X=0,Y=0,E}$  with uniformly chosen inputs. For $G_{AB|X=0,Y=0,E}$ we also show a lower bound from Brown et al.~\cite{BFF2022}. We also demonstrate that the lower bound for $G_{AB|X=0,Y=0,E}$ can be tightened by refining the partitioning of the domain for a specific point (due to the increased computation time, we did not do this throughout).}
\label{fig:rates_conj}
\end{figure}

\begin{figure}[h!]
\centering
\includegraphics[width = 0.5\textwidth]{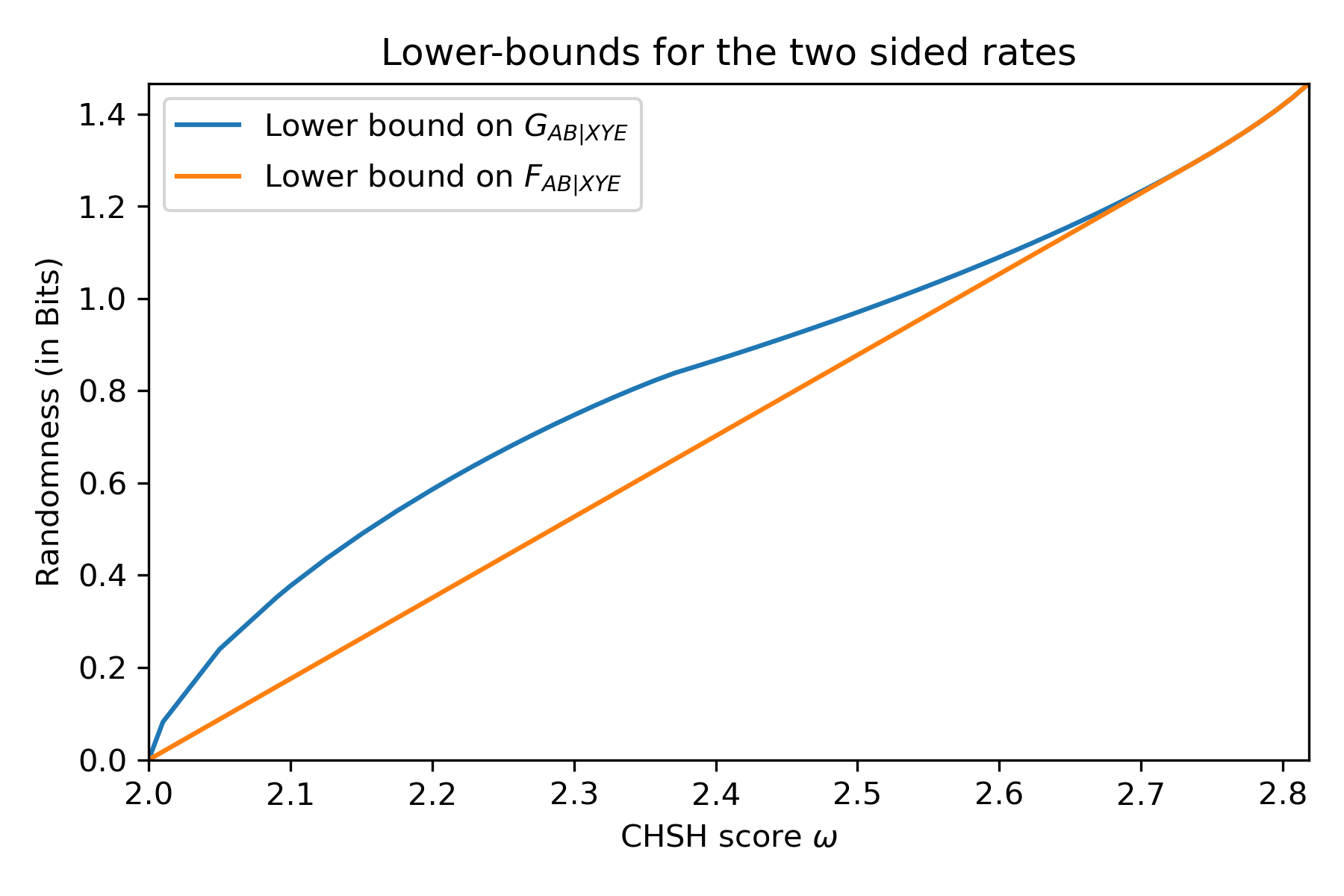}\\
\caption[Graphs for lower bounds on randomness rates for protocols that recycles input randomness.]{Graphs for lower bounds on $G_{AB|XYE}(\score)$ and $F_{AB|XYE}(\score)$  }
\label{fig:HABXYE}
\end{figure}

Lower bounds generated in this way are shown in Fig.~\ref{fig:rates_conj}, and can be seen to be close to the upper bounds. In Fig.~\ref{fig:rates_conj}(b) we also compare with a lower bound on $G_{AB|X=0,Y=0,E}$ from~\cite{BFF2022}. The lower bounds from our technique can be improved by refining the partition of the domain at the expense of increasing the computational time required. As seen in Fig.~\ref{fig:rates_conj}(b), refining the partition moves the lower bound closer to the upper bound. 
The lower bounds for $F_{AB|XYE}$ can be seen in Figure~\ref{fig:HABXYE}
Our numerical evidence suggests that the upper-bounds generated in the previous chapters are tight, as the lower bounds appear to converge to the upper bounds. So, in the remainder of the work, we use upper bounds for computing rates for protocols. 
\begin{conjecture}
The upper bounds in Lemmas~\ref{lem:ABgXYE} and~\ref{lem:AgXYE} are tight.
\end{conjecture}
\chapter{Results and discussion}\label{chap : actual protocols for DI}
In this chapter, we discuss CHSH-based protocols for \RutC{DIRNE} of both the spot-checking and non spot-checking types. We pick specific protocols for concreteness, but there are many possible variations. For instance, the protocols we discuss condense the observed statistics to a single score, but this is not necessary, and in some cases and for some sets of experimental conditions it can be advantageous to use multiple scores~\cite{BRC,TSGPL}.

Before getting to the protocols, we first describe the setup, assumptions and security definition. Although \RutC{DIRNE} requires no assumptions on how the devices used operate, the setup for \RutC{DIRNE} involves a user who performs the protocol within a secure laboratory, from which information cannot leak. Individual devices can also be isolated within their own sub-laboratory and the user can ensure that these devices only learn the information necessary for the protocol (in particular, they cannot learn any inputs given to other devices). The user has access to a trusted classical computer and an initial source (or sources) of trusted randomness.

The quantum devices used for the protocol are only limited by the laws of quantum theory and may share arbitrary entanglement with each other and with an adversary. However, they cannot communicate with each other, or to the adversary after the protocol starts. Furthermore, we assume they are kept isolated after the protocol (see\ the discussion in Appendix~\ref{app:compos}).

For security of the protocols, we use a composable security definition. Consider a protocol with output $Z$ and use $\Omega$ to denote the event that it does not abort. The protocol is $(\epsilon_S,\epsilon_C)$-secure if
\begin{enumerate}
\item $\frac{1}{2}p_{\Omega} || \rho_{ZE|\Omega} - \frac{1}{d_Z}\id_Z \ot \rho_{E|\Omega} ||_1 \leq \epsilon_S$,
  where $E$ represents all the systems held by an adversary and $d_Z$ is the dimension of system $Z$; and
  \item There exists a quantum strategy such that $p_{\Omega}\geq1-\epsilon_C$.
\end{enumerate}
Here $\epsilon_S$ is called the soundness error, and $\epsilon_C$ is the completeness error.

\section{CHSH-based spot-checking protocol for randomness expansion}
We now describe a spot-checking protocol for randomness expansion. It uses a central biased random number generator $R_T$ and two other random number generators, $R_A$ and $R_B$ that are near each of the devices used to run the protocol.

\begin{protocol}\label{prot:spotcheck}  \textbf{(Spot-checking protocol)}
\phantom{a}\\

\noindent\textbf{Parameters:}\\
$n$ -- number of rounds\\
$\gamma$ -- test probability\\
$\score_{\exp}$ -- expected CHSH score\\
$\delta$ -- confidence width for the score
\begin{enumerate}
    \item Set $i=1$ for the first round, or increase $i$ by 1.
    \item Use $R_T$ to choose $T_i\in\{0,1\}$ where $T_i=1$ occurs with probability $\gamma$.
    \item If $T_i=1$ (test round), $R_A$ is used to choose $X_i$ uniformly, which is input to one device giving output $A_i$. Likewise $R_B$ is used to choose $Y_i$ uniformly, which is input to the other device giving output $B_i$. Set $U_i=1$ if $A_i\oplus B_i=X_iY_i$ and $U_i=0$ otherwise.
    \item If $T_i=0$ (generation round), the devices are given inputs $X_i=Y_i=0$, and return the outputs $A_i$ and $B_i$. Set $U_i=\bot$.
    \item Return to Step~1 unless $i=n$.
    \item Calculate the number of rounds in which $U_i=0$ occurred, and abort the protocol if this is larger than $n\gamma(1-\score_{\exp}+\delta)$.
    \item \label{st:7} Process the concatenation of all the outputs with a quantum-proof strong extractor $\Ext$ to yield $\Ext(\B{ AB},\B{ R})$, where $\B{ R}$ is a random seed for the extractor. Since a strong extractor is used, the final outcome can be taken to be the concatenation of $\B{ R}$ and $\Ext(\B{ AB},\B{ R})$ \RutC{(see Section \ref{sec: min-entropy} for details of randomness extraction)}.
\end{enumerate}
\end{protocol}

There are a few important points to take into account when running the protocol. Firstly, it is crucial that each device only learns its own input and not the value of the other input, or of $T_i$. If this is not satisfied it is easy for devices to pass the protocol without generating randomness. Secondly, for implementations in which devices can fail to record outcomes when they should, it is important to close the detection loophole, which can be done by assigning an outcome, say $0$, when a device fails to make a detection.

In order to run the protocol, some initial randomness is needed to choose which rounds are test rounds, to choose the inputs in the test rounds and to seed the extractor. Since the extractor randomness forms part of the final output, it is not consumed in the protocol, so for considering the rate at which the protocol consumes randomness we can work out the amount of uniform randomness needed to supply the inputs. Using the rounded interval algorithm~\cite{HaoHoshi} to make the biased random number generator, $n(H_\bin(\gamma)+2\gamma)+3$ is the expected amount of input randomness required. To achieve expansion, the number of output bits must be greater than this. We use the entropy accumulation theorem (EAT) to lower bound the amount of output randomness. Asymptotically the relevant quantity is $H(AB|X=0,Y=0,E)$. The quantity $H(A|X=0,Y=0,E)$ acts as a lower bound for this, and can be used in its place if convenient, for instance in analyses that are more straightforward with an analytic curve.

\section{CHSH-based protocols without spot-checking}
In this section we discuss two protocols which do not require spot checking. Protocol~\ref{prot:biased} uses two biased local random number generators to choose the inputs on each round. Protocol~\ref{prot:nonspotcheck} eliminates the bias, but also recycles the input randomness. Recycling the input randomness is necessary when unbiased random number generators are used, since otherwise more randomness is required to run the protocol than is generated. Protocol~\ref{prot:nonspotcheck} gives the highest randomness generation rate of all the protocols we discuss.

\begin{protocol}\label{prot:biased} \textbf{ (Protocol with biased local random number generators)}
\phantom{a}\\

\noindent\textbf{Parameters}:\\
$n$ -- number of rounds \\
$\zeta^A$ -- probability of 1 for random number generator $R_A$ (taken to be below $1/2$)\\ 
$\zeta^B$ -- probability of 1 for random number generator $R_B$ (taken to be below $1/2$)\\ 
$\omega_{\text{exp}}$ -- expected CHSH score. \\
$\delta$ -- confidence widths for each score. 
\begin{enumerate}
    \item Set $i=1$ for the first round, or increase $i$ by 1.
    \item Use $R_A$ to choose $X_i\in\{0,1\}$, which is input to one of the devices giving output $A_i\in\{0,1\}$. Likewise use $R_B$ to generate $Y_i\in\{0,1\}$, which is input to the other device giving output $B_i\in\{0,1\}$. Here $X_i=1$ occurs with probability $\zeta^A$ and $Y_i=1$ occurs with probability $\zeta^B$. Set $U_i=(X_i,Y_i,1)$ if $A_i\oplus B_i=X_iY_i$ and $U_i=(X_i,Y_i,0)$ otherwise.
    \item Return to Step~1 unless $i=n$.
    \item \label{st:omega} Compute the value
    \begin{eqnarray}\label{eq:omega}
    \score=\frac{1}{4}\sum_{x,y} \frac{|\{i:U_i=(x,y,1)\}|}{n p_{X}(x) p_Y(y)}
    \end{eqnarray}
    and abort the protocol if $\score<\score_{\exp}-\delta$.
    Here $p_{X}(1)=\zeta^A$, $p_{X}(0)=1-\zeta^A$, $p_{Y}(1)=\zeta^{B}$ and $p_{Y}(0)=1-\zeta^{B}$.
    \item Process the concatenation of all the outputs with a quantum-proof strong extractor $\Ext$ to yield $\Ext(\B{ AB},\B{ R})$, where $\B{ R}$ is a random seed for the extractor. Since a strong extractor is used, the final outcome can be taken to be the concatenation of $\B{ R}$ and $\Ext(\B{ AB},\B{ R})$ \RutC{(see Section \ref{sec: min-entropy} for details)}.
\end{enumerate}
\end{protocol}

Note that the quantity $|\{i:U_i=(x,y,1)\}|/(n p_{X}(x) p_Y(y))$ in~\eqref{eq:omega} is an estimate of the probability of winning the CHSH game for inputs $X=x$ and $Y=y$, and hence the $\score$ computed in Step~\ref{st:omega} is an estimate of the CHSH value that would be observed if the same setup was used but with $X$ and $Y$ chosen uniformly.

The input randomness required per round in this protocol is roughly $H_\bin(\zeta^A)+H_\bin(\zeta^B)$.
To quantify the amount of output randomness (before randomness extraction is performed), in the asymptotic limit similar to the spot checking protocol, the relevant operational quantity is the von Neumann entropy $H(AB|XYE)$. Expansion hence cannot be achieved if $H(AB|XYE)-H_\bin(\zeta^A)-H_\bin(\zeta^B)<0$, which places constraints on the pairs of possible $(\zeta^A,\zeta^B)$. For $\zeta^A$ and $\zeta^B$ smaller than $1/2$, the quantity $H(AB|XYE)-H_\bin(\zeta^A)-H_\bin(\zeta^B)$ increases as $\zeta^A$ and $\zeta^B$ decrease, and hence we want to take these to be small. They only need to be large enough to ensure that $X=1,Y=1$ occurs often enough to give a good estimate of the empirical score.

Since
\begin{eqnarray}
H(AB|XYE) &=&  \sum_{xy} p_{XY}(x,y)H(AB|XYE) \nonumber\\
&\geq&\min_{x,y}H(AB|X=x,Y=y,E) ,  \label{Bakchodi} 
\end{eqnarray}
we can use the bounds formed for $H(AB|X=0,Y=0,E)$ instead, albeit with a loss of entropy (this loss of entropy is small if $\zeta_A$ and $\zeta_B$ are small)\footnote{\RutC{There is nothing special about the choice $X=0$ and $Y=0$ when computing the bounds for $H(AB|X=0,Y=0,E)$.}}.

One reason for using Protocol~\ref{prot:biased} rather than Protocol~\ref{prot:spotcheck} is that the former enables the locality loophole to be closed while expanding randomness. In order to perform the Bell tests as part of a device-independent protocol we need to make inputs to two devices in such a way that neither device knows the input of the other. One way to ensure this is by using independent random number generators on each side of the experiment, and ensuring the outcome of each device is given at space-like separation from the production of the random input to the other. Although space-like separation can provide a guarantee (within the laws of physics) that each device does not know the input of the other\footnote{Provided we have a reasonable way to give a time before which the output of $R_A$ and $R_B$ did not exist.}, in a cryptographic setting it is necessary to assume a secure laboratory to prevent any unwanted information leaking from inside the lab to an eavesdropper. The same mechanism by which the lab is shielded from the outside world can be used to shield devices in the lab from one another and hence can prevent communication between the two devices during the protocol. However, although unnecessary for cryptographic purposes, it is interesting to consider closing the locality loophole while expanding randomness.

This is not possible in a typical spot-checking protocol, where a central random number generator is used to decide whether a round is a test round or not. Considering Protocol~\ref{prot:spotcheck}, the locality loophole can be readily closed during the test rounds, but the use of the central random number generator means that, if one is worried that hidden communication channels are being exploited, there is a loophole that the devices could behave differently on test rounds and generation rounds. For instance, measurement devices that know whether a round is a test or generation round could supply pre-programmed outputs in generation rounds, while behaving honestly in test rounds. Thus, spot-checking protocols do not enable fully closing the locality loophole while expanding randomness.

\begin{figure}
\centering
\includegraphics[width= 0.6\textwidth]{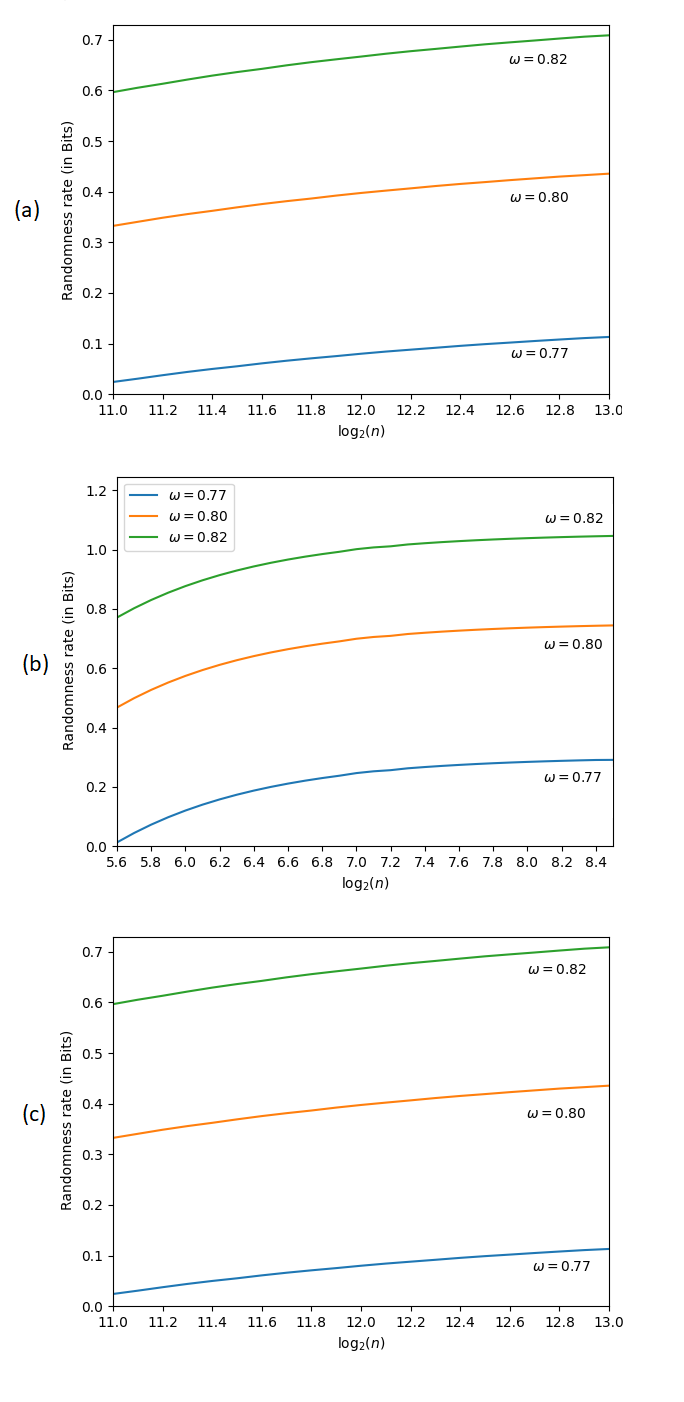}\\
\caption[Graphs of the net rate of certifiable randomness according to the EAT for (a) the spot checking protocol, (b) the protocol with recycled input randomness, and (c) the protocol with biased local random number generators, showing the variation with the number of rounds for three different scores.]{Graphs of the net rate of certifiable randomness according to the EAT for (a) the spot checking protocol (Protocol~\ref{prot:spotcheck}), (b) the protocol with recycled input randomness (Protocol~\ref{prot:nonspotcheck}), and (c) the protocol with biased local random number generators (Protocol~\ref{prot:biased}), showing the variation with the number of rounds for three different scores, $\score$. The error parameters used were $\epsilon_S=3.09\times10^{-12}$ and $\epsilon_C=10^{-6}$. For each point on the curve (a) an optimization over $\gamma$ was performed to maximize the randomness; similarly, the values of $\zeta^A=\zeta^B$ were optimized over to generate the curves in (c).}
\label{fig:EAT_n}
\end{figure}

\begin{figure}
\centering
\includegraphics[width= 0.6\textwidth]{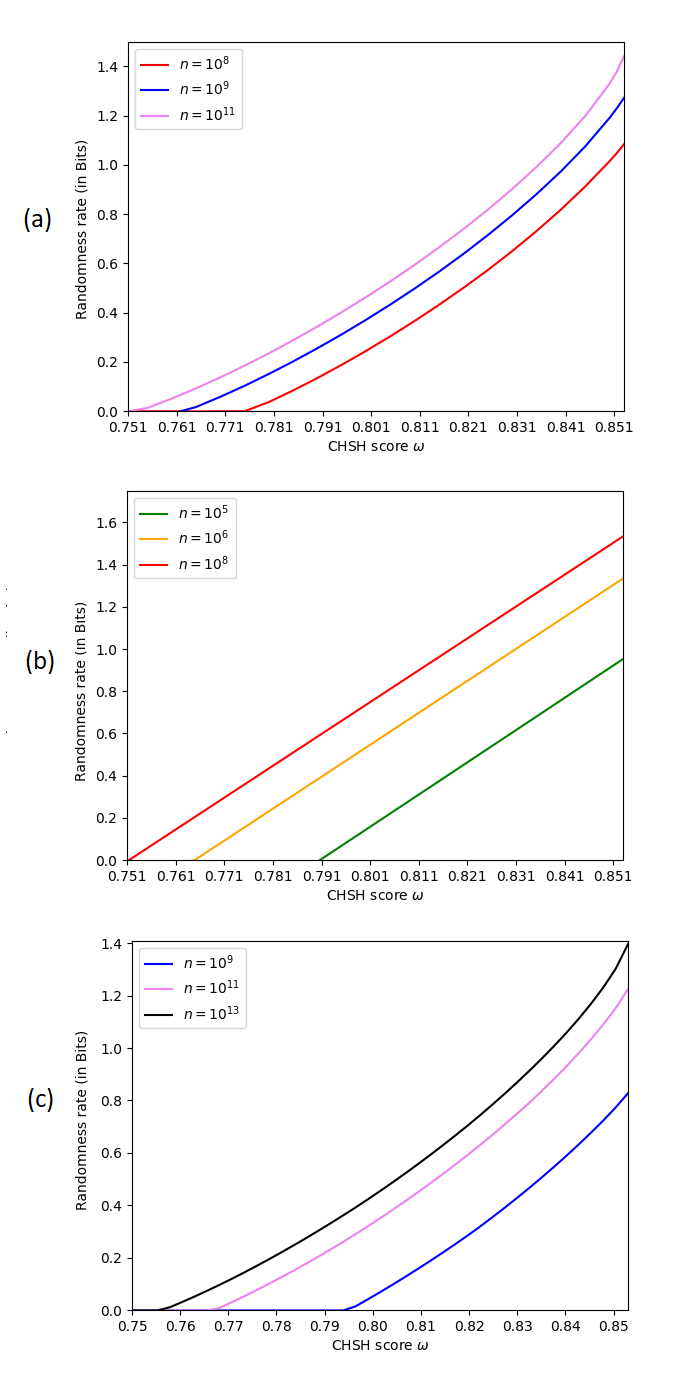} \\
\caption[Graphs of the net rate of certifiable randomness according to the EAT for (a) the spot checking protocol, (b) the protocol with recycled input randomness, and (c) the protocol with biased local random number generators, showing the variation with the CHSH score . The round numbers, are indicated in the legend.]{Graphs of the net rate of certifiable randomness according to the EAT for (a) the spot checking protocol (Protocol~\ref{prot:spotcheck}), (b) the protocol with recycled input randomness (Protocol~\ref{prot:nonspotcheck}), and (c) the protocol with biased local random number generators (Protocol~\ref{prot:biased}), showing the variation with the CHSH score $\score$. The round numbers, $n$, are indicated in the legend. The error parameters used were $\epsilon_S=3.09\times10^{-12}$ and $\epsilon_C=10^{-6}$. As in Figure~\ref{fig:EAT_n}, the values of $\gamma$ (for (a)) and $\zeta^A=\zeta^B$ (for (c)) were optimized over for each point.}
\label{fig:EAT_score}
\end{figure}

When using Protocol~\ref{prot:biased} with $\zeta^A=\zeta^B-\zeta$, the main difference to Protocol~\ref{prot:spotcheck} is that the distribution of $X$ and $Y$ is $((1-\zeta)^2,\zeta(1-\zeta),\zeta(1-\zeta),\zeta^2)$ rather than $(1-3\gamma/4,\gamma/4,\gamma/4,\gamma/4)$. In the analysis this manifests itself in the statistics, and the much lower probability of $X=1,Y=1$ requires an adjustment of $\delta$ to achieve the same error parameters for the protocol. A comparison between the output rates for Protocols~\ref{prot:spotcheck} and~\ref{prot:biased} is shown in Figures~\ref{fig:EAT_n} and~\ref{fig:EAT_score}.\bigskip

\begin{protocol}\label{prot:nonspotcheck} \textbf{ (Protocol with recycled input randomness)}
\phantom{a}\\

\noindent\textbf{Parameters}:\\
$n$ -- number of rounds \\
$\score_{\exp}$ -- expected CHSH score. \\
$\delta$ -- confidence width.
\begin{enumerate}
    \item Set $i=1$ for the first round, or increase $i$ by 1.
    \item Use $R_A$ to choose $X_i\in\{0,1\}$ uniformly, serving as the input to one of the devices giving output $A_i\in\{0,1\}$. Likewise use $R_B$ to generate $Y_i\in\{0,1\}$ uniformly, which is input to the other device giving output $B_i\in\{0,1\}$. Set $U_i=1$ if $A_i\oplus B_i=X_iY_i$ and $U_i=0$ otherwise.
    \item Return to Step~1 unless $i=n$.
    \item Count the number of rounds for which $U_i=0$ occurred and abort the protocol if this is above $n(1-\score_{\exp}+\delta)$.
    \item Process the concatenation of all the inputs and outputs with a quantum-proof strong extractor $\Ext$ to yield $\Ext(\B{ ABXY},\B{ R})$, where $\B{ R}$ is a random seed for the extractor. Since a strong extractor is used, the final outcome can be taken to be the concatenation of $\B{ R}$ and $\Ext(\B{ ABXY},\B{ R})$.
\end{enumerate}
\end{protocol}

An important difference in this protocol compared to Protocols~\ref{prot:spotcheck} and~\ref{prot:biased} is in the extraction step, which now extracts randomness from the input strings $\B{ X}$ and $\B{ Y}$ as well as the outputs. Without recycling the inputs, expansion would not be possible in Protocol~\ref{prot:nonspotcheck}. With this modification, the relevant quantity to decide the length of the output is $H(ABXY|E)$, and so $H(AB|XYE)=H(ABXY|E)-H(XY)=H(ABXY|E)-2$ is the relevant quantity for calculating the rate of expansion. Note that in order to reuse the input in a composable way, it also needs to be run through an extractor~\cite{CK2} (for a discussion of why it is important to do so and a few more composability-related issues, see Appendix~\ref{app:compos}).

We could also consider an adaptation of Protocol~\ref{prot:spotcheck} in which the input randomness is recycled, forming Protocol~\ref*{prot:spotcheck}$'$ from Protocol~\ref{prot:spotcheck} by replacing Step~\ref{st:7} by
\begin{enumerate}
\item[\ref*{st:7}$'$.] Process the concatenation of all the inputs and outputs with a quantum-proof strong extractor $\Ext$ to yield $\Ext(\B{ ABXY},\B{ R})$, where $\B{ R}$ is a random seed for the extractor. Since a strong extractor is used, the final outcome can be taken to be the concatenation of $\B{ R}$ and $\Ext(\B{ ABXY},\B{ R})$.
\end{enumerate}
In this case, as the number of rounds, $n$, increases the advantage gained by this modification decreases, becoming negligible asymptotically. This is because as $n$ increases, the value of $\gamma$ required to give the same overall security tends to zero, and hence the amount of input randomness required becomes negligible. Note that recycling the input randomness in Protocol~\ref{prot:biased} in the case where $\zeta^A=\zeta^B=1/2$ is equivalent to Protocol~\ref{prot:nonspotcheck}. 

Like Protocol~\ref{prot:biased}, Protocol~\ref{prot:nonspotcheck} also allows the locality loophole to be closed if on each round $i$, the random choice $X_i$ is space-like separated from the output $B_i$ and the random choice $Y_i$ is space-like separated from the output $A_i$.

In each of the protocols, the parameter $\delta$ should be chosen depending on the desired completeness error. For the spot-checking protocol, the relation between the two is discussed in~\cite[Supplementary Information I~D]{LLR&}. The analysis there can be applied to the protocol with recycled input randomness by setting $\gamma=1$ and the protocol with biased local random number generators is discussed in Appendix~\ref{app:comp}.

Figures~\ref{fig:EAT_n} and~\ref{fig:EAT_score} show how the amount of certifiable randomness varies with the score, $\score$, and round number, $n$. Note that in the cases where the rate curves are linear, they are linear for most of their ranges. Extending the linear part to the full range of quantum scores makes it easier to use the EAT while only resulting in a small drop in rate for scores close to the maximum quantum value. \RutC{Note that, as mentioned above, strictly speaking, the numerical curves we provided for the von Neumann entropy are upper bounds; the curves in Figures~\ref{fig:EAT_n} and~\ref{fig:EAT_score} are generated under the assumption that these upper bounds are tight}. [We could also use our lower bound instead. This would result in a small down-shifting of the curves, but means that the bounds are provably reliable.]

To demonstrate the increased practicality of the two-sided curves, we use the parameters from a recent experiment~\cite{LLR&} with Protocol~\ref{prot:spotcheck}. There a score of just over $0.752$ was obtained, for which it would require about $9\times10^{10}$ rounds to achieve expansion using Protocol~\ref{prot:spotcheck} with $\gamma=3.383\times10^{-4}$, $\epsilon_S=3.09\times10^{-12}$, $\epsilon_C=10^{-6}$ and taking the one-sided randomness~\cite{LLR&}. Using Protocol~\ref{prot:nonspotcheck} instead, and taking the two-sided randomness for the same score and error parameters allows expansion for $n\gtrsim 8\times10^7$, significantly increasing the practicality. For instance, the main experiment of~\cite{LLR&} was based on a spot-checking protocol and took $19.2$ hours; the use of Protocol~\ref{prot:nonspotcheck} instead would allow the same amount of expansion in about $60$ seconds (this time holds under the assumption that the same repetition rate of the experiment can be met in the non-spot checking protocol\footnote{In some experiments, the rate at which we can switch between the two measurements is relatively slow, and hence when using Protocol~\ref{prot:nonspotcheck}, where switching is required on most rounds, the switching rate dominates, slightly increasing the time.}). Protocol~\ref{prot:biased}, however, produces lower randomness rates compared to the spot-checking protocol. This is partly because more input randomness is required, and also because the completeness error has a worse behaviour. Protocol~\ref{prot:biased} is hence useful when inputs are not recycled and when closing the locality loophole is desirable. 

When discussing randomness expansion we have considered the figure of merit to be the amount of expansion per entangled pair shared. An alternative figure of merit is the ratio of the final randomness to the initial randomness, i.e., here we are considering how much randomness we can get from a given amount of initial randomness. For the latter figure of merit, Protocol~\ref{prot:nonspotcheck} is no longer optimal, since the amount of expansion cannot exceed the amount of input randomness. For the other two protocols the ratio of output randomness to input randomness can be made much higher by taking either $\gamma$ or $\zeta^A\zeta^B$ to be small.

\section{Discussion} 
In chapters~\ref{chap: DI protocols}, \ref{chap: DI upper bounds}, \ref{chap: DI lower bounds}, and \ref{chap : actual protocols for DI} we discussed CHSH based protocols for randomness expansion. We have given numerical bounds on various conditional von Neumann entropies that are relevant for CHSH-based device-independent protocols and discussed when each can be applied. We have investigated their implications using explicit protocols, comparing the finite statistics rates using the EAT, showing use of two-sided randomness has the potential to make a big difference. We also looked at protocols beyond the usual spot checking type. The first removes the spot checking to allow expansion while closing the locality loophole, and the second recycles the input randomness, so allowing expansion while performing a CHSH test on every round.

It remains an open question to find an analytic form for $F_{AB|X=0,Y=0,E}$, $F_{A|E}$ and $F_{AB|E}$. Since the curves $F_{A|E}$ and $F_{AB|E}$ are linear for all but the very highest (experimentally least achievable) scores, in these cases not much is lost by extending the line to all scores forming a lower bound that tightly covers all of the experimentally relevant cases. On the other hand $F_{AB|X=0,Y=0,E}$ is a convex curve throughout and hence a tight analytic form would be particularly useful in this case. Our initial analysis suggests that \RutC{the} form of the parameters achieving the \RutC{optimal values} for these \RutC{functions} is sufficiently complicated that any analytic expression would not be compact. A reasonably tight analytic lower bound for $F_{AB|X=0,Y=0,E}$ could also be useful for theoretical analysis. Note also that the bound $F_{A|E}\geq F_{A|XYE}$ appears to be fairly tight (see Figure~\ref{fig:rates}(a)) so the analytic form for $F_{A|XYE}$ can be used to bound $F_{A|E}$ with little loss. Another open problem is to find a concrete scenario in which $F_{AB|E}$ is directly useful.

The use of Jordan's lemma in this work prevents the techniques used being extended to general protocols, and finding improved ways to bound the conditional von Neumann entropy numerically in general cases remains of interest. For example, protocols that use three inputs for one party can allow up to 2 bits of randomness per entangled pair (see, e.g.~\cite{BRC}), and a way to tightly lower bound the von Neumann entropy in this case would further ease the experimental burden required to demonstrate DIQKD in the lab.\bigskip

\cleardoublepagewithnumber
\part{Semi Device Independent Protocols}
\cleardoublepagewithnumber

\chapter{Introduction to semi-Device Independent Protocols}\label{chap: semi-DI}
\section{Introduction}
As outlined in the introduction, the Device Independent protocols of randomness expansion make minimal assumptions for certifying randomness. However, as they currently stand, these protocols are extremely difficult to implement in a laboratory setting, even with the most advanced technology available. The primary obstacle in this regard is the performance of a loophole-free Bell test. There are two main loopholes that have traditionally caused problems in performing the Bell test. The first is the locality loophole, which refers to the requirement that the devices are spacelike separated when carrying out the Bell test. The second is the detection loophole, resulting from inefficient measurements.

For protocols of randomness expansion, the locality loophole might not be as critical to close, assuming we have control over the laboratory. Shielding mechanisms can be used between the devices to ensure that no communication occurs among them. Theoretically, this could be the same shielding that prevents any data leak from the lab to the outside world. The remaining challenge is the detection loophole, the closure of which requires very high-efficiency detectors. While such detectors are available, they tend to be expensive, and maintaining good-quality entanglement over a relatively large distance presents a significant practical problem.

As noted in the introduction, progress is being made in this direction, but until these challenges can be overcome in a more practical way, Device Independent protocols \RutC{for} randomness expansion can only be used for a select few top-security applications. On a practical level, however, the security of random numbers is desired. Therefore, there is a need for protocols that are not fully Device Independent but rather allow for a certain degree of trust on the components. 

By imposing well-chosen assumptions on the devices or \RutC{the} system's underlying process, semi-Device Independent (semi-DI) approaches can achieve most of the security benefits of a fully Device Independent (DI) protocol \RutC{whilst} bypassing the need for challenging experimental implementations. This intermediate scenario strikes a balance between security and practicality, making it a promising approach for implementing quantum random number generators.

Many semi-DI protocols have been introduced in the scientific literature. A typical setup for these protocols involves having a source that prepares a quantum state, and then sends it to a measurement device that performs a measurement on the prepared quantum state. One advantage of this method is that it doesn't rely on sharing entangled states over a distance, which is a requirement \RutC{of} the Device Independent protocols. This type of experimental setup is often referred to as a ``prepare and measure'' setup (or scenario).

The assumptions underlying a semi-DI scheme can vary based on specific needs. For example, one might consider source-DI or measurement-DI cases, where either the source device or the measurement device is trusted, respectively.

In this section of the thesis, we work on protocol based \RutC{on a} ``prepare and measure'' scenario where the source and measurement devices are both uncharacterized. Thus, it is both source-DI and measurement-DI. This method was first introduced in~\cite{VSoriginal} and been subsequently been studied in  \cite{teb_cv,Brask2017,Rusca2020,avesani2020,Tebyanian_2021, VS}. These works discuss semi-DI Quantum Random Number Generators (QRNGs) that employ a physical system with a unique ground state -- i.e. the lowest eigenvalue state of the system Hamiltonian is unique. The outlined protocols are based on the source preparing the states with  low energy or high overlap\footnote{Overlap here is defined as the fidelity between the state and the ground state.} with the unique ground state. The component of the protocol responsible for the verification of these energy and overlap constraints is trusted, hence making these protocols semi-Device Independent. 

The idea for the protocol is as follows: Recall from section~\ref{sec: sdps} that two quantum states $\rho^{0}$ and $\rho^{1}$, when prepared with a uniform probability distribution, cannot be perfectly distinguished if the distance between them, represented as $||\rho^{0} - \rho^{1} ||_1$, is small. Now consider the \RutC{following} scenario: we randomly select $X \in \{0,1\}$ and prepare $\rho^{x}$ ($x = 0$ or $1$) depending on the outcome of $X = x$, and then task our measurement device with distinguishing between the states $\rho^{0}$ and $\rho^{1}$ to produce a bit $Y$ with the intent that $Y = X$. If the states $\rho^{0}$ and $\rho^{1}$ are sufficiently close\footnote{Here, two states are close to each other if the trace distance between them is small.}, then the probability that $Y = X$ is strictly less than $1$, insinuating that $Y$ must possess randomness, even if $X$ is later revealed. This forms the basis of our protocol. 

The advantage of such a protocol is that it has the features of Device Independent protocols, in the sense that the security of the protocol solely relies on the input-output statistics, namely the probability that  the outputs and inputs \RutC{are} different - i.e. $P(Y \neq X)$. Furthermore, observe that the protocol does not also depend upon which states are prepared, all that matters is the quantity $||\rho^{0} - \rho^{1}||_1$. \RutC{Intuitively, given that the protocol hinges on the inability to perfectly distinguish sufficiently close quantum states, one would anticipate that the randomness generated in the protocol increases as the distance between the states decreases. This intuition will be confirmed when we compute the rates of the protocols in Chapter \ref{chap: semi-DI protocols}.} Thus, if there is an experimental way to ensure the distance $||\rho^{0} - \rho^{1}||_1$ is small, then we have a good protocol for randomness expansion.

\RutC{To ensure the distance} between the generated states is small, one strategy involves seeking systems with a unique ground state, meaning the system's lowest energy state is non-degenerate. If the system's energy is observed to approximate the vacuum energy (i.e., the energy of the ground state), then both states must be nearly equivalent to the ground state and, by extension, to each other. Provided that the measured energy is low enough, it is possible to find bounds on the distance between the states simply by knowing the energies of the states produced using simple arguments. Note that theoretically, the ground state isn't inherently special; the key lies in generating nearly identical states. In practical terms, when the state source is a laser pulse —- common in many prepare-and-measure situations -— the source might produce two coherent states, $\ketbra{\alpha_{0}}{\alpha_0}$ and $\ketbra{\alpha_{1}}{\alpha_1}$, both with minimal mean photon numbers, rendering them close to the vacuum state.

Unfortunately, measuring the energy of the system is not possible in a Device Independent fashion. Thus, \RutC{we need to have a trusted power meter}, whose role is to measure the energy of the states, making this protocol semi-Device Independent. We use the same principle for our protocol of randomness, however, we do not measure energy of the states. \RutC{Rather}, we assume that we have access to a power-meter, which can determine if the prepared states have more than a threshold energy - so it is a yes/no machine (similar to an on/off photo diode). The details of this will \RutC{become} clear in the next section. We can also include an additional component such as a variable attenuator, that can help by reducing the energy of the emitted states. 

For this semi-DI protocol, the primary assumption is that the power meter is a trusted component and has not been tampered with by any adversary nor is it damaged. However, depending on the context, the studies mentioned above have required one or more additional assumptions:
\begin{itemize}
\item Each protocol round is independent of the previous one, and the initial conditions are identical before each round (i.i.d. assumption). This condition implies that the source or the measurement devices do not have an internal memory set by the eavesdropper. 
\item Only classical side-information is considered, which can be caused by device imperfections or classical correlations. This assumption limits the eavesdropper from being entangled with the quantum state prepared by the source. Furthermore, this also supposes that the eavesdropper has no quantum memory. 
\item The source and measurement devices are not entangled with each other. 
\end{itemize}
\RutC{The assumptions above can be rather strong in many scenarios.  For instance,} the assumption that all the rounds are identical and that the eavesdropper only has a classical side information \RutC{is difficult to verify in the experimental setting}. \RutC{The above assumptions also} explicitly rule out the case, in which, the adversary shares entanglement with the source and the measurement device, \RutC{which cannot be easily justified when the RNG is purchased from a untrustworthy party}. Furthermore, device imperfections, environmental changes, and experimental errors make it impossible to achieve identical experiment rounds.
Furthermore, suppose the adversary has a quantum correlation with the devices, such as generated states being entangled with her states. In that case, it becomes easier for her to predict the QRNG measurement outcome.

Our analysis discards all these assumptions, except for the last one, in which the source and the measurement devices share some entanglement. Though we do allow for pre-shared randomness between the source and the measurement device. We permit the eavesdropper to share entanglement with the source and the states prepared by the source. We also allow for a fully uncharacterized measurement device known to the adversary and permit the eavesdropper, source, and measurement device to pre-share arbitrary classical randomness, thereby accounting for a quantum adversary.


In quantum state preparation and measurement, the memory effect means that earlier measurements can influence subsequent ones, and earlier prepared states may impact those prepared later. These memory effects have implications for security, as they introduce correlations and dependencies in the outcomes. To mitigate this issue, we consider the memory effect and other potential sources of correlation in the security estimation stage.

The main advantage is that our protocol is formulated in a way that the Entropy Accumulation Theorem (EAT) can be readily applied. Though this has not been done in the thesis, it can be done in a relatively straightforward manner, thus exploiting several benefits of EAT. The primary advantage is the relaxation of the i.i.d. assumption in the protocol, and accounting for memory effects. Furthermore, the EAT also accounts for the adversary holding quantum side information. Using EAT, the problem of computing randomness rates (randomness per round) is reduced to computing the lower bound on the single round \RutC{von Neumann} entropy of a representative round of a protocol. The problem of computing lower bounds on the single round \RutC{von Neumann} entropies \RutC{for the} semi-DI protocol described above is one of the main aims of \RutC{this section of the thesis}.

One of the challenging aspects of the protocol is that no assumptions have been made on the dimensions of the states $\rho^{x}$. Fortunately, similar to DI protocols, we show that we can leverage Jordan's Lemma for such semi-DI protocols, significantly reducing the complexity of computing the rates of the protocols. Jordan's Lemma allows us to relax the problem to a scenario where the source generates qubit states, and the measurement device performs projective measurements, simplifying the process into an optimization problem involving less than eleven variables. Further simplifications, along with reliable numerical techniques for obtaining lower bounds on polynomial optimization problems as discussed in Chapter~\ref{chap: maths}, enable us to reliably compute the rates for the protocol. The problem of computing these rates will be discussed in the next chapter (Chapter~\ref{chap: semi-DI lower bounds}).

In this work, we present two types of protocols: one for randomness expansion, similar to protocol~\ref{prot:nonspotcheck} in the DI setting, where input randomness is recycled, and another for converting public randomness to private randomness, analogous to protocol~\ref{prot:biased} in the DI setting. The protocols are detailed in Chapter~\ref{chap: semi-DI protocols}, along with a discussion on the asymptotic rates for these protocols.

\RutC{In the remainder of the chapter, we do a brief literature survey of different Quantum Random Number Generators (QRNGs) and then proceed to giving a sketch of the protocol that we study in this thesis}.
\RC{\section{A brief review of different QRNGs}
Before delving into the main semi-DI protocol of randomness expansion discussed in this thesis, we deliver a very brief and non-exhaustive survey of the literature on various Quantum Random Number Generators (QRNGs). There is a vast literature on a range of different protocols for building QRNGs. As discussed in the previous section, there is generally a trade-off between the ease of experimental implementation of a protocol and the security of the protocol. Here, we call a protocol more ``secure'' if it requires fewer experimental assumptions.
Broadly speaking, QRNGs fall into one of these categories:
\begin{enumerate}
\item Fully Device-dependent protocols (DD) (for example see \cite{DD1,DD2,DD3});
\item Source DI protocols (for example see \cite{SDI,SDIone}), wherein the source is untrusted but the measurement apparatus is reliable;
\item MDI protocols (for example see \cite{MDI,MDI2,MDI-pirandola}), where the measurement device is untrusted, but not the source;
\item Semi-DI protocols;
\item DI protocols.
\end{enumerate}
The DD protocols are the easiest to implement since all components are trusted, and no characterization of the device is needed. However, this simplicity might come at the expense of security. On the other hand, the DI protocols are the most difficult to implement with the current technology but promise the highest possible security. The source-DI, measurement-DI, and semi-DI protocols are in the middle with security somewhere in between DD and DI and the practicality also somewhere in between DD and DI. Source and measurement DI protocols have also been studied widely in the literature. These protocols are useful when the user can either characterize the source, which can prepare desired states with minimal noise, or characterize the measurement device that comes equipped with highly efficient detectors.

The exploration of semi-DI protocols in the quantum information theory literature can be traced back to Liang et al. \cite{semiDI-LVTB}. Their work focused on semi-DI protocols for entanglement detection. Building upon these concepts, Paw{\l}owski et al. \cite{semiDI-PB} extended this framework to include semi-DI protocols for Quantum Key Distribution (QKD). This work was further extended by Li et al.~\cite{semiDI-Li} to construct a semi-DI protocol for randomness expansion. These semi-DI protocols operate within the prepare-and-measure scenario and do not make any assumptions about the internal workings of the device, other than the assumption that the dimension of the Hilbert space of the produced states is both bounded and known.

The primary challenge with randomness expansion protocols that depend on a dimension bound for the prepared states is the inability to experimentally verify such an assumption. While a lower bound on the Hilbert space dimension can be certified without trusting the devices \cite{BNV, BQB}, an upper bound cannot be similarly certified. This limitation stems from the fact that the dimension of the Hilbert space is not a directly measurable quantity in quantum theory, leaving no reliable method to validate assumptions about the Hilbert space dimensions of prepared states.

The protocol by Van Himbeeck et al. addresses this issue~\cite{VSoriginal}. Their idea is that the dimension of the Hilbert space can be indirectly deduced through energy measurements. In many quantum systems, both the Hamiltonian and its spectrum are known. The essence of the protocol is that the Hilbert space dimension can be inferred by measuring the energy of the system. For example, if the source prepares quantum states in a system with a known, non-degenerate energy spectrum $\{ E_i \}_{i = 0}^{\infty}$, and the highest observed energy after numerous measurements is $E_{k}$, we can deduce, with high confidence, that the Hilbert space dimension is $k + 1$. This method remains applicable to a degenerate spectrum, as long as the full spectrum, including degeneracies, is known. Importantly, Van Himbeeck et al.~\cite{VSoriginal} realized that if the measured energy stays below a specific threshold, the system can effectively be seen as two-dimensional, with one ground state and all other states can be effectively combined as a single excited state. Subsequent research has further developed and expanded upon these findings \cite{teb_cv, Brask2017, Rusca2020,avesani2020, Tebyanian_2021, VS}.

The method described above for building semi-DI protocols is by no means exhaustive. There are other semi-DI protocols based on different physical or information-theoretic principles (for example, see references \cite{semi-DI-other, semiDIother2, QRAC}). Moreover, just as quantum behavior can be certified using Bell non-locality, another fundamental concept, contextuality, can also certify whether the experimentally observable statistics admit a quantum behaviour or not. Unfortunately, unlike non-locality, non-contextuality cannot be tested in a Device Independent manner. However, without delving into any details, if we place appropriate (partial) trust in the source or the measurement device, it is possible to devise semi-DI protocols of randomness expansion based on the distinction between contextual and non-contextual behaviors. The groundwork for such protocols has been established for the QKD setting in \cite{contextDIQKD} and later extended to randomness expansion in \cite{contextDIRNE}.}
\section{General semi-Device Independent protocol}\label{sec: general_semi-DI protocol}
\begin{figure}\label{fig: protocol-semi-DI}
    \centering
    \includegraphics[width=\textwidth]{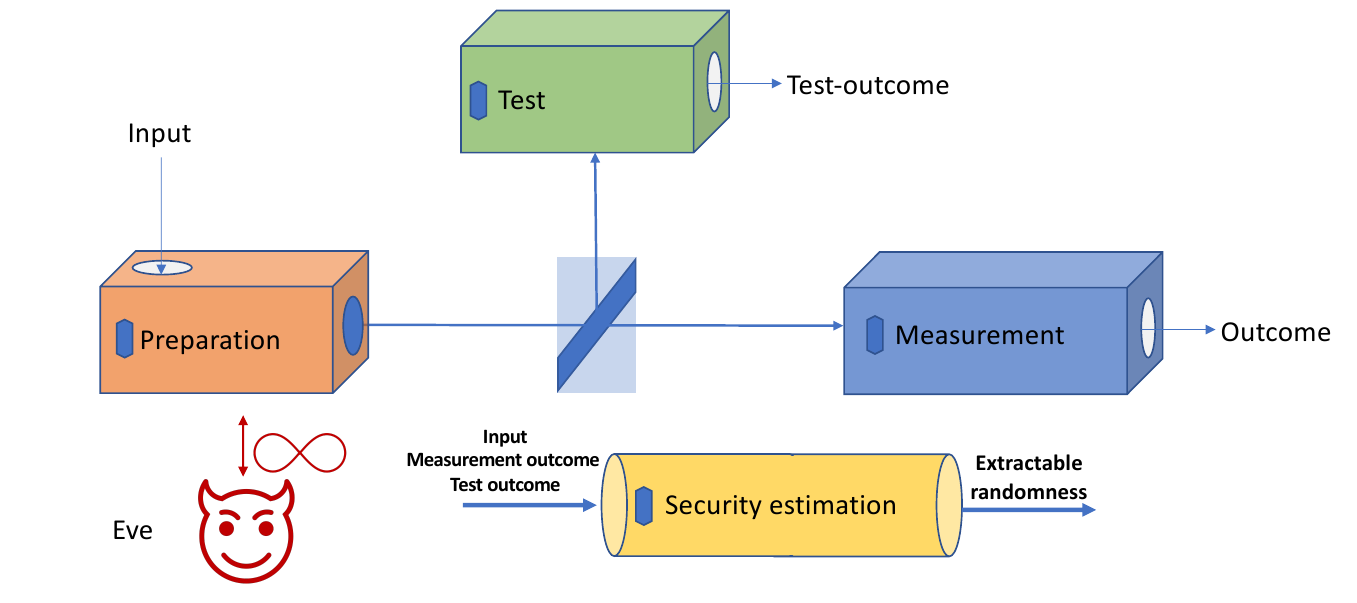}
    \caption{A schematic diagram of our semi-DI protocol.}
\end{figure}
Figure~\ref{fig: protocol-semi-DI} offers a schematic illustration of the protocol, which is split into three primary stages: preparation, testing, and measurement. It's crucial to note that full characterization is necessary only for the component in the testing phase. Both the source and measurement stages remain uncharacterized.

For the protocol to operate effectively, two separate input seeds are essential. The initial seed guides the source in state preparation, while the secondary seed dictates whether the prepared states undergo testing. \RutC{Keeping these seeds hidden from potential adversaries before the protocol's initiation is vital for protocol's security} .

Here is a brief overview of the protocol:

\textbf{Preparation Phase:} 
The source accepts input \RutC{of a}  random variable \( X \in \{ 0 , 1 \} \) that is generated with a probability distribution \( p_{X}(x) \). Based on this input, the source prepares either \(\rho^{0}\) or \(\rho^{1}\) contingent on whether \( X \) equals 0 or 1. The state then gets relayed to component \( BS \), which determines if the state should proceed to the testing or measurement phase. Essentially, device \( BS \) functions \RutC{as a switch}. Utilizing a random number \( T \) produced by the random number generator \( R_{\gamma} \), \(BS\) sends the signal to either testing or measurement (for example, if it receives $T = 0$, then it will send the signal to measurement, otherwise it sends the signal for testing). The bias of \( R_{\gamma} \) is set to ensure most signals reach the measurement phase. If \( BS \) channels the state for testing, we denote that round a test round; otherwise, we call the round ameasurement round. 

\RutC{In practice, if the source is a laser, the device $BS$ can be a mechanical switch that controls a mirror. Depending on the random variable $T$, it directs the signal to the testing phase or the measurement phase. However, such a switch can be lossy and therefore may not very suitable for practical implementation. Alternatively, the device $BS$ can be a half-silvered mirror or a beam splitter, which is a passive optical device that transmits the signal to the measurement device with probability $p_{T}(0) = \gamma$ or reflects the signal with probability $1 - \gamma$. If such a half-silvered mirror is used, it also needs to be fully characterized and assumed not be tampered with by any adversary.}

\textbf{Testing Phase:} 
This is the protocol's fully characterized segment. Recall that the protocol derives its randomness if the states \( \rho^0 \) and \( \rho^1 \) are near identical. This is ensured by ensuring both states are close\footnote{Here closeness of two states is measured in terms of them having either high fidelity or low trace distance.} to the system's unique ground state, symbolized by \( \ketbra{0}{0} \) or \( \Pi_0 \), often referred to as the vacuum state. Essentially, the idea is that if both states are reasonably close to the vacuum state (in the state space), they are close to each other (in the state space). Formally, the definition of closeness that we use here is overlap of two states with vacuum state. The overlap for state \( \rho \) with the vacuum is defines as:

\begin{eqnarray} 
\tr \left( \rho \Pi_0 \right)  = \bra{0} \rho \ket{0}.
\end{eqnarray}

From a theoretical viewpoint, measuring this overlap requires an on-off device that can execute a two-outcome measurement $\{ \Pi_0 , \id - \Pi_0 \}$, with \( \Pi_0 \) representing the projection onto the vacuum state. Alternatively, the overlap can be deduced from the state's energy, given that this measured energy remains below a certain threshold. Therefore, experimentally, this mandates the use of a power-meter or photodiode. For simplicity, this on-off device is often termed a 'power meter', symbolized by \( PM \). If we define the vacuum's energy as zero, this device's primary function becomes the detection of any existing positive energy.

Considering \( PM \) could receive states \( \rho^0 \) or \( \rho^1 \) depending on the source's preparation, there are two separate overlaps with the ground state in a protocol. In our context, the term ``overlap'' (of the protocol) is defined as the average of these individual overlaps of \( \rho^{0} \) and \( \rho^{1} \) with the ground state. Specifically, the overlap \( \overlap \) for a protocol is:

\begin{equation}
    \overlap = \frac{1}{2} \mathrm{tr}(\rho^{0} \Pi_0) + \frac{1}{2} \mathrm{tr} \left(\rho^{1} \Pi_0 \right).
\end{equation}

Ideally, this value should be at its maximum to ensure that the source \( S \) consistently produces states with substantial individual overlaps.

\textbf{Measurement Phase:} 
Here, states \( \rho^0 \) or \( \rho^1 \) are sent to the measurement apparatus \( M \). This device's primary role is to determine whether \( \rho^0 \) or \( \rho^1 \) has been sent. Essentially, it should produce a bit \( Y \) aiming for \( Y \) to equal \( X \). If the overlap is notable, then \( M \) cannot perfectly distinguish between the states, ensuring that the output \( Y \) contains private randomness even if \( X \) is later revealed.

During the generation round, a ``win'' is declared if \( Y \) equals \( X \). Theoretically, the measurement device performs a two outcome POVM $\{M_0 , M_1 \equiv \id - M_0 \}$. The protocol's score is the mean probability of securing a win, defined as:

\begin{equation}
    \score := \frac{1}{2}\tr(M_0 \rho^{0}) + \frac{1}{2} \tr(M_1 \rho^0) ,
\end{equation}

Here, \(  \tr(M_x \rho^x) \) is the probability of winning a generate round when $X = x$ is prepared.

A high score is desirable for our protocol. As we have discussed earlier achieving a high overlap is crucial for the protocol's optimal function. Yet, even with a specific overlap, a high score remains preferable. A low score can arise from strategies like the following: the source prepares two states ($\rho^{x}$ depending upon the inputs $X = x$) that are best distinguished given a fixed overlap. The measurement device can act as follows: with a very high probability, it may ignore the incoming states and output using pre-shared randomness, which is accessible to the adversary. With a smaller probability, it might perform the optimal measurement that distinguishes $\rho^0$ and $\rho^1$. Such a strategy will produce negligible private randomness, as for most rounds, some pre-shared randomness is outputted.

Various protocols can be conceived considering the process of input and output strings. Future chapters will probe two such protocols, focusing on discussions about computing the randomness rate for these protocols.

\chapter{Optimizing the von Neumann entropy}\label{chap: semi-DI lower bounds}
As with the Device Independent (DI) protocols, we can use the Entropy Accumulation Theorem (EAT) to compute the randomness rate (randomness per round) of the semi-DI protocol. Recall that informally, EAT states that \RutC{to compute the randomness generated in a protocol,} it is sufficient to calculate the von Neumann entropy of a single round that is representative of the full protocol. As we shall see in the next chapter, in the context of the protocols considered here, it suffices to compute a lower bound on single round of von Neumann entropy conditioned on the same score $\score$ and the same overlap $\overlap$. The main aim of the chapter is to perform this optimization problem. To do so we make this formalism more rigorous.
\section{Strategies} 
We begin by defining a strategy for our semi-Device Independent protocol: 
\begin{definition}[Strategy] Let $\rho_{AE}^{x} \in \M{S}(\M{H}_{A} \otimes \M{H}_{E})$  and $\{M_{0} , M_{1} \equiv \id - M_1 \}$ be a POVM. A tuple $\M{C} = (\rho^{0}_{AE} , \rho^{1}_{AE} , M_0 )$ is called a strategy.
\end{definition} 
In the definition of a strategy, we allow the states to be sub-normalized as well. However, if a strategy defines  our protocol, then the states $\rho^{x }_{AE}$ should be normalized. Similar to the Device Independent scenario, this strategy is chosen by the adversary Eve. \\
Each strategy has an associated CQ state. Let $p_{X}$ be the input probability distribution, then the CQ state associated with the strategy above is given by 
\begin{equation*}
    \rho_{\M{C}} := \sum_{x} p_{X}(x) \ketbra{x}{x}_{X}\otimes \ketbra{y}{y}_{Y} \otimes \tr\left( M_{y} \otimes \id_{E}  \rho^{x}_{AE} \right). 
\end{equation*}
We introduce a short hand notation of $H(Y|X E)_{\M{C}} \equiv H(Y| XE)_{\rho_{\M{C}}}$ when referring to the conditional von Neumann entropy of the CQ state $\rho_{\M{C}}$. \\
Given a strategy, it is also possible to determine the score it achieves and the overlap that shall be observed, provided the strategy is used in an i.i.d. fashion. Thus it is useful to define the following: 
\begin{definition}[Score of a strategy] 
The score of a strategy $\M{C} = (\rho^{0}_{AE} , \rho^{1}_{AE} , M_0 ) $ is defined as:
\begin{eqnarray}
    \scorefunction(\M{C}) = \frac{1}{2} \sum_{x} \tr\left( \rho^{x}_{AE}  M_{x}\otimes \id_{E} \right). 
\end{eqnarray}
Note that $M_1 = \id - M_0$ is implicitly assumed when defining a strategy. 
\end{definition} 
Similarly, we can define the overlap of the strategy (where $\Pi_0$ is the projector onto the ground state)
\begin{definition}[Overlap of a strategy]
The overlap of a strategy $\M{C} = ( \rho^0_{AE} , \rho^{1}_{AE} , M_0 )$ is given by
\begin{eqnarray}
    \overlapfunction_{\Pi_0}(\M{C}) = \sum_{x} \frac{1}{2} \tr \left( \rho^x_{AE} \Pi_0 \otimes \id_{E} \right).
\end{eqnarray}
\end{definition} 
There are certain special classes of strategies that may be interesting to consider: 
\begin{definition}[Pure state Strategy] A strategy $\M{C} = (\rho^{0}_{AE} , \rho^{1}_{AE} , M_{0})$ is a pure-state strategy if $\rho^{0}_{AE}$ and $\rho^{1}_{AE}$ are pure states.
\end{definition}

\begin{definition}[Projective Strategy] A strategy $\M{C} = (\rho^{0}_{AE} , \rho^{1}_{AE} , P_{0})$ is a projective strategy if $P_{0}$ is a projection operator.
\end{definition}
In the next chapter, it will be shown that, in the asymptotic limit, the rate of the protocol can be  found by computing (or by finding an appropriate lower bound for) of the function $F_{p_{X}}( \score , \overlap)$. This function is given by the optimization problem: 
$$ \inf_{\M{C} \in \Gamma[\score, \overlap]}   H(Y|XE)_{\rho_{\M{C}}} ,$$
where $\Gamma[\score , \overlap]$ \RutC{are} the strategies that achieve a fixed score $\score$ and have an overlap $\overlap$. - i.e. 
\begin{equation*}
    \Gamma := \{ \M{C}: \quad \scorefunction(\M{C}) = \score, \overlapfunction_{\Pi_0}(\M{C}) = \overlap \} .
\end{equation*} 
The optimization problem above can be simplified right-away by showing that it is sufficient to restrict to cases when the states $\rho^{x}_{AE}$ are pure.
\begin{lemma}
For every strategy $\M{C} = (\rho^{0}_{AE} , \rho^{1}_{AE} , M_{0})$ there exists a pure state strategy  $\M{C}' = (\tilde{\rho}^{0}_{AE} , \tilde{\rho}^{1}_{AE} , M_{0})$ such that 
\begin{eqnarray}
H(Y|XE)_{\M{C}} \geq H(Y|XE)_{\M{C}'} \nonumber 
\end{eqnarray}
\end{lemma}
\begin{proof}
\noindent Suppose $\tilde{\rho}^{x}_{AE}$ are not pure states, then let $\rho^{x}_{AEE'}$ be any purification of $\rho^{x}_{AE}$. Let $\M{C}' = (\tilde{\rho}^{0}_{AEE'} , \tilde{\rho}^{1}_{AEE'} , M_{0})$. Then simple computation shows that $\tr_{E'}\rho_{\M{C}'} = \rho_{\M{C}}$. Thus
\begin{eqnarray}
H(Y|X E E')_{\M{C}'} \leq H(Y | X E)_{\M{C}}
\end{eqnarray}
follows from strong subadditivity of \RutC{the} von Neumann entropy. 
\end{proof}
As we restrict to the set of pure states for the rest of the analysis, it shall be understood that the tuple $(\rho^{0}_{A} , \rho^{1}_{A} , M_0)$ is the short-hand for any strategy $(\rho^{0}_{AE} , \rho^{1}_{AE} , M_0 )$, where $\rho^{x}_{AE}$ is any purification of $\rho_{A}^{x}$.

\
\section{Incorporating pre-shared randomness and reduction to projective strategies}  
When executing the protocol, it is necessary to consider potential attacks by the eavesdropper, who may possess pre-shared randomness with the source and measurement devices. During a round of the protocol, the eavesdropper could have complete knowledge of this pre-shared randomness. Additionally, the eavesdropper may instruct the devices to prepare a state based on this pre-shared randomness and can also instruct the measurement device to perform a specific measurement depending on the value of the pre-shared randomness. 

To account for such an attack, we assume that the state $\rho_{AE}^{x}$ can take the following most general form: 

\begin{eqnarray}\label{eqn: General state}
\rho_{AE}^{x} = \sum_{\lambda} p(\lambda) \rho^{\lambda}_{A\tilde{E}} \otimes \ketbra{\lambda} {\lambda}_{\Lambda}, 
\end{eqnarray}

where $E  = \Lambda \tilde{E}$, is the system accessible to $E$. Here, $\Lambda$ represents an additional classical register held by Eve, the source, and the measurement device. It is important to note that the system $\Lambda$ does not possess any information about the input $X$, as it solely represents pre-shared randomness. Since the system $\Lambda$ is not transferred from the source to the device (and the power meter) during each round, the overlap constraint for the protocol is expressed using the projector $\Pi_0 \otimes \id_{\Lambda} \otimes \id_{\tilde{E}} $. 

In the strategy involving pre-shared randomness, the measurement device acts based on the pre-shared randomness. The most general measurement operator for our protocol can be defined as follows 
\begin{eqnarray}\label{eqn: General measurement}
  \hat{M}_{0} =    \sum_{\lambda}  M_{0}^{\lambda} \otimes \ketbra{\lambda}{\lambda}_{\Lambda} \otimes \id_{\tilde{E}}  , \quad    \hat{M}_{1} =    \sum_{\lambda} \left( \id -  M_{0}^{\lambda} \right)  \otimes \ketbra{\lambda}{\lambda}_{\Lambda} \otimes \id_{\tilde{E}}.
\end{eqnarray}

The definition of the measurement operator has been slightly altered from what was presented in the previous section. In this context, the measurement operator can act on the system \( \Lambda \), which the \RutC{Eavesdropper} can access. This change is simply an artefact of our short hand notation. A more accurate representation of the protocol would segregate it into three separate yet perfectly correlated random variables: \( \Lambda_{S} \) for the source, \( \Lambda_{M} \) for the measurement device, and \( \Lambda_{E} \) for the eavesdropper. The states that \RutC{the} source produces depend on the value of $\Lambda_{S}$ (which is a local random variable), and the measurement operator determines the value of the local random variable $\Lambda_{M}$ and performs a measurement $\{ M_{\lambda} , \id - M_{\lambda} \}$, depending upon the value of $\Lambda_{M} = \Lambda_{S} = \lambda$. To streamline the notation, we merged these into a single classical register, \( \Lambda \). As a consequence of this simplification, we effectively allow \RutC{the }measurement apparatus to perform a measurement on a system that is accessible to the eavesdropper.

\RutC{To summarize the discussion above}: our strategy is given by a tuple \( \mathcal{C} = (\rho^{0}_{AE} , \rho^{1}_{AE} , \hat{M}_{0}) \) where \( \rho^{x}_{AE} \)  is of the form \eqref{eqn: General state}. Meanwhile, the measurement operator \( \hat{M}_{0} \) is of the form \eqref{eqn: General measurement}. The CQ state relevant to the most generalized strategy is given by 
\begin{equation}\label{eqn: general CQ state in the generate round}
\rho_{\M{C}} = \sum_{\lambda = 1}^{n} \sum_{x,y} p(\lambda ) p_{X}(x) \ketbra{x}{x}_{X} \otimes \ketbra{y}{y}_{Y} \otimes \tr_{A} \left( M^{\lambda}_{y} \otimes \id_{E} \rho^{x,\lambda}_{A\tilde{E}}  \right) \otimes \ketbra{\lambda}{\lambda}_{\Lambda}. 
\end{equation}
The score and overlap of a strategy are defined via: 
\begin{eqnarray*}
        \score(\mathcal{C}) &=& \frac{1}{2} \sum_{x} \mathrm{tr} \left( \hat{M}_{x} \rho^{x}_{AE} \right) \\
        \overlap_{\Pi_0}(\mathcal{C}) &=& \frac{1}{2} \sum_{x} \mathrm{tr} \left( \Pi_{0} \otimes \id_{E} \rho^{x}_{AE} \right) .
\end{eqnarray*}
The sets $\Gamma[\score, \overlap]$ are defined identically as in the previous section.\par 

We now argue that we can assume the measurement $\{ M_{y}^{\lambda} , \id - M_{y}^{\lambda}\}$ to be a projective measurement without losing any generality. This is because any effect, $M^{\lambda}_0$, can be expressed in terms of convex combination of extremal effects. In quantum theory, these extremal effects are projections \footnote{Note any operator $P$ satisfying $P^2 = P$ is called a projector. Importantly, this also includes the operators $\id$ and zero operator $O$ for our case.}. Therefore, we can write
\begin{align*}
M^{\lambda}_0 = \sum_{\mu} q_{\mu} M^{\lambda, \mu}_{0} , \quad
M^{\lambda}_1 = \sum_{\mu} q_{\mu} M^{\lambda, \mu}_1, 
\end{align*}

for some probability distribution $\{ q_{\mu } \}$. This means, if there are strategies that use non-projective measurements, they can be executed using an extra classical register that only the measurement device can access. Note that having this extra register does not change the score and the overlap of the strategy. 

Because of the strong subadditivity of the conditional \RutC{von Neumann} entropy, letting Eve and the source use this extra classical register cannot worsen the strategy from the point of view of Eve. Considering that the measurement device is not fully characterized, Eve might also have this register in the form of pre-shared randomness instead. Therefore, this classical register can also be integrated into the register $\Lambda$.

We now show that \RutC{the} problem for computing the rate in the most general attack can be reduced to the problem of computing the rate when \RutC{the} eavesdropper does not pre-share any randomness. To facilitate this, we denote the set $\mathfrak{C}_1^{P}[\score , \overlap] \subset  \Gamma[\score , \overlap]$ to be the set of all projective strategies which do not allow for any pre-shared randomness , i.e. the strategies for which $E = \tilde{E}$. In the following, \RutC{$\text{convenv}(.)$ represents the convex envelope (or convex lower bound) as introduced in Chapter \ref{chap: DI upper bounds} (see \ref{app: LF tranform} for a more detailed explanation).}
\begin{lemma}\label{lem: reduction to single stategy} $F_{p_{X}}( \score , \overlap)= \text{convenv}\big( G_{p_{X}} ( \score , \overlap) \big)$, where the function $G_{p_{X}}( \score , \overlap)$ is computed using the optimization problem 
\begin{eqnarray}
    \inf_{\M{C} \in \mathfrak{C}_{1}[\score , \overlap]} H(Y|XE)_{\M{C}} ,
\end{eqnarray}
where the set $\mathfrak{C}_{1}^{P}[\score, \overlap] \subset \Gamma[\score , \overlap]$ represents the set of all pure state and projective strategies that do not allow for any pre-shared randomness - i.e. the collection of the strategies in the set $\Gamma[\score, \overlap]$, for which the classical register $\Lambda$ is trivial. 
\end{lemma}

\begin{proof}
Let $\M{C}$ be any strategy. Then the CQ state $\rho_{\M{C}}$ is of the form 
\begin{equation*}
\rho_{\M{C}} = \sum_{\lambda = 1}^{n} \sum_{x,y} p(\lambda ) p_{X}(x) \ketbra{x}{x}_{X} \otimes \ketbra{y}{y}_{Y} \otimes \tr_{A} \left( M^{\lambda}_{y} \otimes \id_{E} \rho^{x,\lambda}_{A\tilde{E}}  \right) \otimes \ketbra{\lambda}{\lambda}_{\Lambda}. 
\end{equation*} 
The above motivates us to define strategies $\M{C}^{\lambda} = (\rho^{0 , \lambda}_{AE} , \rho^{1 , \lambda}_{AE} , M^{\lambda}_0 )$. By  construction, every strategy $\M{C}^{\lambda}$, 
is in the set $\mathfrak{C}_1[\score_{\lambda} , \overlap_{\lambda}] $ for some $\score_{\lambda}$ and $\overlap_{\lambda} \in [0 , 1]$. \\ 
It is easy now to see that 
\begin{eqnarray}
    \overlapfunction_{\Pi_0}(\M{C}) &=& \frac{1}{2} \sum_{x} \tr \left( \Pi_0 \otimes \id_{E}\left( p(\lambda) \rho^{x, \lambda}_{A\tilde{E}} \otimes \ketbra{\lambda}{\lambda} \right) \right) \nonumber \\ 
    &=& \frac{1}{2} \sum_{x} p(\lambda) \tr \left( \Pi_0 \rho^{x, \lambda}_{A} \right) \nonumber \\ 
    &=& \sum_{\lambda} p(\lambda) \overlapfunction_{\Pi_0}(\M{C}_{\lambda}). 
\end{eqnarray}
Similarly, we can show that $\scorefunction(\M{C}) = \sum_{\lambda} p(\lambda) \scorefunction(\M{C}_{\lambda} )$. \\ 
Finally, 
\begin{eqnarray}
    H(Y|X\Lambda E)_{\M{C}} &=& \sum_{\lambda} p(\lambda) H(Y|X, \Lambda = \lambda , E)_{\M{C}} \nonumber \\
    &=& \sum_{\lambda} p(\lambda) H(Y| XE)_{\M{C}_{\lambda}}. \nonumber
\end{eqnarray} 
Combining all the above gives the objective function 
\begin{eqnarray}
    H(Y| X \Lambda \tilde{E}) =  \sum_{\lambda}p(\lambda) H(Y|X \tilde{E})_{\M{C}_{\lambda}},
\end{eqnarray}
and the constraints are given by 
\begin{eqnarray}
    \sum_{\lambda} p(\lambda) \overlapfunction_{\Pi_0} (\M{C}_{\lambda}) = \overlap \\ 
    \sum_{\lambda} p(\lambda) \scorefunction(\M{C}_{\lambda}) = \score.
\end{eqnarray}
Thus, solving the for the function
\begin{eqnarray}
\begin{aligned}
G_{p_{X}}(\score , \overlap) :=    \quad &\inf  H(Y|X \tilde{E})_{\M{C}_{\lambda}} \\ 
   \quad &    \overlapfunction_{\Pi_0} (\M{C}_{\lambda}) = \overlap \\ 
    \quad &  \scorefunction(\M{C}_{\lambda}) = \score,
\end{aligned}
\end{eqnarray}
and taking the convex lower bound of the function, will give the function $F_{p_{X}}(\score , \overlap)$.
\end{proof}
\noindent In section \ref{app: LF tranform}, we have examined the methods used to calculate the lower bounds over arbitrary probability distributions $p(\mu)$. The fundamental idea of the Lemma above is that when computing rates, it is possible to restrict the calculation to a scenario in which no prior randomness has been shared, by deferring the consideration of lower bounds across all convex combinations at this stage. 

\section{Reduction to qubit strategies}
The main issue with the optimization problem $G_{p_{X}}(\score , \overlap)$ is that no assumption has been made on the dimensions of the state $\rho_{AE}^{x}$. We saw in Chapter~\ref{chap: DI upper bounds} that the complexity of these optimization problems is reduced using Jordan's Lemma. Jordan's Lemma shall be useful for our protocols here as well. We had stated Jordan's Lemma in Chapter~\ref{chap: DI upper bounds} (see lemma~\ref{lem:Jordan}), however, we state it here again for convenience.
\begin{lemma}[Jordan's Lemma Extended]Let $A, B \in S(\mathcal{H})$ be two projections. Then 
\begin{equation}
    \mathcal{H}  = \bigoplus_{\alpha} \mathcal{H}_{\alpha}  
\end{equation}
such that each $\mathcal{H}_{\alpha}$ is an invariant subspace of $\mathcal{H}$ under the action of $A$, $B$, $\id - A $ and $\id - B$ . Moreover, the dimension of each subspace $\mathcal{H}_{\alpha}$ is at most 2. 
\end{lemma}
We will now prove the following technical result:
\begin{lemma}\label{lem: unqquness of lambda0} Let $P_0$ and $\Pi_{0}$ be two projectors on the Hilbert space $\M{H}$. Further let $\Pi_{0}$ be a rank one projector. Consider the Jordan decomposition of $\M{H} = \bigoplus_{\lambda}\M{H}_{\lambda}$ defined by the operators $\{ P_0 , \Pi_0 , \id - P_0 , \id - \Pi_{0} \}$. Among these subspaces, all spaces $\M{H}_{\lambda}$ are contained within the null space of $\Pi_0$, except for a single subspace denoted as $\M{H}_{\lambda_0}$. Furthermore, the projector onto $\M{H}_{\lambda_0}$ takes the form:
\begin{eqnarray}
P_{\lambda} = \Pi_0 + \bar{P},
\end{eqnarray}
where $\bar{P}$ is any projection, up to rank 1, onto the null space of $\Pi_0$
\end{lemma}
\begin{proof}
From Lemma \ref{lem: linear-algebra}, we can deduce that $[\Pi_0, P_{\lambda}] = 0$, indicating that $P_{\lambda}$ and $\Pi_0$ share common eigenvectors. Likewise, $[\id - \Pi_0, P_{\lambda}] = 0$ implies that $\id - \Pi_0$ shares eigenvectors with $P_{\lambda}$. Since $P_{\lambda}$ is a projection onto a subspace of dimension at most 2, it must take the form $P_{\lambda} = \alpha_{\lambda} \Pi_0 + \beta_{\lambda} \bar{P}_{\lambda}$, where $\bar{P}_{\lambda}$ is any projection onto the null space of $\Pi_0$, and $\alpha_{\lambda}, \beta_{\lambda} \in \mathbb{R}$. Considering any two subspaces, $\M{H}_{\lambda_1}$ and $\M{H}_{\lambda_2}$, with orthogonal supports, we observe that $\alpha_{\lambda_1} \alpha_{\lambda_2} = \alpha_{\lambda_1}$, implying that $\alpha_{\lambda} \in [0, 1]$ for every $\lambda$. Furthermore, due to the orthogonality of the supports of $\M{H}_{\lambda}$, $\alpha_{\lambda} = 1$ can only hold for a single subspace.
\end{proof}
We employ Jordan's Lemma to achieve a result that significantly simplifies the problem. This result lets us effectively transform the problem into one concerning only the convex combination of qubits. In essence, the approach is to employ Jordan's Lemma to express the states $\rho^{x}_{AE}$ in a block diagonal form of size $2 \times 2$, denoted by a direct sum $\bigoplus_{\lambda} \rho^{x, \lambda}_{AE}$. Here $\rho^{x,\lambda}_{A} \in \M{S}(\M{H}_{\lambda}) $ is a qubit state. Moreover, the measurement operators $\{ M_{0} , \id - M_{0}\}$ also act on the a projective measurement on each subspace $\M{H}_{\lambda}$ - i.e. $M_{y} =  \bigoplus_{\lambda} M_{y}^{\lambda}$. This transformation effectively simplifies our problem to an optimization problem over qubits, thereby significantly reducing its complexity. Moreover, as we shall see the lemma above can be used to show that the overlap of the strategy arises solely from a single Jordan block, implying that only one qubit block is relevant for analysis. As a result, the optimization problem can be effectively viewed as a scenario in which the source shares a single pair of qubits, rather than a state with unrestricted dimensions.
\begin{lemma}\label{lemm: Jodran's decomposition corollary} $G_{p_{X}}( \score , \overlap) \geq G^{(2)}_{p_{X}} ( \score , \overlap)$, where 
\begin{eqnarray}\label{eqn: Gopt}
\begin{aligned}
    G_{p_{X}}^{(2)}(  \score , \overlap) = \inf \quad &  \sum_{x \in \{ 0 , 1\}}    \eta_{x} p_{X}(x) H(Y| E)_{\rho^{x}}  \\
\textrm{s.t.} \quad & \forall x \in \{ 0 , 1 \}: \eta_{x} \in [0 , 1] \\ 
              \quad & \forall x \in \{ 0 , 1 \}: \rho^{x}_{A} \in S(\M{H}_2) \\ 
              \quad & \forall y \in \{ 0 , 1 \}: M_{y} = \ketbra{\phi}{\phi} \text{ for some } \ket{\phi} \in \M{H}_2   \\ 
              \quad & M_{0} + M_{1} = \id_{2}  \\ 
              \quad & \rho_x = \sum_{y \in  \{ 0 , 1 \}} \ketbra{y}{y}_{Y} \otimes \tr \left((M_{y} \otimes \id_{E}) \rho^{x}_{AE} \right)   \\
\quad & \sum_{x \in \{ 0 , 1\}} \left( \frac{1}{2} \eta_{x} \tr_{A}\left( M_{x} \rho^{x}_{A} \right) \right)   \in  [\score - \sum_{x} \frac{1}{2}(1 - \eta_{x}) , \score]  \\
              \quad &   \sum_{x \in \{ 0 , 1\} } \frac{1}{2}\eta_{x} \tr_{A} \left( \Pi_{0} \rho^{x}_{A} \right) = \overlap ,\\ 
\end{aligned}
\end{eqnarray}
where $\M{H}_{2}$ is a two dimensional Hilbert space and $\id_2$ is the identity on the space. 
\end{lemma}
\begin{proof}
Let $\M{C} = (\rho^{0}_{AE} , \rho^{1}_{AE} , M_{0}) \in \mathfrak{C}_{P}^{1}$. Let $P_{\lambda}$ be the projection onto $2$ dimensional sub-space $\M{H}_{\lambda}$, where $\M{H}_{\lambda}$ is any subspace that is invariant under the actions of the projectors $\Pi_0, M_{0} , \id - M_{0}$ and $\id - \Pi_{0}$. Now for the strategy $\M{C}$ consider the following: 
\begin{eqnarray}
\rho_{\M{C}} &=& \sum_{y,x} p_{X}(x)\ketbra{y,x}{y,x} \otimes \left(\rho_{AE}^{x} M_{y} \otimes \id_{E}    \right) \nonumber\\ 
             &=& \sum_{y,x} p_{X}(x)\ketbra{y,x}{y,x} \otimes \tr_{A}\left(\rho_{AE}^{x} \left( \sum_{\lambda} P_{\lambda}^{2} \otimes \id_{E} \right) M_{y} \otimes \id_{E}    \right) \nonumber\\
             &=& \sum_{y,x,\lambda} p_{X}(x)\ketbra{y,x}{y,x} \otimes \tr_{A}\left( \left(P_{\lambda} \otimes \id_{E} \rho_{AE}^{x} P_{\lambda} \otimes \id_{E}\right) \left( P_{\lambda} M_{y} P_{\lambda}\right) \otimes \id_{E}    \right). \nonumber
\end{eqnarray}
We introduce the quantities $\eta_{x}^{\lambda} \rho^{x, \lambda}_{AE} = P_{\lambda} \otimes \id_{E} \rho^{x}_{AE} P_{\lambda} \otimes \id_{E}$ and $M^{\lambda}_{y} = P_{\lambda} M_{y} P_{\lambda}$, where $\eta_{x}^{\lambda} \in [0, 1]$ (normalization constant) and $\rho^{x, \lambda}_{AE}$ are any normalized states. Furthermore, it is easy to verify that $P_{\lambda} M_{y} P_{\lambda}$ is a projector and $\rho^{x, \lambda}_{A}$ are qubits. Further define $\rho^{\lambda}_{x}$ to be the following CQ state 
$$ \rho^{\lambda}_{x} = \sum_{y} \ketbra{y}{y}_{Y} \otimes \ketbra{x}{x}_{X } \otimes  \tr\left( M_{y} \otimes \id_{2} \rho^{x, \lambda}_{AE}  \right) .$$
Using the concavity of the \RutC{von Neumann} entropy, we have that 
\begin{eqnarray}
H(Y|XE)_{\M{C}} &=& \sum_{x} p_{X}(x) H(Y|X=x,E)_{\rho_{\M{C}}}  \nonumber \\ 
                &\geq&  \sum_{x} p_{X}(x)\sum_{\lambda} \eta_{x}^{\lambda}  H(Y|X=x,E)_{\rho^{\lambda}_{x}} \nonumber \\ 
                &\geq& \sum_{x} p_{X}(x)\eta_{x}^{0}  H(Y|X=x,E)_{\rho^{0}_{x}} \nonumber .
\end{eqnarray}
From Lemma~\ref{lem: unqquness of lambda0}, let $\M{H}_{0}$ be the unique subspace which is not entirely in the nullspace of $\Pi_0$. 
The score function for the strategy is given by:
\begin{eqnarray}
\scorefunction(\M{C}) &=& \sum_{\lambda , x} \frac{1}{2} \tr_{A} \left(\eta^{\lambda}_{x} \rho^{x, \lambda}_{A} M^{\lambda}_{x} \right)  \nonumber\\
&=&  \sum_{ x} \frac{1}{2} \tr_{A} \left(\eta^{0}_{x} \rho^{x, 0}_{A} M^{0}_{x} \right) +  \sum_{\lambda \ne 0 , x} \frac{1}{2} \tr_{A} \left(\eta^{\lambda}_{x} \rho^{x, \lambda}_{A} M^{\lambda}_{x} \right)  \nonumber,
\end{eqnarray}
with $\score_{x}^{\lambda} := \tr \left(M_{x}^{\lambda} \rho^{x,\lambda}_{A} \right)$. The expression above allows us to split the contribution of the strategy into a term that depends only on the single set of qubit states $\rho^{x,0}_{A}$ and projective measurements $M_{y}^{0}$, and a term that depends on the other possible qubit states and measurements. To obtain lower bound and upper bound on the score function $\scorefunction(C)$, we can set $\score^{\lambda}_{x} \in [0 , 1]$ . As $\sum_{\lambda} \eta_{x}^{\lambda}  = 1$, we get the following bounds: 
\begin{eqnarray}
\sum_{x} \frac{1}{2} \left(\eta_{x}^{0} \tr_{A}  \left(\rho^{x, 0}_{A} M_{x} \right) \right) \leq \scorefunction(\M{C}) \leq \sum_{x} \frac{1}{2} \left( \eta_{x}^{0}\tr_{A} \left(\rho^{x, 0}_{A} M_{x} \right) + (1 - \eta_{x}) \right) \nonumber.
\end{eqnarray}
The most importantly, for the overlap conditions, we have that
\begin{eqnarray}
    \overlapfunction_{\Pi_0}(\M{C}) &=& \sum_{x} \frac{1}{2} \eta_{x} \tr \left(\Pi_{0} \rho^{x}_{A} \right)  \nonumber \\ 
    &=& \sum_{x,\lambda} \frac{1}{2} \eta_{x}^{\lambda}  \tr \left( \Pi_0 P_{\lambda}\rho^{x}_{AE} P_{\lambda}\right) \nonumber \\ 
    &=&  \sum_{x,\lambda} \frac{1}{2} \eta_{x}^{0}  \tr \left( \Pi_0 \rho^{x,0}_{AE} \right) \nonumber \\ 
    &=& \overlap \nonumber .
\end{eqnarray}
Now, replacing $\eta_{x}^{0} \rightarrow \eta_{x}$, $\rho^{x , 0}_{AE} \rightarrow \rho^{x}_{AE}$ and $M_{y}^{0} \rightarrow M_{y}$ proves the lemma.
\end{proof}

\section{Converting the optimization problem to a traditional form}
In the previous section, we showed that it is sufficient to restrict to qubit strategies. The aim now is to simplify the optimization problem obtained in the previous section to a more traditional optimization problem that can be solved using numerical techniques. We begin by simplifying the expression of the \RutC{von Neumann} entropy as
\begin{eqnarray}
H(Y|E) &=& H(Y, E) - H(E) \nonumber\\
&=& H(Y) + H(E |  Y) - H(E) \nonumber\\
&\geq& H(Y) - H(E) \nonumber .
\end{eqnarray}
We used the chain rule for conditional von Neumann entropy, as well as the fact that $H(E|Y)_{\rho} \geq 0$ if $\rho = \sum_{y} p_{Y}(y) \ketbra{y}{y} \otimes \rho^{y}_{E}$. In fact, the inequality above becomes tight when $\rho^{y}_{E}$ is a pure state, which is the case in our setting.

\RutC{Next, we express the CQ state of our protocol when $X = x$ is observed:} 
\begin{eqnarray}
\sum_{Y} p_{Y}^{x}(y) \ketbra{y}{y} \otimes \tr_{A} \left( \frac{M^{x}_{y} \otimes \id_{E} \rho^{x}_{AE}}{\tr_{A}\left( M^{x}_{y} \rho^{x}_{A} \right)} \right) ,
\end{eqnarray}
where $p_{Y}^{x}(y) := \tr_{A}\left(\rho^{x}_{A} M_{y} \right)$. By noting that $p_{Y}^{x}(y) = 1 - p_{Y}^{x}(y \oplus 1 )$, we can compute the expression for $H(Y)$ as
\begin{eqnarray}
H(Y) = -\sum_{y} p_{Y}^{x}(y)\log_2\left(p_{Y}^{x}\right) = H_{\bin}\left( \tr \left(\rho^{x}_{A} M_{0} \right)\right).
\end{eqnarray}
Similarly, we can compute $H(E)$ by noting that $\rho^{x}_{AE}$ is a purification of $\rho^{x}_{A}$. Specifically, note that
\begin{eqnarray}
\rho_{E}^{x} &=& \sum_{y} p_{Y}^{x}(y) \tr_{A} \left( \frac{M_{y} \otimes \id \rho^{x}_{AE}} {\tr_{A}\left(\rho^{x}_{A} M_{0} \right)} \right) \nonumber \\ 
&=& \tr_{A} \left( (\sum_{y} M_{y}) \otimes \id_{E} \rho^{x}_{AE} \right) \nonumber \\
&=& \tr_{AE} \left(\rho^{x}_{AE} \right). \nonumber 
\end{eqnarray}
\noindent By applying Lemma \ref{lemm: No-signalling corollary}, we can conclude that the entropies $H(\rho^{x}_{E})$ and $H(\rho^{x}_{A})$ are equivalent if $\rho^{x}_{AE}$ is a purification of $\rho^{x}_{A}$. 

The discussion above leads to the following result
\begin{lemma} The optimization problem~\eqref{eqn: Gopt} can be expressed as
 \begin{eqnarray}\label{eqn: Ggrt in terms of probabilities}
\begin{aligned}
    G_{p_{X}}^{(2)}(  \score , \overlap) = \inf \quad &  \sum_{x \in \{ 0 , 1\}}    \eta_{x} p_{X}(x) \left(H_{\bin}\left(\tr\left(\rho^{x}_{A} M_{0} \right)\right) 
 - H(\rho^{x}_{A})\right)   \\
\textrm{s.t.} \quad & \forall x \in \{ 0 , 1 \}: \eta_{x} \in [0 , 1] \\ 
              \quad & \forall x \in \{ 0 , 1 \}: \rho^{x}_{A} \in S(\M{H}_2) \\ 
              \quad & \forall y \in \{ 0 , 1 \}: M_{y} = \ketbra{\phi}{\phi} \text{ for some } \ket{\phi} \in \M{H}_2   \\ 
              \quad & M_{0} + M_{1} = \id_{2}  \\ 
              \quad & \rho_x = \sum_{y \in  \{ 0 , 1 \}} \ketbra{y}{y}_{Y} \otimes \tr \left((M_{y} \otimes \id_{E}) \rho^{x}_{AE} \right)   \\
\quad & \sum_{x \in \{ 0 , 1\}} \left( \frac{1}{2} \eta_{x} \tr_{A}\left( M_{x} \rho^{x}_{A} \right) \right)   \in  [\score - \sum_{x} \frac{1}{2}(1 - \eta_{x}) , \score]  \\
              \quad &   \sum_{x \in \{ 0 , 1\} } \frac{1}{2}\eta_{x} \tr_{A} \left( \Pi_{0} \rho^{x}_{A} \right) = \overlap. \\
\end{aligned}
\end{eqnarray}
\end{lemma}

\section{Optimization problem in terms of bounded real variables}
 The problem with Lemma \ref{eqn: Ggrt in terms of probabilities} is that it is expressed in terms of the states $\rho^{x}_{A}$ and the measurement $M_{y}$. We would like to re-express this optimization problem in terms of a standard optimization problem on a bounded domain $\mathcal{D} \subseteq \mathbb{R}^{n}$. To achieve this, we without any loss of generality we parameterize
\begin{align}
\rho^{x}_{A} = \frac{\id}{2} + \sum_{i=1}^{3} \frac{a_i^{x}}{2} \sigma_i , \quad  M_{0} = \ketbra{\psi_0}{\psi_0} = \frac{\id}{2} + \sum_{i=1}^{3} \frac{b_i}{2} \sigma_i \text{ and } \Pi_0 = \ketbra{0}{0} = \frac{\id}{2} + \frac{\sigma_1}{2}. 
\end{align}
Here $\{ \sigma_{i} \}_{i = 1}^{3}$ are Pauli $x,y,z$ operators. The constraints $\sum_{i} (a_i^{x})^2 \leq 1$ and $\sum_{i} b_{i}^2 = 1$ must be satisfied, as they ensure that the parameters $\{a_i^{x}\}_{i = 1}^{3}$ and $\{b_i \}_{i = 1}^{3}$ represent a valid density operator and a projective measurement, respectively. We now have an optimization problem over $11$ variables- 3 parameters for each state  $\rho^{x}_{A}$, three parameters describing the measurement operator and $2$ parameters $\eta_0$ and $\eta_1$ that give the probability of the Jordan Block $\M{H}_{0}$ occurring. Thus, the optimization problem can be cast as an optimization problem over a compact subset of $\mathbb{R}^{d}$, where $d=11$ at this stage. We will now reduce the value of $d$ to simplify the optimization problem - i.e. get rid of some redundant variables. \\
Before we simplify this optimization problem any further, we introduce the following function that will help keep the expressions compact
\begin{eqnarray}
  \Phi(x) := H_{\bin} \left(\frac{1}{2} + \frac{x}{2} \right).
\end{eqnarray}
Several properties of $\Phi(x)$ including its monotocity etc. can be easily inferred from the properties of the binary entropy $H_{\bin}$. Some other useful properties of $\Phi(x)$ that are relevant to our proof are discussed in Appendix \ref{app: useful claims}. Now consider the following result
\begin{lemma}
The optimization problem \eqref{eqn: Ggrt in terms of probabilities} is equivalent to the following optimization problem
\begin{eqnarray}\label{eqn: Ggrt in terms of real variables}
 \begin{aligned}
     \inf \quad &  \sum_{x}   \eta_{x} p_{X}(x) \left(\Phi\left( a_{x} \cos(\theta_x - \phi)\right) - \Phi \left(a_{x} \right)\right)  \\
\textrm{s.t.}
\quad & \forall x \in \{ 0 , 1\} : \eta_{x} , a_{x} \in [0 , 1] \\
\quad & \forall x \in \{ 0 , 1 \}: \theta_{x} \in [0 , 2 \pi] \\
\quad & \phi \in [0 , 2 \pi] \\
\quad &  \sum_{x}  \frac{1}{2} \eta_{x} \left( \frac{1}{2} + \frac{(-1)^{x} a_{x} \cos(\theta_x - \phi)}{2}\right) \in [\score - \sum_{x} \frac{1}{2}(1 - \eta_{x}), \score]   \\
              \quad &   \sum_{x} \frac{1}{2}\eta_{x}  \left( \frac{1}{2} + \frac{a_{x} \cos(\theta_{x})}{2}\right) = \overlap .
 \end{aligned} 
\end{eqnarray}
\end{lemma}
\begin{proof}
First, note that the objective function and constraints in the optimization problem \eqref{eqn: Ggrt in terms of probabilities} are expressed solely in terms of $\tr_{A}\left(M_{0} \rho^{x}_{A} \right)$, $\tr_{A}\left( \Pi_{0} \rho^{x}_{A} \right)$, and $H(\rho^{x}_{A})$, so it is useful to express them in terms of the Bloch vectors. Define $\B{a}^{x}$ and $\B{b}$ the Bloch vector that denotes the qubit states $\rho^{x}_{A}$ and the projection $M_{0}$. Further notice that $\Pi_0 = \ketbra{0}{0}$ can be equivalently expressed as the vector $\hat{\B{z}}$ in this notation. Thus,
\begin{eqnarray}
\tr_{A} \left(M_{0} \rho^{x}_{A}\right) &=& \frac{1}{2} + \frac{\B{a}^{x}.\B{b}}{2} \nonumber \\
\tr_{A} \left(\Pi_0 \rho^{x}_{A}\right) &=& \frac{1}{2} + \frac{\B{a}^{x}.\hat{\B{z}}}{2} .
 \nonumber 
\end{eqnarray}
Without any loss of generality, we can chose a basis such that $\B{z} = (1 , 0 , 0)$ and $\B{b} = (\cos(\phi) , \sin(\phi) , 0)$. Furthermore, we write that $\B{a}^{x} = (a_{x} \cos(\theta_{x}) , a_{x} \sin(\theta_{x}) , \tilde{a}_{x})$. In this notation, this gives: 
\begin{eqnarray}
\tr_{A} \left(M_{0} \rho^{x}_{A}\right) &=& \frac{1}{2} + \frac{a_{x} \cos(\theta_{x} - \phi)}{2} \nonumber \\
\tr_{A} \left(\Pi_0 \rho^{x}_{A}\right) &=& \frac{1}{2} + \frac{a_{x} \cos(\theta_{x})}{2}. 
 \nonumber 
\end{eqnarray}
The entropy $H(\rho^{x}_{A})$ can be computed in terms of its eigenvalues $\frac{1}{2} + \frac{|\B{a}^{x}|}{2}$ and $\frac{1}{2} - \frac{|\B{a}^{x}|}{2}$ as
\begin{eqnarray}
H(\rho_A^{x}) = H_{\bin}\left(\frac{1}{2} + \frac{|\B{a}^{x}|}{2} \right) = \Phi\left(|\B{a}^{x}| \right). \nonumber
\end{eqnarray}
We can use the monotonicity of $\Phi(x)$ for $x > 0$, so that 
\begin{eqnarray}
   H(\rho_A^{x})  = \Phi\left(\sqrt{a_{x}^2 + \tilde{a}_{x}^2 }\right) \leq \Phi\left(a_{x}\right). \nonumber
\end{eqnarray}
Note that when optimizing over all strategies, the inequality is tight. Combining all the results above and performing the substitutions in \eqref{eqn: Ggrt in terms of probabilities}, we can prove the lemma.
\end{proof}

\section{Extending the domain} 
To compute the rates, we want to solve the optimization problem \ref{eqn: Ggrt in terms of real variables} using numerical techniques. We find that this optimization problem is resource consuming to solve directly, and therefore, we aim to lower bound this problem, which can be effectively handled numerically. By Lemma \ref{Claim: monotonicity in x} (see Appendix~\ref{app: Semi-DI}), we can lower-bound the objective function by noting that
\begin{eqnarray}
    \eta_{x} \left(\Phi\left( a_{x} \cos(\xi_{x})\right) - \Phi \left(a_{x} \right)\right)  \geq  \left(\Phi\left( \eta_{x} a_{x} \cos(\xi_{x})\right) - \Phi \left(\eta_{x} a_{x} \right)\right) .
\end{eqnarray} 
Furthermore, we can replace the constraints 
\begin{eqnarray}
  \sum_{x}  \frac{1}{2} \eta_{x} \left( \frac{1}{2} + \frac{(-1)^{x} a_{x} \cos(\theta_x - \phi)}{2}\right) &\in& [\score - \sum_{x} \frac{1}{2}(1 - \eta_{x}), \score]  \nonumber  \\
  \sum_{x} \frac{1}{2}\eta_{x}  \left( \frac{1}{2} + \frac{a_{x} \cos(\theta_{x})}{2}\right) &=& \overlap \nonumber
\end{eqnarray}
of the optimization problem \eqref{eqn: Ggrt in terms of real variables} by the relaxed constraints
\begin{eqnarray}
  \sum_{x}  \frac{1}{2} \eta_{x} \left( \frac{1}{2} + \frac{(-1)^{x} a_{x} \cos(\theta_x - \phi)}{2}\right) &\geq& \score - \sum_{x} \frac{1}{2}(1 - \eta_{x})  \nonumber  \\
  \sum_{x} \frac{1}{2}\eta_{x}  \left( \frac{1}{2} + \frac{a_{x} \cos(\theta_{x})}{2}\right) &\geq&  \overlap. \nonumber
\end{eqnarray}
Such a relaxation can only decrease the minimal value, as the feasible set of \eqref{eqn: Ggrt in terms of real variables} is a subset of the relaxed feasible set. As, we shall eventually take the convex envelope of the function later on, this relaxation, does not affect the overall tightness of our rate. \par
It can now be useful to re-label the variables $\eta_{x} a_{x} = \tilde{a}_{x}$. As $a_{x} \in [0 , 1]$, we can add another constraint that $\tilde{a}_{x} \leq \eta_{x}$, completely eliminating the role of $a_{x}$ in the optimization problem. We now make the following substitutions $\theta_x - \phi \rightarrow \xi_{x} $ 
Summarizing all the above leads to the following result:
\begin{lemma}\label{lemm: one more lower bound}
A lower bound on the function $G^{(2)}_{p_{X}}(\score , \overlap) $ can be achieved by computing the optimization problem 
\begin{eqnarray}\label{eqn : Non-polynomial optimization problem}
   \begin{aligned}
  \begin{aligned} 
     \inf \quad &  \sum_{x}  p_{X}(x) \left(\Phi\left( \tilde{a}_{x} \cos(\xi_{x})\right) - \Phi \left(\tilde{a}_{x} \right)\right)  \\
\textrm{s.t.} 
\quad &  \forall x \in \{ 0 , 1 \}: 1 \geq \eta_{x} \geq \tilde{a}_{x} \geq 0 \\
\quad & \forall x: \xi_{x} \in [0 , 1] \\
\quad &  \sum_{x}    \left( -\eta_{x} + (-1)^{x}  \tilde{a}_{x} \cos(\xi_{x})\right)) \geq 4 \score -4  \\
              \quad &   \sum_{x}   \left( \eta_{x} + \tilde{a}_{x} \cos(\xi_{x} + \phi)\right) \geq  4 \overlap. 
 \end{aligned} 
 \end{aligned}   
\end{eqnarray} 
\end{lemma} 

\section{Eliminating one more variable} 

Now, we can immediately simplify the final constraint in terms of overlap by observing that $\phi$ only appears one of the constraints. We should, without any loss of generality, choose the value of $\phi$ that optimizes the constraint (See lemma \ref{lemm: maximizing the constriant} in appendix~\ref{app: Semi-DI}). We observe that:
\begin{eqnarray}
  \sum_{x} \eta_{x} a_{x} \cos(\xi_{x} + \phi) &=&   \left( \sum_{x}\eta_{x} a_{x} \cos(\xi_{x}) \right) \cos(\phi) - \left( \sum_{x} \eta_{x} a_{x} \sin(\xi_{x}) \right)  \sin(\phi) \nonumber\\ 
  &\leq& \sqrt{ \left( \sum_{x}\eta_{x} a_{x} \cos(\xi_{x}) \right)^2 +  \left( \sum_{x} \eta_{x} a_{x} \sin(\xi_{x}) \right)^2} .
\end{eqnarray} 
Note\footnote{It is easy to see that $G_{p_{X}}(\score , \overlap) = 0$ when $\overlap \leq \frac{1}{2}$, so $\overlap \geq \frac{1}{2}$ is the domain of our interest to begin with.} that because $\overlap - \sum_{x} \eta_{x}  \geq 0$ for $\overlap \geq \frac{1}{2}$  and  
\begin{eqnarray}
    \sqrt{ \left( \sum_{x}\eta_{x} a_{x} \cos(\xi_{x}) \right)^2 +  \left( \sum_{x} \eta_{x} a_{x} \sin(\xi_{x}) \right)^2}  + \sum_{x} \eta_{x} \geq 4 \overlap
\end{eqnarray}
implies that 
\begin{eqnarray}
    \left( \sum_{x}\eta_{x} a_{x} \cos(\xi_{x}) \right)^2 +  \left( \sum_{x} \eta_{x} a_{x} \sin(\xi_{x}) \right)^2  \geq \left( 4 \overlap - \sum_{x} \eta_{x}\right)^2 .
\end{eqnarray}
Summarizing all the above long with the results in Lemma \ref{lemm: one more lower bound} leads to the following result:
\begin{lemma} The function $\entropy_{p_{X}}(\score ,\overlap)$ defined by
\begin{eqnarray}
\begin{aligned} 
     \inf \quad &  \sum_{x}    p_{X}(x) \left(\Phi\left(\tilde{a}_{x} \cos(\xi_{x})\right) - \Phi \left(\tilde{a}_{x} \right)\right)  \\
\textrm{s.t.} \quad &  \sum_{x}    \left( -\eta_{x} + (-1)^{x} \tilde{a}_{x} \cos(\xi_{x})\right)) \geq  4 \score -4  \\
              \quad &    \left(  \sum_{x}\tilde{a}_{x} \cos(\xi_{x}) \right)^2 +  \left( \sum_{x} \tilde{a}_{x} \sin(\xi_{x}) \right)^2 - \left( 4 \overlap  - \sum_{x} \eta_{x} \right)^2 \geq 0\\ 
              \quad &  \forall x \in \{0 , 1 \}:  1 \geq \eta_{x} \geq \tilde{a}_{x} \geq 0.
 \end{aligned}
\end{eqnarray}
is a lower bound on the function $G^{(2)}_{p_{X}}(\score, \overlap)$. 
\end{lemma}
In the appendix \ref{app: Semi-DI}, we shall show that the function $\entropy_{p_{X}}(\score, \overlap)$is monotonically increasing in $\score$ and $\overlap$.

\section{Computing the optimization problem over grids}
We now address the problem of computing the convex envelope of the function $\entropy_{p_{X}}(\score, \overlap)$. As the functional form of this is difficult to obtain, we can try to compute this numerically using known optimization techniques. However, in order to compute the convex envelope numerically, we can only compute this on a finite set of points or a finite grid. So, we construct another a lower bound $\entropy^{\M{P}}_{p_{X}}(\score , \overlap)$ of the function $\entropy_{p_{X}}(\score , \overlap)$, that can be only computed using a finite collection of points.
For $\overlap > \frac{1}{2}$ define: 
\begin{eqnarray}
    \entropy^{\M{P}}_{p_{X}}(\score , \overlap)  := \entropy_{p_{X}}(\score_{\M{P}}, \overlap_{\M{P}} ),
\end{eqnarray}
where
\begin{eqnarray}
    \score_{\M{P}}:= \max\{ \score_i  : (\score_i, \overlap_{j}) \in \M{P} , \score_i \leq \score , \overlap_{j}  \leq \overlap\}  \nonumber \\ 
    \overlap_{\M{P}}:= \max\{ \overlap_j  : (\score_i, \overlap_{j}) \in \M{P} , \score_i \leq \score , \overlap_{j}  \leq \overlap\}. \nonumber 
\end{eqnarray} 
Note that $ \entropy^{\M{P}}_{p_{X}}(\score , \overlap)$ is a lower bound on the \RutC{function $\entropy_{p_{X}}(\score_{\M{P}}, \overlap_{\M{P}} )$ as $\entropy_{p_{X}}(\score_{\M{P}}, \overlap_{\M{P}} )$ } 
is also monotonically increasing in $\score$ and $\overlap$; precisely, $\partial_{\score} {\entropy}_{p_{X}} \geq 0$ and $\partial_{\overlap}{\entropy}_{p_{X}} \geq 0$. 

Using the algorithms described in the section \ref{app: LF tranform}, the function $\M{F}_{p_{X}}(\score , \overlap) := \text{convenv}(\entropy^{\M{P}}_{p_{X}}(\score , \overlap))$ can be computed in linear time over the partition $\M{P}$, which can be extended to the entire domain  using the same trick above. 

Note that $ \M{F}^{\M{P}}_{p_{X}}$ can be shown to be a lower bound of the $F_{p_{X}}$ in the section~\ref{sec: lower bounding the randomness rate}.
The optimization problem for $\entropy^{\M{P}}_{p_{X}}(\score , \overlap)$ can be easily relaxed to a polynomial optimization problem by lower bounding the objective function $$ \Phi(\tilde{a}_{x} \cos(\xi_{x})) - \Phi(x) \rightarrow  \Phi_{n}(\tilde{a}_{x} \cos(\xi_{x}) ) - \Phi_{m}(\tilde{a}_{x}) - I_{m+1},$$
and by replacing the variables $\cos(\xi_{x}) \mapsto c_{x} ,\sin(\xi_{x}) \mapsto d_{x}$ with an additional constraint $c_{x}^2 + d_{x}^2 = 1$. \RutC{Here $\Phi_{n}, \Phi_{m}$ are appropriate polynomial approximations of the function $\Phi(x)$ and $I_{m+1} := \sup_{x \in [-1 , 1]} |\Phi(x) - \Phi_{m}(x)|$. In the next section, we shall explicitly construct such polynomial approximations.}
\section{Converting the optimization problem to a polynomial optimization problem}\label{sec: poly-approximations of Phi(x)}
In this section, we approximate the binary entropy function (and its difference) in terms of polynomials. \\ 
We begin by the integral representation of the binary entropy 
\begin{eqnarray} 
\log(x) = \frac{1}{\ln(2)} \int_{0}^{1} \frac{x -1}{t(x - 1) + 1} \dd t .
\end{eqnarray}
From this representation, we can derive the integral representation of the function $\Phi(x) := H_{\bin}(\frac{1}{2} + \frac{x}{2})$ as 
\begin{eqnarray}
    \Phi(x) = \int^{1}_{0} (1 - x^2 ) \frac{(2 - t)}{(2 - t(1 -x)) (2 - t (1 + x)))} \dd t .
\end{eqnarray}
Upon performing some rearrangements, and performing change in variable, we obtain the following expression
\begin{eqnarray}
    \Phi(x) = \int^{1}_{\frac{1}{2}}\frac{1}{z \ln(2)}  \left( \frac{1 - x^2}{1 - \left(\frac{1 - z}{z} \right)^2 x^2} \right) \dd z
\end{eqnarray}
Now as $\frac{1 - z}{z} x \in [0 , 1]$ in the range $z \in [\frac{1}{2} , 1]$ and $x \in [-1 , 1]$, we can expand the integrand to the following convergent infinite sum 
\begin{eqnarray} 
\Phi(z) = \sum_{n = 0}^{\infty} \left(\int_{\frac{1}{2}}^{1}  \frac{1}{z \ln(2)}  \left(\frac{1 - z}{z} \right)^{2n} \dd z\right) x^{2n} (1 - x^2) .
\end{eqnarray}
Note here that the integrals 
\begin{eqnarray}
    I_{n} := \int_{\frac{1}{2}}^{1}  \frac{1}{z \ln(2)}  \left(\frac{1 - z}{z} \right)^{2n} \dd z .
\end{eqnarray}
are analytically solvable, e.g. $I_0 = 1$, $I_1 = 1 - 1/(2\ln(2))$ , $I_3 = 1 - 7/(12\ln(2))$ and so on. Furthermore, using this technique, we obtain lower-bounds on the binary entropy as $I_{n}(x) \geq 0$ can be trivially established. An approximation for the binary entropy can be made by truncating the summation to a certain term 
\begin{eqnarray}
    \Phi(x) \approx \Phi_{n}(x) := \sum_{k = 1}^{n} I_{k} x^{2k} (1 - x^2) .
\end{eqnarray}
The error term is given by 
\begin{eqnarray}
    \Phi(x) - \Phi_n(x) &=& (1 - x^2) \sum_{k = n+1}^{\infty} I_{k} x^{2k} \nonumber\\ 
                        &\leq& (1 - x^2) I_{n+1} \sum_{k = n+1}^{\infty} x^{2k} \nonumber \\ 
                        &=&  I_{n+1} x^{2(n+1)} \nonumber ,
\end{eqnarray}
where we have used the fact that $I_{n} > I_{n + k}$ for every $k \in \mathbb{N}$. We lower bound the objective function in \eqref{eqn : Non-polynomial optimization problem} by introducing the functions
\begin{eqnarray}
    P_{n}(\Tilde{a} , \cos(\xi)  ) := \Phi_{n}(\tilde{a} \cos(\xi)) - \Phi_{n}(\Tilde{a}) - I_{n+1} .
\end{eqnarray}
There is still some work to do in order to lower bound the optimization problem \eqref{eqn : Non-polynomial optimization problem} to a polynomial problem, as $\cos(\xi_{x})$ are not polynomials. However, we can identify the functions $\lambda_{1,x} = \tilde{a}_{x} \cos(\xi_x)$ and $\lambda_{2 , x} = \tilde{a}_{x} \sin(\xi_x)$. Now, if treat $\lambda_{i , x}$ as free variables in our optimization problem by introducing additional constraints 
\begin{eqnarray}
    \sum_{i} \lambda_{i , x}^2 = \tilde{a}_{x}^2, \nonumber    
\end{eqnarray}
then we can lower bound  \eqref{eqn : Non-polynomial optimization problem} to a polynomial optimization problem. Lower bounds can now be obtained to $\entropytwo(\overlap , \score)$ by solving this polynomial optimization problem using a sum-of-squares relaxation by using software such as Ncpol2sdpa. The final optimization problem is now of the form 
\begin{eqnarray}\label{eqn: final optimization problem for semi-DI}
\begin{aligned}  \entropytwo(\score , \overlap) :=
     \inf \quad &  \sum_{x}    p_{X}(x) \left(\Phi_{n}\left(\lambda_{1 , x})\right) - \Phi_{n} \left(\tilde{a}_{x} \right) \right)  - I_{n +1}  \\
\textrm{s.t.} \quad &  \sum_{x}    \left( -\eta_{x} + (-1)^{x} \lambda_{1 , x}\right)) \geq 4 \score -4  \\
              \quad &    \left(  \sum_{x}\lambda_{1 , x} \right)^2 +  \left( \sum_{x} \lambda_{2 , x} \right)^2 \geq \left( 4 \overlap  - \sum_{x} \eta_{x} \right)^2\\ 
              \quad & \lambda_{1, x}^2 + \lambda_{2 , x}^2 = \tilde{a}_{x} \\ 
              \quad &  \eta_{x} \geq \tilde{a}_{x} \\ 
              \quad & 1 \geq \eta_{x} \\ 
              \quad & \tilde{a}_{x} \geq  0 .
 \end{aligned}       
 \end{eqnarray}
Note that this is a constrained polynomial optimization problem over $8$ real variables. Actually, the number of variables is $6$, and we have two dummy variables here. 
Thus, we can summarize the final result of the chapter: 
\begin{theorem}
Let $\entropytwo(\score , \overlap)$ be defined as in \eqref{eqn: final optimization problem for semi-DI}, then 
\begin{eqnarray}
    \inf_{\M{C} \in \Gamma[\score , \overlap]} H(Y|XE)_{\rho_{\M{C}}} \geq \text{convenv}\left( \M{P}_{1}(\score , \overlap) \right)  .
\end{eqnarray}
\end{theorem}
\section{Taking the convex lower-bound of the rate}\label{app: LF tranform}
Upon computing the rate for a single strategy, now we are in the position to compute the rate for the convex combination of such strategies. In this section, we shall discuss a method that allows us to compute the convex envelope of any function $f: \mathbb{R}^{n} \mapsto \mathbb{R}$. We start by defining the following 
\begin{definition}[Convex envelope] 
 Let $f: \mathbb{R}^{n} \mapsto \mathbb{R}$ be any function, then the convex envelope of $f$ given by $\text{convenv}(f)$ is the function 
\begin{eqnarray}
    \mathrm{convenv}(f) := \max \{ g : g \hspace{0.1 cm} \mathrm{ is} \hspace{0.1 cm} \mathrm{ convex} \hspace{0.1 cm}\mathrm{ and } \hspace{0.1 cm} \forall \B{x}: \hspace{0.1 cm} g(\B{x})  \leq f(\B{x})\} .
\end{eqnarray}
Here $\max$ is taken to be point-wise maximum. 
\end{definition}
\noindent In simple words, $\text{convenv}(f)$ is the biggest convex function that is not greater than $f$. In other words, the $\text{convenv}(f)$ is the solution of the optimization problem 
\begin{eqnarray}
\begin{aligned} 
  \text{convenv}(f(\B{x}^*))  =   \inf_{\mu} \quad &  \int  \mathrm{d}\mu(\B{x})  f(\B{x}_{\mu})  \\ 
                    \textrm{s.t.} \quad &   \int  \mathrm{d} \mu(\B{x})   \B{x} = \B{x}^{*} ,\\ 
 \end{aligned}
\end{eqnarray}
where $\inf$ is taken over all the probability measures. 
We now define an important tool in convex analysis that allows us to compute the convex envelope of a function.
\begin{definition}[Legendre-Fenchel transform] Let $f: \mathbb{R}^n \mapsto \mathbb{R}$ be any function. Then $f^*: \mathbb{R}^{n} \mapsto \mathbb{R}$ is its Legendre-Fenchel transform if 
\begin{eqnarray}
    f^*(\mathbf{k}) := \sup_{\mathbf{x} \in \mathbb{R}^{n}} \left(\langle k , x \rangle  - f(\B{x}) \right) .
\end{eqnarray}
\end{definition}
The following can be shown for $f^{*}(\B{x})$
\begin{lemma}
Let $f: \M{D} \mapsto \mathbb{R}$ be bounded. Then the following holds 
\begin{itemize}
    \item $f^{*}$ is convex
    \item $(f^{*})^* = \mathrm{convenv}(f)$  
\end{itemize}
\end{lemma}
\noindent The proof of the lemma above can be found in textbooks of convex analysis such as~\cite{Convex_envelope_book} and shall be omitted in the work. There are algorithms in available in the literature~\cite{lucet1997faster,contento2015linear} to compute the convex envelopes by computing the LF conjugate of the function twice. We compute the convex envelope using the method of ~\cite{contento2015linear}, the code for which was generously provided by the authors to us.\\ 
Finally, we prove the final result about the fact that rate is non-decreasing.
\begin{lemma}
Let $f: \mathbb{R}^2 \rightarrow \mathbb{R}$ be a convex function such that $f \geq 0$. Suppose that $f(x,y) = 0$ for $x \leq x_0$ or $y \leq y_0$ for some $x_0 , y_0 \in \mathbb{R}$. Then $f(x,y)$ is non-decreasing.
\end{lemma}

\begin{proof}
Since $f$ is convex, the functions $g_y := f(\cdot,y)$ and $h_{x} := f(x, \cdot)$ are convex as well.\\
Take $x' > x$ and $\mu \in [0 , 1]$ such that $\mu x' + (1 - \mu) x_0 = x$. Then, by the convexity of $g_y$, we have\\
\begin{align*}
\mu g_{y}(x) + (1 - \mu) g_{y}(x') &\geq g_{y}(\mu x' + (1 - \mu) x_0) = g_{y}(x) ,
\end{align*}
which implies $\mu f(x', y) \geq f(x, y)$ for all $y$. Thus, we have $f(x', y) \geq f(x, y)$ for all $y$, showing that $f$ is non-decreasing in $x$. Similarly, we can show that $f$ is non-decreasing in $y$ using the convexity of $h_{x}$.\\
Therefore, $f(x, y)$ is non-decreasing in both variables.
\end{proof}

\chapter{Results and discussion}\label{chap: semi-DI protocols}
In the previous chapter, we discussed how to find the minimum of the single round \RutC{von Neumann} entropies for strategies that achieve a particular score and have a threshold overlap. In this chapter, we will look into some semi-DI protocols \RutC{and} apply our results to compute randomness rates for these protocols. \par

In  protocols of randomness expansion, we assume that the input randomness is not accessible to any eavesdropper during each round of the protocol. This assumption is crucial \RutC{for} the security of the protocol, \RutC{as if the Eavesdropper knows the input randomness, then they can easily pre-program the devices to produce deterministic outputs}. We design two types of protocols, depending on the source of input randomness: private and public. Before discussing the protocols in detail, we have a discussion regarding the assumptions under which our protocol operates.\par

Beyond the assumption that the source and the measurement device do not share any pre-existing entanglement, and that we have access to a trusted power meter, we have not made any assumptions regarding the functioning of the source and the measurement device. It remains an open question whether having this pre-existing entanglement offers any significant advantage to an eavesdropper's ability to predict the inputs and outputs of the protocol. 

However, it is \RutC{worth noting} that we assume the source and the measurement device can only communicate through the signal $\rho^{x}$ sent by the source \RutC{and cannot communicate by any other means. In order to ensure this, similar to the DI case, the source and measurement devices must be properly shielded}. Much like in Device Independent scenarios, once the protocol commences, neither the source nor the measurement device can convey information to potential adversaries. After the protocol, these devices must remain isolated (see Appendix~\ref{app:compos}) and should not be used again. \par

In the traditional prepare-and-measure scenario, \RutC{it is} crucial to recognize that the source and the measurement device might access alternative communication avenues. These methods could potentially convey the input's information indirectly, rather than encoding it directly within the source. Consider a scenario where the source sends signals $\rho^{x}$ with time lags, $\Delta t$, based on input $X$. This makes it straightforward for the measurement device to identify the input and send outputs using a pre-determined strategy. Ideally, these attacks can be avoided by introducing an additional abort condition which is based on the statistics of the test round of the protocol. In other words, our protocol should  abort if the power meter, our trusted component, detects any anomaly in signal reception timings (for instance, the protocol aborts if the root mean square value of the time lag between consecutive signals is bigger than a threshold). However, this study does not address preventing such attacks. \par

A practical method to circumvent both timing and entanglement-based attacks in semi-Device Independent protocols is to acquire the source and measurement devices from separate manufacturers. While this strategy does not guarantee absolute protection, it is a reasonable precaution in real-world scenarios, based on the security demands of the protocols. \par

For the security of the protocol we use a composable security definition, as detailed in Chapter~\ref{chap: DI protocols}. In the next two sections, we discuss two semi-Device Independent protocols based on the setup described in Section~\ref{sec: general_semi-DI protocol}. These protocols are inspired by the protocols~\ref{prot:biased} and~\ref{prot:nonspotcheck} in Chapter~\ref{chap : actual protocols for DI}. 
\section{Recycled inputs protocol}
As discussed in the Chapter~\ref{chap : actual protocols for DI}, in a standard randomness expansion protocol, it is assumed that the initial randomness is a private source of randomness. If this is the case, there is no incentive to make this initial randomness public. In fact, this input randomness should be recycled, and run through the randomness extractor along with the string of bits generated via the measurement device. This idea forms the basis for the following protocol: 

\begin{protocol}\label{protocol: Semi-DI recycling}
\noindent\textbf{Parameters}:\\
$n$ -- number of rounds \\
$p_{0} $ -- probability of 1 for random number generator $R_A$ (taken to be below $1/2$)\\ 
$\gamma$ -- probability of 1 for random number generator $R_{\gamma}$ (taken to be below $1/2$)\\ 
$\omega_{\text{exp}}$ -- expected score. \\
$\overlap_{\text{exp}}$ -- expected overlap. \\
$\delta_{\overlap}$ -- confidence width for the overlap \\
$\delta_{\score}$ -- confidence with for the score \\
\begin{enumerate}
    \item Set $i=1$ for the first round, or increase $i$ by 1.
    \item Use $R_A$ to choose $X_i\in\{0,1\}$, which is input to the device $S$. Here $X_i=1$ occurs with probability $p_{0}$. The device $S$ prepares a state $\rho^{x}_{i}$ (unknown) and sends it to $BS$. 
    \item Use $R_T$ to choose $T_i \in \{0 , 1\}$, which is input to $BS$. $BS$ sends the state to $M$ if $T_i = 0$ or sends it to $PM$ if $T_i = 1$. 
    \item If $T_i  = 0$ (Generate round): $M$ receives $\rho_{i}^{x}$ and outputs $Y_i \in \{0 ,1 \}$. Set $U_i = (T_i , X_i , Y_i) $.
    \item If $T_i  = 1$ (Test round): $PM$ receives $\rho^{x}_{i}$ and outputs $Y_i$. Set  $U_i = (T_i , X_i , Y_i)$.
    \item Return to Step~1 unless $i=n$. 
        \item Compute the empirical scores $U_{\#}$ and $\score_{\#}$ as 
        \begin{eqnarray}
        U_{\#} &:=& \frac{1}{2}\sum_{i} \frac{|\{i : U_{i} = (0 , x_i , 1) \}|}{n p_{X}(x)\gamma} \nonumber , \\ 
        \score_{\#} &:=& \frac{1}{2}\sum_{i}\frac{|\{i :U_{i} = (1 , x_i , x_i)\}|}{n p_{X}(x)(1 - \gamma) }  \nonumber .
        \end{eqnarray}
    \item Abort the protocol if either of the conditions are not met 
           \begin{enumerate}
               \item $\score_{\#} > \score_{\text{exp}} - \delta_{\score} $. 
               \item $U_{\#} > \overlap_{\text{exp}} - \delta_{\overlap}$.
           \end{enumerate}
    \item\label{st: 9} Process the concatenation of all the outputs with a quantum-proof strong extractor $\Ext$ to yield $\Ext(\B{ XY},\B{ R})$, where $\B{ R}$ is a random seed for the extractor. Since a strong extractor is used, the final outcome can be taken to be the concatenation of $\B{ R}$ and $\Ext(\B{ XY},\B{ R})$.
\end{enumerate}
\end{protocol} 

\section{A protocol to generate private randomness}
There maybe situations in which the initial source of randomness comes from a public source of randomness such as a randomness beacon. Such a source can also be used for randomness expansion as long as the devices are prepared and put in a secure lab before the public randomness is made accessible. In such a scenario, the input randomness cannot be recycled, as any adversary will have access to the input randomness. Nonetheless, the public source of randomness can be converted to a source of private randomness. The only difference in this protocol would be to replace the Step \ref{st: 9} to Step \ref{st: 9}' in the previous protocol (Protocol~\ref{protocol: Semi-DI recycling}), where instead of extracting the key from $\B{XY}$, we only extract the key from $\B{X}$.
\begin{protocol}\label{protcol: private to public}
\noindent\textbf{Parameters}:\\
$n$ -- number of rounds \\
$p_{0} $ -- probability of 1 for random number generator $R_A$ (taken to be below $1/2$)\\ 
$\gamma$ -- probability of 1 for random number generator $R_{\gamma}$ (taken to be below $1/2$)\\ 
$\omega_{\text{exp}}$ -- expected score. \\
$\overlap_{\text{exp}}$ -- expected overlap. \\
$\delta_{\overlap}$ -- confidence width for the overlap \\
$\delta_{\score}$ -- confidence with for the score \\
\begin{enumerate}
    \item Set $i=1$ for the first round, or increase $i$ by 1.
    \item Use $R_A$ to choose $X_i\in\{0,1\}$, which is input to the device $S$. Here $X_i=1$ occurs with probability $p_{0}$. The device $S$ prepares a state $\rho^{x}_{i}$ (unknown) and sends it to $BS$. 
    \item Use $R_T$ to choose $T_i \in \{0 , 1\}$, which is input to $BS$. $BS$ sends the state to $M$ if $T_i = 0$ or sends it to $PM$ if $T_i = 1$. 
    \item If $T_i  = 0$ (Generate round): $M$ receives $\rho^{x}_{i}$ and outputs $Y_i \in \{0 ,1 \}$. Set $U_i = (T_i , X_i , Y_i) $ .
    \item If $T_i  = 1$ (Test round): $PM$ receives $\rho^{x}_{i}$ and outputs $Y_i$. Set  $U_i = (T_i , X_i , Y_i)$.
    \item Return to Step~1 unless $i=n$. 
        \item Compute the empirical scores $U_{\#}$ and $\score_{\#}$ as 
        \begin{eqnarray}
        U_{\#} &:=& \frac{1}{2}\sum_{i} \frac{|\{i : U_{i} = (0 , x_i , 1) \}|}{n p_{X}(x)\gamma} \nonumber  \\ 
        \score_{\#} &:=& \frac{1}{2}\sum_{i}\frac{|\{i :U_{i} = (1 , x_i , x_i)\}|}{n p_{X}(x)(1 - \gamma) }  \nonumber 
        \end{eqnarray}
    \item Abort the protocol if either of the conditions are not met 
           \begin{enumerate}
               \item $\score_{\#} > \score_{\text{exp}} - \delta_{\score} $. 
               \item $U_{\#} > \overlap_{\text{exp}} - \delta_{\overlap}$.
           \end{enumerate}
    \item\label{st: 9'} Process the concatenation of all the outputs with a quantum-proof strong extractor $\Ext$ to yield $\Ext(\B{ X},\B{ R})$, where $\B{ R}$ is a random seed for the extractor. Since a strong extractor is used, the final outcome can be taken to be the concatenation of $\B{ R}$ and $\Ext(\B{ X},\B{ R})$.
\end{enumerate}
\end{protocol} 
\section{Rates of the protocol} 
We will now discuss the computation of rates for the protocol. Similar to the protocols for DIRNE, we can use the Entropy Accumulation Theorem (EAT) to calculate the rates~\cite{DF, DFR}. However, during the course of our project, a more robust extension of EAT called the generalized Entropy Accumulation Theorem was published~\cite{MFR}, which can also be used to compute the finite round key rates. It should be noted that the computation of finite round rates for the protocol is not covered in this thesis and is left as future work.

Roughly speaking, according to the EAT the asymptotic rates for our protocol can be computed by computing the infimum of the single round conditional \RutC{von Neumann} entropy $H(Y|XE)$ over all possible strategies that achieve a score $\omega$ and an overlap $\Theta$. \par

Given that the protocol consists of two distinct round types, the channel describing a single round of a protocol $\M{N}$ can be expressed as 
\begin{equation*}
    \M{N} = (1 - \gamma)\M{N}_{G} + \gamma \M{N}_{T}, 
\end{equation*}
where $\M{N}_{G}$ and $\M{N}_{T}$ are the EAT channels corresponding to the generate and the test rounds and $\gamma$ is the testing probability. 

The EAT channel corresponding to the generate round is given by
\begin{equation*}
    \M{N}_{G}: \rho_{AE} \mapsto \rho_{\M{C}}, 
\end{equation*}
where $\rho_{AE}$ is some quantum state initial quantum state of the form \ref{eqn: General state} and $\rho_{\M{C}}$ is the state of the form \ref{eqn: general CQ state in the generate round}. The channel describing the test round is of the form: 
\begin{eqnarray}
    \M{N}_{T}(\rho_{AE}) = \sum_{x,y} \proj{x} \otimes \proj{y} \otimes \tr_{A} \left( \Pi_{0} \rho^{\lambda}_{x}  \right) \otimes \proj{\lambda}_{\Lambda}. 
\end{eqnarray}

We can use the concavity of the conditional \RutC{von Neumann} entropy to obtain a lower bound 
$$ H(Y|XE)_{\M{N} (\rho_{AE})}  \geq (1 - \gamma) H(Y|XE)_{\M{N}_{G} (\rho_{AE})}.$$ 
In other words, the output randomness in a protocol $\text{rand}_{\text{out}}$ can be computed using the conditional \RutC{von Neumann} entropy from the generate rounds only. This justifies our choice of optimizing $H(Y| XE)_{\rho_{\M{C}}}$ in \RutC{the} previous chapter. \par
Now, let us focus our discussion on the asymptotic rates for individual protocols. In protocol~\ref{protcol: private to public}, the randomness expansion per unit round can be computed by
\begin{eqnarray}
\text{rand}_{\text{out}} - \text{rand}_{\text{in}} = (1 - \gamma)H(Y|XE) - H(X),
\end{eqnarray}
where $H(X)$ denotes the input randomness, which can be computed to be $H_{\bin}(p_{X}(0))$. Note that the entropy $H(Y|XE)$ here is the entropy for generate rounds only. \par
For Protocol \ref{protocol: Semi-DI recycling}, since both the input and output strings are used, the difference in the output randomness (in the asymptotic limit) is given by
\begin{eqnarray}
    \text{rand}_{\text{out}} - \text{rand}_{\text{in}}= (1 - \gamma)H(XY|E) - H(X)  = (1 - \gamma)H(Y|XE) - \gamma H(X) ,
\end{eqnarray}
where we have used the chain rule for the conditional \RutC{von Neumann} entropy. Again, here, the entropy $H(Y|XE)$ is the entropy of generate rounds only.
Thus, in order to compute the rates of the protocols, one should compute the quantity  
$$ F_{p_{X}} ( \score , \overlap) := \inf H(Y|XE), $$
where the infimum is taken over all the single round strategies that achieve a score $\score$ in the generate rounds and have an overlap $\overlap$.
\section{Results} 
\begin{figure}[h!]
    \centering
    \includegraphics[width=\textwidth]{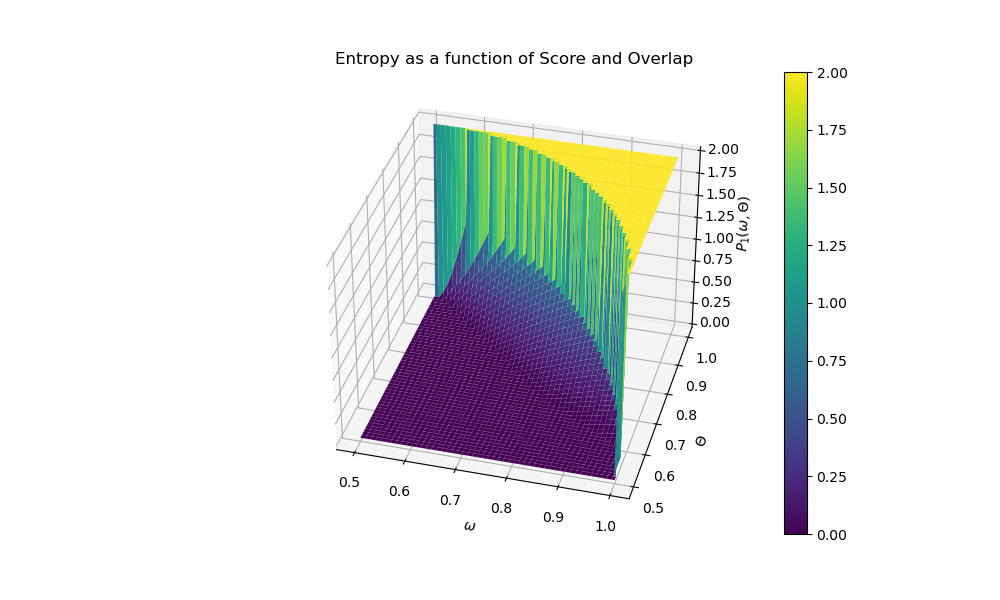}
    \caption[Lower bound of the function G.]{Lower bound of the function $G$ over the extended domain $(\score , \overlap) \in [1/2 , 1]\times [1/2 , 1]$ in the case when $p_{X}(0) = \frac{1}{2}$. The region with $G(\score , \overlap)= 2$, is the region where no quantum strategies are found.}
    \label{fig:Gfunct}
\end{figure}
\begin{figure}[h!]
    \centering
    \includegraphics[width=\textwidth]{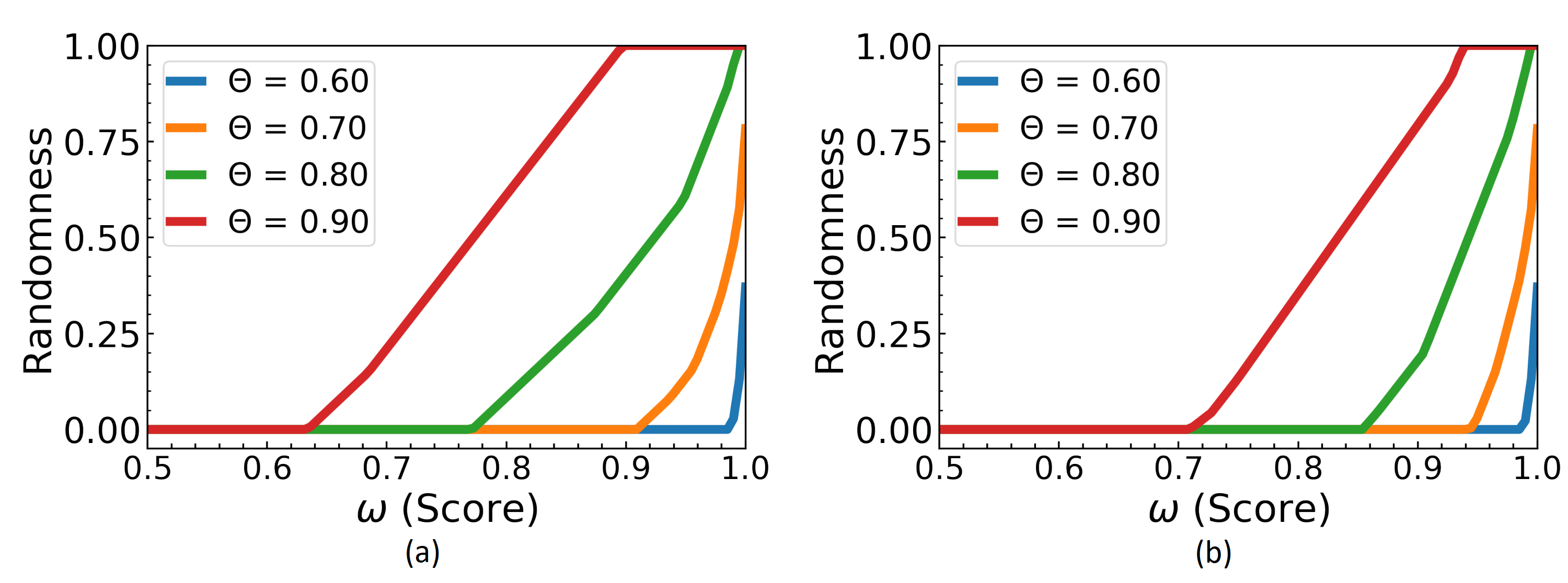}
    \caption[Asymptotic rates in the generate round of our semi-DI protocol.]{Asymptotic rates in the generate round of our protocol. Figure (a) are the rates when $p_{X}(0) = \frac{1}{2}$ and Figure (b) are the rates when $p_{X}(0) = \frac{1}{100}$. Note that \RutC{the rates reported are valid in the limit} $\gamma \rightarrow 0$.}
    \label{fig:asymptotic_rates_Semi_DI}
\end{figure}
In the previous chapter \RutC{(Chapter \ref{chap: semi-DI lower bounds})}, we calculated the asymptotic rates $F_{p_X}(\score, \overlap)$. The rates are shown in Figure~\ref{fig:asymptotic_rates_Semi_DI} as a function of $\score$ for various overlap values. 

As discussed in Chapter~\ref{chap: semi-DI lower bounds}, the function $F_{p_{X}}(\score, \overlap)$ can be obtained by computing the convex envelope (i.e., the smallest convex function below $G$) of $G_{p_{X}}(\score, \overlap)$, which is derived from qubit strategies. Importantly, not all values \RutC{of the tuple} $(\score, \overlap) \in [1/2, 1]\times[1/2 ,1]$ are attainable using a quantum strategy (see~\cite{VSoriginal, VS}). In such situations, our optimization problem for $G_{p_{X}}(\score, \overlap)$ also does not show any feasible solutions. To determine the convex envelope of $G_{p_{X}}(\score, \overlap)$ using numerical algorithms (see~\cite{contento2015linear, lucet1997faster} for fast algorithms for convex envelope), we extend the domain of $G_{p_{X}}(\score, \overlap)$ to the entire range of $[1/2, 1] \times [1/2 , 1]$ and assign it an arbitrarily high value when no feasible quantum strategy is found. This extension results in a function $F_{p_{X}}(\score, \overlap)$ over the domain $[1/2 , 1] \times [1/2 , 1]$. The function reported in Fig~\ref{fig:asymptotic_rates_Semi_DI} as $F_{p_{X}}(\score , \overlap)$ is a lower bound on the randomness rate if there is a quantum achievable strategy. For the regions of the extended domain where no quantum achievable strategy exists, the values of $F_{p_{X}}(\score, \overlap)$ should be disregarded. Nonetheless, this extended domain proves useful when \RutC{EAT is used to get bounds on finite round rates. In particular, it turns out that the min-tradeoff  function needs to be defined on the extended domain,}.

As anticipated, the protocol yields a high randomness rate for a fixed score (assuming quantum strategies can achieve this) when the overlap is significant. Conversely, with minimal overlap, a higher score is essential for generating randomness. This stems from the fact that deterministic strategies, akin to local deterministic strategies in the DI case, can only attain a low score. For instance, achieving a score around \(0.5\) is feasible through mere chance combined with some pre-programming of the devices. To get a large score, a genuine quantum strategy is required, ensuring the presence of randomness in the outputs.

\RutC{For} quantum achievable strategies, our results are promising. For instance, we identify the strategy $\overlap = 0.8$ and $\score = 0.878$ as attainable using quantum theory, achieving approximately $0.319$ bits per round when $p_{X}(0) = \frac{1}{2}$. As shown in the figure, a higher overlap yields better values of randomness rounds. However, it is crucial to consider that the power meter used in the experiment will have a detector efficiency $\eta < 1$. Consequently, for security reasons, we must adopt a pessimistic approach and underestimate the experimental overlap by a factor of $\eta$. Thus, under this pessimistic scenario, an experimental setup (consisting of good power meters with high detector efficiency) would likely result in an overlap value less than $0.8$. Therefore, one should anticipate a randomness expansion of approximately $0.3(1 - \gamma)$ bits per round using this protocol, where $\gamma$ represents the testing probability.

Furthermore, we have plotted the function $F_{p_{X}}(\score , \overlap)$ for the scenario where $p_{X}(0) = \frac{1}{100}$. In this case, the protocol resembles a spot-checking type protocol, where the input randomness is negligible because most of the time $X = 1$ is being sent. Only occasionally, $X = 0$ is sent as a spot check. Since this behavior is unknown to the adversary, for the protocol not to abort, the devices must function honestly during most of the rounds. As expected, this spot-checking protocol provides a lower randomness rate. However, if the figure of merit is the ratio of output to input randomness, then having $p_{X}(0) \ll 1$ may be more suitable. Furthermore, for a finite number of rounds, having $p_{X}(0) \ll 1$, similar to the spot-checking protocol, faces challenges in providing good rates. A significantly large number of rounds is needed to gather enough statistics to be confident in $\score$ and $\overlap$, especially given the scarcity of rounds with $X = 0$. 

\section{Discussion} 
In this work, we have conducted an analysis of the semi-Device Independent protocol based on the energy and overlap bounds. We have proposed a protocol that recycles input randomness for the semi-Device Independent protocols and have also introduced another protocol that converts public randomness to private randomness.

The structure and analysis of these protocols closely resemble Device Independent protocols for randomness expansion. As a result, there are numerous opportunities for applying similar techniques and ideas from Device Independence to enhance these protocols. Just like in Device Independence, alternative \RutC{score} functions other than the CHSH score, as discussed in references~\cite{WAP, WBC}, might yield better performance in our semi-DI protocols. However, determining the optimal functional form of the score remains an open question. Furthermore, the choice of score can significantly impact the performance of the protocols, especially given available experimental statistics.

There is a possibility of exploring connections with concepts like self-testing, as introduced in~\cite{RVMBWPB} that often occur in the discussions of the DI protocols. Self-tests can yield interesting score definitions that could contribute to a partial answer to the question of finding the best score given experimental statistics.

Moreover, developing an NPA-like hierarchy for such protocols ~\cite{NPA}, could be beneficial. Such a hierarchy may allow us to use results for optimizing von Neumann entropies, such as those presented in~\cite{BFF2022}, to analyze these protocols for arbitrary inputs and outputs. The current proof based on Jordan's lemma is limited to protocols with only $2$ inputs and $2$ outputs, so a more generalized approach is desirable.

\chapter{Conclusion and outlook} 

\RutC{In this section of} the thesis, we delved into a comprehensive exploration of randomness expansion protocols based on minimal assumptions regarding the inner workings of the devices. The thesis commenced with an exploration of the need for randomness expansion protocols. Subsequently, we delved into various protocols for achieving randomness expansion. In Chapter~\ref{chap: maths}, we \RutC{discussed} the pivotal role of optimization problems within the context of randomness expansion protocols. We discussed the significance of semi-definite programs in solving polynomial optimization problems. Additionally, we \RutC{identified} the role of min-entropy as a suitable measure for quantifying randomness in a cryptographic scenario. The presentation of the Entropy Accumulation Theorem, as \RutC{an} approach for establishing reliable bounds on min-entropy in randomness expansion protocols, was also discussed at the end of the chapter.

Shifting focus to protocols based on the violation of (generalized) CHSH inequalities in Chapter~\ref{chap: DI protocols}, we explored the concept of Bell's theorem within the context of randomness expansion. Chapters~\ref{chap: DI upper bounds} and \ref{chap: DI lower bounds} delved into strategies for establishing tight lower bounds on conditional von Neumann entropy relevant to DIQKD and DIRNE protocols which rely on violating generalized CHSH inequalities. Our work culminated in Chapter~\ref{chap : actual protocols for DI}, where we introduced and compared three CHSH-based protocols for randomness expansion: the spot-checking protocol (with randomness extracted from both parties), a protocol that recycles input randomness, and a protocol leveraging heavily biased input randomness. Notably, the latter two protocols close the locality loophole. Upon comparative analysis, our results reveal that the protocols that recycle input randomness exhibit exponential enhancements in the finite-round regime, rendering DIRNE protocols notably more practical.

Subsequently, the exploration extended to semi-Device Independent (semi-DI) protocols of randomness expansion. In Chapter~\ref{chap: semi-DI}, the foundational structure of our semi-DI protocol was introduced, focusing on the preparation of states with low energy or high overlap in a system having a unique ground state. The subsequent chapter,~\ref{chap: semi-DI lower bounds}, dealt with the computation of lower bounds on the von Neumann entropic quantity $H(Y|XE)$ for protocols that attain a fixed a overlap with the ground state and a fixed score. Notably, in the final chapter of semi-DI protocols (Chapter~\ref{chap: semi-DI protocols}) we introduced two protocols - one for randomness expansion and another facilitating the conversion of public randomness to private randomness.

Throughout this thesis, we presented significant enhancements in both Device Independent and semi-Device Independent randomness expansion protocols. Our contributions include the development of superior protocols in terms of rates and security. Furthermore, we devised novel approaches to establish tighter bounds on the randomness rate. An important advantage of our work lies in our ability to recycle input randomness for our protocols. The ability to recycle input randomness remarkably yields exponential increases in randomness rates, especially for finite rounds.

While our methodology proficiently computes rates for conventional score definitions, such as the CHSH score in the DI case and the guessing probability in the semi-DI scenario, it can adapt to other linear functions of input-output statistics for generating randomness. We acknowledge the potential of alternative functionals, as demonstrated in~\cite{WBC, WAP}, which might better suit randomness expansion and QKD.

An additional avenue for improvement involves harnessing the potential offered by Multi-partite Bell inequalities. Protocols based on these inequalities offer the promise of generating up to $n$ bits of randomness. While experimentally violating these Multi-partite Bell inequalities proves to be an even greater challenge compared to the CHSH inequality, the theoretical exploration of these scenarios remains invaluable. Such investigations enable us to grasp the true extent of quantum theory's capabilities and equip us with the necessary insights for a future where their implementation becomes more feasible. 

Another avenue of investigation is to for look protocols based on multiple inputs and outputs, rather than restricting to the two input-two output case. It is worth noting that our existing protocols are tailored for binary inputs and outputs due to the reliance on Jordan's Lemma. Unfortunately, no equivalent principle to Jordan's Lemma exists for scenarios involving non-binary inputs or outputs. 

Although there are approaches to bypass the limitations of Jordan's Lemma, these strategies often come with resource-intensive requirements, and/or they do not give good randomness rates. In particular, a computationally feasible way is to obtain bounds on the randomness for multi-partite scenario is using the NPA hierarchy~\cite{NPA} and lower bounding the \RutC{von Neumann} entropy in terms of the min-entropy~\cite{BRC}. However, this method loses tightness. Another way to obtain bounds on the randomness is by adopting the approach from the recently introduced seminal work by Brown et.al.~\cite{BFF}. As it stands, this approach proves challenging for practical execution in multi-partite scenarios, even in tri-partite scenarios where randomness is drawn from all the three parties. However, in simpler cases such as the two-input two-output bipartite scenarios, this method proves to be highly practical. Therefore, it might be worthwhile to look into ways in which this method can be made less resource extensive. 

Interestingly, when it comes to DIRNE protocols, there isn't always a need to close all loopholes in practical implementations. The importance of addressing the locality loophole is somewhat lessened since, in the DIRNE protocols, all devices function within a single lab. Protective shielding methods, implemented to avoid any unwanted external data transmissions, can be also be used to shield the devices themselves and eliminate any potential for device-to-device communications. While obstacles linked to the detection loophole remain, it's worth noting that to exploit this loophole, an eavesdropper would need to craft near-perfect detectors and covertly install them in the devices, which itself is challenging for an adversary to achieve practically.

In this context, the fair sampling assumption~\cite{CH, CHSH} also is essential when performing DI protocols. This assumption forms the foundation for constructing semi-DI protocols based on the violation of a Bell inequality. Consequently, this provides further rationale for developing protocols rooted in different Bell inequalities, even with present day technology. 

To bridge the gap between theory and experimentation, an ideal protocol should align with the best theoretical strategy, thereby achieving optimal rates. Experimental outcomes, however, often diverge from theoretical predictions due to factors like noise, device imperfections, or possible eavesdropping. As a result, statistics gathered from experiments should then be utilized to compute a lower bound on randomness rate. This raises the question: what is the best computed lower bound given the experimental statistics obtained? At first glance, one might consider using all available statistics and employ a method like the one presented in our thesis to determine the best possible rate. \RutC{Intuitively, this seems to be the optimal method, and is indeed the case in the asymptotic limit (i.e. for very large number of rounds)}. Nonetheless, this method has limitations in finite-round cases, primarily because the error term in the Entropy Accumulation Theorem deteriorates when the rate is determined by a very detailed abort condition of the protocol. This translates to the scenario in which if we condition the protocol on the full experimentally achieved input-output distribution \( p_{AB|XY} \), then the error term is significantly worse than if we merely condition on the observation of a CHSH score, which is only a function of the full distribution \( p_{AB|XY} \).

Thus, a trade-off emerges when considering which statistics to condition upon for finitely many rounds. It is therefore crucial to strike the right balance: over-conditioning can worsen the error term while enhancing the primary term. The practicality of various conditioning strategies varies with the finite round regime, and pinpointing the best conditions often requires a case-by-case assessment. It is also worth noting that while the error term of EAT provides a lower bound, it might not be tight. For instance, the error terms of EAT were revised in \cite{DF} and subsequently in \cite{LLR&}, improving the error term from the original EAT statement. Looking ahead, we might see even more refinements to the error term of EAT, which could influence the optimality of how stringent our abort conditions should be to achieve the best rates.

In summary, this thesis delivers advancements in Device Independent and semi-Device Independent randomness expansion protocols. Our contributions span enhanced protocols, tighter bounds, and the inclusion of input randomness recycling. The protocols' practical implementation depends on a delicate balance between theoretical strategies and experimental realities, an aspect that warrants further investigation.

\cleardoublepagewithnumber
\part{Generalized Probability Theories}
\cleardoublepagewithnumber

\chapter{Generalized Probability Theories }\label{chap: GPTs}
\section{Introduction}
The Generalized Probability Theory (GPT) \cite{barrett, plavala2021general} framework provides a description of physical theories based on the experimentally achievable input-output statistics of measurements, given preparations and channels. Each GPT is characterized by its state space, effect space, and set of channels. In this framework, a normalized state is represented by an element $\sigma$ belonging to a convex set $\mathcal{C}$. The state space consists of tuples of the form $\omega = (\lambda, \sigma)$, where $\lambda$ ranges from 0 to 1 and $\sigma$ belongs to $\mathcal{C}$. The value $\lambda \in [0, 1]$ called normalization constant. An effect is defined as a linear map $v$ on the state space satisfying $0 \leq v(\omega) \leq 1$. Physically, an effect represents an outcome of a measurement in the GPT, and its action on the state $v(\omega)$ assigns a probability to the occurrence of the event. A channel, denoted as $\mathcal{M}$, describes the dynamics of the system under specific physical processes.

Quantum theory is a particular GPT where the state space consists of density operators $\rho$, which are positive operators with a trace less than or equal to 1. \RutC{Therefore, for quantum theory}, the convex set $\mathcal{C}$ is identified as the set of \RutC{non-negative} operators with trace 1, and the normalization constant $\lambda = \tr(\rho)$. The effect space in quantum theory includes any \RutC{non-negative} operator $E$ such that $\id - E$ is also \RutC{positive}. The probability of an event occurring is given by $\text{tr}(\rho E)$. The set of channels $\mathcal{M}$ in quantum theory corresponds to Trace-non increasing Completely Positive Maps (TCPMs).

Checking the complete positivity requirement for a channel can be a-priori challenging, as it requires verifying the positivity of $\mathcal{M} \otimes \mathcal{I}_k$ for arbitrary natural numbers $k$. However, there are known results, such as Choi's theorem, that \RutC{provides} a tractable criterion to check whether a given linear map is a channel in quantum theory. Furthermore, it is known that any completely positive map can always be expressed using a Kraus operator decomposition. That is, a map $\mathcal{M}$ is completely positive if and only if there exist operators $\{K_i\}_i^n$ satisfying $\sum_i^n K_i^\dagger K_i \geq 0$, such that $\mathcal{M}(\rho) = \sum_i^n K_i \rho K_i^\dagger$.

In any GPT, similar concepts of complete positivity and positivity come into play. Just like in quantum theory, where effects and channels must adhere to the notion of complete positivity, the same is true for GPTs. Computing the set of ``positive'' maps for most GPTs should be relatively straightforward \RutC{(provided the state space can be embedded in finite dimensional Hilbert spaces)}, and in principle, numerical algorithms can be employed for this purpose, depending on the complexity of the state space. 

However, verifying the condition for complete positivity in GPTs is a significantly more challenging task. Unlike in (finite dimensional) quantum theory, where determining whether a linear map is a channel is in principle \RutC{achievable using computational methods}, there are no equivalent results available for arbitrary GPTs. The main problem is that, \RutC{for an arbitrary GPT, there is no unique method or principle to define composite state spaces.} Furthermore, there are no existing results that allow us to express any completely positive map in an analogous Kraus operator representation. In this section of the thesis, our primary focus is to determine a tractable way to determine the set of channels for Boxworld. 

Boxworld is a well-studied GPT that has attracted considerable attention \cite{gross2010all} in the literature. In  Boxworld, the state space consists of all probability distributions that do not allow super-luminal signaling. Therefore Boxworld stands as the largest theory, in terms of achievable input-output correlations, that is consistent with relativity and includes the quantum theory state space as a strict subset. The extremal points in the $2$-input, $2$-output state space of  Boxworld are known as PR boxes ~\cite{PopescuRohrlich}. These PR boxes represent non-signaling states capable of achieving a maximal CHSH score of $1$, surpassing the Tsirelson bound of $\frac{1}{2} + \frac{1}{2 \sqrt{2}}$ in quantum theory. 

While Boxworld shares some fundamental properties with quantum mechanics, such as the no-cloning and no-broadcasting theorems, as well as the monogamy of correlations \cite{masanes2006general}, it also exhibits distinct features. Notably, Boxworld allows for significantly enhanced communication power compared to quantum mechanics, as demonstrated by van Dam et al. \cite{van2013implausible} and Brassard et al. \cite{brassard2006limit}. Additionally, Linden et al. \cite{linden2007quantum} proved how certain post-quantum theories, including  Boxworld, enable nonlocal computation tasks that are unattainable in quantum mechanics. Consequently, information-theoretic tasks have been proposed as means to distinguish Boxworld from quantum mechanics. \par  

Even if Boxworld is the largest GPT in terms of the achievable input-output correlation, it has very simple mathematical features. This is because Boxworld's larger state space, compared to quantum theory, significantly restricts the dynamics it can exhibit. Barrett~\cite{barrett} demonstrated the absence of joint measurements in Boxworld for two parties. \RutC{This means that any measurement in Boxworld for two parties is essentially a local measurement, and unlike quantum theory, global measurements do not exist.} Further work by Gross et al.~\cite{gross2010all} proved that reversible dynamics within Boxworld are trivial; i.e., any reversible transformation in Boxworld consists solely of local operations and permutations of systems. Notably, the ability to perform joint measurements on two systems is a fundamental aspect of phenomena like teleportation and entanglement swapping~\cite{SPSN,BBCJPW}, and hence these features of quantum theory also cannot be replicated in Boxworld. By studying Boxworld's dynamics, we can understand what sets quantum theory apart from generalized theories and gain a deeper understanding of the phenomena that arise in quantum mechanics. \par

In the next chapter, we carefully define Boxworld in terms of its state space. Through this definition, we develop a linear algebraic framework for Boxworld that can be applied to scenarios involving any number of non-communicating (or space-like separated) parties. Using this framework, we \RutC{prove} that any valid linear map that maps a Boxworld state to another state is a channel in Boxworld. \RutC{Borrowing terminology from quantum theory, this means that any ``positive'' map in Boxworld is automatically ``completely positive''}. This observation indicates that the set of channels in Boxworld is numerically tractable (simply check the action of the channel on the extremal Boxworld states), facilitating an efficient characterization of the dynamics of the theory. \par 

In the remainder of the chapter, we define various aspects of the GPT framework such as states, effects and channels. Furthermore, we shall prove certain fundamental facts within the GPT framework, which apply to all GPTs. 
\section{The GPT framework: A concise literature overview} 
\RC{Before delving into the intricacies of the GPT framework, we provide a brief overview of the literature on GPTs. The GPT framework emerged in the context of axiomatizations of quantum theory. Rather than depending on the conventional ``textbook'' axioms of quantum theory, which are based on abstract mathematical constructs without clear physical motivations, the goal was to formulate quantum theory with axioms from well-motivated physical principles \cite{PR,Hardy,CDP,MM}. To do so, the GPT framework was used to identify the essential features a physical theory should possess and to introduce axioms to derive quantum theory from. The GPT framework appears especially appropriate for such a pursuit, as it operates under minimal requirements -- it describes a physical theory solely in terms of its ability to reproduce the observable statistics of any experiment deemed physical. In other words, the GPT framework only aims to reproduce operational aspects of a theory. 

The pursuit of axiomatizing quantum theory led to the conception of Boxworld. Popescu and Rohrlich, in their seminal paper \cite{PR}, questioned whether the ability to violate the Bell inequality (specifically, the CHSH inequality) was unique to quantum theory. They explored whether the principles of non-locality and relativistic causality (which for the context of this work simply means that faster than speed of light communication is not possible) were sufficient to recover all experimental correlations consistent with the predictions of quantum theory. Surprisingly, they identified non-local correlations beyond the purview of quantum theory, suggesting that merely adopting non-locality and relativistic causality as axioms fails to fully reproduce quantum theory.

Barrett \cite{barrett} constructed ``the generalized non-signaling theory'' keeping non-locality and relativistic causality two as the sole axioms. The generalized non-signaling theory \cite{barrett} subsequently became known as ``Boxworld''. This theory encompasses all non-local correlations, including those identified by Popescu and Rohrlich, which were later called the ``PR box''.

Although the GPT framework was initially introduced to study quantum theory through physically motivated axioms, it has since been applied in a wide variety of contexts. For example, this framework has been extended to investigate computation \cite{GPTcomputer2,GPTcomputer3,GPTcomputer4,GPTcomputer5,GPTcomputer6,GPTcomputer7} and information-theoretic tasks such as bit commitment and communication complexity \cite{GPTcomputer,van2013implausible,brassard2006limit,linden2007quantum,Giorgos1}. An interesting question to address is the extent to which the structure of quantum theory is crucial for proving security in various Device Independent protocols. In this context, Barrett et al. \cite{BHK} illustrated that the standard Device Independent Quantum Key Distribution (DIQKD) protocol remains secure against a spectrum of attacks by post-quantum eavesdroppers, who are limited solely by the non-signaling principle.

The generality of the GPT framework has led to a deep understanding of the ``logical architecture'' of quantum theory. A vast body of research exists that seeks to understand different features of quantum theory that are absent in classical theory. Investigating whether these features are also present in other theoretical constructs not only enhances our understanding of quantum theory but also reveals how multiple features of a theory might be a consequence of the same logical architecture. This research offers insights into the interplay between various quantum features and the ``minimal assumptions'' needed for operational theories to replicate those features. Non-local features such as entanglement, Bell violation, and steering have been studied in various GPTs \cite{non-locality,SB2009,GPT_ent1,GPT-steer2,GPTent2}. Beyond non-locality, other attributes of quantum theory, like non-contextuality, have also been studied within the GPT framework \cite{GPTcontext,GPTcontext2,GPTcontext3}. Additionally, the GPT framework has been used to explore theories compatible with various other physical principles and physical theories, such as causality \cite{HLP,GPTcausality1,Shashaank}, and thermodynamics \cite{thermo1,thermo2,thermo3}. The framework is also instrumental in addressing (apparent) logical paradoxes \cite{Vilasini2} and interpretational issues \cite{Vilasini1} in quantum theory. Furthermore, GPTs have also been used to study theories resulting from the omission of specific axioms of quantum theory \cite{Vicky,Vincenzo}.

In summary, the quest to understand axioms for quantum theory has evolved into a vast and rich domain, allowing to study diverse operational aspects of different mathematically conceivable physical theories.}
\section{State spaces} 
In a GPT, each system  is described  using a mathematical quantity known as a state. Similar to quantum states, these states provide complete information about the probability of any outcome that can occur in any given measurement performed on the system. The collection of all possible states is referred to as the state space of the GPT. 

We start by defining the state space. For each system in the GPT, we associate \RutC{a} Hilbert space, denoted as $\mathcal{H}$. We refer dual space of $\M{H}$ by the notation $\M{H}^*$ (the set of all linear maps $\M{H} \mapsto \mathbb{R}$). The most general state space corresponding to the system is constructed as follows: 
\begin{itemize}
    \item Identify a set $\mathcal{S}^n$ known as the set of normalized states. This set is a convex set that possesses a specific property: there exists a vector $\uniteffect \in \mathcal{H}^*$ such that $\langle \uniteffect, \psi \rangle = 1$ for every $\psi \in \mathcal{S}^n$. The vector $\uniteffect$ is referred to as the identity effect, and it is unique in fulfilling this property. 
    \item Then the set $\M{S}$ is simply defined as 
    \begin{eqnarray}
    \M{S} = \{ (\lambda , \psi) \equiv \lambda \psi  | \hspace{0.2 cm} \lambda \in [0 ,1] , \psi \in \M{S}^{n} \} .
    \end{eqnarray}
\end{itemize}
When $\lambda < 1$, then we say that the state is un-normalized or sub-normalized. Sub-normalized states cannot represent a physical process completely, as they assign an overall probability of less than 1 to the outcomes of any measurement. Nonetheless, they are extremely useful mathematically. Just as in quantum theory, where sub-normalized states arise naturally when considering the action of a quantum channel on a quantum state, they also emerge naturally in the GPT framework when we describe the action of a channel (defined in the next section) on a state. 

In analogy with the quantum theory, for GPTs, we can define the set of pure states as the states that are not ``mixed'' -- i.e. they cannot be a expressed as the statistical mixture of different states. The pure states are on the boundary of the state space and the state space is the convex hull of the set of pure states\RutC{. We} denote the set of pure states by the symbol $\partial \M{S}$ (not to be confused with the boundary of the set $\M{S}$)
\begin{eqnarray}
\partial \M{S} := \{\psi \in \M{S} : \psi = \mu \psi_1 + (1 - \mu) \psi_2  \implies \psi_1 = \psi_2 = \psi\} .
\end{eqnarray}
The GPT framework also allows for a complete description of the experimental statistics of multiple systems. This is done by essentially treating multiple systems as effectively a single system, with a single measurement identified as a collection of measurements on each system\footnote{There may be other types of measurements in such GPTs, \RutC{however}, this form of measurement must always be present.}.

For a multi-system GPT capable of describing interesting phenomena, it must allow for multi-system states which produce correlated outcomes when local measurements are performed on each system. Furthermore, it is also interesting if the multi-system GPT allows for performing non-trivial ``joint measurements'' -- i.e. a measurement that cannot be expressed as a collection of independent measurements on each system. \par

The way we describe a multi-system state space is by first labelling every system in our theory using a unique natural number $n \in \mathbb{N}$. We then label the single or multi-system state space using an index $i \subset \mathbb{N}$. For example, $\mathcal{S}_{\{1, 2, 4\}}$ represents the composite state space describing the systems $1$, $2$, and $4$. Similarly, we use the notation $\M{H}_{i}$ for the Hilbert space that embeds the state space $\M{S}_{i}$.
Therefore, the full description of state spaces in a GPT consists of a collection of these state spaces:

\begin{eqnarray}
\mathcal{T}_{\mathcal{S}} := \{\mathcal{S}_i | i \subset \mathbb{N}\}.
\end{eqnarray}

In order to ensure consistency within our theory, it is necessary for multi-system state spaces to include single-system state spaces. This means that each multi-system state space must contain subsets that are isomorphic to the corresponding single-system state spaces. \RutC{Deferring the discussion on measurements to a later section, as is conventional in the literature,} we assume the validity of local tomography for our theories. Roughly speaking, \RutC{this property} asserts that the global state can be fully characterized through local measurements. Under the assumption of local tomography, the constraint on single-system Hilbert spaces is given by\footnote{Here $\sqcup$ represents the disjoint union of two sets - i.e. it is implicitly understood that sets $i$ and $j$ are disjoint.}~\cite{barrett} 
\begin{equation*}
  \M{H}_{i \sqcup j} \cong \M{H}_i \otimes \M{H}_j. 
\end{equation*}  
\RutC{Here $\otimes$ is the standard tensor product of Hilbert spaces.} Unless stated otherwise, this assumption is made in this thesis. To visually represent multi-system states, we use diagrams with multiple wires, where each wire corresponds to a system's state space. For example, a two-system state \RutC{$\phi \in \M{S}_{\{ 1, 2\}}$} is given by 
\begin{equation}
\begin{quantikz}
& \multiprepareC[2]{\phi \quad}   & \qw^{\M{H}_1} \\ 
&                             & \qw ^{\M{H}_2}    
\end{quantikz} \equiv \begin{quantikz}
& \multiprepareC[1]{\phi}   & \qwbundle{\{1 ,2 \}} 
\end{quantikz} .
\end{equation}
The diagrammatic representation of the state space in GPTs offers a visual approach to understand different aspects of the GPT framework, which is often very useful proving important statements about the GPTs. We shall expand upon the diagrammatic version of GPTs in the last section of this chapter. 
\section{Transformations} 

The description of a physical theory remains incomplete without a means of representing ``dynamics'' in systems. Within the GPT framework, such dynamics are represented using ``channels''.

One of the most basic behaviors a system can exhibit is remaining static, without any evolution. This lack of evolution or change is captured by a map known as the ``identity channel'', denoted by $\idmap$. This channel is defined as follows:
\begin{definition}[Identity channel or map] Let $\M{S}_{k}$ be a state space. The identity channel $\idmap_{k}$ is a unique map that obeys
\begin{eqnarray}
    \forall \psi \in \M{S}_{k}: \idmap_k  \psi = \psi .
\end{eqnarray}
\end{definition}
The identity channel is a valid physical transformation of any GPT~\cite{plavala2021general} by assumption. For majority of the GPTs, the identity map is chosen to be the map $\id$ on the underlying Hilbert space that embeds the state space. 

Consider preparing two states, $\psi_1$ with probability $\mu$ and $\psi_2$ with probability $1 - \mu$. When subjected to the physical process represented by $\M{M}$, the resulting states, $\M{M}\psi_1$ and $\M{M}\psi_2$, should retain their probabilities as $\mu$ and $1 - \mu$, respectively. In other words, a channel must respect the convex structure of the state space i.e. - 
\begin{eqnarray}
\forall \psi_1, \psi_2 \in \M{S}, \forall \mu \in [0 ,1]: \M{M}(\mu \psi_1 + (1 - \mu) \psi_2) = \mu \M{M}(\psi_1) + (1 - \mu) \M{M}(\psi_2) .
\end{eqnarray}
On the level of the Hilbert space $\M{H}$, the above condition can be fulfilled if $\M{M}$ is a linear map on $\M{H}$. Thus, we assume that the channels for our GPTs are linear maps acting on the Hilbert space $\M{H}$. 

Before formally defining a channel, we begin by defining a ``transformation'' or ``positive map'' in the context of GPTs.

\begin{definition}[Transformation or Positive map]
Let $\mathcal{S}_i$ be a state-space embedded in the Hilbert space $\mathcal{H}_i$, and $\mathcal{S}_j$ be a state-space embedded in the Hilbert space $\mathcal{H}_j$. Any linear map $\mathcal{M}: \mathcal{H}_i \rightarrow \mathcal{H}_j$ is called a transformation if:
\begin{equation*}
\forall \psi \in \mathcal{S}_i: \mathcal{M}(\psi) \in \mathcal{S}_j.
\end{equation*}
\end{definition}

For a transformation to be considered a channel, it must satisfy an additional requirement that it consistently acts as a transformation on composite state spaces as well. This is the analogue of the notion ``complete positivity'' in GPTs. We use this requirement to define channels in GPTs. 

\begin{definition}[Channel or completely positive map]
A transformation $\mathcal{M}: \mathcal{S}_i \rightarrow \mathcal{S}_{j}$ is a channel if for any state space $\M{S}_{i \sqcup k}$, the map $\M{M} \otimes \idmap_{k}$ is a transformation. 
\end{definition}
We will adopt the following diagrammatic representation for the transformation. For example, the transformation $\M{M}: \M{S}_{\{1 ,2 \}} \mapsto \M{S}_{\{ 1 , 2 \}}$ is denoted by: 
\begin{equation}
 \begin{quantikz}
& \qw^{\M{H}_1}    &  \gate[wires = 2]{\M{M}} & \qw^{\M{H}_1} \\ 
& \qw^{\M{H}_2}    &                          & \qw^{\M{H}_2} 
\end{quantikz} .
\end{equation}

\section{Effects} 
As the GPT framework aims to describe experimentally achievable statistics in any measurement, it is essential to have the ability to assign probabilities to outcomes of these measurements. These outcomes are represented by mathematical objects known as effects. When an outcome is observed in a given experiment, the probability of the outcome can be computed by action of the effect on the state in which the system is prepared. 
As motivated by the set of transformations, the map $E$ must be linear in order to preserve the convexity of the state space. Thus, the effects are also taken to be linear maps. When a state resides in the space $\mathcal{H}$, the set of effects, being linear maps on the set of states, exists in the dual space $\mathcal{H}^*$. \par

Similar to the subtleties between channels and transformations, caution is required when defining effects. We define a witness as any linear map that assigns a probability to any state within our state space.
\begin{definition}[Witness]
A linear map $\M{H}_i \mapsto \mathbb{R}$ is a witness if $$ \forall \psi \in \M{S}_{i}: \quad E(\psi) \in [0 , 1].$$ 
\end{definition}
\RutC{For a witness $E$ to correspond to an outcome of an experiment in a GPT, $E \otimes \idmap$ also needs to be a witness on all state spaces that contain the state space $\M{S}$. Witnesses that can be assigned to outcomes of an experiment in a GPT are called effects.}
\begin{definition}[Effect]
A witness $E$ is an effect, if 
\begin{itemize}
    \item for every state space $\M{S}_{i \sqcup k}$, the map $E \otimes \idmap_{k}$ is a channel, and
    \item  $\uniteffect - E$ is an effect.
\end{itemize}
\end{definition}
While the first condition of $E \otimes \idmap$ stems from the fact that an effect should act consistently on the composite state space, the condition that $\uniteffect - E$ is also an effect is used in order to ensure an effect must form a part of a measurement in our GPT. We also assume that the set of effects, or the effect space, is a convex set. This means that any convex combination of two effects must also be an effect in the GPT.\par

\RutC{Mathematically, it is significantly easier to check if a particular map serves as a witness than to determine if it qualifies as an effect. This is because, to verify that a linear map \( E \) is an effect, one must confirm that an infinite number of linear maps of the form \( E \otimes \text{id}_{k} \) act as witnesses.} We denote the set of witnesses for the state space $\M{S}_i$ by the notation $\M{S}_{i}^*$. \par  
This duality between states and witnesses is much stronger. Given a state space, there is a unique set of witness, however, the relation goes the other way round too -- given the set of witnesses, the set
\begin{equation*}
(\M{S}^*)^* := \{ \phi \in \M{H} \quad | \quad \forall E \in \M{S}^*: \quad E(\phi) \in [0 , 1]  \} \end{equation*}    
of maps that assign probability to the witnesses is precisely the state space.
\begin{lemma}
Let $\M{S}$ be any state space, then
\begin{equation*}
 (\M{S}^*)^*  = \M{S} .
\end{equation*}
\end{lemma}
\begin{proof}
Let $\phi \in (\M{S}^*)^*$   be such that 
\begin{equation*}
    \forall E \in \M{S}^*: E \phi \in [0 , 1] ,
\end{equation*}
but $\phi \notin \M{S}$. We \RutC{prove that such a vector $\phi$ cannot exist. If such a vector $\phi$ existed, then }a witness that assigns a negative probability to $\phi$ \RutC{would also exist}, which contradicts our assumption. \\   
As $\M{S}$ is a state space, it is a convex set by construction. Thus, by the separating hyperplane theorem, there exists a linear map $F$ such that 
\begin{equation*}
F \phi \leq 0 \text{ and } F \psi \geq 0 \quad \forall \psi \in \M{S} .
\end{equation*} 
Let $\lambda = \max_{\psi \in \M{S}}(F \psi)$, then  clearly $\frac{1}{\lambda} F \in \M{S}^*$. However $\frac{1}{\lambda} F \phi < 0$, which leads to the conclusion that $\frac{1}{\lambda} F \notin \M{S}^*$, which leads to a contradiction. 
\end{proof}
Often in GPTs, the set of states and effects are put on an equal footing. This translates to the demand that given a state space $\M{S}$, the set of effects must be the set of witnesses $\M{S}^*$. We call this property the \textbf{no-restriction hypothesis}  \cite{JL, CDP}. While this symmetry in preparations and measurements may not be crucial for an arbitrary GPT, it should be noted that this holds for \RutC{quantum theory}. \\

We also use diagrams to express the set of effects. For example, \RutC{an} effect \RutC{$E$} for the state space $\M{S}_{\{ 1 ,2\}}$ is given by: 
\begin{equation}
\begin{quantikz}
      \qw^{\M{H}_1}    & \multimeterD[2]{E} \\  
        \qw^{\M{H}_2}                    &                                      
\end{quantikz}.
\end{equation}
Finally, we are able to define a measurement in the GPT framework:
\begin{definition}
We can define a measurement in the GPT framework in terms of measurement  $\{ E_{i} \}_{i}$ such that $\sum_i E_{i} = \uniteffect$.
\end{definition}  
\section{GPTs as diagrams}
The diagrammatic notation (see~\cite{plavala2021general} \RutC{for a full discussion and refer to papers such as \cite{GPTs_diagram1} and \cite{GPTs_diagram2} for examples of usage of the notation)} proves to be highly convenient when dealing with GPTs and demonstrating important claims about them. Its visual nature allows us to easily identify results that might remain obscured in the conventional linear algebraic framework. \RutC{A similar diagrammatic framework for finite dimensional quantum theory was introduced by Coecke et. al. \cite{Gpts_diagram4,Gpts_diagram3}. This graphical framework has been used for various areas, such as quantum error correction, quantum natural language processing, among other areas of quantum information theory and quantum computing \cite{coecke2023zxlectures}. }\\ 

Throughout the chapter, we have \RutC{familiarized} ourselves with the diagrammatic notation, for states, channels and effects. In light of this, let us consider expressing a state $\mu \psi_1 + (1 - \mu) \psi_2$ in terms of two states $\psi_1$ and $\psi_2$, where $\mu \in [0, 1]$. Employing the diagrammatic notation, this can be expressed as follows:
\begin{equation*}
\begin{quantikz}
\multiprepareC[1]{\mu \psi_1 + (1 - \mu) \psi_2}& \qw  
\end{quantikz} = \mu \begin{quantikz}
\multiprepareC[1]{\psi_1}& \qw                      
\end{quantikz}  + (1 - \mu) \begin{quantikz}
\multiprepareC[1]{ \psi_2}& \qw                      
\end{quantikz} .
\end{equation*}
Similarly holds for effects:
\begin{equation*}
\begin{quantikz}
 \qwbundle{n} & \multimeterD[1]{\mu E_1 + (1 - \mu) E_2} 
\end{quantikz} = \mu \begin{quantikz}
\qwbundle{n} &  \multimeterD[1]{E_1}               
\end{quantikz}  + (1 - \mu) \begin{quantikz}
\qwbundle{n} & \multimeterD[1]{E_2}         
\end{quantikz} .
\end{equation*}
Note that above, we have dropped the label for the Hilbert space on the wires. We will omit the explicit Hilbert space label on the wires whenever the interpretation of the diagram remains unambiguous. \\

The primary function of GPT is to compute the probability of outcomes in any given experiment. This probability is determined by the inner product of effects and states. In the diagrammatic notation, the inner products are given by diagrams with closed legs. Let $E$ be an effect and $\psi$ be a state, then the probability $E(\psi)$ is given by: 
\begin{equation*}
E (\psi) = 
\begin{quantikz}
\multiprepareC[1]{\psi}& \qw & \multimeterD[1]{E}         
\end{quantikz} .
\end{equation*}
Representing transformations in the diagrammatic framework is relatively straightforward. The identity channel in the GPT is denote by a wire:
\begin{equation*}
\begin{quantikz}
& \qw & \qw &   \qw       
\end{quantikz}  = 
\begin{quantikz}
 & \gate[1]{\idmap} &   \qw       
\end{quantikz} .
\end{equation*}
To illustrate how all three components of our GPT fit together, let's consider the following example:
\begin{equation*}
E ( \M{M} \psi) = 
\begin{quantikz}
\multiprepareC[1]{\psi}&  \gate[1]{\M{M}}  & \multimeterD[1]{E}. 
\end{quantikz} .
\end{equation*}
As can be seen, this notation makes the linear algebraic expressions visual and thus easy to understand and perform mathematical manipulations. 
\chapter{Channels in Boxworld}\label{chap: Boxworld}
This chapter is solely dedicated to exploring Boxworld, which stands as the largest theory (in terms of achievable measurement statistics) that remains consistent with the principles of relativity, as introduced in the previous chapter. This chapter begins with the definition of the most general Boxworld state space and the corresponding set of channels for Boxworld.

Following this, we delve into redefining the state space in terms of non-signaling vectors, which helps us to construct a linear-algebraic framework for Boxworld. In the final section of this chapter, we employ this framework to prove that, unlike the case with \RutC{quantum theory, every transformation in Boxworld is also a channel}. 
\section{Notations and definitions}
In this section, we present the most general definition of Boxworld state space. Our aim is to establish a clear and well-defined representation of Boxworld, which will serve as the foundation for constructing a linear algebraic framework of the theory. This framework will not only facilitate our analysis but also contribute to the proof of our central result, namely the equivalence of transformations and channels in Boxworld. \par

The physical scenario described by Boxworld involves systems being prepared by a source and subsequently measured by spacelike-separated (or non-communicating) entities such as distinct laboratories or parties. Recall that a state in the GPT framework \RutC{provides complete information about the probability of any outcome that can occur in any given measurement performed on the system. In Boxworld}, a state provides the probability for every possible outcome of all the local measurements performed by the distinct parties. A multi-system or multipartite Boxworld state can be uniquely described by conditional probability distributions of the form:
\begin{equation}
p_{\psi}(a_1, a_2, \ldots, a_n | x_1, x_2, \ldots, x_n). \nonumber
\end{equation}
Here, $x_i \in \{0, 1, \ldots, p_i\}$ represents the \RutC{set} of distinct measurements that the $i^{\text{th}}$ party can perform, and $a_i \in \{0, 1, \ldots, k_i\}$ denotes the \RutC{set} of outcomes \RutC{of} each measurement. For convenience, we assume a fixed number of outcomes for each measurement performed by a party, although our results can be generalized to cases where this assumption does not hold. However, \RutC{as the parties performing measurement are non-communicating,} Boxworld can \RutC{only allow for states that correspond to ``non-signalling'' probability distributions.} For example, in the case of a two partite distribution $p_{\psi}$, the distribution is considered non-signaling \RutC{(NS)} if the following condition holds:
\begin{equation}
NS: \qquad p_{\psi}(a_1 | x_1 , x_2) = p_{\psi}(a_1 | x_1 , x_2') \nonumber
\end{equation}
This condition must hold for all possible values of $a_1$, $x_1$, $x_2$, and $x_2'$. In general, we introduce the set of 
the non-signalling distributions:
\begin{definition}[Non-signalling set] A probability distribution $p $ is considered to be a non-signalling probability distribution if it satisfies the following: 
\begin{eqnarray}
\forall j: \sum_{a_j = 0}^{k_j} p(a_1, a_2, \cdots , a_{j} , \cdots , a_{n} |  x_1 , x_2 , \cdots , x_{j} , \cdots , x_{n}) \text{ is independent of } x_{j}.\nonumber
\end{eqnarray}
Conditions such as the one above are called ``non-signalling conditions''. We denote the collection of all the probability distributions that obey the non-signalling conditions by the set $\M{NS}$. 
\end{definition}
It is worth noting that the set of non-signaling probability distributions is convex, and therefore can be consistently defined to be a state space. The proof of this fact is straightforward, we shall skip the proof here for brevity. \par

Let us now formally define \RutC{the} Boxworld state space. We begin by describing a single system state space. If we label the measurements as $x \in \{0, 1 ,\cdots , p \}$ and assign the labels $a \in \{0, 1 , \cdots , k \}$ to the outcomes of each measurement, an element in the state space of the single system (or party) Boxworld can be expressed as follows:

\begin{equation}\label{Boxworld state}
\psi = \begin{bmatrix}
p_{\psi}(0|0) \\
\vdots \\ 
p_{\psi}(k | 0) \\ 
----\\
p_{\psi}(1|1) \\
\vdots \\
---- \\
p_{\psi}(0 | p) \\ 
\vdots \\
p_{\psi}(k|p)
\end{bmatrix}\RutC{ \in \mathbb{R}^{kp}}.
\end{equation}
Here, $p_{\psi}(a|x)$ represents the conditional probability of obtaining outcome $a$ when measurement $x$ is performed. \RutC{The state in equation \ref{Boxworld state} is a list of  $kp$ non-negative numbers, and therefore, can be treated as an element of the vector space $\mathbb{R}^{kp}$. This means that the state space of a single system Boxworld state space is a convex set in the space $\mathbb{R}^{kp}$ for some $k , p \in \mathbb{N}$}. \par 

To facilitate a more convenient representation of Boxworld states space, we will transition to using the Dirac Notation. This transition involves using the bra-ket notation, such as $\ket{\psi}$, to represent states instead of simple Greek letters like $\psi$. Referring back to the state $\psi$ as in Equation \eqref{Boxworld state}, we express it as: 
\begin{eqnarray}
\ket{\psi} = \sum_{a_1  = 0}^{k_1} \sum_{x_1 = 0}^{p_1} p(a_1 | x_1) \ket{e_{a_1 | x_1}} .
\end{eqnarray}
Here, $\{ |e_{a_1 | x_1} \rangle \}$ represents orthonormal basis in $\mathbb{R}^{kp}$.\\  
This notation proves to be particularly useful as it allows us to express the multi-system state $\psi$ in a concise manner\footnote{Here $\sum_{a}^{n}$ and $\sum_{x}^{n}$ are shorthand notations for $\sum_{a_1  = 0}^{k_1} \cdots \sum_{a_{n} = 0}^{k_n}$ and $\sum_{x_1 = 0}^{p_1} \cdots \sum_{x_n =  0}^{p_n}$ respectively. }:
\begin{eqnarray}\label{eqn: associate probability distribution}
\ket{\psi} = \sum_{a}^{n} \sum_{x}^{n} p_{\psi}(a_1 , a_2 , \cdots a_n|x_1 , x_2 , \cdots x_n) \ket{e_{a_1|x_1}} \otimes \ket{e_{a_2|x_2}} \otimes \cdots \ket{e_{a_n | x_n}} ,
\end{eqnarray}
where, $p_{\psi}(a_1 , \cdots,  a_n|x_1 , \cdots ,  x_n)$ is the probability of obtaining the outcomes $(a_1 , \cdots ,  a_n) $ when performing the measurement labeled by $(x_1 , x_2 , \cdots , x_n)$ are performed. \\ 
The underlying vector space that embeds our Boxworld state space is the tensor product space $\bigotimes_{i} \mathbb{R}^{k_i p_i} $, which, moving forward, will be denoted by $\M{H}_{N}^{\B{k} , \B{p}}$. We can now define the state space for Boxworld as follows: 
\begin{definition}[Boxworld state space]
Let $n \in \mathbb{N}$ denote the number of parties. Let $\mathbf{k} = (k_1, k_2, \ldots, k_n)$ denote the vector corresponding to the number of outputs for each measurement, and let $\mathbf{p} = (p_1, p_2, \ldots, p_n)$ be the vector representing the number of distinct measurements made by each party. The state space $\bnkp$ is defined as the set of all  vectors $\ket{\psi} \in \bigotimes_{i = 1}^{n} \mathbb{R}^{k_i p_i}$  such that the distribution $p_{\psi}$ defined by:\footnote{The notation $\ket{x,y}$ and $\ket{x} \otimes \ket{y}$ will be used interchangeably in this context.}
\begin{eqnarray}
p_{\psi}(a_1 , a_2 , \cdots a_{n} | x_1 , x_2 , \cdots x_{n}) := \bra{e_{a_1|x_1}, e_{a_2|x_2}, \ldots, e_{a_n|x_n}}\ket{\psi}
\end{eqnarray}
is a (sub-normalized) non-signalling distribution i.e. $p_{\psi} = \lambda p$ for some $\lambda \in [0 , 1]$ and $p \in \M{NS}$.
\end{definition}
If the parameter $\lambda  = 1$, then we call the state $\psi$ as a normalized state. For a normalized state $\ket{\psi}$, if $\ket{\psi} \in \partial \bnkp$ ,then we call it a ``pure'' state. This notation is borrowed from quantum theory, in which the pure states are the boundaries of the state space. If $\lambda < 1$, then we call the state un-normalized. In general, this parameter $\lambda$ defines the norm of the state. 
\begin{definition}[Norm]
Let $\ket{\psi} \in \bnkp$, and let $p_{\psi}$ be the distribution defined by \eqref{eqn: associate probability distribution}. Since $p_{\psi} = \lambda p$ for some $p \in \mathcal{NS}$ and $\lambda \in [0,1]$, we define the norm of $\ket{\psi}$ as $\normone {\ket{\psi}} = \lambda$.
\end{definition} 
The norm defined above, when acting as a map $\normone{.} : \bnkp \mapsto [0 ,1]$ is linear. This can be shown by observing that for any $\lambda, \epsilon \in \mathbb{R}$ and $p_1, p_2 \in \M{NS}$, the distribution $\lambda p_1 + \epsilon p_2$ can be expressed as $(\lambda + \epsilon) \left(\frac{\lambda}{\lambda + \epsilon} p_1 + \frac{\epsilon}{\lambda + \epsilon} p_2\right)$. It is evident that $\frac{\lambda}{\lambda + \epsilon} p_1 + \frac{\epsilon}{\lambda + \epsilon} p_2 \in \M{NS}$, which follows from the convexity of $\M{NS}$. Hence, the linearity of the norm map follows from \RutC{the} one to one correspondence of Boxworld states and probability distributions. \\ 
 
Now that we have defined the state space, we can proceed to define transformations in Boxworld. Recall that \RutC{transformations} in GPTs \RutC{are} linear maps between any two state spaces in the GPT, \RutC{and not every transformation is a channel. A channel is a transformation that acts consistently on composite state spaces as well}. However, here we deviate from this terminology, and instead, adopt the terminology inspired from the quantum theory. Consider the following definitions: 
\begin{definition}[Positive maps] A linear map $\M{M}: \M{H}^{\B{kp}}_{n} \mapsto \M{H}^{\B{k'p'}}_{n'}$ is positive if  
\begin{equation*}
\forall \ket{\psi} \in \bnkp:  \quad \M{M} \ket{\psi}  \in \bnkprime .
\end{equation*} 
\end{definition}
\noindent We introduce the notion of \RutC{$l$} positive maps 
\RutC{\begin{definition}[$l$ positive maps] A positive linear map $\M{M}: \M{H}^{\B{kp}}_{n} \mapsto \M{H}^{\B{k'p'}}_{n'}$ is a $l$ positive if the map $\M{M}\otimes \id_{l} : \M{H}^{\B{kp}}_{n} \otimes \M{H}^{\tilde{\B{k}} \tilde{\B{p}} }_{l}  \mapsto \M{H}^{\B{k'p'}}_{n'} \otimes \M{H}^{\tilde{\B{k}} \tilde{\B{p}} }_{l} $ is positive. Here $\id_{l}$ is the identity defined on the space $\M{H}^{\tilde{\B{k}} \tilde{\B{p}} }$. 
\end{definition}}
This inspires the definition of completely positive maps.
\begin{definition}[Completely positive maps]
A linear map $\M{M}$ is completely positive if $\M{M}$ is $k$ positive for every $k \in \mathbb{N}$. 
\end{definition}
We adopted the terminology of `positivity' and `complete positivity' over 'transformations' and 'channels' because it allows us to draw parallels between concepts in Boxworld and those in quantum theory. 
\section{Redefining Boxworld state space }
In this section, our objective is to find an alternative definition of the Box-world state space which is free from the direct reference to the probability distribution associated with the state. To do so, we start by introducing a special set of vectors known as the non-signaling vectors:
\begin{definition}[Non-signalling vectors] Let 
\begin{equation*}
|f_{x_{i}, x_{i}'}\rangle :=  \sum_{a_{i} = 0}^{k_i} \left( |e_{a_{i}| x_{i}}\rangle  - |e_{a_{i}| x_{i}'} \rangle \right) ,
\end{equation*}  then we define the set $v_{\M{NS}}$ as the set of all the vectors of the form
\begin{eqnarray}
 \ket{e_{a_1 | x_1} ,  \cdots  , e_{a_{i} | x_{i}}  , \cdots ,   f_{x_{j}, x_{j}'} , \cdots  , e_{a_{n} |x_{n}}} \nonumber .
\end{eqnarray}
\end{definition}
As we will show next, the set $v_{\M{NS}}$ comprises vectors that have a one-to-one correspondence with the non-signaling constraints. 
\begin{lemma} Let $\ket{\psi} \in \bnkp$, then $\forall \ket{w} \in v_{\M{NS}}$, $\bra{w}\ket{\psi} = 0$ .
\end{lemma}
\begin{proof} Let $p_{\psi}$ be the probability distribution associated with the state $\psi$. Since $p_{\psi}$ obeys the non-signalling constraints, we have that: 
\begin{eqnarray}
 \sum_{a_j = 0}^{k_{j}} \left(   p_{\psi}(\cdots, a_{j} , \cdots  | \cdots , x_{j}, \cdots x_{l}, \cdots   ) -  p_{\psi}(\cdots,  a_{j} , \cdots |\cdots , x_{j}', \cdots x_{l}, \cdots   ) \right)  = 0. \nonumber 
\end{eqnarray}
The above can be re-written as
\begin{eqnarray}
\sum_{a_{j} = 0}^{k_j} \left( \langle e_{a_1|x_1}| \otimes \cdots \otimes \left( \langle  e_{a_{j}|x_{j}}| - \langle e_{a_{j} | x_{j}'} | \right) \cdots \otimes \langle e_{a_{l}|x_{l}} | \otimes \cdots \otimes  \langle e_{a_{n}|x_n} | \right) |\psi   \rangle &=&  0. \nonumber
\end{eqnarray}
This is equivalent to $\bra{w}\ket{\psi} = 0$ for every $\ket{w} \in v_{\M{NS}}$. 
\end{proof}
The non-signalling vectors defines a sub-space 
\begin{eqnarray}
    \mathfrak{NS} := \text{span} \left( \{\ket{w}| \ket{w}  \in v_{\M{NS}}  \} \right) \nonumber 
\end{eqnarray}
of all the vectors orthogonal to the state space. Each vector $\ket{w} \in \mathfrak{NS}$ can be used to define a different non-signalling condition, making this subspace interesting \RutC{to} analyse. The discussion in the section helps us to re-define Boxworld state space.
\begin{lemma} 
An element $\ket{\psi}  \in \M{H}_{n}^{\B{kp}}$ is a state in Boxworld iff it satisfies the following: 
\begin{itemize}
    \item \textbf{Positive:} For every $a_1 \cdots  a_{n}$ and for every  $x_1  \cdots  x_{n}$:  $\langle e_{a_1 | x_1} , \cdots, e_{a_n | x_n} | \psi \rangle \geq 0$.
    \item \textbf{Normalizable:} $\normone{\ket{\psi}} \leq 1$. 
    \item \textbf{Non-signalling:} $\forall \ket{w} \in \mathfrak{NS}: \bra{w}\ket{\psi} = 0 $.
\end{itemize}
\end{lemma}

\section{Some examples and properties of completely positive map}
Now, let's consider some examples and properties of completely positive maps in Boxworld. We will specifically focus on examples that will be useful for proving our results in the next section. Let's consider the following example: 
\begin{lemma} \label{lemm: post-seclection map}
The map $\M{M}_{a_i|x_{i}}$ defined by 
\begin{equation*}
\M{M}_{a_i | x_i} \ket{\psi} = \langle e_{a_i | x_{i}} | \psi \rangle \end{equation*} 
is a completely positive map.
\end{lemma}
\begin{proof}
To simplify the argument, we show that $\M{M}_{a_1 | x_1}$ is a completely positive map, as the reasoning is not explicitly dependent on the specific number of measurement outcomes or the total number of measurements. Let $p_{\psi}$ be the probability distribution associated with the state $\ket{\psi}$ as defined in equation \eqref{eqn: associate probability distribution}. We show that the distribution $q$ defined by the components 
\begin{eqnarray}
q(a_2 \cdots a_{n} | x_2 \cdots x_n ) := p_{\psi}(a_1, a_2 \cdots a_n | x_1 , x_2 \cdots x_n)
\end{eqnarray}
is indeed a (sub-normalized) non-signaling probability distribution - i.e. $q$ is of the form $\lambda p$ for some $p\in \M{NS}$  and $\lambda \in [0 , 1]$. Note that the positivity of the components $q(a_2 \cdots a_{n} | x_2 \cdots x_n )$ follows trivially for the expression above. To show that $q$ is a (sub-normalized) non-signalling probability distribution, we show that $\normone{\M{M}_{a_1|x_1} \ket{\psi}} \leq \normone{\ket{\psi}}$ and that $q$ obeys all the non-signalling constraints. \\  
We first show that $ \normone{\M{M}_{a_1|x_1} \ket{\psi}} \leq \normone{\ket{\psi}} $ as follows:\footnote{The notation \RutC{$\sum_{a \ne 1}^{n} $ is a short hand notation for $ \sum_{a_2 = 0}^{k_2} \cdots \sum_{a_3 = 0}^{k_3} \cdots \sum_{a_{n}}^{k_n}$}. }
\begin{eqnarray}
\sum_{a\ne 1}^{n} q(a_2 , a_3 \cdots a_{n} | x_2 , x_3, \cdots x_{n} ) &=& \sum_{a \neq 1}^{n} p_{\psi}(a_1 , a_2 , \cdots, a_{n} | x_1 , x_2 , \cdots , x_n) \nonumber \\  
&\leq& \sum_{a}^{n} p_{\psi}(a_1 , a_2 , \cdots, a_{n} | x_1 , x_2 , \cdots , x_n) \label{eqn: random inequality}\\ 
&=& \normone{\ket{\psi}}. \nonumber 
\end{eqnarray}
\RutC{The inequality \eqref{eqn: random inequality}} holds as a consequence of $p_{\psi}(a_1 , \cdots , a_{n}|x_{1}, \cdots , x_{n})$ all non-negative. \\ 
We now check if $q$ obeys the non-signalling conditions. Observe that
\begin{eqnarray}
 \sum_{a_{j} = 0}^{k_j} q(\cdots, a_{j} , \cdots | x_2 , \cdots , x_{j} , \cdots ) &=& \sum_{a_{j} =1}^{k_j} p_{\psi}(a_1, \cdots, a_{j} , \cdots  |x_{1} \cdots , x_{j} , \cdots ) \nonumber \\ 
 &=& \sum_{a_{j} = 0}^{k_j}  p_{\psi}(a_{1}, \cdots, a_{j} , \cdots |x_{1} \cdots , x_{j}' , \cdots, ) \nonumber \\ 
 &=& \sum_{a_{j} = 0}^{k_j} q(\cdots, a_{j} , \cdots | x_2 , \cdots , x_{j}' , \cdots). \nonumber
\end{eqnarray}
Importantly, the validity of our argument is independent of the number of parties and the number of measurements each party makes, meaning that $\mathcal{M}$ is completely positive.
\end{proof}

By utilizing distinct positive maps in Boxworld, we can construct other completely positive maps. One straightforward way to achieve this is by composing positive maps. If we have two completely positive maps $\M{M}$ and $\M{N}$, their composition $\M{M} \circ \M{N}$ is also a completely positive map. Consequently, the composition of multiple maps of the form $\M{M}_{a_i|x_i}$ - i.e. maps of the form
\begin{equation}
\M{M}_{a_1| x_1} \circ \M{M}_{a_2 | x_2} \circ \cdots \circ \M{M}_{a_j | x_j} \nonumber
\end{equation}
is a completely positive map. \\

In some cases, summing two completely positive maps can result in the construction of a new completely positive map. \RutC{To see why this is the case, let $\ket{\psi}$ be a Boxworld state. If} $\M{M}$ and $\M{N}$ are both completely positive maps, then the vector $(\M{M} + \M{N})\ket{\psi}$ is orthogonal to the set $\mathfrak{NS}$, which can be easily seen by taking the appropriate inner products. Therefore, to establish whether $(\M{M} + \M{N})\ket{\psi}$ is a valid state, we only need to verify if $\normone{(\M{M} + \M{N})\ket{\psi}} \leq 1$ holds for every state $\ket{\psi}$. \\ 
One such example is the map $\mathcal{M}_{x_1}$ defined by 
\begin{equation*}
\mathcal{M}_{x_i} \ket{\psi} = \langle e_{x_i} | \psi \rangle.\end{equation*}
where $\ket{e_{x_i}} := \sum_{a_i = 0}^{k_i} \ket{e_{a_i | x_i}}$. From the discussion above, to show that this map is completely positive, it suffices to show that $\M{M}_{x_i}$ is norm non-increasing. In fact, we show that this map preserves the norm of the state:
\begin{align*}
\normone{\mathcal{M}_{x_1} \ket{\psi}} &= \sum_{a_1=0}^{k_1} \normone{\mathcal{M}_{a_1|x_1} \ket{\psi}} \\
&= \sum_{a_1 = 0}^{k_1} \sum_{a \neq 1}^{n} p_{\psi}(a_1, a_2 , \cdots , a_n|x_1, x_2, \cdots , x_{n}) \\
&= \normone{\ket{\psi}}. 
\end{align*} 
Here, we have used the linearity of the norm map. Therefore, the map $\mathcal{M}_{x_{1}}$ is completely positive. \\ 
Again, it is worth mentioning explicitly that the composition of such maps such as
$$\M{M}_{x_1} \circ \M{M}_{x_2} \circ \cdots \circ \M{M}_{x_n}$$ 
yields a completely positive map. Moreover, this composition of maps individually preserves the norm of the state. Thus, we can define the norm map in terms of such composition - i.e. for every $\psi \in \bnkp:$
$$ \norm{\ket{\psi}} = \norm{\M{M}_{x_1} \circ \M{M}_{x_2} \circ \cdots \circ \M{M}_{x_n} \ket{\psi}}.$$ 

\RutC{Observe that $\M{M}_{x_i}$ does not increase the norm of the state. We will now prove a more general claim: any transformation (positive map) in a GPT cannot increase the norm of the state. Here, we prove this claim in context of Boxworld, the arguments employed are applicable to any GPT.}
\begin{lemma}\label{lemm: norm decreases}
Let $\M{M}: \bnkp \mapsto \bnkprime$ be a positive map. Then, for all $\ket{\psi} \in \bnkp$, we have $\normone{\ket{\psi}} \geq \normone{\M{M}\ket{\psi}}$.
\end{lemma} 
\begin{proof}
Assume the existence of $\ket{\psi} \in \bnkp$ such that $\normone{\ket{\psi}} < \normone{\M{M}\ket{\psi}} \leq 1$. Choose $\lambda \in \left(\frac{1}{\normone{\M{M}\ket{\psi}}}, \frac{1}{\normone{\ket{\psi}}} \right)$. Clearly, $\normone{\lambda \ket{\psi}} < 1$, indicating that $\lambda \ket{\psi} \in \bnkp$. However, we observe that $\normone{\M{M}(\lambda \ket{\psi})} = \normone{\lambda \M{M}\ket{\psi}} \geq 1$. Therefore, $\M{M}(\lambda \ket{\psi}) \notin \bnkprime$. As a result, $\M{M}$ is not a positive map, leading to a contradiction.
\end{proof}
Note that replacing $\bnkp$ by $\M{S}_{n}$ and $\bnkprime$ by $\M{S}_{m}$ proves the claim for general GPTs.
\section{Positive maps and completely positive maps in Boxworld} 
We are now prepared to prove the central claim of this chapter, which asserts that any positive map in Boxworld is also completely positive. Recall that to establish a vector as a state in Boxworld, it is sufficient to prove that the vector possesses three properties: it has non-negative entries in the canonical bases, it has a norm less than or equal to 1, and it is orthogonal to the set $\mathfrak{NS}$. We will break our proof into three steps, wherein we demonstrate that $\M{M} \otimes \id_{l}\ket{\psi}$, for any $l \in \mathbb{N}$, satisfies all three conditions when $\M{M}$ is a positive map and $\ket{\psi}$ is a Boxworld state (in the domain of $\M{M} \otimes \id_{l}\ket{\psi}$).
Let us begin by proving that $\M{M} \otimes \id_{l} \ket{\psi}$ exhibits non-negative components in all canonical bases for every Boxworld state $\ket{\psi}$ in the domain of $\M{M} \otimes \id_{l} \ket{\psi}$. 
\begin{lemma}\label{lemm}
Let $\M{M}: \M{H}_{n}^{\B{kp}}   \mapsto \M{H}_{n'}^{\B{k'p'}}$ be a positive map, and let $\ket{\psi} \in \M{H}_{n}^{\B{kp}} \otimes \M{H}_{n'}^{\B{\tilde{k} \tilde{p}}}$ be any Boxworld, state then  
\begin{eqnarray}
    \langle e_{a_1|x_1},  \cdots,  e_{a_{n'}| x_{n'}} , \cdots ,  e_{a_{n'+l} | x_{n'+l} } | \M{M} \otimes \id_{l} |  \psi \rangle \geq 0
\end{eqnarray}
\RutC{holds} for every possible of measurement $(x_0 , x_1 , \cdots x_{n'+l})$ and output $(a_1 , a_2 , \cdots a_{n'+l})$ .
\end{lemma}
\begin{proof} 
Let $\ket{\psi} \in \M{H}_{n}^{\B{kp}} \otimes \M{H}_{l}^{\B{k'p'}}$ be a Boxworld state. Notice that the inner product
\begin{eqnarray}
\langle e_{a_1 |x_1} , e_{a_2 | x_2} , \cdots , e_{a_{n'} | x_{n'} }| \otimes \langle e_{a_{n' + 1}| x_{n' + 1}} , \cdots,  e_{a_{n' + l} | x_{n' + l} } | \M{M} \otimes \id_{l} | \psi \rangle \nonumber 
\end{eqnarray}
can be rewritten as
\begin{eqnarray}
\langle e_{a_1 |x_1} , e_{a_2 | x_2} , \cdots , e_{a_{n'} | x_{n'} }| \M{M} \left(\langle e_{a_{n' + 1}| x_{n' + 1}} , \cdots e_{a_{n' + l} | x_{n' + l} } | \psi \rangle \right). \nonumber
\end{eqnarray}
By Lemma \ref{lemm: post-seclection map}, the vector $\ket{\tilde{\psi}}$ defined by $\left(\langle e_{a_{n' + 1}| x_{n' + 1}} , \cdots e_{a_{n' + l} | x_{n' + l} } | \psi \rangle \right)$ is indeed an $n$-partite state, i.e., $\ket{\tilde{\psi}} = \left(\langle e_{a_{n' + 1}| x_{n' + 1}} , \cdots e_{a_{n' + l} | x_{n' + l} } | \psi \rangle \right) \in \bnkp$. This is because \begin{equation*}
 \langle e_{a_{n' + 1}| x_{n' + 1}} , \cdots,  e_{a_{n' + l} | x_{n' + l} } | \psi \rangle = \M{M}
_{a_{n' +1}|x_{n' + 1}} \circ \M{M}_{a_{n' +2}|x_{n' + 2}} \circ \cdots \circ \M{M}_{a_{n'+ l} | x_{n' + l}} \ket{\psi}\end{equation*}  is a vector that is obtained \RutC{by} repeated application of positive maps and thus results in a state in the set $\bnkp$. \\
Now, by definition, $\langle e_{a_1 |x_1} , e_{a_2 | x_2} , \cdots , e_{a_{n'} | x_{n'} }| \M{M} |\tilde{\psi} \rangle \in \bnkprime$ since $\M{M}$ is a positive map. Thus, $$    \langle e_{a_1|x_1},  \cdots,  e_{a_{n'}| x_{n'}} , \cdots ,  e_{a_{n'+l} | x_{n'+l} } | \M{M} \otimes \id_{l} |  \psi \rangle \geq 0. $$
\end{proof}
In the second part of our proof, we face the most challenging aspect of the overall argument. We show that $\M{M} \otimes \id_{l}$ cannot be a signalling map. This means that \RutC{the} resulting vector, obtained by applying $\M{M} \otimes \id_{l}$ to a state in Boxworld, is orthogonal to the subspace $\mathfrak{NS}$. 
\begin{lemma} 
If $\M{M}: \M{H}_{n}^{\B{kp}}  \mapsto \M{H}_{n'}^{\B{k'p'}}$ is positive, then $\forall \ket{w} \in v_{\M{NS}}$, 
\begin{eqnarray}
    \langle w | \M{M} \otimes \id_l | \psi \rangle = 0 \nonumber
\end{eqnarray}
holds for all Boxworld states $\ket{\psi} \in \M{H}_{n}^{\B{kp}} \otimes \M{H}_{l}^{\B{\tilde{k} \tilde{p} }}$. 
\end{lemma} 
\begin{proof}
As we know that $\M{M}$ is a positive map, the following \RutC{holds} for every $\ket{\phi} \in \bnkp$ and for every $\ket{w} \in \mathfrak{NS}$,   
\begin{equation*}
\bra{w} \M{M} \ket{\phi} = 0 .
\end{equation*}
The only way the equation above holds for every $\ket{\phi} \in \bnkp$ is when $\M{M}^{\dagger}$ leaves the space $\mathfrak{NS}$ invariant - i.e. $ \forall \ket{w} \in \mathfrak{NS}:$ $\M{M}^{\dagger} \ket{w} \in \mathfrak{NS}$. \\
We need to check if for every $\ket{w} \in \mathfrak{NS}$, $\bra{w}(\M{M} \otimes \id_{l}) \ket{\psi} = 0$. As the map $\M{M}$ acts on the first $n$ tensor factors of the state $\ket{\psi}$, the vector $\M{M}^{\dagger}$ on $\ket{w}$ acts on the first $n'$ tensor factors in the dual space. This allows us to consider $2$ distinct types of vectors $\ket{w}$ \RutC{(up to} relabeling of parties) 
\begin{eqnarray}
\textbf{Case 1}:    \ket{v_{1}} &=&  \ket{f_{x_1, x_1'} , e_{a_2| x_2}, \cdots , e_{a_{n'} | x_{n'}}}  \otimes \ket{e_{a_{n'+1} | x_{n'+1} } , \cdots, e_{a_{n' + l}| x_{n'+l}}} \nonumber. \\ 
\textbf{Case 2}:    \ket{v_{2}} &=&  \ket{e_{a_1|x_1}  , e_{a_2 | x_2}, .. ,e_{a_{n'} |  x_{n'}}} \otimes \ket{f_{x_{n'+1} , x'_{n'+1} } ,   e_{a_{n' + 2}| x_{n'+2}},  \cdots,  e_{a_{n' + l}|  x_{n'+l}}}. \nonumber
\end{eqnarray}
We now deal with both the cases individually. \\ 
\textbf{Case 1:}  We can write 
\begin{eqnarray}
\M{M}^{\dagger} \otimes \id_{l} \ket{v_1} &=& (\M{M}^{\dagger}  \ket{f_{x_1, x_1'} , e_{a_2| x_2}, \cdots , e_{a_{n'} | x_{n'}}}  ) \otimes \ket{e_{a_{n' + 1}|x_{n'+1}} , \cdots ,e_{a_{n' + l}|x_{n'+l}}}. \nonumber 
\end{eqnarray}
Here, $\M{M}^{\dagger}  \ket{f_{x_1, x_1'} , e_{a_2| x_2}, \cdots , e_{a_{n'} | x_{n'}}} \in \mathfrak{NS}$ as discussed above. Thus, $(\M{M} \otimes \id_l) \ket{v_1}$ must be expressable as a linear combination of vectors of the form 
$$\ket{w} \otimes | e_{a_{n'+1}|x_{n' +1} } , \cdots, e_{a_{n' + l} | x_{n' + l}} \rangle,$$ 
where $\ket{w} \in v_{\M{NS}}$. This shows that $\langle v_1 | \M{M} \otimes \id_{l} |\psi \rangle = 0$.\\ 

Alternatively, we can prove $\bra{v_1} \M{M} \otimes \id_{l} \ket{\psi} = 0$ by observing that $\bra{v_1}\M{M}\otimes \id_{l} \ket{\psi}$ can be written as:
\begin{eqnarray}
     \langle v_{1}|  \M{M} | \psi \rangle = \langle (f_{x_1 , x_1'} , e_{a_2 | x_2} , \cdots , e_{a_{n'}|x_{n'}} | \M{M} \left( \bra{e_{a_{n' + 1}|x_{n' + 1}} ,  \cdots, e_{a_{n' + l}|x_{n' + l}}} \ket{\psi}\right) \nonumber
\end{eqnarray}
However, notice from Lemma \ref{lemm: post-seclection map}, we can infer that
\begin{equation*}
\ket{\tilde{\psi}} := \bra{e_{a_{n' + 1}|x_{n' + 1}} , e_{a_{n' + 2} | x_{n' +2}} , \cdots,  e_{a_{n' + l} | x_{n' + l}}} \ket{\psi} \in \bnkp,    
\end{equation*}
as it is obtained as a composition of positive maps acting on a Boxworld state. Thus,
\begin{equation*}
\bra{v_1} \M{M} \otimes \id_{l} \ket{\psi} = \langle f_{x_1 , x_1'} , e_{a_2 | x_2} , \cdots , e_{a_{n'} | x_{n'}} | \tilde{\psi} \rangle = 0 ,\end{equation*}
since $|f_{x_1 , x_1'} , e_{a_2 | x_2} , \cdots , e_{a_{n'} | x_{n'}}\rangle \in \mathfrak{NS}$.\\ 

\textbf{Case 2}: 
Notice that, vectors of the form $\ket{e_{a_1|x_1}, e_{a_2|x_2}, \cdots, e_{a_{n'}|x_{n'}}}$ span the entire space $\mathcal{H}_{n'}^{\B{k'p'}}$. Thus, we can express $\M{M}^{\dagger} \ket{e_{a_1|x_1}, e_{a_2 | x_2}, \cdots, e_{a_{n'} | x_{n'}}}$  using a linear combination:
\begin{equation*}
\sum_{a}^{n'} \sum_{x}^{n'} \alpha_{a_1, a_2 , \cdots,  a_{n'} ; x_{1} , x_{2 } , \cdots x_{n'} } \ket{e_{a_1|x_1} , e_{a_2 | x_2}, \cdots ,  e_{a_{n}|x_{n'}}},     
\end{equation*} 
where $\alpha_{a_1, a_2 , \cdots a_{n'} ; x_{1} , x_{2 } , \cdots x_{n'} } \in \mathbb{R}$ are some real coefficients. Thus, 
\begin{equation*}
\M{M}^{\dagger} \otimes \id_l \ket{v_2} = \sum_{a}^{n'} \sum_{x}^{n'} \alpha_{a_1, a_2 , \cdots,  a_{n'} ; x_{1} , x_{2 } , \cdots x_{n'} } \ket{e_{a_1|x_1} , e_{a_2 | x_2}, \cdots ,  e_{a_{n}|x_{n'}}} \otimes \ket{w},     
\end{equation*} 
where $\ket{w} \in v_{\M{NS}}$. Thus, $ \M{M}^{\dagger} \otimes \id_l \ket{v_2}  \in \mathfrak{NS}$. This implies that $\bra{v_2} \M{M} \otimes \id_{l} \ket{\psi} = 0$. \\ 
Combining the results for Case 1 and Case 2, we can prove the claim.
\end{proof}
We now end the proof by showing that $\M{M} \otimes \id_{k}$ cannot increase the norm of the states.
\begin{lemma}
If $\M{M}: \M{H}_{n}^{\B{kp}}   \mapsto \M{H}_{n'}^{\B{k'p'}}$ is positive, then $\M{M} \otimes \id_{k}$ is norm non-increasing. 
\end{lemma}
\begin{proof}
Let $\ket{\psi} \in \M{H}_{n}^{\B{kp}} \otimes \M{H}_{l}^{\B{\tilde{k} \tilde{p}}} $ be a state. From our discussion in the previous section, we can express $ \normone{\M{M} \otimes \id_{l} \ket{\psi}}$ as 
\begin{equation*}
\langle e_{x_1} , e_{x_2} , \cdots e_{x_{n'}} | \M{M} (\langle e_{x_{n'+ 1}} , \cdots , e_{x_{n' + l}} | \psi \rangle ) \equiv \langle e_{x_1} , e_{x_2} , \cdots e_{x_{n'}} | \M{M}| \hat{\psi} \rangle )
\end{equation*}
for some Boxworld state $|\hat{\psi}\rangle := \langle e_{x_{n'+ 1}} , \cdots , e_{x_{n' + l}} | \psi \rangle$. Note that $|\hat{\psi}\rangle$ is a Boxworld \RutC{state} as 
$$\langle e_{x_{n'+ 1}} , \cdots , e_{x_{n' + l}} | \psi \rangle  = \M{M}_{x_{n'+ 1}} \circ \cdots 
 \circ \M{M}_{x_{n' + l}} \ket{\psi} = |\hat{\psi}\rangle.$$ 
Using above and discussion from the previous section, we can also infer that 
$\normone{|\hat{\psi}\rangle } = \normone{\ket{\psi}}$. From above, we know that,
\begin{eqnarray}
\normone{\M{M} \otimes \id_{l} \ket{\psi}} &=& \langle e_{x_1} , \cdots , e_{x_{n'}} |\M{M}| \hat{\psi} \rangle \nonumber \\ 
&=& \normone{\M{M} \ket{ \hat{\psi}}} \nonumber \\ 
&\leq&  \normone{\ket{\hat{\psi}}} \nonumber \\ 
&=& \normone{\ket{\psi}} \nonumber .
\end{eqnarray}
The inequality above follows from the Lemma \ref{lemm: norm decreases}. 
\end{proof} 
\section{Conclusion}
In this chapter, we began by introducing a general definition of Boxworld for arbitrary many parties. We briefly discussed what non-signalling probability distributions are and defined an appropriate state-space for both single-party and multi-partite Boxworld state spaces. Subsequently, we re-defined the Boxworld state space using a linear algebraic framework that does not directly reference the probability distributions associated with a state. Instead, it is described using three different linear algebraic constraints that are equivalent to verifying if the underlying probability distribution is a non-signalling probability distribution. We also explored some straightforward examples of completely composite maps in Boxworld and discussed some easily provable properties of completely positive maps. Using these examples and the three alternative criteria, we proved that any positive map in Boxworld is also completely positive.

It remains an open problem to identify all theories in which such a straightforward result is true. Additionally, it's interesting to determine if all the GPTs that abide by such a result also indeed lack non-trivial channels, as is the case in Boxworld. We discuss some of these problems, as well as some (partial) progress in this direction, in the next and final chapter of the thesis. 

\chapter{Discussion and conclusion} 

In this section of the thesis, we delve into the GPT framework, starting with the definitions of states, effects, and channels. Given the analogous concept of complete positivity in a GPT, we highlight the mathematical challenges associated with defining a channel and an effect when a GPT is determined solely by its state space. We then turn our attention to Boxworld, the theory that is capable of producing any multi-partite non-signaling probability distributions. Importantly, we make no assumptions about the number of parties involved, the number of measurements each party can conduct, or the number of outputs for each measurement.

Our definition of Boxworld results in a fully linear algebraic framework for the theory. \RutC{Contrasting with quantum theory, a distinction arises: in Boxworld, any transformation (positive map) is a channel (completely positive map)}. This insight provides a tractable criterion, in principle, for determining whether a particular linear map qualifies as a channel within Boxworld. \par 

Boxworld's limited dynamics have lead to variations like \RutC{Houseworld~\cite{SB2009, SPB2009} and Noisy Boxworld~\cite{JL}} to capture specific quantum theory features absent in classical theory. Houseworld emerged to address the absence of non-locality swapping in Boxworld, an analog of entanglement swapping in quantum theory. In Houseworld, specific effects demonstrate non-locality swapping, illustrating that this phenomenon isn't exclusive to quantum theory. The two-party state space in Houseworld is defined by excluding certain extremal states (PR boxes) from Boxworld's state space.

Noisy Boxworld replicates various quantum theory aspects, especially Tsirelson's bound for maximally entangled states. \RutC{The state space (or the effect space) of this theory is constructed by taking a convex hull of ``noisy'' versions of extremal states (or effects) of Boxworld.}

Both Houseworld and certain Noisy Boxworld instances lack specific relabeling symmetries. Preliminary research (in collaboration with Mr. Kuntal Sengupta) suggests tractable criteria for defining effects in these theories when using the min-tensor product. Notably, the no-restriction hypothesis appears inconsistent with the max tensor product in these contexts.

This indicates a need to delve deeper into discerning sets of channels and effects, as such determinations can be intricate. \RutC{Addressing such questions is important} for understanding non-trivial GPTs, especially those being designed to  \RutC{study interesting  properties of multi-partite} state spaces.

Ultimately, one of the important pursuit of understanding GPTs is to axiomatize quantum theory from an information-theoretical perspective. A good method for defining the effects and channels for GPTs allowing for composite state spaces will play a crucial role if our goal is to understand all \RutC{non-classical} features of quantum theory using the GPT framework.

\cleardoublepagewithnumber

\appendix

\chapter{Appendix for Device Independent Protocols}
\section{Upper bounding the derivatives - H(A|XYE)}\label{appendix: H(A|XYE)} 
In this section, we derive bounds on the functions $\cos(u(\B{x}) \pm v) \sin(2 \theta)$. Since $\sin(2 \theta)$ is always positive and increasing in $\theta$ in our domain, we get the following result for any $(\eta , \theta , v) \in \M{C}_{i,j,k}$. We can lower bound $\cos(u(\B{x}) \pm v) \sin(2 \theta)$ as  
\begin{eqnarray}\label{eqn: Taylors theorem for H(A|XYE)}
  \begin{cases}\max_{\B{x} \in \M{C}_{i,j,k}} \Big (\cos(u(\B{x}) \pm v) \Big)  \sin(2 \theta_{i+1}) & \text{if} \max_{\B{x} \in \M{C}_{i,j,k}} \Big( \cos(u(\B{x}_{i,j,k}) \pm v) \Big) > 0 \\ 
    \max_{\B{x} \in \M{C}_{i,j,k}} \Big(\cos(u(\B{x}) \pm v) \Big)  \sin(2 \theta_{i})    &  \text{if} \max_{\B{x} \in \M{C}_{i,j,k}} \Big( \cos(u(\B{x}_{i,j,k}) \pm v) \Big) < 0.
 \end{cases}
\end{eqnarray}
Let $\B{x}=(\eta,\theta,v) \in \M{C}_{i,j,k}$ and, for brevity, write $g_{\pm,y}(\B{ x})=u_{\pm}(\B{x})+(-1)^yv$ with $y\in\{0,1\}$. Then, by Taylor's theorem (cf.\ Theorem~\ref{thm:Taylor}), there exists $\B{ x'}\in\M{C}_{i,j,k}$ such that
\begin{eqnarray}
\cos(g_{\pm,y}(\B{x})) &=& \cos(g_{\pm,y}(\B{x}_{i,j,k})) + \partial_\eta\cos(g_{\pm,y}(\B{x}))\big|_{\B{ x'}}(\eta-\eta_i)+ \nonumber \\ 
                         & & \hspace{0.1cm} \partial_\theta\cos(g_{\pm,y}(\B{x}))\big|_{\B{ x'}}(\theta-\theta_j^{(i)}) + \partial_v\cos(g_{\pm,y}(\B{x}))\big|_{\B{ x'}}(v-v_k^{(i,j)}).\label{eq:cosg}
\end{eqnarray}
We upper bound this by upper bounding each of the partial derivatives on $\M{C}_{i,j,k}$:
\begin{eqnarray*}
  \partial_\eta\cos(g_{\pm,y}(\B{x}))&=&-\sin(g_{\pm,y}(\B{x}))\partial_\eta u_{\pm}\\
  \partial_\theta\cos(g_{\pm,y}(\B{x}))&=&-\sin(g_{\pm,y}(\B{x}))\partial_\theta u_{\pm}\\
  \partial_v\cos(g_{\pm,y}(\B{x}))&=&-\sin(g_{\pm,y}(\B{x}))(\partial_v u_{\pm}+(-1)^y).
\end{eqnarray*}
We have bounded the derivatives of $u$ on $\M{C}_{i,j,k}$ in~\eqref{eq:uderivs}.

We now consider the different cases. Firstly, suppose $\max_{\B{x}\in\M{C}_{i,j,k}}\left[-\sin(g_{\pm,y}(\B{x}))\right]\geq0$.

Consider the terms in~\eqref{eq:cosg}. Using the bounds in~\eqref{eq:uderivs}, we have
\begin{align*}
  \partial_\eta\cos(g_{+,y}(\B{x}))\big|_{\B{ x'}}(\eta-\eta_i)&\leq2\max_{\B{x}\in\M{C}_{i,j,k}}\left[-\sin(g_{+,y}(\B{x}))\right](\eta_{i+1}-\eta_i)\\                                    \partial_\theta\cos(g_{+,y}(\B{x}))\big|_{\B{ x'}}(\theta-\theta_j^{(i)})&\leq0\\
  \partial_v\cos(g_{+,y}(\B{x}))\big|_{\B{ x'}}(v-v_k^{(i,j)})&\leq\begin{cases}2\max_{\B{x}\in\M{C}_{i,j,k}}\left[-\sin(g_{+,0}(\B{x}))\right](v_{k+1}^{(i,j)}-v_k^{(i,j)})&y=0\\0&y=1\end{cases}.
\end{align*}
Similarly,
\begin{align*}
  \partial_\eta\cos(g_{-,y}(\B{x}))\big|_{\B{ x'}}(\eta-\eta_i)&\leq0\\                                    \partial_\theta\cos(g_{-,y}(\B{x}))\big|_{\B{ x'}}(\theta-\theta_j^{(i)})&\leq0\\
  \partial_v\cos(g_{-,y}(\B{x}))\big|_{\B{ x'}}(v-v_k^{(i,j)})&\leq\begin{cases}2\max_{\B{x}\in\M{C}_{i,j,k}}\left[-\sin(g_{-,0}(\B{x}))\right](v_{k+1}^{(i,j)}-v_k^{(i,j)})&y=0\\0&y=1\end{cases}.
\end{align*}
Combining all of these, and bounding the $-\sin(g_{\pm,y}(\B{x}))$ terms by $1$, we find
\begin{align}
\cos(g_{+,0}(\B{x}))&\leq\cos(g_{+,0}(\B{x}_{i,j,k}))+2(\eta_{i+1}-\eta_i)+2(v_{k+1}^{(i,j)}-v_k^{(i,j)})\nonumber\\
  \cos(g_{+,1}(\B{x}))&\leq\cos(g_{+,1}(\B{x}_{i,j,k}))+2(\eta_{i+1}-\eta_i)\label{eq:gplus1}\\
  \cos(g_{-,0}(\B{x}))&\leq\cos(g_{-,0}(\B{x}_{i,j,k}))+2(v_{k+1}^{(i,j)}-v_k^{(i,j)})\nonumber\\
  \cos(g_{-,1}(\B{x}))&\leq\cos(g_{-,1}(\B{x}_{i,j,k}))\nonumber
\end{align}

Secondly, in the case $\max_{\B{x}\in\M{C}_{i,j,k}}\left[-\sin(g_{\pm,y}(\B{x}))\right]\leq0$, we have
\begin{align*}
  \partial_\eta\cos(g_{+,y}(\B{x}))\big|_{\B{ x'}}(\eta-\eta_i)&\leq0\\                                    \partial_\theta\cos(g_{+,y}(\B{x}))\big|_{\B{ x'}}(\theta-\theta_j^{(i)})&\leq-2\min_{\B{x}\in\M{C}_{i,j,k}}\left[-\sin(g_{\pm,y}(\B{x}))\right](\theta^{(i)}_{j+1}-\theta^{(i)}_j)\\
  \partial_v\cos(g_{+,y}(\B{x}))\big|_{\B{ x'}}(v-v_k^{(i,j)})&\leq\begin{cases}0&y=0\\-\min_{\B{x}\in\M{C}_{i,j,k}}\left[-\sin(g_{+,0}(\B{x}))\right](v_{k+1}^{(i,j)}-v_k^{(i,j)})&y=1\end{cases}.
\end{align*}
and
\begin{align*}
  \partial_\eta\cos(g_{-,y}(\B{x}))\big|_{\B{ x'}}(\eta-\eta_i)&\leq-2\min_{\B{x}\in\M{C}_{i,j,k}}\left[-\sin(g_{\pm,y}(\B{x}))\right](\eta_{i+1}-\eta_i)\\                                    \partial_\theta\cos(g_{-,y}(\B{x}))\big|_{\B{ x'}}(\theta-\theta_j^{(i)})&\leq-2\min_{\B{x}\in\M{C}_{i,j,k}}\left[-\sin(g_{\pm,y}(\B{x}))\right](\theta^{(i)}_{j+1}-\theta^{(i)}_j)\\
  \partial_v\cos(g_{-,y}(\B{x}))\big|_{\B{ x'}}(v-v_k^{(i,j)})&\leq\begin{cases}0&y=0\\-\min_{\B{x}\in\M{C}_{i,j,k}}\left[-\sin(g_{+,0}(\B{x}))\right](v_{k+1}^{(i,j)}-v_k^{(i,j)})&y=1\end{cases}.
\end{align*}

Combining all of these, and bounding the $-\sin(g_{\pm,y}(\B{x}))$ terms by $-1$, we find
\begin{align}
\cos(g_{+,0}(\B{x}))&\leq\cos(g_{+,0}(\B{x}_{i,j,k}))+2(\theta^{(i)}_{j+1}-\theta^{(i)}_j)\nonumber\\
  \cos(g_{+,1}(\B{x}))&\leq\cos(g_{+,1}(\B{x}_{i,j,k}))+2(\theta^{(i)}_{j+1}-\theta^{(i)}_j)+(v_{k+1}^{(i,j)}-v_k^{(i,j)})\label{eq:gplus2}\\
  \cos(g_{-,0}(\B{x}))&\leq\cos(g_{-,0}(\B{x}_{i,j,k}))+2(\eta_{i+1}-\eta_i)+2(\theta^{(i)}_{j+1}-\theta^{(i)}_j)\nonumber\\
  \cos(g_{-,1}(\B{x}))&\leq\cos(g_{-,1}(\B{x}_{i,j,k}))+2(\eta_{i+1}-\eta_i)+2(\theta^{(i)}_{j+1}-\theta^{(i)}_j)+(v_{k+1}^{(i,j)}-v_k^{(i,j)}).\nonumber
\end{align}

\section{Upper bounding the derivatives - H(AB|00E)} \label{appendix: H(AB|00E)} 
For brevity in this section we often use $\bar{\theta}=\pi/4+\theta$. We upper-bound the derivatives for the functions $\hat{\alpha}_0, \tilde{\epsilon}, \hat{R}$. We first upper bound the derivatives for $\alpha$ as 
\begin{eqnarray}
\Big| \partial_{\lambda} \hat{\alpha}_0 \Big| &=&  \Big|\frac{2 \cot \left(\bar{\theta}\right) \csc ^2(\lambda )}{\cot ^2\left(\bar{\theta}\right) \cot
   ^2(\lambda )+1}  \Big|\\
\Big| \partial_v \hat{\alpha}_0 \Big| &=& \Big|\frac{\cot \left(\bar{\theta}\right) \csc ^2(v)}{\cot ^2\left(\bar{\theta}\right) \cot ^2(v)+1} \Big|\\ 
\Big| \partial_{\bar{\theta}} \hat{\alpha}_0 \Big| &=& \Big|\frac{ 2 \csc ^2\left(\bar{\theta}\right) \cot (\lambda )}{\cot ^2\left(\bar{\theta}\right) \cot ^2(\lambda )+1} -  \frac{\csc ^2\left(\bar{\theta}\right) \cot (v)}{\left(\cot ^2\left(\bar{\theta}\right) \cot ^2(v)+1\right)} \Big|.
\end{eqnarray}
Observe that for $x\in\mathbb{R}$
\begin{eqnarray}
    \frac{a \csc^2(x)}{a^2 \cot^{2}(x) + 1} &\leq& \max \{a, \frac{1}{a}\}. 
\end{eqnarray}
Noting that $\cot(\bar{\theta}) \leq 1$ for $\theta \in [0,\frac{\pi}{4}]$. This gives us 
\begin{eqnarray}
|\partial_{\lambda}\hat{\alpha}_0 |  &\leq&  2 \tan(\bar{\theta}) =: \alpha_{\lambda}\\
|\partial_v\hat{\alpha}_0 |          &\leq& \tan(\bar{\theta})  =: \alpha_v.
\end{eqnarray}
The identity $\frac{x}{1 + a^2 x^2}\leq \frac{1}{2|a|}$ can be used to get the following upper bound 
\begin{eqnarray*}
    \Big| \partial_{\bar{\theta}} \hat{\alpha}_0 \Big| &\leq& \Big|\frac{2 \csc ^2\left(\bar{\theta}\right) \cot (\lambda )}{\cot ^2\left(\bar{\theta}\right) \cot ^2(\lambda )+1} \Big|  + \Big|\frac{\csc ^2\left(\bar{\theta}\right) \cot (v)}{\left(\cot ^2\left(\bar{\theta}\right) \cot ^2(v)+1\right)}  \Big|\\ 
    &\leq&  2\frac{\csc^{2}(\bar{\theta}) \tan(\bar{\theta})}{2} + \frac{\csc^2(\bar{\theta})\tan(\bar{\theta})}{2}\\ 
    &=&\frac{3}{2\sin(2\bar{\theta})}:=\alpha_{\bar{\theta}}.
\end{eqnarray*}

Define $z(\lambda, v, \theta)$ to be the denominator in~\eqref{two sided function 2}, i.e., \begin{eqnarray}
z(\lambda, v, \theta)\!&:=&\!\cos(v\!-\!\lambda)\!\left[\sin(\hat{\alpha}_0)\sin(v)\cos(\frac{\pi}{4}\!+\!\theta)\!+\!\cos(\hat{\alpha}_0)\cos(v)\sin(\frac{\pi}{4}\!+\!\theta)\right]\! \nonumber\\ 
& & \hspace{4.5cm} - \quad\frac{\sin(v\!-\!\lambda)}{\sqrt{2}}\sqrt{1\!-\!\cos(2 v)\sin(2 \theta)}.
\end{eqnarray}
We now compute the derivatives of $z$. For the derivative with respect to $\lambda$, we write $\partial_{\lambda} z = (b_1 + b_3) + b_2 \partial_{\lambda} \hat{\alpha}_0$, where 
\begin{eqnarray*}
b_1 &=& \sin(v-\lambda)\left(\sin(\hat{\alpha}_0)\sin(v)\cos(\bar{\theta})+\cos(\hat{\alpha}_0)\cos(v)\sin(\bar{\theta})\right)\\
b_2 &=& \cos(v-\lambda)\left(\cos(\hat{\alpha}_0)\sin(v)\cos(\bar{\theta})-\sin(\hat{\alpha}_0)\cos(v)\sin(\bar{\theta})\right) \\ 
b_3 &=& \frac{\cos(v-\lambda)\sqrt{1-\cos(2v)\sin(2\theta)}}{\sqrt{2}}.
\end{eqnarray*}
We can then bound these by $b_1 \leq \cos(\bar{\theta}) + \sin(\bar{\theta}) +1/\sqrt{2} \leq \sqrt{2} + 1/\sqrt{2}$ and $b_2 \leq \cos(\bar{\theta}) + \sin(\bar{\theta}) \leq \sqrt{2} $, so that 
\begin{eqnarray}
|\partial_{\lambda} z| \leq \sqrt{2}(3/2+\alpha_{\lambda}) =: z_{\lambda}.
\end{eqnarray}

Note that $\partial_v[\cos(v-\lambda)\sin(v)]=\cos(\lambda-2v)$ and $\partial_v[\cos(v-\lambda)\cos(v)]=\sin(\lambda-2v)$. We can hence write the $v$ derivative as 
\begin{eqnarray}
\partial_vz = a_1 + a_2 + a_3 \partial_{v} \hat{\alpha}_0\text{,\quad where}
\end{eqnarray}
\begin{eqnarray*}
a_1&=&\cos(\lambda-2v)\sin(\hat{\alpha}_0)\cos(\bar{\theta})+\sin(\lambda-2v)\cos(\hat{\alpha}_0)\sin(\bar{\theta}) \leq \cos(\bar{\theta}) + \sin(\bar{\theta}) \leq \sqrt{2}\\
   a_2 &=& -\frac{\cos (v-\lambda ) \sqrt{1-\sin (2
   \theta ) \cos (2 v)}}{\sqrt{2}}-\frac{\sin (2 \theta ) \sin(2 v) \sin (v-\lambda )}{\sqrt{2} \sqrt{1-\sin (2 \theta )\cos (2 v)}} \leq \sqrt{2}\\
   a_3 &=& \cos(v-\lambda)\left(\cos(\bar{\theta})\sin(v)\cos(\hat{\alpha}_0)-\sin(\bar{\theta})\cos(v)\sin(\hat{\alpha}_0)\right) \leq \cos(\bar{\theta}) + \sin(\bar{\theta}) \leq \sqrt{2},
\end{eqnarray*} 
and where we obtained the bound on $a_2$ using $|\frac{\sin(2v)\sin(2\theta)}{\sqrt{1 - \sin(2\theta) \cos(2v))}}| \leq \sqrt{2}$. Hence, we can bound
\begin{eqnarray}
|\partial_vz|\leq\sqrt{2}(2+\alpha_v)=: z_v.
\end{eqnarray}
Finally we compute the $\bar{\theta}$ derivative 
\begin{eqnarray}
\partial_{\bar{\theta}} z &=& c_1 + c_2+c_3 \partial_{\bar{\theta}} \hat{\alpha}_0
\end{eqnarray}
where 
\begin{eqnarray*}
c_1 &=&\cos(v-\lambda)\left(\cos(\hat{\alpha}_0)\cos(v)\cos(\bar{\theta})-\sin (\hat{\alpha}_0)\sin(v)\sin(\bar{\theta})\right) \leq \cos(\bar{\theta}) + \sin(\bar{\theta}) \leq \sqrt{2}\\
c_2&=&\frac{\cos(2\theta)\cos(2v)\sin(v-\lambda)}{\sqrt{2}\sqrt{1-\sin(2\theta)\cos(2v)}}\leq\frac{\cos(2\theta)}{\sqrt{2}\sqrt{1-\sin(2\theta)}}=\sqrt{\frac{1+\sin(2\theta)}{2}}\leq1\\
c_3 &=&\cos(v-\lambda)\left(\cos(\hat{\alpha}_0)\sin(v)\cos(\bar{\theta})-\sin(\hat{\alpha}_0)\cos(v)\sin(\bar{\theta})\right)\leq \cos(\bar{\theta}) + \sin(\bar{\theta}) \leq \sqrt{2}
\end{eqnarray*}
We hence obtain
\begin{eqnarray}
|\partial_{\bar{\theta}} z| &\leq&  \sqrt{2} + 1 + \sqrt{2}\alpha_{\bar{\theta}} =: z_{\theta}.
\end{eqnarray}
We now compute the derivatives of $\tilde{\epsilon}$:
\begin{eqnarray*}
   \partial_{\lambda} \tilde{\epsilon} &=& -\partial_{\lambda}\hat{\alpha}_0 (\cos (\theta ) \sin
   (\hat{\alpha}_0+\lambda-2 v)+\sin (\theta )
   \sin (\hat{\alpha}_0 -\lambda +2 v)) \\ 
   & &\hspace{4.5cm}+\sin (\theta
   ) \sin (\hat{\alpha}_0-\lambda +2 v)-\cos
   (\theta ) \sin (\hat{\alpha}_0+\lambda-2 v)\\ 
\partial_{v}\tilde{\epsilon} &=& -\partial_{v}\hat{\alpha}_0 (\cos (\theta)\sin(\hat{\alpha}_0+\lambda-2v)+\sin(\theta)
   \sin (\hat{\alpha}_0-\lambda +2 v))\\ 
   & & \hspace{4.5cm}-2 \sin
   (\theta ) \sin (\hat{\alpha}_0-\lambda +2 v)+2
   \cos (\theta ) \sin (\hat{\alpha}_0+\lambda-2
   v)\\
   \partial_{\theta}\tilde{\epsilon} &=&-\partial_{\theta} \hat{\alpha}_0 (\cos (\theta ) \sin(\hat{\alpha}_0+\lambda-2 v)+\sin (\theta )
   \sin (\hat{\alpha}_0 -\lambda +2 v))+\cos (\theta
   ) \cos (\hat{\alpha}_0 -\lambda +2 v) \\ 
   & & \hspace{4.5 cm}-\sin
   (\theta ) \cos (\hat{\alpha}_0+\lambda-2 v).
\end{eqnarray*}
Using the same techniques as above, we find the following bounds 
\begin{eqnarray*}
|\partial_{\lambda}\tilde{\epsilon}| &\leq& \alpha_{\lambda} (\cos(\theta) + \sin(\theta )) + \cos(\theta) + \sin(\theta) \\
                                      &\leq&  \sqrt{2}(\alpha_{\lambda}+1)=: \epsilon_{\lambda}\\
|\partial_{v}\tilde{\epsilon}| &\leq& \alpha_{v} (\cos(\theta) + \sin(\theta )) + 2\cos(\theta) + 2\sin(\theta) \\ 
                               &\leq&   \sqrt{2}(\alpha_{v}+2)=:\epsilon_v\\
|\partial_{\theta}\tilde{\epsilon}| &\leq& \alpha_{\theta} (\cos(\theta) + \sin(\theta )) + \cos(\theta) + \sin(\theta)   \\ 
                                &\leq &   \sqrt{2}(\alpha_{\theta}+1) =: \epsilon_{\theta}.
\end{eqnarray*}
\section{Appendix for H(AB|XYE)}\label{app: Two sided rates on grids}
\subsection{General result on Grids} 
Suppose that we want to optimize the function: 
\begin{equation}\label{eqn: general result on grids}
\begin{aligned}
\min_{\mathbf{x}\in \M{C} ,\B{y} \in \M{D} } \quad & g(\B{x} , \B{y}) + h(\B{x}) \\
\textrm{s.t.} \quad & \forall i: \quad f^{i}(\B{x} , \B{y}) \geq 0.
\end{aligned}
\end{equation}
where $\B{x} = (x_0 , x_1 , \cdots , x_{n}) \in \mathbb{R}^{n} $ and $\B{y} = (y_0 , y_1 , \cdots, y_{m}) \in \mathbb{R}^{m}$. Also, we assume that $f_i , g , h$ are differentiable functions in the respective domains. \\ 
Our goal is to eliminate certain parameters in the optimization problem, making it suitable for numerical optimizations. We further assume that there exists $\B{x}_0 $ such that 
\begin{equation}\label{eqn: what is x0?}
    \inf_{\B{x} \in \M{C}}: h(\B{x}) \geq h(\B{x}_0).
\end{equation} 
Let $\Delta_{\B{x}_0} $ be the difference  vector defined by
$$ \Delta_{\B{x}_0}(\B{x})  :=  \B{x} - \B{x}_0.$$ 
and the gradient restricted in $\mathbb{R}^{n}$ given by
$$ \nabla_{x} g(\B{x}, \B{y}) = \left(\partial_{x_0}g(\B{x}, \B{y}) , \partial_{x_1}g(\B{x}, \B{y}), \cdots , \partial_{x_{n}}g(\B{x} , \B{y})   \right) $$
We can lower bound the objective function in terms of the function using the Taylor's theorem
\begin{eqnarray} 
 g(\B{x} , \B{y}) + h(\B{x}) &\geq& g(\B{x}_0 , y) + h(\B{x}_{0}) -  \nabla_{x} g(\Tilde{\B{x}} , \tilde{\B{y}}).\Delta_{\B{x}_0}(\tilde{\B{x}})  \nonumber\\ 
                            &\geq& g(\B{x}_0 , y) + h(\B{x}_{0}) - g_{\max} \Delta_{\max} \label{eqn: teri baari aa hi gayi}
\end{eqnarray}
for some $\tilde{\B{x}} \in \M{C}$, $\tilde{\B{y}} \in \M{D}$ and $g_{\max}$ and $\Delta_{\max}$ are any positive numbers satisfying: 
\begin{eqnarray}\label{eqn: what is g and delta?}
g_{\max} \geq \max_{\mathbf{x}\in \M{C} ,\B{y} \in \M{D} } || \nabla_{x} g(\B{x}_0 , \B{y}) || \text{ and }  \Delta_{\max} \geq \max_{\mathbf{x} \in \M{C}} || \Delta_{\B{x}_0} ||.
\end{eqnarray}
For the sake of completion, we also define the quantity 
\begin{eqnarray}\label{eqn: what is f?}
f_{\max}^{i} \geq  \max_{\mathbf{x}\in \M{C} ,\B{y} \in \M{D} } || \nabla_{x} f^{i}(\B{x}_0 , \B{y}) ||.
\end{eqnarray}

Now consider the following result that shall enable us to compute the lower bound of the optimization problem \eqref{eqn: general result on grids}:  
\begin{lemma}\label{lemm: gazab book-keeping}
Consider the optimization problem defined as follows:
\begin{eqnarray}\label{eqn: nayi optimization problem}
\begin{aligned}
\min_{\B{y} \in \M{D}} \quad & g(\B{x}_0 , \B{y}) + h(\B{x}_0) - g_{\max} \Delta_{\max}\\ 
\textrm{s.t.} \quad & \forall i: \quad f^{\max}(\B{x}_0, \B{y}) \geq -f^{i}_{\max} \Delta_{\max},
\end{aligned}
\end{eqnarray}
where $\B{x}_0$ is defined according to \eqref{eqn: what is x0?}, and $g_{\max}$, $\Delta_{\max}$, and $f_{\max}$ are defined as in \eqref{eqn: what is g and delta?} and \eqref{eqn: what is f?}, respectively. This optimization problem serves as a lower bound on the problem \eqref{eqn: general result on grids}.
\end{lemma}
\begin{proof}
In the discussion above, we showed that the objective function in \eqref{eqn: general result on grids} can be bounded from below using \eqref{eqn: teri baari aa hi gayi}. It remains to show that the feasible set of the optimization problem  \eqref{eqn: general result on grids} is subset of that of \eqref{eqn: nayi optimization problem}. This can also be shown using Taylor's Theorem. Let $(\B{x}, \B{y})$ be any point in the feasible set of  \eqref{eqn: general result on grids}, then 
\begin{eqnarray*}
     f^{i}(\B{x} , \B{y})  =   f^{i}(\B{x}_0 , \B{y}) -   \nabla_{x} f^{i}(\tilde{\B{x}_0}  , \tilde{\B{y}}).\Delta_{\B{x}_0}(\tilde{\B{x}}) \nonumber
\end{eqnarray*}
for some $\tilde{\B{x}_0} \in \M{C}  , \tilde{\B{y}} \in \M{D}$. Now, it is straightforward to see that,
\begin{eqnarray*}
     f^{i}(\B{x}_0 , \B{y}) -   \nabla_{x} f^{i}(\tilde{\B{x}_0}  , \tilde{\B{y}}).\Delta_{\B{x}_0}(\tilde{\B{x}})  \geq 0 \implies   f^{i}(\B{x}_0 , \B{y}) +  f_{\max} \Delta_{\max} \geq 0.
\end{eqnarray*}
Thus, the set of feasible points of the problem \eqref{eqn: general result on grids}, remain feasible for the optimization problem \eqref{eqn: nayi optimization problem} as well. 
\end{proof}
\section{Usage of EAT for different protocols}\label{appendix: EAT}
\subsection{Protocol with recycled input randomness (Protocol~\ref{prot:nonspotcheck})}
For this protocol we want to extract randomness from the inputs and outputs. We hence set $C_i=A_iB_iX_iY_i$ and take $D_i$ to be trivial. When running a protocol, we do not generally know the set of EAT channels being used (these are set by the adversary), but instead only know that they have the no-signalling form, i.e., we have
$$\cM(\rho_{A'B'})=\sum_{abxy}\proj{a}\ot\proj{b}\ot\proj{x}\ot\proj{y}\ot\proj{u(a,b,x,y)}\ot\cM^{abxy}(\rho_{A'B'})\,,$$
where $\cM^{abxy}(\rho_{A'B'})=p_{XY}(x,y)(\cE^{x,a}\ot\cF^{y,b})(\rho_{A'B'})$ and $\{\cE^{x,a}\}_a$ and $\{\cF^{y,b}\}_b$ are instruments on $A'$ and $B'$ respectively (cf.~\eqref{EAT channel}). Henceforth, the set $\mathfrak{G}$ will refer to all channels of this type.

In the CHSH protocol without spot-checking the rate function should be a lower bound on $H(ABXY|E)=2+H(AB|XYE)$ and we can form our rate function via $\rate(\{1-s,s\})=2+F_{AB|XYE}(s)$ or $\rate(\{1-s,s\})=2+F_{A|XYE}(s)$, the former being preferred as it is larger. A min-tradeoff function can then be obtained by taking the tangent at some point. Since $F_{AB|XYE}(s)$ is linear for $3/4\leq s\leq\score_{AB|XYE}^*\approx0.847$, for experimentally relevant scores we can form the min-tradeoff function using the extension of this line to the domain $[0,1]$, i.e., we can take $f(\{1-s,s\})=2+G'_{AB|XYE}(\score^*)(s-3/4)$ in Theorem~\ref{thm:EAT} when applying to Protocol~\ref{prot:nonspotcheck}, and in this case $d_C=d_Ad_Bd_Xd_Y=16$ and we get a bound on $H_{\min}^{\epsilon_h}(\B{ ABXY}|E)$. The theorem holds for all $\alpha\in(1,2)$ and we can optimize over $\alpha$ to increase the bound.

\subsection{Spot-checking CHSH protocol (Protocol~\ref{prot:spotcheck})}
To use the EAT in the spot-checking CHSH protocol (Protocol~\ref{prot:spotcheck}) we set $C_i=A_iB_i$ and $D_i=X_iY_i$ in Theorem~\ref{thm:EAT}. The channels again have the no-signalling form mentioned above, and we can use either $F_{AB|00E}$ or $F_{A|00E}$ as the basis of our rate function. Since the two-sided version is larger, it is better to work with $F_{AB|00E}(s)$, and the related min-tradeoff function based on taking its tangent at some point. Modification is required to account for the spot-checking structure. If we let $g_t(\{1-s,s\})$ be the tangent of $F_{AB|00E}(s)$ taken at $t$ then we can form the spot-checking min-tradeoff functions
\begin{align*}
f_t(\delta_u)=\begin{cases}\frac{1}{\gamma}g_t(\delta_u)+(1-\frac{1}{\gamma})g_t(\delta_1)&u\in\{0,1\}\\
    g_t(\delta_1)&u=\bot\end{cases}\,.
\end{align*}
where $t$ can be chosen (see e.g.~\cite[Section~5]{DF} for the argument behind this). Using this construction the following theorem can be derived (this is an adaptation of Theorem~3 in~\cite{LLR&}).
\begin{theorem}[Entropy Accumulation Theorem for spot-checking CHSH protocol]\label{thm:EATspot} Let $\rho_{\B{ ABXYU}E}$ be a CQ state obtained using the spot-checking CHSH protocol (Protocol~\ref{prot:spotcheck}). Let $\Omega$ be the event $\left|\{i:U_i=0\}\right|\leq n\gamma(1-\score_{\exp}+\delta)$ with $p_\Omega$ being the probability of this event in $\rho_{\B{ ABXYU}E}$, and let $\rho_{\B{ ABXYU}E|\Omega}$ be the state conditioned on $\Omega$. Let $\epsilon_h\in(0,1)$ and $\alpha\in(1,2)$. Then for any $r$ such that $f_t(\Freq_\B{ U})\geq r$ for all events in $\Omega$ we have
\begin{eqnarray*}
H^{\epsilon_h}_{\min}(\B{ AB}|\B{ XY}E)_{\rho_{\B{ ABXY}E|\Omega}} > && n r - \frac{\alpha}{\alpha -1} \log\left(\frac{1}{p_\Omega (1- \sqrt{1 - \epsilon_h^2} )}\right) \\
&&+ n\inf_{p \in\cQ^\gamma_{\mathfrak{G}}}(\Delta (f_t , p) - (\alpha - 1) V(f_t,p) - (\alpha -1)^2 K_{\alpha}(f_t))\,,
\end{eqnarray*}
where 
\begin{eqnarray*}
\Delta(f_t,p)&:=&F_{AB|XYE}(p(1)/\gamma)-f_{t}(p) \\ 
V(f_t, p) &=& \frac{\ln 2}{2} \left(\log(9)+\sqrt{\Var_p(f_t)+2}\right)^2 \\
K_{\alpha}(f_t)&=& \frac{1}{6 \log(2 - \alpha)^3 \ln2 }2^{(\alpha -1)(2 + \Max(f_t) -\Min_{\M{Q}^\gamma_{\mathfrak{G}}}(f_t))} \ln^3(2^{2 + \Max(f_t) - \Min_{\M{Q}^\gamma_{\mathfrak{G}}}(f_t)}+\e^2).
\end{eqnarray*}
\end{theorem}
To use this theorem we can take $r=(F_{AB|00E}(t)+(\score_{\exp}-\delta-t)F'_{AB|00E}(t))$ (cf.\ the discussion in~\cite{LLR&}), and since the theorem holds for any $t$ and $\alpha$ these can be optimized over.

\subsection{Protocol with biased local random numbers (Protocol~\ref{prot:biased})}
To derive the randomness rates, we use Theorem~\ref{thm:EAT} with $C_i=A_iB_i$ and $D_i=X_iY_i$, as in the previous subsection. What remains is to derive the min-tradeoff function and error terms. In this section, we compute these quantities and derive the expression for the completeness error in terms of the biasing parameters $\zeta_A,\zeta_B$ and statistical error $\delta$. 

\subsubsection{Deriving the min-tradeoff function} 
We seek a min-tradeoff function suitable for using with Protocol~\ref{prot:biased}. To construct it we write the EAT channel in a slightly different way that is explicit in the input distribution $p_{XY}$:
\begin{eqnarray}
\cM_{p_{XY}}(\rho)&=& \sum_{abxy}p_{XY}(x,y)\proj{a}_A\ot\proj{b}_B\ot\proj{x}_X \nonumber \\
                  & &\hspace{2cm}  \ot\proj{y}_Y\ot\proj{(x,y,w)}_U\ot\cM^{x,y}_{a,b}(\rho)\,,\label{EAT channel2}
\end{eqnarray}
where $\cM^{x,y}_{a,b}$ are subnormalized channels. We can also consider the analogous channel where the $U$ register only stores $w$ (we use $\tilde{\cM}$ to indicate this case). Next consider the entropy $H(AB|X=0,Y=0E)$, this entropy is calculated for the normalization of the state
$$(\proj{0}_X\ot\proj{0}_Y\ot\id_{ABUE})(\cM_{p_{XY}}\ot\cI_E)(\rho_{RE})(\proj{0}_X\ot\proj{0}_Y\ot\id_{ABUE})\,.$$
For fixed $\{\cM^{x,y}_{a,b}\}$, this is independent of $p_{XY}$ (it is defined provided $p_{XY}(0,0)\neq0$).

We next note that for $q$ as the distribution on the score ($U$) register
\begin{eqnarray*}
(\cM_{p_{XY}}\ot\cI_E)(\rho_{RE})_U&=&\sum_{abxyw}p_{XY}\tr(\cM^{x,y}_{a,b}(\rho_R))\proj{(x,y,w)} \\ 
                                    &=&\sum_{xyw}q(x,y,w)\proj{(x,y,w)}\\
(\cM_{1/4}\ot\cI_E)(\rho_{RE})_U&=&\sum_{abxyw}\frac{1}{4}\tr(\cM^{x,y}_{a,b}(\rho_R))\proj{(x,y,w)} \\ 
                                &=& \sum_{xyw}\frac{q(x,y,w)}{4p_{XY}}\proj{(x,y,w)}\,,
\end{eqnarray*}
and hence
$$(\tilde{\cM}_{1/4}\ot\cI_E)(\rho_{RE})_U=\sum_{xyw}\frac{q(x,y,w)}{4p_{XY}}\proj{w}\,.$$

It follows that
\begin{align*}
  &\biggl\{H(AB|X=0,Y=0,E)_{(\tilde{\cM}_{1/4}\ot\cI_E)(\rho_{RE})}:(\tilde{\cM}_{1/4}\ot\cI_E)(\rho_{RE})_U=(1-s)\proj{0}+s\proj{1},\ \\ 
  & \hspace{12cm}s=\sum_{xy}\frac{q(x,y,1)}{4p_{XY}}\biggr\}\\
  &= \biggl\{H(AB|X=0,Y=0,E)_{(\cM_{p_{XY}\ot\cI_E)}(\rho_{RE})}:(\cM_{p_{XY}}\ot\cI_E)(\rho_{RE})_U \\ 
  & \hspace{8cm}=\sum_{xyw}q(x,y,w)\proj{(x,y,w)}\biggr\}\,.
\end{align*}

Let $\mathfrak{G}_{\zeta^A,\zeta^B}$ be the set of channels for which $X$ and $Y$ are independent, $X$ is 1 with probability $\zeta^A$ and $Y$ is 1 with probability $\zeta^B$.

\begin{lemma}
  The function $F_{AB|00E}$ as defined in the main text can be used to define a rate function for $\mathfrak{G}_{\zeta^A,\zeta^B}$ by taking
  $\rate_{\zeta^A,\zeta^B}(q)=F_{AB|00E}(\score(q))$ for $q\in\cQ_{\mathfrak{G}_{\zeta^A,\zeta^B}}$
  where
  \begin{equation}\label{eq:sc}
    \score(q)=\frac{1}{4}\sum_{xy}\frac{1}{p_X(x)p_Y(y)} q((x,y,1))\,.
    \end{equation}
\end{lemma}
\begin{proof}
  We have
  \begin{align*}
F_{AB|00E} (\score(q))& :=\!\!\!\inf_{(\tilde{\cM},\rho_{RE})} \Big\{H(AB|X=0,Y=0,E)_{(\tilde{\cM}_{1/4}\ot\cI_E)(\rho_{RE})}\!:\!(\tilde{\cM}_{1/4}\ot\cI_E)(\rho_{RE})_U \\ & \hspace{7cm}=(1-\score(q))\proj{0}+\score(q)\proj{1}\Big\}\\
              &=\!\!\!\inf_{(\cM,\rho_{RE})}\!\biggl\{\!H(AB|X=0,Y=0,E)_{(\cM_{p_Xp_Y}\ot\cI_E)(\rho_{RE})}\!:\!(\cM_{p_Xp_Y}\ot\cI_E)(\rho_{RE})_U \\  &\hspace{7cm}=\!\sum_{xyw}q(x,y,w)\proj{(x,y,w)}\!\biggr\}\\
    &\leq\inf_{(\cM,\rho_{RE})}\Big\{H(AB|XYE)_{(\cM_{p_Xp_Y}\ot\cI_E)(\rho_{RE})}:(\cM_{p_Xp_Y}\ot\cI_E)(\rho_{RE})_U \\ 
    & \hspace{6cm}=\sum_{xyw}q(x,y,w)\proj{(x,y,w)}\Big\},
  \end{align*}
and hence $F_{AB|00E}(\score(q))$ is a rate function for $q\in\cQ_{\mathfrak{G}_{\zeta^A,\zeta^B}}$.
\end{proof}
We can hence form min-tradeoff functions suitable for using with Protocol~\ref{prot:biased} by taking affine lower bounds to $F_{AB|00E}$. Taking the tangent to $F_{AB|00E}$ at $t$ we have min-tradeoff function
$$f_t(q):=F_{AB|00E}(t)+F'_{AB|00E}(t)\left(\frac{1}{4}\sum_{x,y}\frac{1}{p_X(x)p_Y(y)} q((x,y,1))-t\right),$$
or, in other words, considering deterministic distributions on $U=(x,y,w)$
$$f_t(\delta_{(x,y,w)})=\begin{cases}
    \frac{F'_{AB|00E}(t)}{4p_X(x)p_Y(y)}+F_{AB|00E}(t)-tF'_{AB|00E}(t) & \text{if}\ w=1\\
    F_{AB|00E}(t)-tF'_{AB|00E}(t)& \text{if}\ w=0
    \end{cases}$$

    We have
    \begin{align*}
      \Max(f_t)&=\frac{F'_{AB|00E}(t)}{4\zeta^A\zeta^B}+F_{AB|00E}(t)-tF'_{AB|00E}(t)\\
      \Min_{\cQ_{\mathfrak{G}_{\zeta^A,\zeta^B}}}(f_t)&=F_{AB|00E}(t)-F'_{AB|00E}(t)\left(t-\frac{1}{2}\left(1-\frac{1}{\sqrt{2}}\right)\right)
      \end{align*}
    We now find a bound on $\Var_q(f_t)$ using the Bhatia-Davis bound~\cite{BhatiaDavis}.
\begin{lemma}[Bhatia-Davis bound]
Let $X$ be a real-valued random variable with $\max(X) = M$, $\min(X) = m$ and $\mathbb{E}(X)=\mu$, then
\begin{eqnarray}
\Var_X\leq(M-\mu)(\mu-m)\,.
\end{eqnarray}
\end{lemma}
In our case, $M=\Max(f_t)$, $m=F_{AB|00E}(t)-tF'_{AB|00E}(t)$ and $\mu=\mathbb{E}_q(f_t)=F_{AB|00E}(t)+F'_{AB|00E}(t)\left(\score(q)-t\right)$, where $\score(q)$ is defined in~\eqref{eq:sc}. Thus,
\begin{eqnarray*}
  \Var_q(f_t) &\leq&(F'_{AB|00E}(t))^2\score(q) \left(\frac{1}{4\zeta^A\zeta^B}-\score(q)\right)\\
              &\leq&\begin{cases}(F'_{AB|00E}(t))^2\left(\frac{1}{4\zeta^A\zeta^B}-1\right)&\text{if}\ \zeta^A\zeta^B<1/8\\\left(\frac{F'_{AB|00E}(t)}{8\zeta^A\zeta^B}\right)^2&\text{if}\ \zeta^A\zeta^B\geq1/8\end{cases}
\end{eqnarray*}
where we have optimized over $\score(q)\in[0,1]$ for the second inequality.

\subsubsection{Completeness error}\label{app:comp}
We can form a bound on the completeness error using Hoeffding's inequality~\cite{Hoeffding}.
\begin{lemma}[Hoeffding's inequality] 
Let $X_i$  be $n$ i.i.d.\ random variables with $a \leq X_i \leq b$, $a, b \in \mathbb{R}$. If $S  = \sum_{i} X_{i}$ and $\mu=\mathbb{E}(S)$. Then for $t > 0$
\begin{eqnarray}
\mathbb{P}(S - \mu \geq t) \leq \e^{-\frac{2t^2}{n (b - a)^2 }}\,.
\end{eqnarray}
\end{lemma}
\begin{theorem}
Suppose Protocol~\ref{prot:biased} is run using honest devices that behave in an i.i.d.\ fashion and that have an expected CHSH score $\score_{\exp}$. The probability that the protocol aborts is no greater than
\begin{eqnarray}
 \e^{-32n(\delta\zeta^A\zeta^B)^2}\,.
\end{eqnarray}
\end{theorem}
\begin{proof}
Recall the abort condition in the protocol, which states that $\score<\score_{\exp}-\delta$ where
$$\score=\frac{1}{4}\sum_{x,y} \frac{|\{i:U_i=(x,y,1)\}|}{n p_{X}(x) p_Y(y)}\,.$$
We can write this as $\sum_iJ_i$, where 
\begin{eqnarray}
J_i((x,y,w)) &=& \begin{cases} 0& \text{ if } w=0\\
1/(4np_X(x)p_Y(y)) & \text{ if } w=1
\end{cases}
\end{eqnarray}
This construction gives $\mathbb{E}\left[\sum_i J_i\right]=n\mathbb{E}[J_i]=\sum_{xy}\frac{1}{4p_X(x)p_Y(y)}\mathbb{P}(U=(x,y,1))$.
In an honest implementation of the protocol, the distribution on the register $U$ takes the form
\begin{eqnarray}
\mathbb{P}(U=(x,y,w)) = \begin{cases} p_X(x)p_Y(y) (1-\score_{xy}) & \text{ if } w=0 \\
p_X(x)p_Y(y) \score_{xy} & \text{ if } w=1
\end{cases}
\end{eqnarray}
where $\sum_{xy} \score_{xy} = 4 \score_{\exp}$, and hence $\mathbb{E}\left[\sum_i J_i\right]=\score_{\exp}$.
The abort condition can be expressed as $\score_{\exp}-\sum_i J_i>\delta$. We have
\begin{align*}
    \mathbb{P}(\score_{\exp}-\sum_i J_i>\delta)&=\mathbb{P}(\sum_i(-J_i)-(-\score_{\exp})>\delta)\\
    &\leq\e^{-32n(\delta\zeta^A\zeta^B)^2}\,,
\end{align*}
where we have used Hoeffding's inequality for the random variable $-J_i$ with $a=-1/(4n\zeta^A\zeta^B)$ and $b=0$.
\end{proof}

\subsection{Error parameters}

Both Theorems~\ref{thm:EAT} and~\ref{thm:EATspot} are stated in terms of the probability that the protocol does not abort, $p_\Omega$, which is unknown to the users of the protocol. However, if we replace $p_\Omega$ by $\epsilon_{\text{EAT}}$, then if $p_{\Omega}\geq\epsilon_{\text{EAT}}$ we have a correct bound on the entropy. On the other hand, if $p_{\Omega}<\epsilon_{\text{EAT}}$ then the protocol aborts with probability greater than $1-\epsilon_{\text{EAT}}$. In other words, prior to running the protocol the probability that it will both not abort and that the entropy is not valid is at most $\epsilon_{\text{EAT}}$. The soundness error of the protocol is $\epsilon_S=\max(\epsilon_{\text{EAT}},2\epsilon_h+\epsilon_{\text{EXT}})$, where $\epsilon_{\text{EXT}}$ is the extractor error (essentially the probability that the extraction fails). A summary of the aspects of extraction relevant to the present discussion and in the same notation as used here can be found in~\cite[Supplementary Information I~C]{LLR&}.

\subsection{Application to $H(AB|E)$ and $H(A|E)$}
Note that the EAT as stated in Theorem~\ref{thm:EAT} cannot be directly used in conjunction with $H(AB|E)$ and $H(A|E)$. The basic reason is that the event $\Omega$ should be an event on $\B{ U}$, which in turn should be a deterministic function of $\B{ C}$ and $\B{ D}$. To use $H(AB|E)$ and $H(A|E)$ we need $\B{ D}$ to be empty and $\B{ C}$ to be $\B{ AB}$. This means the register $\B{ U}$ cannot depend on the inputs, $\B{ XY}$, but without a score that depends on the inputs we cannot certify non-classicality let alone randomness.

Since we do not have strong use cases for $H(AB|E)$ and $H(A|E)$ , we do not consider possible extensions of the EAT in this work.

An alternative, which loses tightness, is to use an idea from~\cite[Appendix~B.3]{MDRHW}. Applying to the present case this would mean taking $\B{ D}$ to be empty and $\B{ C}$ to be either $\B{ ABV}$ or $\B{ AV}$, where $V_i$ records whether the CHSH game was won on the $i$th round, with $U_i=V_i$. Then, proceeding with the former, because $H(ABV|E)\geq H(AB|E)$ we can base our min-tradeoff function on $H(AB|E)$, and we can use a chain rule to recover a bound on the smooth min entropy of $\B{ AB}$ given $E$ from that of $\B{ ABV}$ given $E$. The bounds used in this approach are tightest when $V$ has low entropy, so we expect better performance with spot-checking protocols. 

\section{Discussion of composability}\label{app:compos}
Throughout this work we consider a composable security definition, which means that, except with some small probability, the output randomness can be treated as a perfect random string as part of any larger protocol. Composable security definitions involve a distinguisher who tries to guess whether the real protocol or a hypothetical ideal protocol is being run. This distinguisher is allowed access to all the systems an eavesdropper has access to and is also assumed to learn whether or not the protocol was successful. 

The main purpose of this appendix is to briefly discuss composability for protocols that recycle the input randomness. [The discussion here is not relevant for protocols without such recycling.] In general, input randomness (the strings $\B{ X}$ and $\B{ Y}$) is not directly reusable without processing~\cite{CK2}. For instance, the devices could be set up such that the protocol aborts unless $X_1=0$ and so if the protocol passes it is known that $X_1=0$. If $\B{ X}$ directly forms part of the output, then with probability $1/2$ one bit of the final output is known, which contradicts the security claim that the probability of distinguishing the output from perfect randomness is at most the soundness error.

Hence, in order to recycle the input randomness, it also has to undergo extraction to remove possible information that has leaked about it. The only mechanism for information leak allowed by the correct running of the protocol is whether it aborted or not, which constitutes at most one bit. This is easily remedied by compressing the concatenation of all inputs and outputs by one extra bit during the extraction step (the argument mirrors the more general case below).

One could also imagine more general protocols in which the length of the string output by the protocol is a variable (that depends on the actual score observed in the experimental run), in contrast to the protocols we use in the present paper whose outputs are all a pre-determined fixed length that depends on the expected CHSH score. Note that allowing varying output length requires a modified security definition with a different ideal state, as well as a modified analysis, both of which are beyond the scope of the present paper (see, e.g.,~\cite{PortRen,BHLMO}). The main point we wish to make here is that if additional information $L$ potentially leaks during the protocol then this could convey information about the input string. In protocols that recycle the input randomness, we can deal with this by additional compression in the extraction step. More precisely, Equation~(3.21) of~\cite{Renner} implies $H_{\min}^{\epsilon_h}(R|SL)\geq H_{\min}^{\epsilon_h}(R|S)-\log d_L$, where $d_L$ is the dimension of $L$.

In a protocol with variable output length, the distinguisher can get potentially useful information from the length $L$ of the final random string. Given a bound on $H_{\min}^{\epsilon_h}(\B{ ABXY}|E)$ (to reiterate: obtaining this bound and the correct security definition are beyond the scope of this work), we can account for the information potentially conveyed by the final string length as follows. Conservatively, in a CHSH-based protocol with $n$ rounds, the concatenation $\B{ ABXY}$ comprises at most $4n$ bits. Therefore, the length of the final output should satisfy
$$H_{\min}^{\epsilon_h}(\B{ ABXY}|EL)\geq H_{\min}^{\epsilon_h}(\B{ ABXY}|E)-\log d_L\geq H_{\min}^{\epsilon_h}(\B{ ABXY}|E)-\log (4n)\,.$$
Hence, if we reduce the length of the extractor output by $\log(4n)$ bits we can recycle the input randomness. Since the leading order term in $H_{\min}^{\epsilon_h}(\B{ ABXY}|E)$ is proportional to $n$, this reduction is minor.

Because we are working with device-independent protocols, the ongoing security of any randomness generated can be compromised if the devices used for one instance of the protocol are subsequently reused~\cite{bckone}. Hence, our discussion of security assumes devices are not reused (possible modifications to protocols that aim to allow restricted reuse are also discussed in~\cite{bckone}).
 
\chapter{Appendix for semi-Device Independent protocols}\label{app: Semi-DI}
\section{Useful claims}\label{app: useful claims}
This section is comprised of a series of standalone proofs that substantiate the claims presented in the main text. These proofs are independent of one another unless explicitly linked by a reference.
\begin{lemma}\label{Claim: montonicity in theta}
Let $\theta \in [0 , \frac{\pi}{2}]$ and $x , h >0$ are any reals (such that the functions below are defined), consider the functions 
\begin{eqnarray}
\Lambda_0(x , h , \theta) &:=& \Big( \Phi((x + h) \cos(\theta))  - \Phi(x + h) \Big) -  \Big( \Phi(x \cos(\theta)) - \Phi(x) \Big) \\
\Lambda_1(x , h , \theta) &:=& \Big( \Phi((x + h) \cos(\theta))   -   \Phi(x \cos(\theta))  \Big)
\end{eqnarray}
Then $\Lambda_{0}(x , h , \theta_1) \geq \Lambda(x , h , \theta_2)$ and $\Lambda_{1}(x , h , \theta_1) \geq \Lambda_{1}(x , h , \theta_2)$ if $\theta_1 > \theta_2$. 
\end{lemma}
\begin{proof}
\begin{eqnarray}
\partial_{\theta} \Lambda_{0}(x , h , \theta) = \partial_{\theta} \Lambda_{1}(x , h , \theta) &=& \big(-\Phi'(x\cos(\theta)+ h\cos(\theta))(1 + h) + \Phi'(x\cos(\theta)) \big)  \sin(\theta) \nonumber\\ 
                                          &\geq& \big(-\Phi'(x\cos(\theta)+ h\cos(\theta))  + \Phi'(x\cos(\theta)) \big)  \sin(\theta) \nonumber \\
                                          &\geq& 0   \nonumber
\end{eqnarray}
Where the last line follows from the fact that $-\Phi(x) > -\Phi(y)$ if $x > y$ and $x , y > 0$.
\end{proof}

\begin{lemma}\label{Claim: monotonicity in x}
Let $\theta \in [0 , \frac{\pi}{2}]$. The function 
\begin{eqnarray}
f(x) := \Phi(x \cos(\theta)) - \Phi(x)
\end{eqnarray}
is increasing for all $x \in [0, 1]$. 
\end{lemma}
\begin{proof}
Consider the function 
\begin{eqnarray}
\Lambda(x , h , \theta) := \Big( \Phi((x + h) \cos(\theta))  - \Phi(x + h) \Big) -  \Big( \Phi(x \cos(\theta)) - \Phi(x) \Big)
\end{eqnarray}
We need to show that $\Lambda(x , h , \theta) > 0$ . We first start by proving this claim for $\theta \leq \cos^{-1}\Big(\frac{x}{x +h}\Big)$. The choice of such $\theta$ shall be made clear in the second part of the proof. Suppose, that we are able to show that $\Lambda(x , h , \cos^{-1}\Big(\frac{x}{x+h}\Big)) \geq 0$, then we prove the $\Lambda(x , h , \theta) \geq 0$ for $\theta \geq \cos^{-1}\Big(\frac{x}{x+h}\Big)$ by using \ref{Claim: montonicity in theta} and the obervation 
\begin{eqnarray}
\Lambda(x , h , \theta) \geq \Lambda(x , h , \cos^{-1}(\frac{x}{x+h})) \geq 0
\end{eqnarray}
We now prove that $\Lambda(x , h ,\theta) \geq 0$ for every $\theta \in [0 , \cos^{-1}\big(\frac{x}{x+h} \big) ]$. For any $\theta$ in this range, the sets $[x\cos(\theta) , x]$ and $[(x+h) \cos(\theta) , x+h]$ are disjoint. Using Taylor's theorem, we know that there exists $c_1 \in [x\cos(\theta) , x]$ such that 
\begin{eqnarray}
\Phi(x \cos(\theta)) - \Phi(x) &=& - \Phi'(c_1) x (1 - \cos(\theta))\nonumber
\end{eqnarray}
Thus, $\exists c_2 \in [(x + h) \cos(\theta) , x+h] > c_1$ such that
\begin{eqnarray}
\Lambda(x , h , \theta) &=& -\Phi'(c_2) (x + h) \cos(\theta) + \Phi'(c_1) x \cos(\theta) \nonumber \\
                        &\geq& \big( -\Phi'(c_2) + \Phi'(c_1) \big) x \cos(\theta) \nonumber \\ 
                        &\geq& 0
\end{eqnarray}
The last line here follows from the fact that $c_2 > c_1 > 0$ implies $-\Phi'(c_2) > -\Phi'(c_1) $.
\end{proof}

\begin{lemma}\label{lem: linear-algebra}
If $\M{H}_{\lambda}$ is invariant under the action of $\hat{O}$. If $P_{\lambda}$ is the projection onto the subspace $\M{H}_{\lambda}$ then 
\begin{equation}
   P_{\lambda} \hat{O} \in \M{H}_{\lambda} \quad \mathrm{and} \quad [P_{\lambda} , \hat{O}] = 0
\end{equation}
\end{lemma}
\begin{proof}
Let $\ket{v} \in \M{H}_{\lambda}$. As $\M{H}_{\lambda}$ is invariant under the action of $\hat{O}$, $\hat{O}\ket{v} = \ket{w}$ for some $\ket{w} \in \M{H}_{\lambda}$. Then $P_{\lambda} \ket{w} \in \M{H}_{\lambda} \implies P_{\lambda} \hat{O} \ket{v} \in \M{H}_{\lambda}$. \\ 
Finally $P_{\lambda} \hat{O} \ket{v} = \ket{w} =  \hat{O}  \ket{v} = \hat{O} P_{\lambda} \ket{v}$. As $\ket{v}$ is any-arbitrary element of $\M{H}_{\lambda}$ and $\forall \ket{u} \in \M{H}: \quad P_{\lambda} \hat{O} \ket{u} \in \M{H}_{\lambda}$, we must have $[P_{\lambda} , \hat{O} ] = 0$. 
\end{proof}

Another result that we use is the corollary of the no-signalling for our protocol 
\begin{lemma}\label{lemm: No-signalling corollary}
Let $\rho^{x}_{AE}$ be any purification of $\rho^{x}_{A}$. Then, the von Neumann entropy of the reduced state on system $E$, denoted by $H(\rho^{x}_{E})$, is equal to the von Neumann entropy of $\rho^{x}_{A}$, denoted by $H(\rho^{x}_{A})$.
\end{lemma}

\begin{proof}
First, note that any two purifications of $\rho^{x}_{A}$, $\rho^{x}_{AE}$ and $\tilde{\rho}^{x}_{AE}$, are related by a unitary $U$ acting on system $E$, i.e., $\tilde{\rho}^{x}_{AE} = (\id_A \otimes U)\rho^{x}_{AE}(\id_A \otimes U^{\dagger})$. Using the properties of partial trace, we have
\begin{align*}
\tr_{A}(\tilde{\rho}^{x}_{AE}) &= \tr_{A}\left[ (\id_A \otimes U)\rho^{x}_{AE}(\id_A \otimes U^{\dagger}) \right] \\
&= (\id_A \otimes U) \tr_{A}(\rho^{x}_{AE}) (\id_A \otimes U^{\dagger}) \\
&= U \tr_{A}(\rho^{x}_{AE}) U^{\dagger}.
\end{align*}
Since $H(U \rho U^{\dagger}) = H(\rho)$ holds for any state $\rho$ and unitary $U$, it suffices to construct any specific purification of $\rho^{x}_{A}$ and compute its entropy.\\
Let ${\ket{i}}$ be the eigenbasis of $\rho^{x}_{A}$, and let ${\lambda_i}$ be the corresponding eigenvalues. Consider the following purification of $\rho^{x}_{A}$:
\begin{equation*}
\rho^{x}_{AE} = \sum_{i,j} \sqrt{\lambda_i \lambda_j} \ketbra{i}{j}_A \otimes \ketbra{i}{j}_E.
\end{equation*}
It is easy to verify that $\rho^{x}_{AE}$ is indeed a purification of $\rho^{x}_{A}$. Moreover, it is easy to check that
\begin{equation*}
\tr_{A}(\rho^{x}_{AE}) = \sum_{i} \lambda_i \ketbra{i}{i}_{E}.
\end{equation*}
Thus, we have
\begin{equation*}
H(\rho^{x}_{E}) = H\left(\tr_{A}(\rho^{x}_{AE})\right) = H(\rho^{x}_{A}),
\end{equation*}
as desired.
\end{proof}

\section{Elminating some parameters} 
The proofs below help in reduction of the parameter space for optimization problem for computing the rate functions. 
\begin{lemma}\label{lemm: maximizing the constriant}
Suppose we have the optimization problem:
\begin{eqnarray}\label{eqn: general optimization : 1}
 \begin{aligned}
     \inf \quad &  f(x_1 ,x_2 , \cdots , x_{n} , y) \\
\textrm{s.t.}
\quad & \forall i: h_i(x_1 , x_2 , \cdots , x_{n}) \geq 0\\
\quad &  g(x_1 , x_2 , \cdots , x_{n} , y) \geq 0  
\end{aligned}. 
\end{eqnarray}
The optimization problem \ref{eqn: general optimization : 1} is equivalent to the following problem: 
\begin{eqnarray}\label{eqn: general optimization}
 \begin{aligned}
     \inf \quad &  f(x_1 ,x_2 , \cdots , x_{n} , y) \\
\textrm{s.t.}
\quad & \forall i: h_i(x_1 , x_2 , \cdots , x_{n}) \geq 0\\
\quad &  g(x_1 , x_2 , \cdots , x_{n} , y) \geq 0 \\
\quad &  y \in Y^*
\end{aligned} 
\end{eqnarray}
where $Y^*$ is the following set: 
\begin{equation*}
    Y^* := \{ y \in \mathbb{R}: \quad \partial_{y} g(x_1 , x_2 , \cdots , x_{n} , y) = 0 \}
\end{equation*}
\end{lemma}
\begin{proof}
We can solve the optimization problem \eqref{eqn: general optimization : 1} by writing the lagrangian: 
\begin{equation*}
    \M{L}(x_1, \cdots , x_{n} , y ) := f(x_1 , \cdots , x_n , y) - \sum_{i} \lambda_i h_i(x_1 , \cdots, x_{n} ) - \mu g(x_1 , \cdots, x_n , y)  
\end{equation*}
where $\lambda_i , \mu \in \mathbb{R}$ are some KKT multipliers. To determine the optimal parameter $y$, we must have that $\partial_{y} \M{L} =  0$, which is equivalent to  
\begin{equation*}
    \mu \partial_{y }g(x_1 , \cdots , x_n , y) = 0. 
\end{equation*}
Which gives the result unless $\mu = 0$. If $\mu = 0$, then the optimal solution for the \eqref{eqn: general optimization : 1} does not depend upon the value $y$, as $y$ only appears in the constraint $g(x_1, \cdots ,x_n , y) = 0$. Thus, we can choose $y \in Y^*$ without any loss of generality.
\end{proof}

\section{Monotonicity of rates} 
This section consists of proofs for the monotonicity of the rate function $ \entropy_{p_{X}}$. 
\begin{lemma}
The function $g_{\score}(\overlap) := \entropy_{p_{X}}(\score , \overlap)$  is increasing for all $\score \in [\frac{1}{2} , 1]$ and $\overlap \geq \frac{1}{2}$. 
\end{lemma}
\begin{proof}
It suffices to show that the set of feasible points for the optimization problem $\entropy_{p_{X}}(\score , \overlap)$ is a strict subset of the feasible set for the optimization problem for $\entropy_{p_{X}}(\score , \overlap - \delta \overlap)$ for all $\delta \overlap \in [0 , \overlap - \frac{1}{2}]$. \\ 
The parameter $\overlap$ appears in only one constraint denoted by: 
\begin{equation}\label{eqn:  monotonicity of theta 1}
     \left(  \sum_{x}\tilde{a}_{x} \cos(\xi_{x}) \right)^2 +  \left( \sum_{x} \tilde{a}_{x} \sin(\xi_{x}) \right)^2 - \left( 4 \overlap   - \sum_{x} \eta_{x} \right)^2 \geq 0. 
\end{equation}
Thus we aim to show that the set of points that obey 
\begin{equation}\label{eqn: monotonicity of theta 2}
     \left(  \sum_{x}\tilde{a}_{x} \cos(\xi_{x}) \right)^2 +  \left( \sum_{x} \tilde{a}_{x} \sin(\xi_{x}) \right)^2 - \left( 4 (\overlap - \delta \overlap)   - \sum_{x} \eta_{x} \right)^2 \geq 0, 
\end{equation}
also obey \eqref{eqn: monotonicity of theta 1}. This can be easily proved by observing that: 
\begin{eqnarray*}
 - \left( 4 (\overlap - \delta \overlap)   - \sum_{x} \eta_{x} \right)^2 &=&  - \left( 4 \overlap   - \sum_{x} \eta_{x} \right)^2 + 8 \delta \overlap \left(\left( 4 \overlap   - \sum_{x} \eta_{x} \right) - 2 \delta \overlap \right). 
\end{eqnarray*}
Furthermore, it is easy to check that  $\forall \delta \overlap \in [0 , \overlap - \frac{1}{2}]$, the following holds: 
$$8 \delta \overlap \left(\left( 4 \overlap   - \sum_{x} \eta_{x} \right) - 2 \delta \overlap \right) \geq 0 . $$
Thus, we can see that the set of points that obey \eqref{eqn: monotonicity of theta 1}, must automatically obey \eqref{eqn: monotonicity of theta 2}.
\end{proof}
\begin{lemma}
The function $g_{\overlap}(\score) := \entropy_{p_{X}}(\score , \overlap)$  is increasing for all $\overlap \in [\frac{1}{2} , 1]$ and $\score \geq \frac{1}{2}$. 
\end{lemma}
\begin{proof}
As in the previous proof, it suffices to show that the set of feasible points for the optimization problem $\entropy_{p_{X}}(\score , \overlap)$ is a strict subset of the feasible set for the optimization problem for $\entropy_{p_{X}}(\score  - \delta \score , \overlap)$ for all $\delta \score \in [0 , \overlap - \frac{1}{2}]$. This is even more straightforward in this case as 
$$  \sum_{x}    \left( -\eta_{x} + (-1)^{x} \tilde{a}_{x} \cos(\xi_{x})\right)) \geq  4 \score -4 \geq 4(\score - \delta \score) - 4  $$
holds for all $\delta \score \in [0 , 1]$. 
\end{proof}

\backmatter


\bibliographystyle{unsrt}
\bibliography{Bibfiles/firstbib} 


\end{document}